%% file: thesis.tex
\documentclass[
PhD, 
DIC=off,
12pt,
bibliography=totoc, 
listof = totoc, 
index= totoc, 
onehalfspacing, 
a4paper 
]
{icldt}

\usepackage{graphicx}
\usepackage{amsmath}
\usepackage{amssymb}
\usepackage{amsfonts}
\usepackage{epsfig}
\usepackage{graphicx,epsf}
\usepackage{color}
\usepackage{bm}
\usepackage{psfrag}
\usepackage{mathtools}
\usepackage{siunitx}
\usepackage{adjustbox}
\usepackage{enumitem}
\usepackage{afterpage}
\usepackage{currfile}
\usepackage{tensor}
\usepackage{cancel}
\usepackage[square, numbers, sort&compress]{natbib}
\DeclareMathAlphabet{\mathpzc}{OT1}{pzc}{m}{it}
\DeclareMathOperator{\order}{\mathcal{O}}

\newcommand{\infrac}[2]{#1/#2}
\newcommand{\arm}{a_\text{eq}}
\newcommand{\Hrm}{{\cal{H}}_\text{eq}}

\renewcommand{\vec}[1]{\mathbf{#1}}

\newcommand*\nc[1]{\tikz[baseline=(char.base)]{
    \node[shape=circle,draw,inner sep=2pt] (char) {#1};}}
\newcommand{\balg}{\begin{enumerate}[label=\protect\nc{\arabic*}]}
\newcommand{\ealg}{\vspace{6 pt}\end{enumerate}}

\newcommand{\smax}{s_\text{max}}

\newcommand{\tauc}{\tau_\text{c}}
\newcommand{\rhor}{\rho_\text{r}}
\newcommand{\rhom}{\rho_\text{m}}
\newcommand{\kunit}{\infrac{\Hrm}{\sqrt{2}}}
\newcommand{\tunit}{\infrac{\sqrt{2}}{\Hrm}}

\newcommand{\drs}{\delta_{\text{r}(s)}}

\newcommand{\hmatch}{h_\text{rm}}

\newcommand{\hr}{h_\text{r}}
\newcommand{\tr}{\tau_\text{r}}
\ExplSyntaxOn
\NewDocumentCommand{\countlist}{m}
 {
  \clist_count:n { #1 }
 }
\ExplSyntaxOff
\newcommand{\rfcite}[1]{\ifnum\countlist{#1}=1  Ref.~\cite{#1}\else Refs.~\cite{#1}\fi}
\newcommand{\eref}[1]{Eq.~(\ref{#1})}

\renewcommand{\eqref}[1]{(\ref{#1})}
\def\bea{\begin{eqnarray}}
\def\eea{\end{eqnarray}}
\def\bi{\begin{itemize}}
\def\ei{\end{itemize}}
\def\be#1\ee{\begin{equation}#1\end{equation}}
\def\ba#1\ea{\begingroup
\addtolength{\jot}{9pt}\begin{align}#1\end{align}\endgroup}
\def\bas#1\eas{\begingroup
\addtolength{\jot}{9pt}\begin{align}\begin{split}#1\end{split}\end{align}\endgroup}
\def\bfa#1\efa{\begingroup
\addtolength{\jot}{9pt}\begin{flalign}#1\end{flalign}\endgroup}
\def\bml#1\eml{\begingroup
\addtolength{\jot}{9pt}\begin{multline}#1\end{multline}\endgroup}
\def\nt#1\ent{{\texttt#1 }} 
\newcommand\eq[1]{Eq.~(\ref{#1})}
\newcommand\eqs[1]{Eqs.~(\ref{#1})}
\def\be{\begin{equation}}
\def\ee{\end{equation}}
\def\bea{\begin{eqnarray}}
\def\eea{\end{eqnarray}}
\def\bi{\begin{itemize}}
\def\ei{\end{itemize}}
\def\p{\partial}

\def\dd{\text{d}}
\def\drf{\delta^{(1)} \rho}
\def\drs{\delta^{(2)} \rho}
\def\dkf{\delta^{(1)} k}

\def\dRf{\delta^{(1)} R}
\def\dRs{\delta^{(2)} R}
\def\dzf{\delta^{(1)} z}
\def\dzs{\delta^{(2)} z}
\def\dVf{\delta^{(1)} V}
\def\dVs{\delta^{(2)} V}
\def\daf{\delta^{(1)} d_{A}}
\def\das{\delta^{(2)} d_{A}}
\def\E{{\cal E}}
\def\Ef{\delta^{(1)}{\cal E}}
\def\Es{\delta^{(2)}{\cal E}}
\def\dS{\delta^{(1)}\Sigma}
\def\dnuf{{\delta^{(1)} \nu}}
\def\dnf{{\delta^{(1)} n}}
\def\dnus{{\delta^{(2)} \nu}}
\def\dns{{\delta^{(2)} n}}
\def\ff{{\Phi_{1}}}
\def\pf{{\Psi_{1}}}
\def\fs{{\Phi_{2}}}
\def\ps{{\Psi_{2}}}

\def\P{{{\cal{P}}}}
\def\H{{{\cal{H}}}}
\def\G{\mathcal{G}}
\def\H{\mathcal{H}}

\usepackage{hyperref}
\usepackage[title]{appendix}
            
\hypersetup{colorlinks = true, linkcolor=blue}
            
\makeatletter
\newcommand\appendix@numberline[1]{}
\g@addto@macro\appendix{%
  \addtocontents{toc}{
    \let\protect\numberline\protect\appendix@numberline}%
}
\makeatother

\begin{document}

\input{tex/title}

\include{tex/abstract}

\include{tex/acknowledgements}

\tableofcontents

\onehalfspacing

\include{tex/chapter_1}

\include{tex/chapter_2}

\include{tex/chapter_3}
\include{tex/chapter_4}
\include{tex/chapter_5}
\include{tex/appendix}

\begin{singlespace}

\bibliography{bib/mybib2}
%


\end{singlespace}

\listoffigures 

\end{document}

%% file: tex/title.tex
\setlength{\tabcolsep}{0in}
\newcommand{\isep}{-2 pt}
\newcommand{\lsep}{-0.5cm}
\newcommand{\psep}{-0.6cm}
\renewcommand{\labelitemii}{$\circ$}
\college{Queen Mary, }
\department{Physics and Astronomy}
\supervisor{Karim A Malik and Timothy Clifton}
\title{Applications of Cosmological Perturbation Theory in the Late Universe}
\author{Jorge Luis Fuentes Venegas}
\date{September 2018}
%
%
%

\declaration{%
I, Jorge Luis Fuentes Venegas, confirm that the research included within this thesis is my own work or that where it has been carried out in collaboration with, or supported by others, that this is duly acknowledged below and my contribution indicated. Previously published material is also acknowledged below.
\newline
\newline
I attest that I have exercised reasonable care to ensure that the work is original, and does not to the best of my knowledge break any UK law, infringe any third party's copyright or other Intellectual Property Right, or contain any confidential material.
\newline
\newline
I accept that the College has the right to use plagiarism detection software to check the electronic version of the thesis.
\newline
\newline
I confirm that this thesis has not been previously submitted for the award of a degree by this or any other university. The copyright of this thesis rests with the author and no quotation from it or information derived from it may be published without the prior written consent of the author.
\newline
\newline
Details of collaboration and publications:
\begin{itemize}
\item \underline{Gravitational waves in a flat radiation-matter universe including anisotropic stress} \\
	Andrew J.~Wren, Jorge L.~Fuentes and Karim A.~Malik \\
	e-Print: 1712.10012 [gr-qc]
\item \underline{Linear cosmological perturbations in almost scale-invariant fourth-order gravity} \\
	Jorge L.~Fuentes, Usman A.~Gillani and Karim A.~Malik \\
	e-Print: 1812.00938 [gr-qc]
\item \underline{Galaxy number counts at second order: an independent approach} \\
	Jorge L.~Fuentes, Juan Carlos Hidalgo and Karim A.~Malik \\
	Published in: Class.Quant.Grav. 38 (2021) 6, 065014 \\e-Print: 1908.08400 [astro-ph.CO]
\item \underline{Galaxy number counts at second order in perturbation theory:}\\ \underline{a leading-order term comparison} \\
	Jorge L.~Fuentes, Juan Carlos Hidalgo and Karim A.~Malik \\
	e-Print: 2012.15326 [astro-ph.CO]
\end{itemize}
\vfill
\noindent Signature: Jorge Luis Fuentes Venegas
\newline
Date: 17/07/2020}

\maketitle

%% file: tex/abstract.tex
\chapter*{Abstract}
\label{ch:abstract}
\addcontentsline{toc}{chapter}{Abstract}
\section*{}
\singlespacing

In this thesis, we discuss some of the applications of cosmological perturbation theory in the late universe. We begin by reviewing the tools used to understand the standard model of cosmology theoretically and to compute its observational consequences, including a detailed exposition of cosmological perturbation theory. We then describe the results in this thesis; we present novel analytical solutions for linear-order gravitational waves or tensor perturbations in a flat Friedmann-Robertson-Walker universe containing two perfect fluids --radiation and pressureless dust-- and allowing for neutrino anisotropic stress. One of the results applies to any sub-horizon gravitational wave in such a universe. Another result applies to gravitational waves of primordial origin (for example, produced during inflation) and works both before and after they cross the horizon. These results improve on analytical approximations previously set out in the literature. Comparison with numerical solutions shows that both these approximations are accurate to within 1\% or better, for a wide range of wave-numbers relevant for cosmology. We present a new and independent approach to computing the relativistic galaxy number counts to second order in cosmological perturbation theory. We also derive analytical expressions for the full second-order relativistic observed redshift, for the angular diameter distance and the volume spanned by a survey. We then compare our result with previous works which compute the general distance-redshift relation, finding that our result is in agreement at linear and leading nonlinear order. Lastly, we briefly study a class of almost scale-invariant Gauss-Bonnet modified gravity theory and derive the Einstein-like field equations to first order in cosmological perturbation theory in longitudinal gauge.

%% file: tex/acknowledgements.tex
\chapter*{Acknowledgements}
\label{ch:acknowledgements}
\addcontentsline{toc}{chapter}{Acknowledgements}

I would like to thank my supervisor Karim \scshape{Malik} \normalfont for his guidance and his patience, and express my gratitude to Juan Carlos \scshape{Hidalgo} \normalfont for all his help with this work.

I must also thank the Cosmology Group for all the constant inspiration and useful discussions.

I want to thank all my fellow PhD students, former and present, for making the PhD life more bearable and making the Astronomy Unit great again. I would particularly like to thank my friend Pedro \scshape{Carrilho} \normalfont for all his guidance and help since the very beginning. I also want to thank former Cosmology Group members Charalambos \scshape{Pittordis}\normalfont, Shailee \scshape{Imrith}\normalfont, Viraj \scshape{Sanghai}\normalfont, and John \scshape{Ronayne} \normalfont for great insights on what being a cosmology PhD student is. I also want to thank Domenico \scshape{Trotta} \normalfont for his friendship and support through the years, and I additionally would like to thank my fellow fourth year colleagues Christopher \scshape{Gallagher}\normalfont, Jack \scshape{Skinner}\normalfont, Louis \scshape{Coates}\normalfont, Sanson \scshape{Poon}\normalfont, Clark \scshape{Baker}\normalfont, and Sandy \scshape{Shaoshan Zeng}\normalfont; we made it! 

Additionally I would like to thank Rebeca \scshape{Carrillo}\normalfont, Paula \scshape{Soares}\normalfont, Alice \scshape{Giroul}\normalfont, Jesse \scshape{Coburn}\normalfont, Eline \scshape{de Weerd}\normalfont, Francesco \scshape{Lovascio}\normalfont, Jessie \scshape{Durk}\normalfont, George \scshape{Turpin}\normalfont, Julian \scshape{Adamek}\normalfont, Fraser \scshape{Kennedy}\normalfont, Gavin \scshape{Coleman}\normalfont, Maritza \scshape{Soto}\normalfont, Joe \scshape{Hayling}\normalfont, Eddie \scshape{Thorpe}\normalfont, Zolt\'{a}n \scshape{Laczk\'{o}}\normalfont, Antony \scshape{Frey}\normalfont, Steve \scshape{Cunnington}\normalfont, Miguel \scshape{Quetzeri}\normalfont, Anna \scshape{Tolmarsh}\normalfont, Alex \scshape{Umeda-Pelling}\normalfont, Jasmin \scshape{K\"{o}hnken-Sawall}\normalfont, William \scshape{Megone}\normalfont, Matt \scshape{Dibble}\normalfont, Nelson \scshape{Cano}\normalfont, Nicole \scshape{Hudgins}\normalfont, Adrien \scshape{Ycard}\normalfont, and Gemma \scshape{Munday} \normalfont for the good times, pints and all the football and rugby matches.

I would particularly like to thank Massissilia \scshape{Hamadouche} \normalfont for reading through this thesis and helping to get rid of typos and small idiomatic errors.

I would like to thank my long-time friends Eduardo \scshape{Serna}\normalfont, Germ\'{a}n \scshape{Ram\'{i}rez}\normalfont, Oscar \scshape{Salmer\'{o}n}\normalfont, Andr\'{e}s \scshape{Arroyo}\normalfont, Roberto \scshape{Tapia}\normalfont, Mariano \scshape{Renter\'{i}a}\normalfont, Aar\'{o}n \scshape{Becerra}\normalfont, David \scshape{Rangel}\normalfont, Leonardo \scshape{Ram\'{i}rez}\normalfont, and Hugo \scshape{Trejo} \normalfont for their continued friendship, even living in different countries.

I would not be here without my parents Jorge and M\'{o}nica, whose support and love helped me go through it all. I also want to extend my gratitude to my sisters Carolina and M\'{o}nica for believing in me.
\vfill
I acknowledge financial support from a Queen Mary University of London Studentship and the Consejo Nacional de Ciencia y Tecnolog\'{i}a (CONACYT) grant No.~603085 from 2016 to 2020.

%% file: tex/chapter_1.tex
\chapter{Introduction}
\label{chapter:intro}


In recent years technology has advanced at a rapid pace, giving scientists the computational power to test theoretical and analytical solutions for specific cosmological models through simulations, perform state of the art experiments and undertake new observations to answer long standing questions in cosmology. 

A good example of a theoretical model that is particularly useful in cosmology is Albert Einstein's theory of gravity which was proposed in 1905. It passed several tests through observations like the anomalous perihelion shift in the orbit of Mercury, which had been seen using Newtonian gravity but was measured with outstanding accuracy with relativity between 1966 and 1990 using radio telescopes. 

According to general relativity, light does not travel along straight lines when it propagates in a gravitational field. Instead, it is deflected in the presence of massive bodies. In particular, the light from a star is deflected as it passes near the Sun, leading to apparent shifts of up $1.75$ arc seconds in the stars' positions in the sky (an arc second is equal to $1/3600$ of a degree). In the framework of Newtonian gravity, a heuristic argument can be made that leads to light deflection by half that amount. The different predictions can be tested by observing stars that are close to the Sun during a solar eclipse. In this way, a British expedition to West Africa in 1919, directed by Arthur Eddington, confirmed that Einstein's prediction was correct, and the Newtonian predictions wrong, via observation of the May 1919 eclipse. Eddington's results were not very accurate; subsequent observations of the deflection of the light of distant quasars by the Sun, which utilise highly accurate techniques of radio astronomy, have confirmed Eddington's results with significantly better precision.

Another prediction of general relativity, the gravitational redshift was first measured in a laboratory setting in 1959 by Pound and Rebka. It is also seen in astrophysical measurements, notably for light escaping the white dwarf Sirius B. The related gravitational time dilation effect has been measured by transporting atomic clocks to altitudes of between tens and tens of thousands of kilometres.

General relativity had passed plenty of tests through observations but one; it predicted the existence of a gravitational counterpart to electromagnetic waves, namely, \textit{gravitational waves}, remaining undetected through direct means until 2015. Gravitational waves are `ripples' in space-time caused by some of the most violent and energetic processes in the Universe. These cosmic ripples travel at the speed of light, carrying with them information about their origins, as well as clues to the nature of gravity itself.


Indirect detection of gravitational waves had been made, through energy and angular momentum measured from binary stars orbiting each other, a good example is the Hulse-Taylor pulsar, whose orbit has decayed since the binary system was initially discovered, in precise agreement with the loss of energy due to gravitational waves described in general relativity; but no direct detection was made until September 2015. 

On 11 February 2016, the Laser Interferometer Gravitational-Wave Observatory\footnote{www.ligo.caltech.edu}  (LIGO)\cite{2016PhRvL.116f1102A} announced the detection of GW150914, a gravitational wave signal coming from a binary blackhole collision demonstrating that Einstein's theory appears to withstand the passage of time. 

The Theory of General Relativity (GR), relates the amount of matter (or energy) in the universe to the geometry of space, meaning that one affects the other. Matter tells the universe how to curve and the geometry of space tells matter how to move \cite{weinberg}. Precise experiments such as LIGO demonstrate that a metric theory such as GR is currently the best description for our universe.


Other than GR, there are many other theories that aim to describe the state of the universe by modifying gravity. The idea of modifying gravity on cosmological scales has really taken off over the past decades triggered by theoretical developments involving higher dimensional theories, as well as new developments in constructing renormalizable theories of gravity. More phenomenologically, Bekenstein's relativistic formulation of Milgrom's Modified Newtonian Dynamics (MoND) has provided a fresh impetus for new study: what was a rule of thumb for how weak gravitational fields might behave in regions of low acceleration, was suddenly elevated to a theory that could be used to study cosmology. Insights such as Bertschinger's realisation that large-scale perturbations in the universe can be directly related to the overall expansion rate have also made it possible to characterise large classes of theories simply in terms of how they make the universe evolve. Finally, there has been tremendous progress observationally. A key step here has been the measurement of the growth of structure at higher redshifts, by Guzzo and his collaborators \cite{TCMG}. 

Along with MoND \cite{CP1, CP2} there are extensions to GR such as $f(R)$ which work by adding functions to the action that describes the theory.
Effective field theory descriptions of the universe also exist, Horndeski's theory being a popular example \cite{TCMG, Nojiri:2017ncd}. Horndeski's theory is the most general theory of gravity in four dimensions whose Lagrangian is constructed out of the metric tensor and a scalar field and leads to second order equations of motion. The theory was first proposed by Gregory Horndeski in 1974 \cite{horn} and has found numerous applications, particularly in the construction of cosmological models of Inflation and dark energy \cite{TCMG}. Horndeski's theory contains many theories of gravity, including General relativity, Brans-Dicke theory, Quintessence, Dilaton, Chameleon and covariant Galileon \cite{esposito} as special cases.

In this thesis we mostly use the GR description of the universe, and in Appendix \ref{chapter:mgb} we explore one modified theory of gravity called Gauss-Bonnet gravity, which has been a topic of discussion at the time of writing this thesis \cite{PGS1, PGS2}.


The progress of technology does not end with probing GR predictions, as it also has helped cosmologists to measure cosmological parameters to unprecedented precision. At the present time the Sloan Digital Sky Survey\footnote{https://www.sdss.org/} (SDSS) is creating the most detailed three dimensional map of the universe ever made, with deep multi-colour images, and spectra for more than three million astronomical objects. 

With easily accessible data, SDSS has created an era of Big Data for cosmology, where at the moment there is so much data that it would take years for the quarter-million Milky Way stars' spectra to be processed, and there will even be more with SDSS-IV running now (2020) with its final data release scheduled for July 2021 and SDSS-V starting in summer 2020, with its first data release expected two years later.

One of the ongoing surveys within SDSS is the Extended Baryon Oscillation Spectroscopic Survey (eBOSS), which combined with previous phases of SDSS, precisely measures the expansion history of the universe throughout eighty percent of cosmic history, back to when the universe was less than three billion years old. It is providing improving constraints on the nature of ``Dark Energy'', the observed phenomenon that the expansion of the universe is currently accelerating.

\textit{Dark energy} is an unknown form of energy that affects the universe on the largest scales. The first observational evidence for its existence came from supernovae measurements, which showed that the universe does not expand at a constant rate; rather, the expansion of the universe is accelerating \cite{RevModPhys.75.559}. Understanding the evolution of the universe requires knowledge of its starting conditions and its composition. Prior to these observations, the only forms of matter-energy known to exist were ordinary matter, dark matter, and radiation. Measurements of the cosmic microwave background suggest the universe began in a hot Big Bang, from which general relativity explains its evolution and the subsequent large scale motion. Without introducing a new form of energy, there was no way to explain how an accelerating universe could be measured. Since the 1990s, dark energy has been the most accepted premise to account for the accelerated expansion. Two proposed forms of dark energy are the cosmological constant \cite{Carroll_2001}, representing a constant energy density filling space homogeneously, and scalar fields such as quintessence or moduli, dynamic quantities having energy densities that can vary in time and space. Contributions from scalar fields that are constant in space are usually also included in the cosmological constant.

As a small scale example, another ongoing survey is the Apache Point Observatory Galactic Evolution Experiment (APOGEE-2) that observes the ``archeological'' record embedded in hundreds of thousands of stars to explore the assembly history of the Milky Way Galaxy. The details as to how the galaxy evolved are preserved today in the motions and chemical compositions of its stars.

Some of the other surveys that complement SDSS are Mapping Nearby Galaxies at APO (MaNGA), the Multi-object APO Radial Velocity Exoplanet Large-area Survey (MARVELS) and Sloan Extension for Galactic Understanding and Exploration (SEGUE).

Along with these ongoing surveys, there are several future experiments that will help cosmologists look further into the past and into the unknown, tightening known constraints and finding interesting new answers as well as questions for generations of new scientists. One of these future surveys is Euclid\footnote{www.euclid-ec.org}, an European Space Agency (ESA) satellite to map the geometry of the universe and better understand the mysteries of dark energy and dark matter, which make up most of the energy budget of the cosmos. This mission will investigate the distance-redshift relationship and the evolution of cosmic structures by measuring shapes and redshifts of galaxies and clusters out to redshift $\sim2$, or equivalently to look back in time 10 billion years. In this way, Euclid will cover the entire period over which dark energy played a significant role in accelerating the expansion of the universe  \cite{euclid}. Its launch is planned for 2022. 

Scheduled to begin in October 2022, the Vera C. Rubin Observatory is to conduct the 10-year Legacy Survey of Space and Time\footnote{www.lsst.org} (LSST), previously known simply as the Large Synoptic Survey Telescope \cite{LSST}. LSST will deliver a 500 petabyte\footnote{A \textit{petabyte} is a measure of memory or data storage capacity that is equal to $2^{50}$ of bytes. There are $1024$ terabytes (TB) in a petabyte -- or 1 million gigabytes (GB).} set of images and data products that will address some of the most pressing questions about the structure and evolution of the universe and the objects in it, such as understanding dark matter and dark energy, hazardous asteroids and the remote Solar System, the Transient Optical Sky and the formation and structure of the Milky Way galaxy. Due to the amount of data it will be generating (20 terabytes each night), its software is one of the most challenging aspects of the Rubin Observatory. 

A different kind of observatory that will be completed in the near future; the Square Kilometre Array\footnote{www.skatelescope.org} (SKA) \cite{SKA}, which is an international effort to build the world's largest radio telescope, with eventually over a square kilometre of collecting area. The SKA will be able to conduct ground breaking discoveries in astronomical observations. From challenging Einstein's seminal theory of relativity to the limits using pulsars to test general relativity in extreme conditions, for example, close to black holes. It will explore the very first stars and galaxies formed just after the Big Bang, with a precision that have not been accomplished before, helping scientists understand the nature of dark energy. The SKA's is different to previous experiments of its kind due to the resolution it can accomplish. The SKA's sensitivity stems from the huge number of radio receivers at low, mid and high frequencies, which will combine in each frequency range from the locations in Africa and Australia to form a collecting area equivalent to a single radio telescope one kilometre wide, as mentioned above. The combined factors of sensitivity and resolution will dwarf all existing telescopes currently in operation, and give the SKA unparalleled view of the early formation of the universe. Early science observations are expected to start in the mid-2020s with a partial array. 

These surveys have a major factor in common, which is that they all depend on very well developed technology, that has made the era of precision cosmology possible. However, lest we forget that most of these experiments are, in essence, trying to falsify a theory or model of the universe by explaining observations with mathematical tools. One of these tools that is exceptionally useful today as it has been since the 19th century (when it was developed) is perturbation theory. In particular, perturbation theory allows us in principle to deduce from observations the power spectrum of the curvature perturbation at Hubble-exit during two-field slow-roll inflation, in single field inflation, observations directly constrain the tensor to scalar ratio ($r$) and from perturbation theory, we can then compute the slow-roll parameter $\epsilon$. Lastly, another application for perturbation theory directly connected with observations would be the non-linear evolution and non-Gussianity using the primordial curvature perturbation $\zeta$.


Perturbation theory was first devised to solve otherwise intractable problems in the calculation of the motion of planets in the solar system. For example, Newton's law of universal gravitation explained the gravitation between two astronomical bodies, but when a third body is added, the problem becomes much harder to analyse. Like today, the increase in precision in astronomical observations led to the need for more accurate solutions to Newton's gravitational equations. Lagrange and Laplace extended and generalised the solutions with the use of the first methods of perturbation theory, and these well developed methods were then adopted and adapted to solve new problems from quantum mechanics and subatomic physics to large scale structure in cosmology.

In physical cosmology, cosmological perturbation theory (CPT) is used to study the evolution of structure. It uses GR to compute the gravitational forces causing small perturbations to grow and eventually seed the formation of structures like stars, galaxies and clusters. It is useful in situations in which the universe is predominantly homogenous, such as cosmic inflation and large parts of the Big Bang. Since the universe is believed to be homogeneous, CPT is a good approximation on the largest scales. On smaller scales, the power of CPT may be suppressed and other techniques may be needed, such as N-body simulations. The formulation of CPT can be subtle, so we follow the approach of Ref.~\cite{malik} throughout this thesis. We will discuss perturbation theory and its applications in detail in Chapter \ref{chapter:cpt}. 


Cosmological perturbation theory is widely used within the scientific community and, similar to the situation with Newton's gravitational equations, more precision is needed in these solutions to keep up with the advances being made with the ongoing and upcoming surveys. Previously mentioned experiments such as eBOSS, Euclid and LSST will be able to measure cosmological parameters up to a percentage level ($\sim 0.01\%$), and an analytical counterpart is needed to explain and model the observations, hence the need for higher order perturbation theory.

Higher order perturbation theory can be used to explain slight perturbations on top of existing ones, modelling nonlinearities with precision that could not be modelled before using simple linear perturbation theory, making the computation of cosmological observables such as the bispectrum possible. 

Since perturbation theory is very powerful, we make use of it to solve Einstein equations and in cosmology, this helps us to constrain theoretically models of the universe. One of these models is the concordance $\Lambda$CDM model, which is a parametrisation of the Big Bang and contains three major components: first, the cosmological constant denoted by the greek letter Lambda ($\Lambda$) normally associated with dark energy; second, the postulated cold dark matter (CDM); and third, ordinary matter. It is frequently referred to as the \textit{standard model} of cosmology because it is the simplest and most widely used model that provides a reasonably good account of the properties of the universe. The existence of the cosmic microwave background (CMB), the large scale structure in the distribution of galaxies, the observed abundances of hydrogen, helium and lithium and the accelerating expansion of the universe, all contribute to the acceptance of the model. This model appeared in the late 1990s, and up until today, it is the most accepted model of cosmology there is.

Within the standard model there are several parameters normally divided into three categories. The first one being \textit{independent parameters}, which are mostly not predicted by current theory; the parameter values and uncertainties are estimated using large computer searches to locate the region of parameter space providing an acceptable match to cosmological observations. These parameters are comprised of the physical baryon density parameter ($\Omega_{b}h^{2}$), the physical dark matter density parameter ($\Omega_{c}h^{2}$), the age of the universe ($\sim 14\times10^{9}$ years), the scalar spectral index ($n_{s}$), the curvature perturbation amplitude ($k_{0} \sim 0.002 \text{Mpc}^{-1}$, $\Delta^{2}_{R}$) and the reionisation optical depth ($\tau$).

Other parameters are the fixed at ``natural'' values, e.g. total density parameter is equal to one. The \textit{fixed parameters} are the total density parameter ($\Omega_{\text{tot}}$), the equation of state of dark energy ($w$), the tensor to scalar ratio ($r$), the running of the spectral index, the sum of three neutrino masses and the effective number of relativistic degrees of freedom ($N_{\text{eff}}$). In the literature, from time to time the temperature of the CMB photons can be taken as a fixed parameter, although it is tightly constrained it can be taken as a fixed parameter in say, the $\lambda$-CDM model. The same happens with the motion of the solar system with respect to the CMB, which establishes the class of comoving observers.

The last group are the \textit{calculated values}; the Hubble constant ($H_{0}$), the baryon density parameter ($\Omega_{b}$), the dark matter density parameter ($\Omega_{c}$), the matter density parameter ($\Omega_{m}$), the dark energy parameter ($\Omega_{\Lambda}$), the critical density ($\rho_{\text{crit}}$), the present root-mean-square matter fluctuation averaged over a sphere of radius $8h^{-1}$ Mpc ($\sigma_{8}$), the redshift at decoupling ($z_{*}$), the age of the universe at decoupling ($t_{*}$) and the redshift of reionisation ($z_{\text{re}}$).

The parameters of this model have been determined to great precision. That which enables this determination is that a modern-day cosmological observation typically maps out the distribution of an observable quantity (e.g., temperature and polarisation of the CMB, photons, the number density of galaxies and clusters, etc.) as a function of redshift and/or angular position in the sky. To extract any meaningful information about the universe from this data, we require a set of free model parameters associated with the cosmological model and a theoretical framework to compute and predict the observable quantities given.


Observables are given as a function of redshift, and in CPT some of the widely studied observables are Redshift Space Distortions (RSD) \cite{kaiser, chen, hamilton, kaiser2, saito}, which are an effect where the spatial distribution of galaxies appears to be squashed and distorted when their positions are plotted as a function of their redshift. This effect is due to the peculiar velocities of the galaxies causing a Doppler shift in addition to the redshift caused by the expansion of the universe. RSD manifest in two particular ways; the ``Fingers of God'' effect, where the galaxy distribution appears to be elongated pointing towards the observer, and the ``Kaiser effect'', in which the galaxy distribution appears to flatten, though this effect is much smaller than the fingers of God effect and can be distinguished by the fact that it occurs on larger scales. 

Other observational effects include gravitational lensing \cite{yoo6}. Predicted by Einstein's general theory of relativity, the effect consists of having a light source, a massive object and an observer. If they all lie in a straight line, the light will be deflected by the massive object and the source will appear as a ring around the massive lensing object to the observer. Other lensing effects include multiple distorted images of the same source around the massive lens, e.g. roulettes \cite{roulettes}. There are three classes of gravitational lensing: strong lensing (Einstein rings, arcs and multiple images), weak lensing (stretching of the background perpendicular to the direction to the centre of the lens) and microlensing (no distortion in shape can be seen but the amount of light received from the source object changes in time). It was not until 1979 that this effect was confirmed by observations \cite{1979Nature}. 

Two of the most widely studied observables in the literature are the Sachs-Wolfe effect and the Integrated Sachs-Wolfe (ISW) \cite{sachs, sachs2, sachs3}, named after Rainer K. Sachs and Arthur M. Wolfe, these effects are a property of the CMB, in which photons from the CMB are gravitationally redshifted, causing the CMB spectrum to appear uneven. This effect is the predominant source of fluctuations in the CMB for angular scales above about ten degrees.  The non-integrated Sachs-Wolfe effect is caused by gravitational redshift occurring at the surface of last scattering\footnote{Set of points in space at the right distance from us so that we are now receiving photons originally emitted from those points at the time of photon decoupling.}. The effect is not constant across the sky due to differences in the matter and energy density at the time of last scattering. The ISW is also caused by gravitational redshift, but it occurs between the surface of last scattering and Earth, and is not part of the primordial CMB.

An observable which has been of particular interest to study using cosmological perturbation theory is the galaxy number count. This observable is of crucial importance recently due to it being one of the main observables in the upcoming surveys. Galaxy surveys measure the number of galaxies per unit of solid angle in redshift bins, and this is precisely the galaxy number count. Refs.~\cite{antony, jeong1, durrer1} have computed this at linear order, but the high precision of the future galaxy surveys makes higher order perturbation theory absolutely necessary.


Nonlinearities arise when working at higher orders in cosmological perturbation theory and the complicated calculations and length of the resulting description of observables makes this a highly demanding and laborious task. Various attempts have been made to compute the galaxy number counts at second \cite{cc1,cc2,durrer2,yoo1,yoo2,jeong2,yoo5} and at third order \cite{didio3}, finding new relativistic corrections to known effects such as the Sachs-Wolfe, ISW and RSD, as well as other effects that arise from the combination of such and sourcing coming from previously neglected potentials.

The lengthy results have made the different groups question whether or not their results match each other, and there have been attempts to compute the difference between different descriptions of the same observable, but only for the dominating terms, not the full expressions \cite{nielsen}. 


Starting from first principles, as given by Einstein's theory and clearly explained in Refs.~\cite{weinberg, misner}, the theory of null geodesics \cite{ellis}, and the computational power of \texttt{xPand} \cite{xact, brizuela, toolkit} is required to perform the perturbations needed to compute an expansion of the galaxy number counts up to second order in perturbation theory. We present a general result for the galaxy number counts in terms of the affine parameter of the geodesic equation that can be then compared to previous works. 

In order to do this, we use techniques ranging from differential geometry to perturbation theory, passing through numerical methods and the analysis of analytical descriptions of observables. We start by describing cosmological perturbation theory in chapter \ref{chapter:cpt}, since it is essential for the understanding of the concordance model. This technique is indeed crucial to solving the differential equations of General Relativity, as a completely non-perturbative formulation is still far from reach by even the most powerful computers available. We show these solutions, since they will be necessary for the rest of the thesis. 

In chapter \ref{chapter:wkb}, we use these solutions and consider gravitational waves with sufficiently small amplitudes to be modelled as linear tensor perturbations travelling through a Friedmann-Lema\^{i}tre-Robertson-Walker (FLRW) universe filled with radiation and matter, allowing for a source term due to neutrino anisotropic stress.

In chapter \ref{chapter:gnc}, we study the galaxy number counts observable, and all the different observable contributions, from redshift, to the relation between luminosity and angular diameter distance, and the geometric description of the geodesic deviation equations, the physical volume and the energy that take part in the full expansion of the observable. This chapter is where the most intensive research that was completed is presented. Most of the second order quantities are left in the appendices so the text is easier to read, but the main results have been left in the main body. 

Finally, in chapter \ref{chapter:conclusions} we discuss the conclusions reached in this thesis and point towards future research and work.

%% file: tex/chapter_2.tex
\chapter{Cosmological Perturbation Theory}
\label{chapter:cpt}

\section{Introduction}
\label{sec:intropt}

Perturbation theory is a widely used technique in physics \cite{malik}. It helps to simplify equations and calculations and to study problems which would otherwise be impossible to solve. In order for perturbation theory to work, it requires the existence of a small quantity around which all the analysis is made and, in a first approximation, may be neglected. These quantities may be parameters of the theory or dynamical variables, like in cosmology, thus, cosmological perturbation theory (CPT). In general, one starts by expanding a quantity into a homogeneous background and an inhomogeneous perturbation
\be
T(x^{\mu}) = T^{(0)}\left(x^{0}\right) + \delta T(x^{\mu}).
\ee
The background part is a time-dependent quantity, $T^{(0)}\equiv T^{(0)}\left(x^{0}\right)$, whereas the perturbations depend on time and space coordinates $x^{\mu} = \left[ x^{0}, x^{i}\right]$. The smallness of a dynamical variable studied in cosmology is often related to a small parameter $\epsilon$. In linear perturbation theory for example, we only consider first-order terms, $\epsilon^{1}$, and can neglect the product of two perturbations, which in turn would be of order $\epsilon^{2}$ or higher. We can continue this until an $n$-th order is reached, as desired. The perturbation can be further expanded as a power series in the smallness parameter $\epsilon$,
\be
\label{eq:deltat}
\delta T\left( x^{0}, x^{i}\right) = \sum_{n=1}^{\infty} \frac{\epsilon^{n}}{n!} \delta T^{(n)}\left( x^{0}, x^{i}\right),
\ee
in which we use the conventional factor $1/n!$ inspired by the Taylor expansion and defined the notation $\delta T^{(n)}$ to distinguish between the perturbations of order $n$ and the background part variable, $T^{(0)}$ or $\bar{T}$. For some particular quantities the background part vanishes, but in general and in this thesis in particular, it does not vanish and it represents a solution of the system of equations under study in a simple case where the problem allows an exact solution.

Although useful, this technique does have its caveats. It works only if the desired level of accuracy ``is only a small correction to the previous order non-zero variable'', meaning if the new correction to $\delta T^{(n)}$ is $\delta T^{(n+1)}$, then $\epsilon \delta T^{(n+1)}\ll \delta T^{(n)}$, for all values of $n$. If the small quantities under study are the dynamical variables themselves, this may not always be verified, as these variables may grow beyond the size which verifies the previous condition.

One of the advantages of perturbation theory is that, in plenty of cases, it makes it possible to linearise a system of equations, thus simplifying it considerably. The first thing to do is to solve the background equations and to find $T^{(0)}$, then to solve the linearised system to find the solution $\delta T^{(1)}$, which should depend on the background solution. To find the next order solution, in this case $\delta T^{(2)}$, another linear equation needs to be solved, depending now on the first order term $\delta T^{(1)}$ for which we have a solution for already. This procedure, although time consuming, can be used to model more complicated phenomena that describe reality better. This can continue up to arbitrary orders to improve accuracy of the results to a desired level, as well as to study new effects not present at lower orders.

In CPT we can run into the problem that our small quantities are the dynamical variables themselves, as mentioned above, but only when studying the late universe on small scales. On large scales or in the early universe, this problem does not usually occur, at least for the most common models. We assume perturbation theory is valid throughout this thesis.

The study of perturbations in cosmology has been a case of study for a long time. The first studies were done by Lifshiftz in Refs.~\cite{Lifshitz:2017aa, lif} calculating the evolution of density perturbations at linear level --first order in perturbation theory. Another study by Tomita in Ref.~\cite{tomita} demonstrated the second order computation for the density evolution. One of the most cited and important works that set the foundation of the formalism for cosmological perturbations was done by Bardeen in Ref.~\cite{Bardeen80}, who defined the first gauge invariant perturbations, which now carry his name. Kodama and Sasaki generalised the gauge invariant formalism in Ref.~\cite{KS}, deriving the equations for different cosmological scenarios. Many works have contributed to the development of CPT. More information can be found in the reviews \cite{MFB,MM2008}.

In this chapter we review cosmological perturbation theory. We present first the more general relativistic perturbation theory in Section \ref{section:rel} and then its application on the cosmological setting with a Friedmann-Lema\^{i}tre-Robertson-Walker background in Section \ref{section:flrw}. This is so we can introduce the notation used in the rest of the thesis and the background necessary to understand the following chapters.

Many calculations shown in this chapter and in the rest of the thesis were performed using the \textit{Mathematica}\footnote{www.xact.es/xPand} package \texttt{xPand} \cite{xpand}, which is built into the tensor calculus package \texttt{xAct}\footnote{www.xact.es} \cite{xact}.

\section{Relativistic perturbation theory}
\label{section:rel}

In the context of relativistic theories of gravity, relativistic perturbation theory is widely used to solve the nonlinear equations posed by the theory. It is particularly useful since it takes into account the fundamental symmetry of the theory in its formulation, and is therefore the correct perturbative treatment to study tensor fields on Riemaniann manifolds. In this section we focus only on general relativity, however, most of the information presented is also valid in more general scenarios.

\subsection{General relativity}

The theory of general relativity (GR) was developed by Albert Einstein in 1915 and describes spacetime as a manifold with its curvature determined by the matter content present in the spacetime \cite{weinberg, misner}. It is based on Riemaniann geometry and we will review some of its main points here. 

The fundamental dynamical variable in GR is the metric tensor, $g_{\mu \nu}$, which defines infinitesimal distances between points in spacetime
\be
\label{eq:metrictensor}
\dd s^{2} = g_{\mu \nu} \dd x^{\mu}\dd x^{\nu}.
\ee

The metric tensor is symmetric, invertible and we use the $(-, +, +, +)$ convention for its signature \cite{wald}. The curvature of spacetime can be calculated from the metric tensor by defining the Levi-Civita connection, $\nabla_\mu$, which is a natural generalisation of the connection in the classical differential geometry of surfaces. We use the fundamental theorem of (pseudo-)Riemannian geometry, which states that on a (pseudo-)Riemannian manifold $(M, g)$ there exists a unique symmetric connection which is \textit{compatible} with the metric $g$. This connection is called the Levi-Civita connection, or \textit{symmetric connection} \cite{nakahara}, which means
\be
\Gamma\indices{^\lambda_{\mu\nu}} = \Gamma\indices{^\lambda_{\nu\mu}},
\ee 
and for $X^{\alpha}\in M$ it is given by
\be
\nabla_{\mu} X^{\alpha} = \partial_{\mu}X^{\alpha} + \Gamma\indices{^\alpha_{\mu\lambda}} X^{\lambda} .
\ee

Thus the curvature of spacetime compatible with the metric is related to the Riemann tensor given by
\be
R\indices{^\alpha_{\beta\mu \nu} } = \Gamma^{\alpha}_{\beta \nu, \mu} - \Gamma^{\alpha}_{\beta\mu,\nu} + \Gamma^{\alpha}_{\mu\rho}\Gamma^{\rho}_{\beta_\nu} - \Gamma^{\alpha}_{\nu\rho}\Gamma^{\rho}_{\beta\mu},
\ee
which has components determined by the Christoffel symbols
\be
\Gamma_{\mu\nu}^{\alpha} = \frac{1}{2}g^{\alpha \beta}\left( g_{\beta\mu, \nu} + g_{\beta\nu,\mu} - g_{\mu\nu, \beta} \right).
\ee
The Ricci tensor, $R_{\mu\nu}$, and the Ricci scalar, $R$, are contractions of the Riemann curvature tensor
\be
R_{\mu\nu} = R\indices{^\alpha_{\mu\alpha\nu}}, \qquad R = R\indices{^\mu_\mu}=g^{\mu\nu}R_{\mu\nu},
\ee
and their combination defines the Einstein tensor by
\be
\label{eq:ETensor}
G_{\mu\nu} = R_{\mu\nu} - \frac{1}{2} R g_{\mu\nu}.
\ee
There are at least three ways to compute the Einstein tensor from the Riemann tensor: (i) by successive contractions of the Riemann tensor, which is the one used in this thesis, and for completeness; (ii) by forming the dual of the Riemann tensor and then contracting, 
\begin{align}
\cancel{G}\indices{_{\alpha\beta}^{\gamma\lambda}}&\equiv (*R*)\indices{_{\alpha\beta}^{\gamma\lambda}} = \epsilon_{\alpha\beta\mu\nu} R\indices{^{|\mu\nu|}_{|\rho\sigma|}} \epsilon^{\rho\sigma\gamma\lambda} = - \delta\indices{^{\lambda\rho\sigma}_{\beta\mu\nu}} R\indices{^{|\mu\nu|}_{|\rho\sigma|}} , \\
G\indices{_\beta^{\lambda}} &= \cancel{G}\indices{_{\alpha\beta}^{\alpha \lambda}},
\end{align}
where $\epsilon_{\alpha\beta\mu\nu}$ is the Levi-Civita symbol, and $(*X*)$ denotes the dual of $X$. The third (iii) consists of combining the equations obtained in method (ii) obtaining
\be
G\indices{_\beta^\lambda} = G\indices{^\lambda_\beta} = - \delta\indices{^{\lambda\rho\sigma}_{\beta\mu\nu}}R\indices{^{|\mu\nu|}_{|\rho\sigma|}},
\ee
where
\be
\delta\indices{^{\lambda\rho\sigma}_{\beta\mu\nu}} = \delta\indices{_{\beta\mu\nu}^{\lambda\rho\sigma}} = 
	\begin{cases}
	+1 & \text{if $\lambda\rho\sigma$ is an even permutation of $\beta\mu\nu$}, \\
	- 1 & \text{if $\lambda\rho\sigma$ is an odd permutation of $\beta\mu\nu$}, \\
	0 & \text{otherwise}.
	\end{cases}
\ee

The Einstein tensor is symmetric and has the important property of being divergence free, i.e. $\nabla_{\mu}G^{\mu\nu} = 0$ due to the Bianchi identities
\be
R\indices{^a_{\beta[\lambda\mu;\nu]}} = 0.
\ee

This is the reason that the Einstein tensor is used in the Einstein field equations
\be
\label{eq:efield}
G_{\mu\nu} = 8\pi G T_{\mu\nu},
\ee
where the constant $G$ is the Newtonian constant of gravity and we choose units for which the speed of light, $c$, is set to unity. Since the stress-energy tensor of matter, $T_{\mu\nu}$, must also be divergence-free to preserve local conservation of energy and momentum, it obeys
\be
\label{eq:energytensor}
\nabla_{\mu}T^{\mu\nu} = 0.
\ee
The field equations can also be derived from the action
\be
\label{eq:gr-s}
S = \int \dd^{4}x\sqrt{-g}\left[ \frac{R}{16 \pi G} + \mathcal{L}_{m}\right],
\ee
where $g$ is the determinant of the metric, which defines an invariant volume measure and $\mathcal{L}_{m}$ is the matter Lagrangian. The first term of the action in Eq.~\eqref{eq:gr-s} is the so-called Einstein-Hilbert action, and describes the gravitational dynamics. In the standard metric formulation of gravity, the Einstein equations are derived by varying this action with respect to the metric tensor. Other formulations exist, which give rise to the different equations in modified gravity like $f(R)$ or Gauss-Bonet theory, which is discussed in Appendix \ref{chapter:mgb}. Both the Einstein-Hilbert action and the field equations are invariant under diffeomorphisms, which means this theory can be described by tensors, independently of the choice of coordinate system or basis. The fact that this symmetry exists is of extreme importance and has consequences in the development of a consistent perturbation theory, as we show in the next section.

From the action, Eq.~\eqref{eq:gr-s}, it can be seen that the matter Lagrangian, $\mathcal{L}_{m}$, is related to the stress-energy tensor by,
\be
\label{lmn}
T_{\mu\nu} = -2 \frac{\delta \mathcal{L}_{m}}{g^{\mu\nu}} + g_{\mu\nu}\mathcal{L}_{m}.
\ee
The energy-momentum tensor of a fluid with density $\rho$, isotropic pressure $P$ can be decomposed by choosing a set of observers represented by a time-like unit vector field, $u^{\mu}$, defined in the next section. The energy-momentum tensor is then defined as (\cite{malik,ellis,durrer3}) 
\be
\label{tmunu}
T_{\mu\nu} = \left( \rho + P\right) u_{\mu}u_{\nu} + P g_{\mu\nu} + q_{\mu}u_{\nu} + q_{\nu}u_{\mu}+\pi_{\mu\nu}.
\ee
where the remaining variables are the energy flux, $q_{\mu}$, and the anisotropic stress, $\pi_{\mu\nu}$. Note that the energy-momentum tensor defined in Eq.~\eqref{tmunu} does not necessarily follow Eq.~\eqref{lmn}, and it is not derived from it. 

Each of the variables in Eq.~\eqref{tmunu} can also be obtained from the energy-momentum tensor itself, if we define the projection tensor orthogonal to the velocity $u^{\mu}$ as $h_{\mu\nu} = g_{\mu\nu} + u_{\mu}u_{\nu}$, then the observer-dependent physical quantities are \cite{KS}
\begin{align}
\text{Energy density}		&\quad \rho = T_{\mu\nu} u^{\mu}u^{\nu},\\
\text{Pressure} 			&\quad P = T_{\mu\nu}h^{\mu\nu}/3, \\
\text{Energy flux}		&\quad q^{\alpha} = -T_{\mu\nu}h^{\mu\alpha}u^{\nu}, \\
\text{Anisotropic stress}	&\quad \pi^{\alpha\beta} = h^{\mu\alpha}h^{\nu\beta}T_{\mu\nu} - P h^{\alpha\beta},
\end{align}
with the constraints
\be
\label{pi-cons}
q_{\alpha}u^{\alpha} = 0, \quad \pi_{\alpha\beta}u^{\alpha} = 0, \quad \pi\indices{^\alpha_{\alpha}} = 0,
\ee
which follow from their definitions. In most cases, the frame is chosen such that the energy flux vanishes, i.e. $q^{\mu} = 0$ \cite{carroll}. This is called the \textit{energy frame}, and it will be used throughout this thesis. In this frame all the degrees of freedom that described the energy flux are now transferred to the observer's 4-velocity, $u^{\mu}$, since it is now constrained to follow the flow of the matter to conserve the vanishing energy flux. Any other frame can be chosen as this procedure is covariant, meaning that the quantities generated are the same in every coordinate system. Note that this definition does depend on the observer, which implies that all quantities defined with respect to a particular observer must be known in order to make predictions about observables related to that particular observer.

The equations of motion for the fluid quantities can also be derived from the conservation of the stress-energy tensor. Eq.~\eqref{eq:energytensor} gives rise to four equations, one for each value of the free index $\nu$, but in total there are ten functional degrees of freedom in the energy-momentum tensor. To completely describe the evolution of the system, the typical approach is to work in the perfect fluid scenario, wherein this case, the anisotropic stress vanishes and the number of degrees of freedom of the system is reduced to five. To reduce this to the four equations we do have, an equation of state is normally used that relates the pressure of the system with the other variables. A barotropic equation of state, $P=P(\rho)$, is often used and can successfully describe many fluids relevant for cosmology. Alternatively, if the fluid is not perfect, one can take an approach where the anisotropic stress depends only on the other fluid parameters, such as when it is well represented by shear viscosity.

If matter is composed of particles that can be approximated as test particles, its dynamical evolution can also be described using the geodesic equation \cite{ellis,misner}
\be
\label{eq:geointro}
p^{\mu}\nabla_{\mu}p^{\nu} = 0,
\ee
where $p^{\mu}$ is the 4-momentum tensor. This is useful for computing the dynamical evolution of matter if the system under study is composed only of a few particles or if the calculations are to be performed numerically, like in the case of N-body simulations. The usefulness of the geodesic equation is far more general than to describe the matter degrees of freedom, as we will see in Section \ref{chapter:gnc}. The geodesic equation is mostly applied to study the geometry of a spacetime by analysing the trajectories of test particles, such as massless particles that obey
\be
p^{\mu}p_{\mu} = 0,
\ee
which will be studied further in Section \ref{chapter:gnc}, or massive ones with mass $m$,
\be
p^{\mu}p_{\mu} = -m^{2}.
\ee
\break\break\break\break
The Einstein equations, Eq.~\eqref{eq:efield}, these are nonlinear partial differential equations for the metric tensor and for that reason, they are very difficult to solve in general scenarios. Some general solutions do exist for simple systems with particular symmetries, one of which would be the De Sitter universe, which is a completely symmetric space, without matter or curvature whatsoever. Using a spherical mass with spherical or axial symmetry then we get the exact solutions for black holes, such as Schwarzschild. Taking isotropy and homogeneity but no curvature, we are left with Minkowski spacetime and allowing for curvature we then have the Friedmann-Lema\^{i}tre-Robertson-Walker (FLRW) family of spacetimes.

The exact solution that we focus on in this thesis is part of the FLRW family of spacetimes. These spacetimes have homogenous and isotropic spatial slices. The FLRW line element is, in spherical coordinates, given by
\be
\label{eq:flrw-spherical}
\dd s^{2} = a^{2}(\eta)\left[ -\dd \eta^{2} + \frac{\dd r^{2}}{1-K r^{2}} + r^{2}\left(\dd \theta^{2} + \sin^{2}\theta \dd\varphi^{2}\right) \right],
\ee
where $\eta$ is the conformal time, and the function $a(\eta)$ is the scale factor, which has to obey the Friedmann evolution equations, \eqref{F1} and \eqref{F2} below, that are derived from the Einstein equations; $K$ represents the constant curvature of the homogenous spatial slices. In the next subsection, we solve perturbatively the Einstein equations when no exact solution can be found.

\subsection{Perturbed spacetime}

As discussed in Section \ref{sec:intropt}, we will assume that the physical quantities can usefully be decomposed into a homogeneous background, where quantities depend only on cosmic time, and inhomogeneous perturbations. These perturbations ``live'' on the background spacetime and it is this background spacetime which is used to split the four-dimensional spacetime into spatial 3-hypersurfaces and a temporal component that labels each spatial slice, using a (3+1) decomposition \cite{hartle}. The metric tensor can be split up as
\be
g_{\mu\nu} = g_{\mu\nu}^{(0)}+ \delta g_{\mu\nu},
\ee
where $g^{(0)}_{\mu\nu}$ is the background metric and $ \delta g_{\mu\nu}$ is the perturbed metric. The background metric $g_{\mu\nu}$ or $\bar{g}_{\mu\nu}$ is given by
\be
\label{eq:bgmet}
g_{\mu\nu}^{(0)} = a^{2}(\eta)  \begin{pmatrix}
							-1 & 0 \\
							0 & \gamma_{ij}		
							\end{pmatrix},
\ee
and $\gamma_{ij}$ is the metric on the 3-dimensional space with constant curvature $K$, which in spherical coordinates, is given by Eq.~\eqref{eq:flrw-spherical}.

The metric tensor has 10 independent components in 4 dimensions. For linear perturbations, it turns out to be very useful to split the metric perturbation into different parts that are labelled scalars, vectors or tensors according to their transformation properties on spatial 3-hypersurfaces \cite{Bardeen80}. The reason for splitting the metric perturbation into these three types is that they all decouple in the linear perturbation equations.

\textit{Scalar} perturbations can always be constructed from a scalar quantity, or its derivatives, and any background quantities such as the 3-metric $\gamma_{ij}$. We can construct any first-order scalar metric perturbation in terms of four scalars $\phi$, $B$, $\psi$ and $E$, where
\begin{align}
\label{pertg00}
\delta g_{00} &= -2 a^{2}\phi, \\
\label{pertg0i}
\delta g_{0i} &= a^{2}B_{i}, \\
\label{pertgij}
\delta g_{ij} &= 2 a^{2} C_{ij}.
\end{align}
The $0 - i$ and the $i - j$ components of the metric tensor can be further decomposed into scalar, vector and tensor parts
\begin{align}
\label{bi}
B_{i} &= B_{,i} - S_{i}, \\
\label{cij}
C_{ij} &= -\psi \delta_{ij} + E_{,ij} + F_{(i,j)} + \frac{1}{2} h_{ij},
\end{align}
where $S_{i}$ and $F_{i}$ are \textit{vector} metric perturbations, and $h_{ij}$ is a \textit{tensor} metric perturbation.

Any 3-vector, such as $B_{,i}$, constructed from a scalar is necessarily curl-free, i.e. $B_{,[i,j]}=0$. 

\textit{Vector} perturbations are divergence-free. We can distinguish a pure vector part of the metric perturbation $\delta g_{0i}$, which is denoted by $-S_{i}$, which gives a non-vanishing contribution to $\delta g_{0[i,j]}$. Similarly, we define the vector contribution to $\delta g_{ij}$ constructed from the derivative of a vector $F_{(i,j)}$.

Lastly, there is a \textit{tensor} contribution to $\delta g_{ij} = a^{2}h_{ij}$ which is both transverse, $h\indices{_{ij,}^j}=0$, meaning divergence-free, and trace-free, $h\indices{^i_i} = 0$, which therefore cannot be constructed from inhomogeneous scalar or vector perturbations.

We have introduced four scalar functions, two spatial vector valued functions with three components each, and a symmetric spatial tensor with six components. After taking into account the constraints above mentioned, we are left with 10 degrees of freedom, the same number as the independent components of the perturbed metric.

Reminding ourselves that the reason for splitting the metric perturbation into these three types is that the governing equations decouple at linear order, therefore we can solve each perturbation type separately. At higher order this is no longer true, as there is mixing of the different types. In general, at higher orders, $n>1$, the different types of perturbations of order $n$ decouple, but are sourced by terms comprising perturbations of lower order \cite{malik}.

We follow the notation of Malik and Wands \cite{malik} and Mukhanov et al. \cite{MFB}, so that the metric perturbation, $\psi$, can be identified with the intrinsic scalar curvature of spatial hypersufaces at first order. There are different choices for the way the spatial metric is split into different perturbation variables at second and higher orders in the perturbations, however, we will not be investigating these further in this thesis.

Note that the metric perturbations in Eqs.~\eqref{pertg00}, \eqref{pertg0i}, and \eqref{pertgij} include all orders. If we write down the complete metric tensor, up to and including second-order in perturbations we have
\begin{align}
\label{eq:metricsec2}
g_{00} &= -a^2 \left[1 + 2 \phi_{1} + \phi_{2}\right], \\
g_{0i} &= a^{2} \left[ B_{1i} + \frac{1}{2} B_{2i}\right],\\
g_{ij} &= a^{2}\left[\delta_{ij} + 2 C_{1ij} + C_{2ij} \right], 
\end{align}
where the first and second order perturbations $B_{1i}$ and $C_{1ij}$, and $B_{2i}$ and $C_{2ij}$, can be split further following Eqs.~\eqref{bi} and \eqref{cij}.

The metric tensor obeys the constraint,
\be
g_{\mu\nu} g^{\nu\lambda} = \delta\indices{_\mu^\lambda},
\ee
and can be used to compute the contravariant metric tensor, which up to second order is given by
\begin{align}
\label{g00}
g^{00} &= -a^{-2}\left[ 1- 2\phi_{1}- \phi_{2} + 4{\phi_{1}}^{2} -B_{1k}B\indices{_1^k}\right],\\
\label{g0i}
g^{0i} &= a^{-2}\left[ B\indices{_{1}^k} + \frac{1}{2} B\indices{_2^k} - 2 \phi_{1} B\indices{_1^i} - 2 B_{1k} C\indices{_1^{ki}} \right], \\
\label{gij}
g^{ij} &= a^{-2}\left[ \delta^{ij} - 2 C\indices{_1^{ij}}-C\indices{_2^{ij}} + 4 C\indices{_1^{ik}}C\indices{_{1k}^{j}} -B\indices{_1^i}B\indices{_1^j} \right].
\end{align}

We are interested in how spacetime geometry, described by the metric tensor, $g_{\mu \nu}$, is affected by the perturbed matter content, described by the energy-momentum tensor. The four velocity shown before, $u^{\mu}$, is defined by
\be
u^{\mu} = \frac{\dd x^{\mu}}{\dd \tau},
\ee
where $\tau$ is the proper time comoving with the fluid, defined as $\dd \tau = \dd s$, and it obeys the constraint,
\be
\label{uu1}
u^{\mu}u_{\mu} = -1.
\ee
Using Eqs.~\eqref{g00}, \eqref{g0i}, \eqref{gij}, and \eqref{uu1}, we find that the components of the 4-velocity up to and including second order are given by
\begin{align}
\label{eq:vel}
u_{0} &= -a \left[ 1 + \phi_{1} + \frac{1}{2} \phi_{2}- \frac{1}{2}\phi_{1}^{2}+\frac{1}{2}v_{1 k}v_{1}^{k} \right], \\
u_{i} &= a \left[ v_{1 i} + B_{1i}+ \frac{1}{2}\left( v_{2 i}+ B_{2i}\right)-\phi_{1}B_{1i}  - 2 C_{1ik} v_{1}^{k}\right], \\
u^{0} &= a^{-1} \left[1-\phi_{1} - \frac{1}{2}\phi_{2}+\frac{3}{2} \phi_{1}^{2}+\frac{1}{2}v_{1 k}v_{1}^{k}+v_{1k}B\indices{_1^k} \right],\\
u^{i} &= a^{-1} \left[ v_{1}^{i} + \frac{1}{2} v_{2}^{i} \right],
\end{align}
where 
\be
\label{velsplit}
v^{i} = \partial^{i} \text{v} + v^{i}_{\text{v}},
\ee
with $\text{v}$ the velocity potential, and $\nabla^{i}X = \partial^{i}X = {X_{,}}^{i}$ is the spatial part of the covariant derivative. For the 4-scalars, we simply write the perturbations by
\begin{align}
\label{rho}
\rho &= \rho_{0} + \delta \rho, \\
\label{pre}
P &= P_{0} + \delta P,
\end{align}
where $\rho_{0} = \rho^{(0)}$ and $P_{0} = P^{(0)}$, are the background quantities. 

The energy-momentum tensor of a fluid defined in Eq.~\eqref{tmunu}, taking a vanishing energy flux, can be split into first and second order parts in the usual way, and thus the anisotropic stress tensor, $\pi_{\mu\nu}$ is
\be
\pi_{\mu\nu} \equiv \pi_{(1)\mu\nu} + \frac{1}{2}\pi_{(2)\mu\nu},
\ee
following the constraints given in Eq.~\eqref{pi-cons}. The anisotropic stress vanishes for a perfect fluid or a minimally coupled scalar field, but may be non-zero in the presence of free-streaming neutrinos or a non-minimally coupled scalar field. The decomposition of the anisotropic stress tensor is complicated due to its constraints and for this reason its components also depend on the velocity fluctuations, as well as the metric. The anisotropic stress up to and including second order is given by,
\begin{align}
\pi_{00} &= 0, \\
\pi_{i0} &= -2 \pi_{ij} v^{j}, \\
\label{pij}
\pi_{ij} &= a^{2}\left[ \Pi_{ij} + \Pi_{(i,j)} + \Pi_{,ij}-\frac{1}{3}\delta_{ij}\nabla^{2}\Pi \right] + \frac{4}{3} \delta_{ij}\pi_{kl}C^{kl},
\end{align}
where we have defined the scalar, $\Pi$, vector, $\Pi_{i}$, and tensor, $\Pi_{ij}$, parts of the anisotropic stress.

From Eqs.~\eqref{tmunu}, \eqref{g00}, \eqref{g0i}, \eqref{gij}, \eqref{eq:vel}, \eqref{rho}, and \eqref{pre}, the components of the stress energy tensor in the background
\be
T\indices{^0_0} = -\rho_{0}, \quad T\indices{^0_i}=0, \quad T\indices{^i_j} = \delta\indices{^i_j}P_{0},
\ee
at first order,
\begin{align}
{}^{(1)}\delta T\indices{^0_0} &= -\delta \rho_{1}, \\
{}^{(1)}\delta T\indices{^0_i} &= \left( \rho_{0} + P_{0} \right) \left( v_{1i} + B_{1i} \right), \\
{}^{(1)}\delta T\indices{^i_j} &= \delta P_{1} \delta\indices{^i_j} + a^{-2}\pi\indices{_{(1)}^{i}_{j}},
\end{align}
and at second order
\begin{align}
{}^{(2)}\delta T\indices{^0_0} &= -\delta \rho_{2} - 2\left( \rho_{0} + P_{0} \right)v_{1 k}\left( v\indices{_1^k}+B\indices{_1^k}\right), \\
{}^{(2)}\delta T\indices{^0_i} &= \left( \rho_{0} + P_{0}\right)\left[ v_{2i} + B_{2i} + 4C_{1ik}v\indices{_1^k}-2\phi_{1}\left( v_{1i}+ B_{1i}\right) \right]\\
&\qquad +2\left( \delta\rho_{1}+\delta P_{1}  \right)\left( v_{1i} + B_{1i} \right) + \frac{2}{a^{2}}\left( B\indices{_{1}^k} + v\indices{_1^k} \right)\pi_{1 i k} , \notag \\
{}^{(2)}\delta T\indices{^i_j} &= \delta P_{2} \delta\indices{^i_j} + a^{-2} \pi\indices{_2^i_j} - \frac{4}{a^2}C\indices{_1^{ik}}\pi_{1jk} + 2\left( \rho_{0} + P_{0}\right) v\indices{_1^i} \left(v_{1j} + B_{1j} \right).
\end{align}
Note, that for compactness we have not split the above perturbations into their scalar, vector and tensor parts; these are given for $v^{i}$, $B_{i}$, $C_{ij}$ and $\pi_{ij}$ in Eqs.~\eqref{velsplit}, \eqref{bi}, \eqref{cij}, \eqref{pij}, respectively.

\section{Perturbations in Friedmann-Lema\^{i}tre-Robertson-Walker}
\label{section:flrw} 

The most general homogeneous and isotropic line element is the FLRW line element, given in Eq.~\eqref{eq:flrw-spherical}. In the background, the components of the metric depend only on time, which is represented by the conformal time coordinate, $\eta$. The conversion to cosmic time, $t$, is given by
\be
t = \int a \dd \eta.
\ee 

In this section, we begin by describing the evolution of the scale factor $a(\eta)$ and the other matter variables at the background level. We then introduce perturbations to this background in the flat case, i.e. $K=0$ in Eq.~\eqref{eq:flrw-spherical}. We show the perturbed evolution equations for both metric and matter perturbations and we introduce the concept of gauge, discuss the gauge issue and give some examples of gauge transformations.

\subsection{Background}
The Einstein equations give rise to only two independent evolution equations for the scale factor, $a(\eta)$, of which only one is dynamical. Before discussing the evolution equations, we introduce the Hubble rate, $H$, which quantifies the rate of change of the scale factor and is given by
\be
H = \frac{\dot{a}}{a}, 
\ee
in which a dot over the quantity denotes the derivative with respect to the cosmic time, $t$. The conformal Hubble rate, that we use throughout this thesis, is similarly given by
\be
\H = \frac{a'}{a} = a H,
\ee
where the prime denotes the derivative with respect to conformal time. Having introduced the conformal Hubble rate, using Eqs.~\eqref{eq:efield} and \eqref{eq:flrw-spherical} we can derive the Friedmann equations, which are constraints on the conformal Hubble rate and are given by
\be
\label{F1}
\H^{2} = \frac{8 \pi G}{3}a^{2} \rho - K.
\ee
The only other independent part of the Einstein field equations can be found from their trace and is
\be
\label{F2}
\H' = -\frac{4 \pi G}{3} a^{2} \left( \rho + 3P \right).
\ee
Note that, to simplify the notation in this case we have used the symbol of the variable to denote its background value, i.e. $\rho = \rho^{(0)}$ and $P = P^{(0)}$. We assume that matter is well described by a perfect fluid at background level and that the frame used to project the stress-energy tensor is the energy frame, as mentioned above. The conservation of the stress-energy tensor gives another dynamical equation which is not independent of the two Einstein equations and is given by
\be
\rho' = -3\H\left(\rho + P\right).
\ee
Solutions to these equations have been found in simple cases, such as when a single fluid dominates the energy density and has the simple equation of state
\be
P = w \rho,
\ee
with $w$ constant. Solving for the density, $\rho$, we find
\be
\rho(a) = \rho_{0} a^{-3(1+w)},
\ee
where $\rho_{0}$ is an integration constant, usually set to the value of $\rho$ today. The particular cases widely studied in the literature are those with zero curvature, $K=0$ and with specific equations of state for radiation ($w=1/3$), matter ($w=0$) and vacuum energy ($w=-1$) \cite{ellis, durrer3}. The corresponding solutions for $a(\eta)$ are
\begin{align}
a(\eta) &= \sqrt{\frac{8 \pi G \rho_{0}}{3}}\eta, \qquad w=\frac{1}{3}, \\
a(\eta) &= \frac{2\pi G\rho_{0}}{3}\eta^{2}, \qquad w=0, \\
a(\eta) &= \sqrt{\frac{3}{8\pi G\rho_{0}}}\frac{1}{\eta}, \qquad w=-1, 
\end{align}
in which we have assumed expanding initial conditions ($a'>0$). Many more solutions exist; with fluid mixtures, as discussed in Section \ref{sec:linear-tensor-perturbations}, or with scalar fields, and some that cannot be found in a closed form but many more exist numerically \cite{tong}.

\subsection{Gauges}

\subsubsection{Perturbing spacetime}

Perturbation theory in a relativistic setting gives rise to interesting issues related to the fact that spacetime itself is perturbed. We have to make sure that the formalism is adapted to the geometric nature of the problem and is covariant. We therefore follow Refs.~\cite{penrose,malik2008,malik,malik2}.

Usually, the first step is to identify the exact solution of the Einstein field equations, Eq.~\eqref{eq:efield}, that approximates the system under study. For cosmology, this is the FLRW solution, but here we will attempt to be fully general and call that solution ``the background solution'' with the background metric $g_{\mu\nu}^{(0)}$, as given in \eqref{eq:metrictensor}. The physical manifold $\mathcal{M}$ is split into a background manifold $\mathcal{M}_{0}$ and perturbed manifold, which is part of a one-parameter family of manifolds $\mathcal{M}_{\epsilon}$, with $\epsilon$ being the small parameter defining the perturbative scheme, as seen in Section \ref{sec:intropt}. All of these 4-dimensional manifolds are embedded in a 5-dimensional manifold $\mathcal{N}$. We can then define a diffeomorphism $\phi_{\epsilon} : \mathcal{M}_{0} \rightarrow \mathcal{M}_{\epsilon}$, which identifies points in $\mathcal{M}_{0}$ with those in $\mathcal{M}_{\epsilon}$. It is also useful to define a vector field $X$ in the tangent bundle of $\mathcal{N}$, whose integral curves, $\gamma$, intersect each of the manifolds of the family $\mathcal{M}_{\epsilon}$, generating the diffeomorphism $\phi_{\epsilon}$, by identifying each interception in $\mathcal{M}_{\epsilon}$ to a point in the background manifold $\mathcal{M}_{0}$.

Given a tensor field $T$, its Taylor expansion around any point in $\mathcal{M}_{0} \subset \mathcal{N}$, along the integral curve, $\gamma$, is given by
\be
\label{eq:gaugeexp}
T_{\phi} \equiv \phi_{\epsilon}^{*} T_{\epsilon} = e^{\epsilon \pounds_{X}}T\big|_{0} = T_{0} + \epsilon \left( \pounds_{X}T\right)\big|_{0} +  \frac{\epsilon^{2}}{2} \left( \pounds_{X}^{2} T \right)|_{0} + \mathcal{O}\left( \epsilon^{3}\right),
\ee
in which $T_{\epsilon}$ is the tensor field $T$ evaluated at the manifold $\mathcal{M}_{\epsilon}$, $\pounds_{X}$ is the Lie derivative along the vector field $X$ and $\phi^{*}_{\epsilon}$ denotes the pull-back of the diffeomorphism $\phi_{\epsilon}$, which is used to evaluate the result on $\mathcal{M}_{0}$. The use of the exponential of the Lie derivative is simply a shorthand for the Taylor expansion, and it is useful to simplify certain calculations. Labelling perturbations as $\delta T$, as before, we can separate the full result order by order as (omitting pull-backs)
\be
T_{\phi} = T_{\phi}^{(0)} + \delta T_{\phi}^{(1)} + \frac{1}{2} \delta T_{\phi}^{(2)} + \dots ,
\ee
so that $\delta T_{\phi}^{(n)} = \epsilon^{n} \left( \pounds_{X}^{n} T \right)|_{0}$, in which we use a similar notation to Eq.~\eqref{eq:deltat}, but the $\epsilon$ parameters are absorbed into the perturbations, for simplicity. Note that all quantities are evaluated in $\mathcal{M}_{0}$ and will therefore be written in terms of the coordinates of the background manifold.

\subsubsection{Gauge choice}
\label{sub:gauge}

The choice of vector field in Eq.~\eqref{eq:gaugeexp} and corresponding diffeomorphism $\phi_{\epsilon}$ is not unique, since there is no unique way to identify points between the background and the perturbed manifolds. This choice is called \textit{the gauge choice} and $X$ is called the generator of that gauge. As we will see, perturbed quantities defined in different gauges will not be equal. This is not surprising, as quantities in one gauge are evaluated at different points from quantities in another gauge. It is useful, therefore, to relate quantities in different gauges and to establish ways to fix the chosen gauge. Two approaches exist for deriving the relation between the quantities in different gauges, or \textit{gauge transformations}, called the active and passive approaches \cite{malik}. They differ by the choice of manifold on which to focus. The active approach focuses on each point in the perturbed manifold $\mathcal{M}_{\epsilon}$ and compares tensors in the corresponding points in $\mathcal{M}_{0}$ using different gauge generators. The passive approach does the opposite. It begins with points on $\mathcal{M}_{0}$ and evaluates tensors at different points in the perturbed manifold. They are equivalent and lead to the same transformation equations and, for that reason, we only use the active approach here.

The process of fixing a gauge is of great importance, since most cosmological quantities are not gauge invariant, i.e. they depend on which gauge was chosen to start with. However, once a gauge is fixed, all quantities in that gauge are well defined. 

The process of fixing a gauge is often based on choosing an appropriate number of tensor fields and giving some constraints on their perturbations. As an example, suppose we have a scalar field $\varphi$, whose background value $\varphi^{(0)}$ is not constant. We can (partially) fix a gauge by deciding that its perturbations vanish, i.e. by forcing $\varphi$ to obey the symmetries of the background manifold. In the language described above, we are choosing to map the points in $\mathcal{M}_{\epsilon}$ to points in $\mathcal{M}_{0}$ for which the value of $\varphi$ is the same, which is certainly possible. Making this choice, along with similar ones for three other variables (in 4 dimensions), eliminates the freedom in choosing the gauge generators. The mapping between the background and perturbed manifolds is completely determined, and thus all perturbations are uniquely defined.  

The Einstein field equations \eqref{eq:efield} are invariant no matter the gauge choice, as is any equation relating tensors, since it can always be rewritten as
\be
\label{eq:gmt}
G_{\mu\nu} - 8 \pi G T_{\mu\nu} = 0,
\ee
and the right-hand-side, zero, is obviously invariant. This has the consequence that the Einstein field equations can always be written equivalently in any gauge or with any choice of gauge-invariant variables. Another consequence of a symmetry like the one given in Eq.~\eqref{eq:gmt}, is that all quantities that can possibly be observed must be gauge-invariant, because there is no way for the equations to have information about the gauge in which they were used.

\subsubsection{Poisson Gauge}
Many gauges have become popular in the literature and in this thesis we make use of the Longitudinal Gauge and Poisson Gauge. We now describe their definitions and compute some of the gauge-invariant quantities that arise from them.

A very popular gauge for studying the post-inflationary universe is the longitudinal gauge. This gauge is also called conformal Newtonian gauge, and is defined by the following condition, 
\be
\label{eq:poisson-scalar}
B = E = 0.
\ee
If the problem under study only involves scalars, this choice is sufficient and turns out to diagonalise the metric, making many calculations simpler. When extended to include vector degrees of freedom, this gauge is often denoted as Poisson gauge. Two possible definitions exist in the literature, with the choice
\be
F^{i} =0,
\ee
being the most common \cite{cc1,cc2,cc3, durrer2, pedro}. It is motivated by the similarity with the Coulomb gauge of electromagnetism ($\nabla \cdot A = 0$), since its gauge conditions are equivalent to
\be
B\indices{^i_{,i}} = 0, \qquad C\indices{_{T}^{ij}_{,i}} = 0,
\ee
where $B_{i}$ and $C_{ij}$ were defined in Eqs.~\eqref{bi} and \eqref{cij}, and with $C\indices{_{T}^{ij}}$ being the traceless part of the perturbations of the spatial metric, $C^{ij}$. The alternative choice is made by taking
\be
S^{i} = 0,
\ee
is also used and is based on the requirement that the contravariant vector orthogonal to spatial hypersurfaces has a vanishing spatial part \cite{malik, MFB}. The gauge-invariant quantities arising in this gauge are the Bardeen potentials \cite{Bardeen80, durrer1}, given by
\begin{align}
\Phi &\equiv \phi + \H\left( B-E'\right) + \left( B - E'\right)', \\
\Psi &\equiv \psi - \H \left( B - E'\right). 
\end{align}
These quantities were the first gauge-invariants to be explicitly calculated and have been used in the literature for a very long time \cite{cc1, cc2, cc3, cc4, cc5, durrer1, durrer2, durrer3, malik, ellis}. They have the property of simplifying one of the equations of motion considerably as one can verify by substituting them into the Einstein equations. This gauge also has the advantage of nearly mimicking the evolution equations of Newtonian cosmology on short scales, at least at first order.

\subsection{Matter velocity field and peculiar velocities}
The components of the 4-velocity $u^{\mu}=\dd x^{\mu}/\dd \eta$ defined in general in Eq.~\eqref{eq:vel}, in the longitudinal gauge up to and including second order using the perturbed metric and using only scalar perturbations are given by
\begin{align}
\label{eq:pertu0}
u_{0} &= -a \left[ 1 + \ff + \frac{1}{2} \fs- \frac{1}{2}\ff^{2}+\frac{1}{2}v_{1 k}v_{1}^{k} \right], \\
\label{eq:pertui}
u_{i} &= a \left[ v_{1 i} + \frac{1}{2} v_{2 i}  - 2 \pf v_{1 i}\right], \\
u^{0} &= a^{-1} \left[1-\ff - \frac{1}{2}\fs+\frac{3}{2} \ff^{2}+\frac{1}{2}v_{1 k}v_{1}^{k} \right],\\
u^{i} &= a^{-1} \left[ v_{1}^{i} + \frac{1}{2} v_{2}^{i} \right],
\end{align}
where $v_{i} = \partial_{i} \text{v}$, with $\text{v}$ the velocity potential.

\subsection{Evolution equations}

\subsubsection{Einstein equations}
We show the perturbed Einstein field equations, Eq.~\eqref{eq:efield}, up to linear order. We split the equations into their scalar, vector and tensor parts and write them without specifying any gauge. We begin with the time--time Einstein equation:
\be
3\H\left( \H \phi + \psi' \right) - \nabla^{2}\left( \psi + \H\sigma \right) = -4 \pi G a^{2}\delta \rho,
\ee
where $\sigma = E'-B$.

The space--time equation results in a scalar equation,
\be
\H \phi + \psi' = -4\pi G a^{2}\left( \rho + P\right)\left( v+B\right),
\ee
and a vector equation,
\be
\nabla^{2}\left( F'_{i} + S_{i} \right) + 4\left( \H^{2} - \H' \right)\left( v_{\text{v} i} - S_{i} \right) = 0.
\ee
The off-diagonal is given by
\be
\sigma' + 2\H \sigma + \psi-\phi = 8 \pi G a^{2} \Pi.
\ee
The spatial part of the Einstein equation, like the spatial metric, is composed of two scalar parts
\begin{align}
\psi'' + 2\H \psi' + \H \phi' + \left( 2\H' + \H^{2} \right)\phi &= 4 \pi G a^{2} \left( \delta P + \frac{2}{3}\nabla^{2}\Pi \right), \\
E'' - B' + 2 \H \left( E' - B \right) +\psi - \phi &= 8\pi G a^{2} \Pi,
\end{align}
one vector part
\be
F_{i}'' + S_{i}' + 2\H\left( F_{i}' +S_{i} \right) = 8\pi G a^{2} \Pi_{i},
\ee
and a tensor part
\be
{h\indices{^i_j}}'' + 2 \H{h\indices{^i_j}}' - \nabla^{2}{h\indices{^i_j}} = 8\pi G a^{2}\Pi\indices{^i_j}.
\ee

We can clearly see in all equations above, one of the advantages of the SVT decomposition -- the scalars, vectors and tensors do not couple to each other at first order, implying that one can solve their respective equations independently of the others. At second order, this is no longer exactly true, as the second-order equations are sourced by combinations of first-order scalars, vectors and tensors. This can be seen in the nonlinear part of the time--time equation, which is
\begin{align}
{\Xi_{\text{NL} }}_{0}^{0} &= \left( \H^{2} - \H' \right) v_{i} \left( v^{i} + B^{i} \right) + \frac{3}{2} \H^{2} B^{i}B_{i} - 6 \H^{2} \phi^2 + 2\H C' \phi \\
& \quad+ B^{i,j}\left( \frac{1}{2}\left( \delta_{ij} C - C'_{ij} \right) - 2 \H\left( C_{ij} + \phi \delta_{ij}\right) +\frac{1}{4} \left( B_{(i,j)} - \delta_{ij} B\indices{^k_{,k}} \right) \right) \notag \\
& \quad+B^{i} \left( {C^{j}_{[j,i]}}' + \H C_{,i} - 2\H C^{j}_{i,j} + \H \phi_{,i} + \frac{1}{4} \H B^{,j}_{[i,j]}\right) \notag \\
& \quad+ \frac{1}{4}\left( C'_{ij}{C^{ij}}' - (C')^{2}\right) + 2\H C_{ij}{C^{ij}}' - C^{ij}C_{,ij} + 2 C^{ij} C^{k}_{i, j k} - C^{ij} C\indices{_{ij,k}^{,k}} \notag \\
& \quad+ \frac{1}{4}C_{,j}C^{,j} + C^{j}_{i,j}C\indices{^i_k^{,k}} -C_{,i}C\indices{^i_{k}^{,k}} + \frac{1}{2} C_{ij,k}C^{ik,j} - \frac{3}{4} C_{ij,k} C^{ij,k} , \notag
\end{align}
where $C\equiv C^{k}_{k}$, and we can see that all types of couplings exist. However, the second-order parts of variables continue not to mix, so one can still evolve them independently.

\subsubsection{Conservation of the stress-energy tensor}
We now show the equations derived from the covariant conservation of the stress-energy tensor, Eq.~\eqref{eq:energytensor}. The time component is given by
\be
\delta \rho' + 3 \H\left( \delta \rho + \delta P \right) + \left( \rho+P \right)\Big[ \nabla^{2}\left( \sigma + v + B \right) - 3\psi' \Big] = 0,
\ee
where the the perturbations are `full', meaning \textit{all orders} are enclosed in the perturbed variables for simplicity, i.e. $\psi = \psi^{(1)} + \psi^{(2)} + \cdots$ .

The spatial component gives the generalisation of the Euler equation, which can be further split into a scalar and a vector component. The scalar equation is
\be
\Big[ \left( \rho+P \right)\left( v + B \right) \Big]' + \left( \rho + P \right)\Big[ 4\H\left( v + B \right)+\phi \Big] + \delta P + \frac{2}{3}\nabla^{2}\Pi = 0.
\ee
and the vector equation is given by
\be
\left( v_{\text{v}}^{i} - S^{i} \right)' + \left(\rho - 3P\right)' \H \left( v_{\text{v}}^{i}-S^{i} \right) + \frac{1}{2\left( \rho + P \right)} \nabla^{2} \Pi^{i} = 0.
\ee

%% file: tex/chapter_3.tex
\chapter{Gravitational waves in a flat radiation--matter universe}
\label{chapter:wkb}

\section{Introduction}
\label{sec:intro}

In 2016, LIGO \cite{2016PhRvL.116f1102A} directly detected gravitational waves for the first time. The origin of these waves was astrophysical: the merger of two large stellar mass black holes.
In contrast, primordial gravitational waves originating from inflation have tightly constrained upper limits, but have not been detected, by \textsc{Planck} satellite observations combined with data from BICEP2/Keck Array \cite{PhysRevLett.114.101301} and with other data sources \cite{2015arXiv150201589P}.  The simplest inflationary models consistent with these results tend to favour inflationary scenarios which generate gravitational waves of relatively low, but not negligible ($r\ge0$), amplitude compared with scalar perturbations.

In this chapter, we consider gravitational waves with sufficiently small amplitudes to be modelled as linear tensor perturbations travelling through a flat Friedmann--Lema\^{i}tre--Robertson--Walker (FLRW) universe filled with radiation and matter, as seen in Section \ref{section:flrw}.  The matter component is pressureless dust, and our perturbation equations treat both radiation and matter as perfect fluids. We allow for a source term due to neutrino anisotropic stress.

As mentioned in \rfcite{1993PhRvD..48.4613T,1994PhRvD..50.3713A,1995PhRvD..52.2112N,1996PhRvD..53..639W,2004PhRvD..69b3503W,2005AnPhy.318....2P,2005astro.ph..5502B,2008cosm.book.....W,2013PhRvD..88h3536S}, there are well known exact analytical expressions for the evolution of linear tensor perturbations in a flat universe dominated by either radiation only or matter (pressureless dust) only. These approximations do not work particularly well when used for a universe containing both radiation and matter. The next simplest approximation is to use the radiation--only solution before the time of radiation--matter equality, and the matter--only solution after that time. A slightly more sophisticated approach (see e.g.~\rfcite{1994PhRvD..50.3713A,1995PhRvD..52.2112N}) is to match these two solutions together at the time of radiation--matter equality, termed the ``sudden approximation''. Reference~\cite{1995PhRvD..52.2112N} shows that these approaches do not provide particularly good approximations to a numerical solution of the linear tensor perturbation equation, finding that for large wavenumbers the standard scale-invariant wavefunction for the gravity wave severely \textit{underestimates} the actual size of the gravity wave and there is a definite shift in the \textit{phase} of the gravity wave as it crosses the radiation-matter phase transition. In \rfcite{1996PhRvD..53..639W}, a smoother transition is sought between the radiation and matter solutions by fitting parameters to numerical solutions, a method which is most suited to dealing with wavenumbers much lower (corresponding to much larger wave--lengths) than those we consider.

Later work in \rfcite{2004PhRvD..69b3503W,2005AnPhy.318....2P,2005astro.ph..5502B,2008cosm.book.....W,2013PhRvD..88h3536S} is aimed primarily in approximating the effects of neutrino free--streaming as expressed via an integro--differential equation derived in \rfcite{PhysRevD.50.2541,2004PhRvD..69b3503W}, rather than at improving accuracy in the perfect fluid model of radiation and matter.
In this chapter, we are also improving the accuracy of the solutions also in the case of non-zero anisotropic stress.

We now briefly survey the accuracy of the approximations of \rfcite{1995PhRvD..52.2112N,2005AnPhy.318....2P,2005astro.ph..5502B,2008cosm.book.....W,2013PhRvD..88h3536S}. References~\cite{1995PhRvD..52.2112N,2005AnPhy.318....2P}, \rfcite{2008cosm.book.....W} and \rfcite{2005astro.ph..5502B} pursue three different forms of the Wentzel-–Kramers-–Brillouin (WKB) method for finding approximate analytical solutions for ordinary differential equations.

The methods of \rfcite{1995PhRvD..52.2112N,2005AnPhy.318....2P} and  \rfcite{2008cosm.book.....W} provide results which, when restricted to the perfect fluid case, are accurate for primordial gravitational waves to within about $10\%$ for inverse wavenumbers, evaluated at the present day, of around \num{10}--\SI{15}{Mpc}. Reference~\cite{2005astro.ph..5502B} presents another form of WKB approximation for a flat radiation--matter universe. Restricted to the perfect fluid case, this provides a good sub--horizon approximation for gravitational waves of primordial origin. However, it does not retain such good accuracy outside the horizon, and covers only one of the two independent solutions of the governing equation for tensor perturbations. 

A different kind of approach is pursued in \rfcite{2013PhRvD..88h3536S}, using sums of spherical Bessel functions of successive orders to approximate tensor perturbations.  However, \rfcite{2013PhRvD..88h3536S}'s approach is best used for numerical calculations based on expansions of very high order~--- they exemplify solutions to 20th and 100th order. 

We derive improved results by employing a method set out, in a non--cosmological context, by Feshchenko, Shkil' and Nikolenko (FSN) in 1966 \cite{FSN}, and used in a recent paper \cite{DPS} to approximate the solutions of the ordinary differential equations governing  scalar density perturbations. The FSN method is well suited for solving linear second order ordinary differential equations, that also depend on a small parameter, which here we take to be the inverse wavenumber.  In effect, this approach extends the WKB method of \rfcite{2008cosm.book.....W}.

In Section~\ref{sec:linear-tensor-perturbations} of this chapter, we present the differential equation which governs tensor metric perturbations which were derived in Section \ref{section:flrw}, focusing on a flat radiation--matter universe, and also noting the simpler differential equation for a flat radiation--only model and its analytical solution with and without neutrino anisotropic stress. In Section~\ref{sec:approximating-tensor-perturbations}, we then use the method first set out in~\rfcite{FSN} to find an approximate analytical solution to the governing equation of Section~\ref{sec:linear-tensor-perturbations}. 

Our approximation method is based on the \emph{sub--horizon} assumption that the tensor perturbation has a wavenumber larger than the Hubble parameter.  However, using numerical solutions, we can see that, for a wide range of wavenumbers, it is possible to extend the approximation back to earlier times when the wavenumber is similar to the Hubble parameter. We find the approximation of third order in the inverse wavenumber is accurate to within $1\%$ or better for gravitational waves with a present day inverse wavenumber less than, or equal to, around  $\SI{17}{Mpc}.$ 
 
Section~\ref{sec:modelling-gravitational-waves-through-the-whole-radiation--matter-epoch} extends the analytical approximation to cover gravitational waves of primordial origin, from early times when their wavelength is larger than the horizon through their subsequent sub--horizon evolution. Our approximation to second order in the inverse wavenumber is accurate to within $1\%,$ or better, for gravitational waves corresponding to those with an actual present day inverse wavenumber less than, or equal to, around  $\SI{35}{Mpc}.$  This range includes wavenumbers which represent scales of key relevance for structure formation in the Universe, and broadly corresponds  to CMB multipoles $l\gtrsim 120$. 


A \emph{Mathematica} notebook which executes the approximation method automatically is available online at \rfcite{GWGithub}.  It enables all the approximations presented in this paper to be calculated in less than a second on a standard PC. The template can be easily adapted for use with other, similar, second order differential equations.

Throughout this chapter, we work in conformal time $\tau,$ with a dash indicating the derivative of a function with respect to conformal time.  We set the speed of light $c=1$, comoving spatial co-ordinates are denoted $x^i$ and Latin indices $i,j,k$ range from $1$ to $3$.

\section{Linear tensor perturbations}
\label{sec:linear-tensor-perturbations}

We now study linear perturbations in a flat Friedmann--Lema\^{i}tre--Robertson--Walker (FLRW) model with two non--interacting perfect fluids~--- radiation and matter, the latter being pressureless dust. We begin with the background equations, with the conformal time Friedmann equation studied in Section \ref{section:flrw}, and shown for a general fluid in Eq.~\eqref{F1}, taking the following form
\be\label{eq:Friedmann}
\H^2=\frac{8\pi G}{3} a^2 \left(\rhor+\rhom \right)\,,
\ee
where $\H$ is the conformal Hubble factor, $G$ is the gravitational constant, $a$ is the scale factor, $\rhor$ the homogeneous radiation density, and $\rhom$ the homogeneous matter density. As usual, the radiation and matter densities obey
\ba \label{eq:densities}
\rhor\propto a^{-4} & \text{\qquad and\qquad }  \rhom\propto a^{-3}\,,
\ea
and, as
noted in, for example, \rfcite{DPS,2006AIPC..843..111P}, substituting Eqs.~\eqref{eq:densities} in Eq.~\eqref{eq:Friedmann} and integrating, the scale factor is given by
\ba \label{eq:rmscalefactor}
a(\tilde{\tau})=\arm\left(\frac{\tilde{\tau}}{\tauc}+\frac{\tilde{\tau}^2}{4\tauc^2}\right)\,,
\ea
where $\arm$ is the time of radiation--matter equality, $\tilde{\tau}$ is the conformal time, and $\tauc=\infrac{\sqrt{2}}{\Hrm}.$ In the following, we will use a normalised conformal time co--ordinate and replace $\infrac{\tilde{\tau}}{\tauc}$ by $\tau,$ giving us the simpler expression
\ba \label{eq:rmscale}
a(\tau)=\arm\left(\tau+\frac{\tau^2}{4}\right)
\,.
\ea
We also normalise the comoving wavenumber $k$ by using the corresponding inverse units $\tauc^{-1}=\infrac{\Hrm}{\sqrt{2}},$ and we note that, in those units, the conformal Hubble parameter is then given by
\ba\label{eq:h-in-tau}
\H(\tau)=\frac{2\left(2+\tau\right)}{\tau\left(4+\tau\right)}\,.
\ea

For future reference, using the cosmological parameters of \rfcite{2015arXiv150201589P}, we find that we have
\ba \label{eq:phys-unit} \kunit=\SI[group-digits = false]{0.00727}{a_0.Mpc^{-1}}
\,,
\ea
where $a_0$ is the value of the scale factor at the present time, \emph{assuming the universe contains only radiation and matter}.  To compare inverse wavenumbers with \emph{actual} present day distances, we use a value of $a_0$ which takes account of  the  expansion of the Universe due to dark energy.      From for example  \rfcite{2006PASP..118.1711W}, we find that the Universe is currently some $16\%$ bigger than it would have been without dark energy. If, as used implicitly in \rfcite{2015arXiv150201589P}, we take the actual present day value of $a_0$ to be $1,$ then, in the universe with only radiation and matter assumed in  \eref{eq:phys-unit}, we have $a_0\approx\infrac{1}{1.16}\approx 0.86,$ giving us
\ba \label{eq:actual}
\kunit=\SI{0.0063}{Mpc^{-1}}\,.
\ea

For use below, in drawing figures directly comparable with those of \rfcite{2005AnPhy.318....2P}, we note the conformal time of recombination, $\tr.$  Using the redshift of $z_\star=\num{1089.90}$ from \cite{2015arXiv150201589P}, we find $\tr=\num[group-digits = false]{2.56069}.$ \\

As discussed in Section \ref{section:rel}, to linear order, tensor, vector and scalar perturbations decouple.  We therefore can focus solely on tensor perturbations, and the metric then has a line element
\ba 
ds^2=a^2\big(-d\tau^2+\left[\delta_{ij}+h_{ij}\right]dx^i dx^j\big)\,,
\ea
where $h_{ij}$ are transverse, traceless metric tensor perturbations which depend on both $\tau$ and the comoving spatial co-ordinate $\vec{x}.$ At this linear order, transverse, traceless tensor perturbations do not depend on any choice of gauge.\\

The tensor perturbations have two independent  polarisations,  $+$ and $\times,$
\bas
\label{eq:dof}
h^+_{ij}=h^+e^+_{ij}\qquad\text{and}\qquad
h^\times_{ij}=h^\times e^\times_{ij}\,,
\eas
where the polarisation matrices $e^+_{ij}$ and $e^\times_{ij}$ represent eigenmodes of the spatial Laplacian, each satisfying
\ba
\nabla^2\, e_{ij}= -k^2\,e_{ij}\,,
\ea
and the Laplacian's derivatives are with respect to the comoving co--ordinate $\vec{x},$  with $k$ being the perturbation's comoving wavenumber.\\

The governing equation for the tensor perturbations with neutrino anisotropic stress is given, in general, by (see e.g.~\cite{malik,PhysRevD.50.2541})
\ba
\label{eq:gov-general}
h_{ij}''(\tau)+2\H h_{ij}'(\tau)+\nabla^2 h_{ij}(\tau)=16 \pi G {\pi}^{(\nu)}_{ij}\,,
\ea
where ${\pi}^{(\nu)}_{ij}$ denotes neutrino anisotropic stress, which obeys the constraints
\ba
\label{eq:stress}
\pi^{i}_{i} = 0\,, \qquad\text{and}\qquad \partial_{i}\pi_{ij} = 0\,.
\ea
For a perfect fluid $\pi_{ij} = 0$, but this is not true in general.

\section{Governing equations}
\label{gov_equ_sect}

We can now rewrite the general evolution equation for the tensor perturbations, \eref{eq:gov-general} above, for the particular cases we would like to find solutions. We start with simplest case, in which the anisotropic stress is vanishing.

\subsection{Perfect fluid}
\label{subsec:perfect-fluid}

\subsubsection{Radiation and dust}

As set out in, for example, \rfcite{malik}, for a model where the energy density is in the form of perfect fluids, the tensor perturbation parameters, $h^+$ and $h^\times,$ from \eref{eq:dof} each are separately governed by the conformal time equation
\ba\label{eq:gov}
h''(\tau)+2\H h'(\tau)+k^2h(\tau)=0\,,
\ea
where $\H$ is again the conformal Hubble parameter and the dashes denote differentiation with respect to $\tau.$ Using \eref{eq:h-in-tau}, we can therefore write \eref{eq:gov} explicitly in terms of $\tau$ as
\ba
\label{eq:gov-tau}
h''(\tau)+\frac{4\left(2+\tau\right)}{\tau\left(4+\tau\right)}\, h'(\tau)+k^2\, h(\tau)=0\,.
\ea
This is the governing equation for linear tensor perturbations in a flat universe filled with radiation and pressureless matter, without a cosmological constant, and with no source term due to anisotropic stress.

\subsubsection{Radiation only}
\label{subsec:radonly-perfect-fluid}

Similarly, we can derive the governing equation for linear tensor perturbations in flat universes which contain either only radiation or only matter.  Unlike, \eref{eq:gov-tau},  the governing equations in each of these simpler models has  tractable, and well known, analytical solutions.

\begin{figure}
\centering
\includegraphics[width=0.85\linewidth]{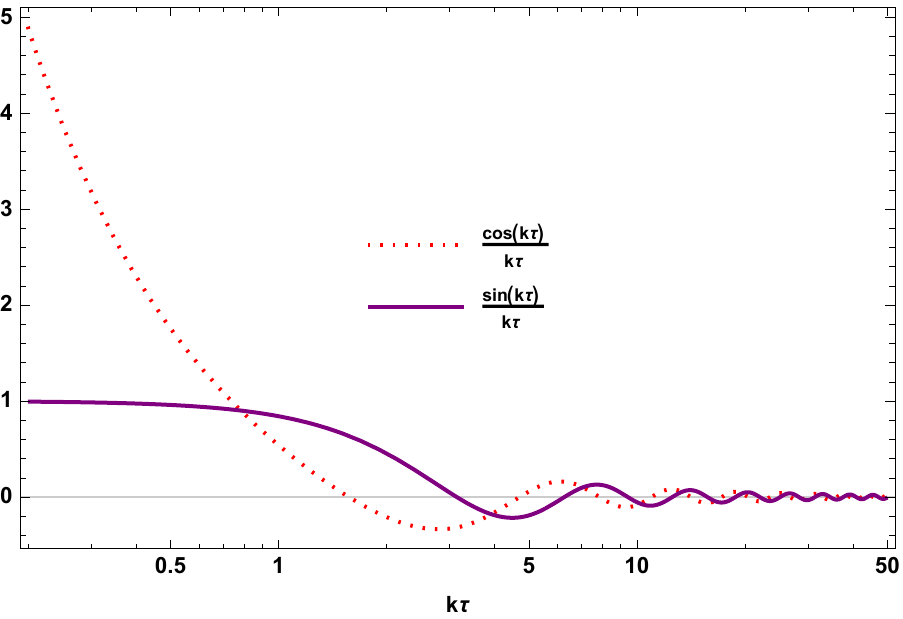}
\caption{Components of the analytical solution for linear tensor perturbations in a flat radiation--only universe, as set out in~\eqref{eq:flat-rad-only-soln} for a perfect fluid. The $k\tau$ axis is logarithmic.}
\label{fig:Fig1-sin-cos}
\end{figure}

For example, with only radiation, the Friedmann equation can be solved to show that $a\propto \tau.$ We then have $\H=\infrac{1}{\tau},$ giving us the governing equation for our perturbation, $\hr,$ as
\ba
\label{eq:gov-tau-rad-only}
\hr''(\tau)+\frac{2}{\tau}\, \hr'(\tau)+k^2\, \hr(\tau)=0\, .
\ea
As recalled in, for example, \rfcite{1993PhRvD..48.4613T,1996PhRvD..53..639W,1994PhRvD..50.3713A,1995PhRvD..52.2112N,2005AnPhy.318....2P,2005astro.ph..5502B}, it is well--known that this can easily be solved analytically to give
$\hr$ as a linear combination
\ba \label{eq:flat-rad-only-soln}
\hr(\tau)=A_\text{s}\,\frac{\sin(k\tau)}{k\tau}+A_\text{c}\,\frac{\cos(k\tau)}{k\tau}\, ,
\ea
where $A_\text{s}$ and $A_\text{c}$ are real constants set by  initial conditions.

This solution also describes the evolution of gravitational waves which originate very early in the history of a flat radiation--matter universe, when the matter density can be neglected relative to the radiation density. Any primordial $A_\text{c}$ component may be neglected because it increases very rapidly in early times, as can bee seen from Figure~\ref{fig:Fig1-sin-cos} and it is preferable to avoid any singularities. In Section~\ref{sec:modelling-gravitational-waves-through-the-whole-radiation--matter-epoch}, we will use this property to allow us to neglect the $A_\text{c}$ component for gravitational waves which have a very early cosmological origin, such as inflation.

For use in the following sections, we recall that a perturbation with wavenumber $k$ is said to cross the horizon at the time $\tau_k$ when $\H=k.$  From \eref{eq:h-in-tau}  we get 
\ba \label{eq:horizon}
\tau_k=\frac{\sqrt{4 k^2+1}-2 k+1}{k}=\frac{1}{k}+\order\left({\frac{1}{k^2}}\right),
\ea
where the $\order\left({\infrac{1}{k^2}}\right)$ means that the remaining part of the expression is of order $\infrac{1}{k^2}$.

\subsection{Including anisotropic stress}
\label{subsec:including-stress}

\subsubsection{Radiation and dust, including anisotropic stress}

From \ref{subsec:radonly-perfect-fluid} we have an analytical solution for the tensor perturbations in a perfect fluid. Here, we denote how we can extend this to include neutrino anisotropic stress and we first recall the expression for the anisotropic stress given in \rfcite{2004PhRvD..69b3503W,PhysRevD.50.2541},
\ba \label{eq:stress-int}
\pi_{ij}(\tau) = -4 \bar{\rho}_{\nu}(\tau)\H^{2} \int_{0}^{\tau}K(\tau-t)h'_{ij}(t)dt, 
\ea
where $K$ is the kernel defined as
\ba
\begin{split}
\label{eq:kernel}
K(\tau) &\equiv \frac{1}{16}\int_{-1}^{1}dx(1-x^2)^2 e^{i x \tau}, \\
&= -\frac{\sin{\tau}}{\tau^{3}} -\frac{3\cos{\tau}}{\tau^{4}}+\frac{3\sin{\tau}}{\tau^{5}}.
\end{split}
\ea
and $\bar{\rho}_{\nu}$ is the unperturbed neutrino energy density. Eq~\eqref{eq:stress-int} explains, for instance, in any imperfect fluid with shear viscosity $\xi$, we have $\pi_{ij}=-\xi h'_{ij}$, even when $h_{ij}$ becomes time-independent as the wavelength of a mode leaves the horizon, and remains time-independent until horizon re-entry. All modes of cosmological interest are stillfar outside the horizon at the temperature $\approx 10^{10}$ K where neutrinos are going out of equilibrium with electrons and photons, so $h_{ij}$ can be affected by anisotropic inertia only later, when neutrinos are freely streaming, and this equations explain the contribution of freely streaming neutrinos exactly. We then use \eref{eq:stress-int} in \eref{eq:gov-general} which gives the integro-differential equation for $h_{ij}(\tau)$ firstly derived in \rfcite{PhysRevD.50.2541} and popularised by \rfcite{2004PhRvD..69b3503W}
\ba
\label{eq:gov-tau-stress}
h''(\tau)+\frac{4\left(2+\tau\right)}{\tau\left(4+\tau\right)}\,h'(\tau)+k^2\, h(\tau)=-24 f_{\nu}(\tau) \left[\frac{4\left(2+\tau\right)}{\tau\left(4+\tau\right)}\right]^2\ \int_{0}^{\tau}K(\tau-t)h'(t)dt.
\ea
This is the governing equation for linear tensor perturbations in a flat  universe filled with radiation and pressureless matter, without a cosmological constant, and including a source term due to anisotropic stress, where $f_{\nu} = \bar{\rho}_{\nu}/\bar{\rho}$.

\subsubsection{Radiation only with anisotropic stress}
\label{subsec:radonly-stress}

For a radiation only universe \eref{eq:gov-tau-stress} reduces to
\ba
\label{eq:gov-tau-stress-radonly}
\hr''(\tau)+\frac{2}{\tau}\,\hr'(\tau)+k^2\, \hr(\tau)=-24 f_{\nu}(\tau) \left[\frac{4}{\tau^2}\right]\ \int_{0}^{\tau}K(\tau-t)\hr'(t)dt.
\ea
We proceed to use the method from \cite{DPS} (explained further in detail in Appendix \ref{sec:calculation-of-the-anisotropic-stress-solution}) to get the solution for the first order $h$ sourced by free--streaming neutrinos to compare with the analytic result \eref{eq:flat-rad-only-soln}. 

We consider wavelengths short enough to have re-entered the horizon during the radiation-dominated era (though after neutrino decoupling), and then compare with the analytical result. We take the initial time to be early enough so that it can be approximated as $\tau~ 0$, with the zero of time defined so that during the radiation-dominated era we have $a\sim \sqrt{\tau}$. Then we have $a'/a \sim 1/\tau$, while for three neutrino flavours $f_{\nu}$ takes the constant value $f_{\nu} = 0.40523$. Taking the approximation this gives
\ba \label{eq:flat-rad-only-soln-stress}
{\hr}_{(1)}(\tau)= \left( -\frac{i}{k}\right)\left( \frac{e^{i k \tau - f_{\nu}(0)\tau}}{\tau}\right),
\ea
which is less than $1\%$ away from the analytic solution without anisotropic stress when the perturbation enters the horizon. For wavelengths that enter the horizon after electron-positron annihilation and well before radiation-matter equality, all quadratic effects of the tensor modes in the cosmic microwave background, such as the whole of the ``B-B'' polarisation multipole coefficients $C_{\ell B}$, are $35.6\%$ less than they would be without the damping due to free-streaming neutrinos \cite{2004PhRvD..69b3503W}. Photons also contribute to $\pi_{ij}$, but this effect is much smaller because at last scattering photons contribute much less than $40\%$ of the total energy.

\section{Analytical solutions for the  tensor perturbations}
\label{sec:approximating-tensor-perturbations}

  In this section, we will derive an approximate analytical solution to the governing equation \eref{eq:gov-tau} valid in  the \emph{sub--horizon} case, $k\gg\H.$  We begin by recalling the method of \rfcite{FSN} as applicable to a single second order differential equation, having a small parameter. Here, the small parameter is the comoving inverse wavenumber, $k^{-1}.$

As described in \rfcite{DPS}, the double power series method consists of finding an approximation for $h$ of the form
\ba \label{eq:FSN-approx}
h(\tau)= \exp\left[{\sum_{s=0}^\infty \int k^{-s+1}\,\omega_s(\tau)\,d\tau}\right],
\ea
  and for  convenience, we write
\ba\label{eq:omega}
\omega&=\sum_{s=0}^\infty k^{-s+1}\,\omega_s(\tau),
\ea
where $\omega$ depends on both $\tau$ and $k,$ and we will regard $k$ as a fixed parameter. We then have
\bas
h'&=\omega h,\\
\mathrm{and\quad\quad} h''&=\left(\omega'+\omega^2\right)h\,  . 
\eas

\subsection{Approximating perfect fluid tensor perturbations}
\label{sec:approximating-tensor-perturbations}

\begin{figure}
\centering
\includegraphics[clip, trim= {0 6.5cm 0 0} , width=\linewidth]{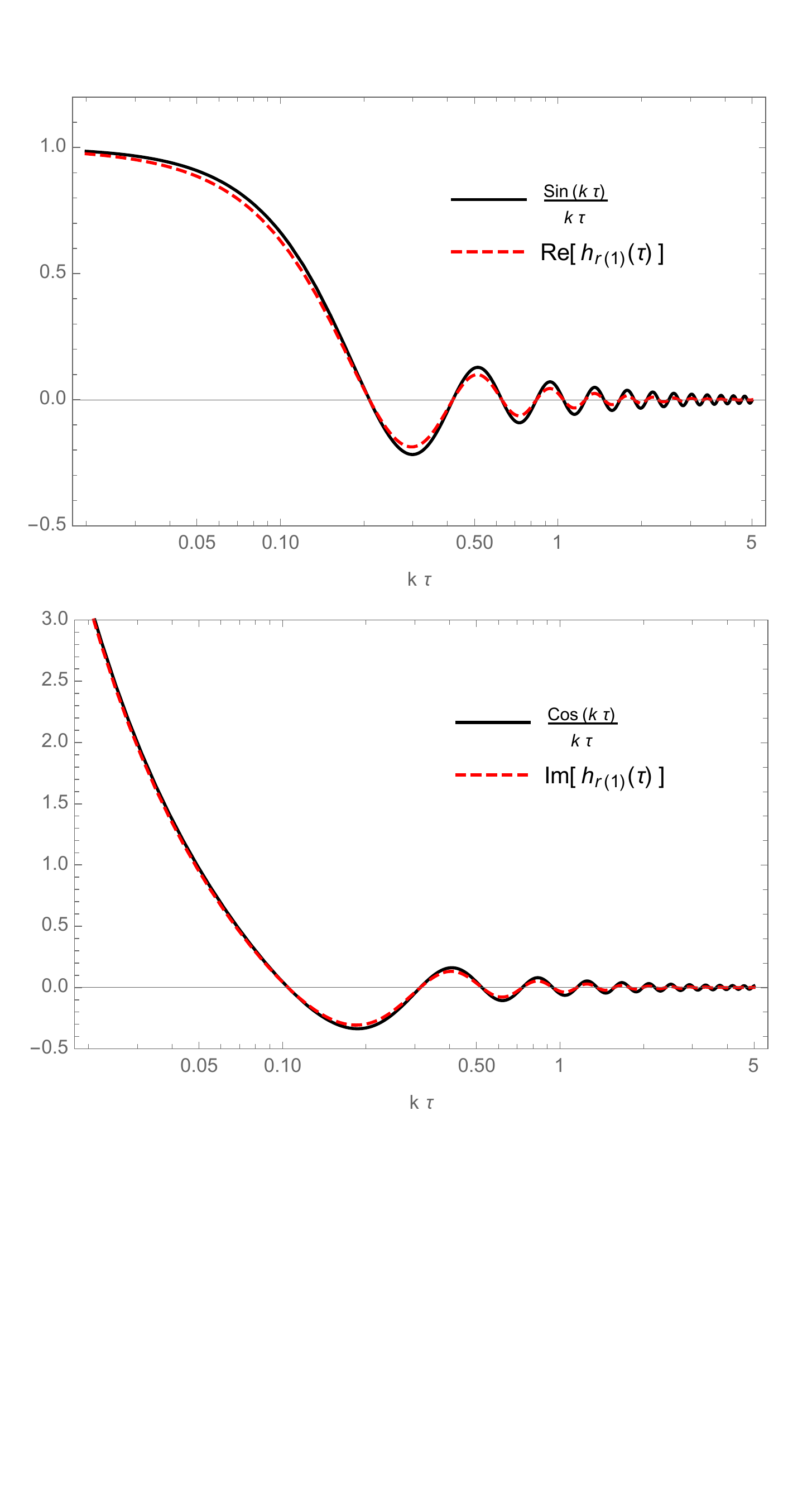}
\caption{\emph{Upper panel:} Real components of the analytical and approximate solution for linear tensor perturbations in a flat radiation--only universe without (solid black line red dashed line) and with the effect of free streaming neutrinos anisotropic stress with $f_{\nu}(0)=0.5$  (red dashed line), as set out in Eqs.~(\ref{eq:flat-rad-only-soln}) and (\ref{eq:flat-rad-only-soln-stress}). \emph{Lower panel:} Imaginary components of Eqs.~(\ref{eq:flat-rad-only-soln}) and (\ref{eq:flat-rad-only-soln-stress}). The $k\tau$ axis is logarithmic.}
\label{fig:Fig2-sin-cos-stress}
\end{figure}

Using the approximation of \eref{eq:FSN-approx} in our governing equation \eref{eq:gov-tau}, and dividing through by $h,$ we get  the ``characteristic'' equation from which we will derive all orders of our approximation, 
\ba \label{eq:approxequation}
\left(\omega'+\omega^2\right)+\frac{4\left(2+\tau\right)}{\tau\left(4+\tau\right)}\,\omega+k^2=0\, .
\ea
We  can now  substitute \eref{eq:omega} into  \eref{eq:approxequation} and equate coefficients of like powers of the inverse wavenumber, $k^{-1}$, meaning we can write the $\omega_{s}$ in terms of previously calculated ones, i.e. $\omega_{s+1} \propto \omega_{s}$ and so build a better approximation for $h$.  The calculation is similar to that set out in a Bessel type differential equation in \rfcite{DPS}. It can also be carried out using the \emph{Mathematica} notebook available online at \rfcite{GWGithub}.

We can form successive order $\smax$ approximations 
\ba \label{eq:smax}
h_{(\smax)}
=\exp\left[\sum_{s=0}^{\smax}
\int k^{-s+1}\,\omega_s(\tau)\,d\tau\right].
\ea
We find that $h_{(\smax)}$ has complex values.   Taking the real and imaginary parts of $h_{(\smax)}$ provides approximations for two independent solutions of \eref{eq:gov-tau}.

For clarity, we should point out that, in this chapter, when we describe an approximation as being of a particular order, we are not referring to the order of the perturbation theory (as described in, for example, \rfcite{malik}). We use the term order in this context to refer to the value of $\smax$ in the approximation of \eref{eq:smax}. We only use expressions of first order in perturbation theory, and, to minimise confusion, we consistently describe these as being of linear order.

The largest value of $\smax$ we shall use is $\smax=3.$  This is because  trial against numerical solutions\footnote{  In Section~VII of \rfcite{DPS} the authors set out a method for estimating which order approximation is the most accurate for a given value of $\tau$ without employing numerical solutions. } shows that  larger choices of $\smax$ do not uniformly improve the approximation for all sub--horizon values of $\tau.$ 

This third order approximation is to take any linear combination of the real and imaginary parts of
\ba \label{eq:third-order}
h_{(3)}(\tau)=\frac{1}{\tau(4+\tau )}\left(\frac{4+\tau }{\tau }\right)^{\infrac{i}{4 k}}\exp\left[i k \tau+\frac{1}{2k^2\tau \left( 4+\tau  \right)}\right].
\ea
Recall, as set out following \eref{eq:rmscalefactor}, that $\tau$ and $k$ have been normalised in terms of $\tunit$ and are therefore dimensionless numbers.

We want to compare our approximations with numerical solutions.\footnote{As usual, our numerical solutions are derived from initial conditions.  In \rfcite{DPS} the authors had to use final conditions to manage a particular numerical instability.} 
Figure~\ref{fig:Fighappr3k10} shows the $h_{(3)}$ approximation for wavenumber $k=10\kunit,$ plotted against the corresponding numerical solution.

We want now to quantify the error in the approximation $h_{(3)}.$ One way to try to do this would be to take the numerical solution $h$ and calculate the ratio $\infrac{\left|\left(h_{(3)}-h\right)}{h}\right|.$  However, this runs into a difficulty.  Since $h$ is oscillating and repeatedly takes zero values, unless there is no error at all at these zeros, the ratio $\infrac{\left|\left(h_{(3)}-h\right)}{h}\right|$ will repeatedly become infinite.  

We therefore adapt our approach in order to avoid this problem.  Broadly speaking, we estimate the typical error compared with the amplitude of oscillation.  To do this we first decide a target degree of accuracy, $1\%$ say.  We then compare a plot of the error $\left|h_{(3)}-h\right|$ with a plot of the target accuracy, here $1\%\cdot \left|h\right|.$  Both plots will usually spike repeatedly downwards towards zero.  We consider that the approximation is accurate to within $1\%$ if, spikes when $h=0$ aside, $\left|h_{(3)}-h\right|\le 1\%\cdot \left|h\right|.$  This will show on the graph as the plot of $\left|h_{(3)}-h\right|$ being level with, or below, the plot of $1\%\cdot \left|h\right|$ (except near $h=0$ spikes). 

As shown in Figure~\ref{fig:Fighappr3k10error},  the $h_{(3)}$  approximation is accurate to within $1\%,$ or better from  horizon--crossing onwards $(\tau\ge\tau_k).$  In fact, this degree of accuracy applies for wavenumbers $k\gtrsim 9\kunit\sim\SI{0.07}{a_0.Mpc^{-1}}.$ Using \eref{eq:actual}, this corresponds to actual present day distances of {$\SI{17}{Mpc}.$ 

\begin{figure}
\centering
\includegraphics[width=\linewidth]{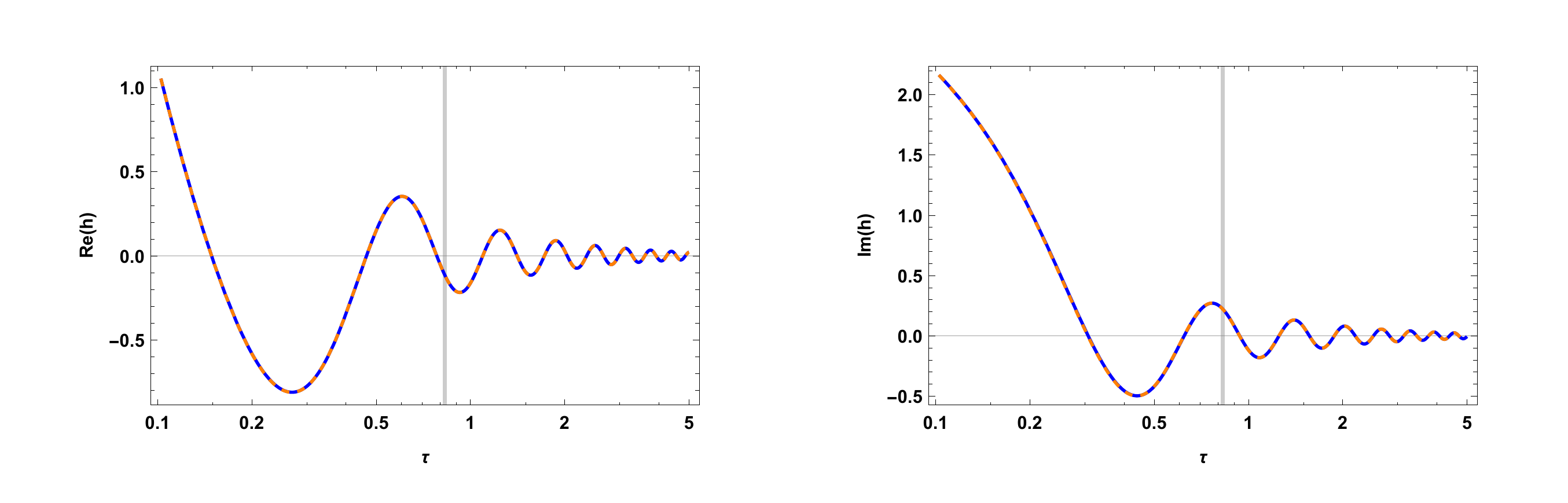}
\caption{  The real and imaginary parts of the approximate third order solution for a perfect fluid given by~\eref{eq:third-order} (solid blue line) and the numerical solution (dashed orange line) for \eref{eq:gov-tau} with $k=10\kunit$. The horizontal axes show $\tau$ in units of $\tau_\text{r},$ the conformal time of recombination, calculated using \rfcite{2015arXiv150201589P}. The curves are plotted for values of $\tau$ after horizon--crossing. The solid vertical line corresponds to radiation--matter equality, as defined by \eref{eq:rmscale}. }
\label{fig:Fighappr3k10}
\end{figure}

\begin{figure}
\centering
\includegraphics[width=\linewidth]{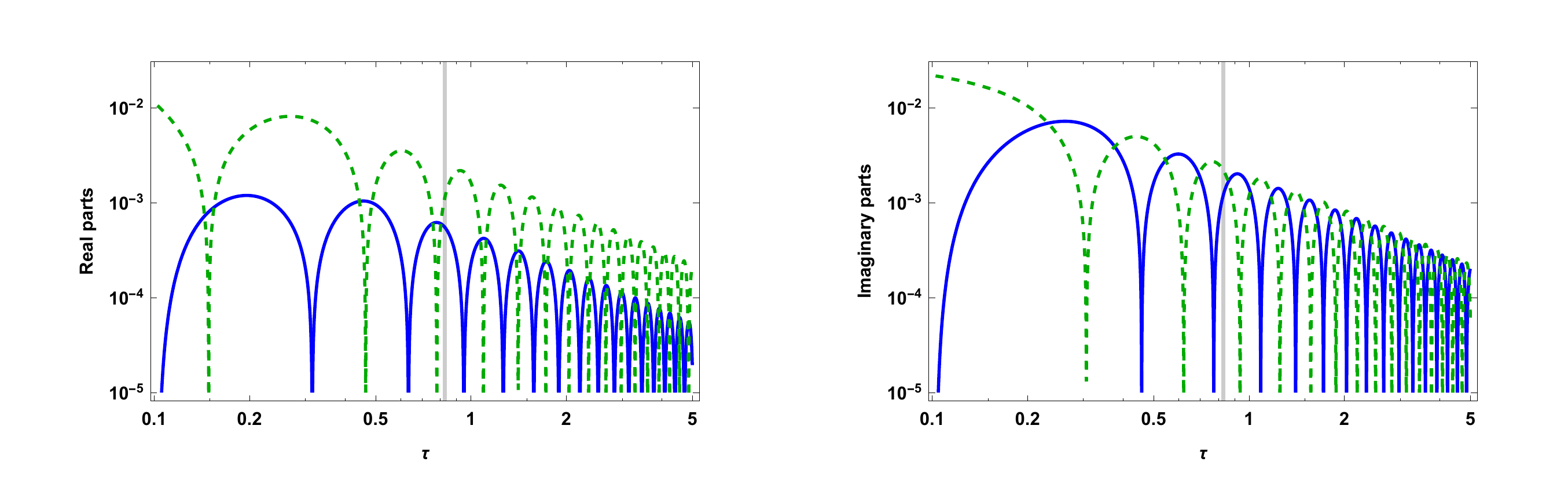}
\caption{  The solid lines show the absolute errors in the approximation of Figure~\ref{fig:Fighappr3k10} as $\left|f_{(3)}(\tau)-f(\tau)\right|,$ where $f_{(3)}$ is the real or imaginary part, as appropriate, of $h_{(3)}$ from \eref{eq:third-order} and $f$ is the real or imaginary relevant part of the numerical solution for a perfect fluid.  For comparison, the dashed line shows 
$\left|1\%\times f(\tau)\right|.$ Downward spikes in the lines represent points where the value becomes zero. The blue solid line being below or level with the green dashed line indicates an error of less than or equal to $1\%.$  See the text for further explanation. }
\label{fig:Fighappr3k10error}
\end{figure}

 The second order and first order approximations constructed by our method will also be of use in the next section, and we set them out here.   The second order approximation is to take any linear combinations of the real and imaginary parts of
\ba \label{eq:second-order}
h_{(2)}(\tau)=\frac{1}{\tau(4+\tau )}\left(\frac{4+\tau }{\tau }\right)^{\infrac{i}{4 k}}\exp\left[i k \tau\right].
\ea
For use in Section~\ref{sec:modelling-gravitational-waves-through-the-whole-radiation--matter-epoch}, we also write this more explicitly as 
\ba \label{eq:second-order-trig}
h_{(2)}(\tau)=\frac{B_\text{s}\sin \Big(k \tau +\frac{1}{4k}\ln \left(1+\frac{4}{\tau }\right)\Big)+B_\text{c}\cos \Big(k \tau +\frac{1}{4k}\ln \left(1+\frac{4}{\tau }\right)\Big)}{\tau  (\tau +4)},
\ea
with $B_\text{s}$ and $B_\text{c}$ real constants. This second order approximation is accurate to within $1\%,$ or better, from horizon-crossing onwards,  when we have $k\gtrsim 17\kunit\sim\SI{0.12}{a_0.Mpc^{-1}}.$

The first order, or leading order, approximation~--- as set out in \cite{2008cosm.book.....W}~--- is to take any linear combination of the real and imaginary parts of
\ba \label{eq:leading}
h_{(1)}(\tau)=\frac{1}{\tau(4+\tau )}\exp\left[i k \tau\right]
\propto
\frac{\exp\left[i k \tau\right]}{a(\tau)}
=\frac{\cos\left( k \tau\right)+i\sin\left( k \tau\right)}{a(\tau)},
\ea
  which  is accurate to $1\%,$ or better, from horizon-crossing onwards, when we have $k\gtrsim 120\kunit\sim\SI{0.87}{a_0.Mpc^{-1}}.$

\subsection{Approximating tensor perturbations with anisotropic stress}
\label{sec:approximating-tensor-perturbations-with-stress}

In this section, we compute an approximate analytical solution to the governing equation \eref{eq:gov-tau-stress}. The fraction of the total energy density in neutrinos is \cite{2004PhRvD..69b3503W} 
\ba
\label{eq:f-neutrinos}
f_{\nu}(\tau) =
\frac{\Omega(\infrac{a_{0}}{a})^{4}}{\Omega_{\textrm{m}}(\infrac{a_{0}}{a})^{3}+\left(\Omega_{\gamma} + \Omega_{\nu}\right) (\infrac{a_{0}}{a})^{4}} = \frac{f_{\nu}(0)}{1+\tau},
\ea

We follow the method described in \rfcite{DPS} along with (\ref{eq:kernel}) and (\ref{eq:f-neutrinos}) to get an approximate solution to \eref{eq:gov-tau-stress}. The third order approximation is to take any linear combination of the real and imaginary parts of 
\ba 
\label{eq:third-order-stress}
\begin{split}
h_{(3)}(\tau)&= \left( -\frac{i}{k} - \frac{2(2+\tau)}{k^{2} \tau (4+\tau)} + \frac{i[16 f_{\nu}(0)(2+\tau)^{2} + 5(16+28\tau+15\tau^{2}+3\tau^{3})]}{5 k^{3} \tau^{2}(1+\tau)(4+\tau)^{2}} \right) \\
&\qquad \times \tau^{-(1 + (i/k)\alpha)} (1+\tau)^{(i/k) \beta} (4+\tau)^{-1+(i/k) \gamma} \\
&\qquad \times \exp\Bigg[ i k \tau - f_{\nu}(0)\tau- \frac{2 i [-15(2+\tau)+4 f_{\nu}(0)(6+\tau)]}{15 k \tau (4+\tau)} +\\
&\qquad\qquad\qquad\qquad\qquad\qquad\frac{16 f_{\nu}(0)(2+\tau)^{2} + 5(16+28\tau+15\tau^{2}+3\tau^{3})]}{10k^{2} \tau^{2}(1+\tau)(4+\tau)^{2}}\Bigg] .
\end{split}
\ea
where
\ba
\alpha = \frac{5+8 f_{\nu}(0)}{20}, \quad \beta = \frac{16}{45}f_{\nu}(0), \quad \text{and} \quad \gamma = \frac{45+8 f_{\nu}(0)}{180} .
\ea
\begin{figure}
\centering
\includegraphics[width=\linewidth]{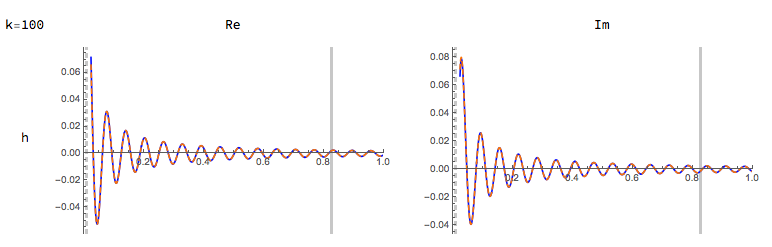}
\caption{The real and imaginary parts of the approximate third order solution with free neutrino anisotropic stress given by~\eref{eq:third-order-stress} (solid blue line) and the numerical solution (dashed orange line) for \eref{eq:gov-tau-stress} with $k=100\kunit$ and $f_{\nu}(0)=0.40523$. The curves are plotted for values of $\tau$ after horizon--crossing. The solid vertical line corresponds to radiation--matter equality, as defined by \eref{eq:rmscale}.}
\label{fig:Fighappr5k100}
\end{figure}

\section{Modelling primordial gravitational waves}
\label{sec:modelling-gravitational-waves-through-the-whole-radiation--matter-epoch}

In this section, we model the propagation of primordial gravitational waves from early times when the waves are super--horizon, through to later times when they are sub--horizon.  We do this by matching an approximate solution for early times with a different one for late times.  This gives us a solution valid for all times in the radiation--matter model, not just when the waves are sub--horizon.

\begin{figure}
\centering
\includegraphics[width=\linewidth]{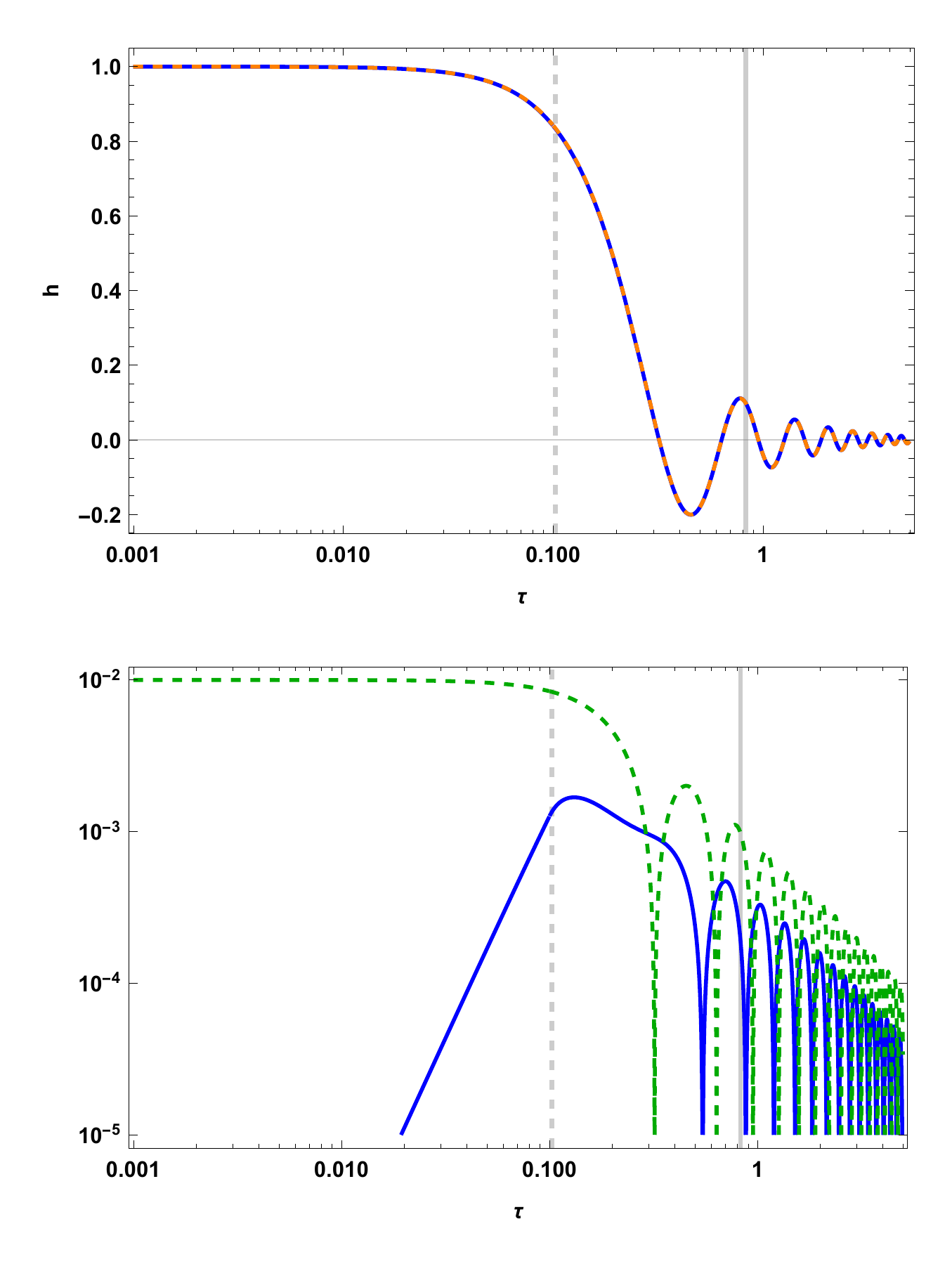}
\caption{\emph{Upper panel:} The approximate solution, $\hmatch,$ of
  ~\eref{eq:match} (solid blue line) and the numerical solution for a perfect fluid
  (dashed orange line) for \eref{eq:gov-tau} with $k=10\kunit$.
  \emph{Lower panel:} The error in the matched solution (solid blue
  line) plotted against $1\%$ of the numerical solution (dashed green
  line). \emph{In both panels:} The vertical dashed line is
  horizon--crossing, $\tau_{k=10}\approx \num{0.102},$ which is very
  close to the matching point, $\tau=\num{0.1}.$ The solid vertical
  line is radiation--matter equality.  }
\label{fig:Fig-match}
\end{figure}

For early times, we use the radiation--only solution, \eref{eq:flat-rad-only-soln}, while for late times we use the previous section's approximate solution $h_{(2)}$ from \eref{eq:second-order-trig}.  We choose the time at which we match these solutions to be $\tau=\infrac{1}{k}.$ We note from \eref{eq:horizon} that for relevant values of $k,$ the matching time $\tau=\infrac{1}{k}$ will be close to the horizon--crossing time $\tau_k.$ The form of the resulting equations will, however, be much simpler than if we had done the matching at the horizon--crossing time itself.

The derivation of the approximations set out in the previous section depended on the sub--horizon assumption, $k\ll\H.$  However, below we show that, for $k\gtrsim 4.5\kunit,$  our matching approach gives an approximation, $\hmatch,$ which is accurate to $1\%,$ or better, through the whole radiation--matter epoch.  We use the second order solution $h_{(2)}$ in the matching because trial and error shows it is the order of approximate solution which works to within $1\%$ for the widest range of wavenumbers. 

From~\eref{eq:flat-rad-only-soln}, we have that the early time solution is 
\ba \label{eq:flat-rad-only-soln2}
\hr(\tau)=A_\text{s}\,\frac{\sin(k\tau)}{k\tau}+A_\text{c}\,\frac{\cos(k\tau)}{k\tau}.
\ea
As discussed above following \eref{eq:flat-rad-only-soln}, for gravitational waves with primordial origin, we can set $A_\text{c}=0.$ We therefore take
\ba \label{eq:flat-rad-only-soln3}
\hr(\tau)=\frac{\sin(k\tau)}{k\tau}
\ea
to be our solution for $\tau\le\infrac{1}{k}$.

 From \eref{eq:second-order-trig}, we have our solution for $\tau\ge\infrac{1}{k}$  as
\ba \label{eq:second-order-trig2}
h_{(2)}(\tau)=\frac{B_\text{s}\sin \left(\lambda(k,\tau)\right)+B_\text{c}\cos \left(\lambda(k,\tau)\right)}{\tau  (\tau +4)},
\ea
where we have written 
\be \label{eq:lamdba}
\lambda(k,\tau)=k\tau+\frac{1}{4 k}\ln \left(1+\frac{4}{\tau}\right).
\ee

We then need to choose the real constants $B_\text{s}$ and $B_\text{c}$ in \eref{eq:second-order-trig2}  to be such that $\hmatch$ and $\hmatch'$  are continuous. In other words, we ensure the values of the two functions and the values of their first $\tau$ derivatives each match at $\tau=\infrac{1}{k}.$ The calculation is set out in the \hyperref[sec:calculation-of-the-matching-conditions-for-approximating-primordial-gravitational-waves]{Appendix}.  From \eref{eq:match-a}, we have the resulting matching approximation 
\begin{equation}\label{eq:match}
\hmatch(\tau)=
\begin{dcases}
\frac{\sin(k\tau)}{k\tau}
&\mbox{if } \tau\le\frac{1}{k}
\\[9pt]
\frac{4 k+1}{4 k^3 \tau  (4+\tau)}
\bigg[
\,4 k\, \sin \Big(k\tau + L(k,t)\Big)
+ \mu
\sin \Big(k\tau + L(k,\tau)-1\Big)
\bigg]
 &\mbox{if } \tau\ge\frac{1}{k},
\end{dcases}
\end{equation}
where 
\be
L(k,\tau)=\frac{1}{4 k}\ln \left(\frac{1+4\,\tau^{-1} }{1+4\,k\hfill}\right) 
\qquad
\text{
and }\qquad
\mu
=
\sin\left(1\right)+\cos\left(1\right)
=
1.38177...\, .
\ee

This is our approximate analytical solution to \eref{eq:gov-tau}, the governing equation for tensor perturbations (gravitational waves) in a flat radiation--matter universe.  As we now show, it is a good approximation for both sub--horizon ($\tau\lesssim\infrac{1}{k}$) and super--horizon ($\tau\gtrsim\infrac{1}{k}$) for wavenumbers $k$ over a wide range of values relevant for structure formation in the Universe.

Figure~\ref{fig:Fig-match} shows this matched solution, $\hmatch,$ for $k=10\kunit$ in its upper panel and in the lower panel plots the error using the approach also used in Figure~\ref{fig:Fighappr3k10error}. The lower panel shows that the error is consistently less than $1\%,$ both before and after horizon--crossing.  We found that $\hmatch$ has errors of $1\%,$ or better for all $\tau$, when we $k\gtrsim 4.5\kunit=\SI{0.033}{a_0.Mpc^{-1}}.$

Using \eref{eq:actual},   this corresponds to $\hmatch$ meeting   our $1\%$ accuracy test for all times, when we consider inverse wavenumbers with actual present day values of less than, or equal to, around $\SI{35}{Mpc}.$  We can also express this in multipoles, as used in CMB calculations, via the broad correspondence ~--- see for example \rfcite{2003itc..book.....R,2008cosm.book.....W,2009pdp..book.....L} 
~---   that a multipole, $l,$ receives its main contributions from present day physical wavenumbers $k_\text{ph}\simeq {c H_0} \,l.$ Using this rule, our range of $1\%$ accuracy, $k\gtrsim 4.5\kunit,$ broadly corresponds to multipoles $l\gtrsim 120.$

We found that matching using $h_{(1)}$ of \eref{eq:leading}, in place of $h_{(2)},$ obtains errors of $1\%,$ or better for all $\tau$, only when we have  $k\gtrsim 180\kunit=\SI{1.3}{a_0.Mpc^{-1}}.$ This corresponds to actual present day inverse wavenumbers of less than or equal to around $\SI{0.9}{Mpc},$ and, broadly, to multipoles $l\gtrsim\num{5000}.$


The WKB solutions of \rfcite{1995PhRvD..52.2112N,2005AnPhy.318....2P}, \rfcite{2005astro.ph..5502B} and \rfcite{2008cosm.book.....W} aim to derive approximations for primordial gravitational waves. We now compare these with our approximation~--- not restricting our attention only to the perfect fluid case, but considering the anisotropic stress due to free--streaming neutrinos as well like \rfcite{2005AnPhy.318....2P,2008cosm.book.....W,2005astro.ph..5502B} .

Turning first to the WKB approximation of  \rfcite{2008cosm.book.....W}, we note that this is the same as our leading order approximation $h_{(1)}$ of  \eref{eq:leading}.  For primordial gravitational waves, along the lines discussed following \eref{eq:flat-rad-only-soln}, we are interested in only the imaginary (sine) part of \eref{eq:leading}, which is proportional to
\ba \label{eq:leading-s}
h_{(1),\text{s}}=\frac{\arm \sin(k\tau)}{k\, a(\tau)}=\frac{4\sin(k\tau)}{k\tau(4+\tau)},
\ea
where we have chosen the constant of proportionality so that $h_{(1),\text{s}}(0)=1.$
Figure~\ref{fig:Fig-match-happr1} shows that, for $k=10\kunit,$ this gives errors of around $10\%.$   

\begin{figure}
\centering
\includegraphics[width=0.85\linewidth]{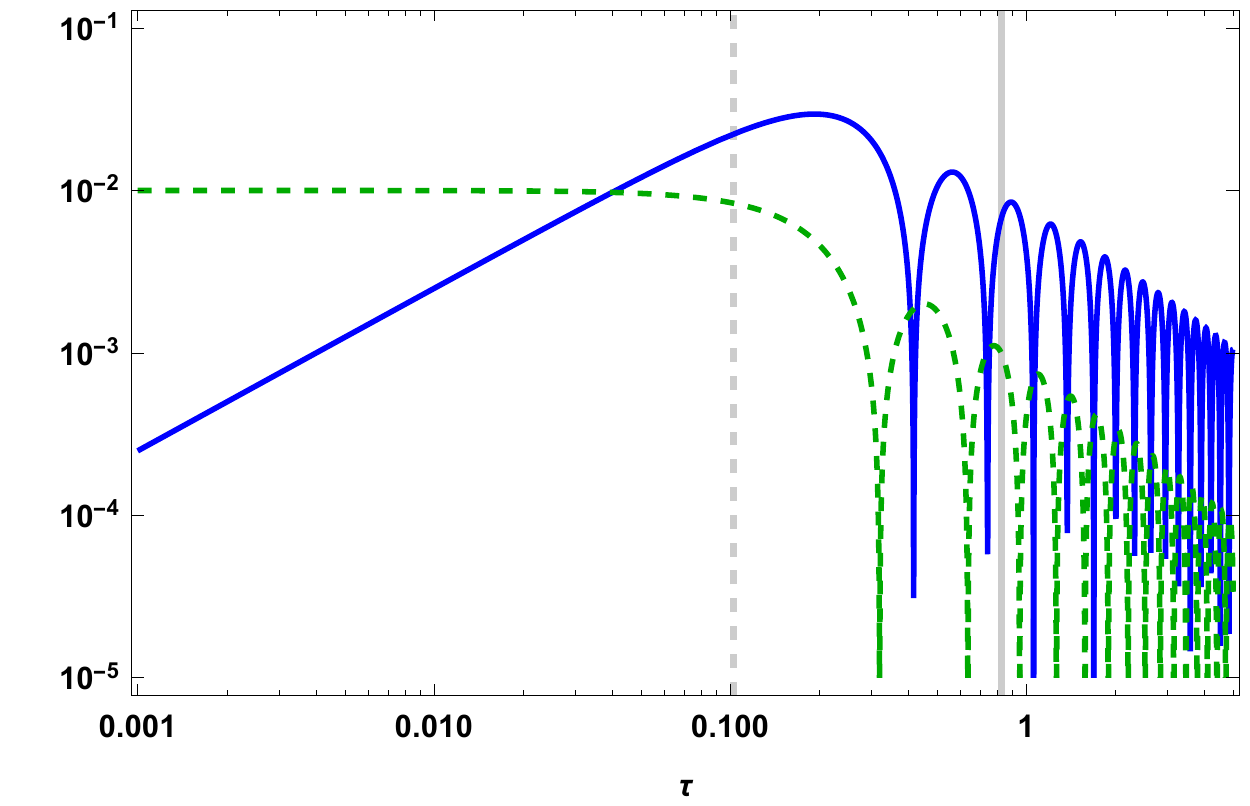}
\caption{As for the lower panel of Figure~\ref{fig:Fig-match}, but using the leading order approximation, $h_{(1),\text{s}}$ of \eref{eq:leading-s}, instead of $\hmatch.$}
\label{fig:Fig-match-happr1}
\end{figure}

References~\cite{1995PhRvD..52.2112N,2005AnPhy.318....2P} presented another less straightforward form of WKB approximation.  Figures in those papers show that, for $k\approx 16.4\kunit$ in Fig.~1 of \rfcite{1995PhRvD..52.2112N}, and for $k\approx 12.1\kunit$ in Fig.~2 of \rfcite{2005AnPhy.318....2P}, there are errors of around $10\%,$ or greater.  Their figures also show that alternative simpler approximations~--- using a radiation--only solution, a matter--only solution, or the two matched together at $\tau=\tau_\text{eq}$ in a ``sudden approximation''~--- are all less accurate than the WKB approximation of \rfcite{1995PhRvD..52.2112N,2005AnPhy.318....2P}, and so also much less accurate than our approximation $\hmatch$, given in \eref{eq:match}.

Our calculations suggest that the best approximation we have found in the literature for sub--horizon primordial gravitational waves is the WKB solution explored in passing in \rfcite{2005astro.ph..5502B}. That approximation is constructed for only one of the two independent solutions of the governing equation \eref{eq:gov-tau}.  As can be seen from Figure~\ref{fig:Fighappr3k10error}, typically one of a pair of approximate WKB solutions will be better than the other.  For $k=10\kunit,$ and for sub--horizon values of $\tau,$ the approximation of \rfcite{2005astro.ph..5502B} is accurate to within about $0.8\%,$ which compares with the accuracies for the two approximations of Figure~\ref{fig:Fighappr3k10error} of around $0.25\%$ (real part) and $1\%$ (imaginary part). Starting outside the horizon, \rfcite{2005astro.ph..5502B}'s approximation becomes considerably less accurate than our matching approximation, $\hmatch.$ 

The Bessel function approach of \rfcite{2013PhRvD..88h3536S} also approximates primordial gravitational waves.  As indicated in  Section~\ref{sec:intro}, a comparison with numerical solutions shows that many terms are needed to derive  accurate  approximations.

\section{Conclusion}
\label{sec:conclusion}

In this chapter, we have derived new analytic solutions to the gravitational wave or tensor evolution equation at linear order in cosmological perturbation theory, using the method presented in \rfcite{DPS} and including neutrino anisotropic stress. The solutions depend on time, wavenumber, and the ratio of the neutrino background energy density and the total energy density at $\tau=0$, $f_{\nu}(0)$, which allows us to vary the damping of the gravitational waves due to the neutrino anisotropic stress.

We show how the method of \rfcite{DPS} can be applied to an integro-differential equation, and our solutions present a simple way to analyse the evolution of the tensor perturbations in the presence of anisotropic stress. The analytic solutions save computational time compared to solving the equations numerically, for example if we are interested in a wide range of different parameter values $f_\nu$ (the ratio of neutrino to total background energy density).
%
We have compared our analytical approximation to numerical solutions with anisotropic stress and find that the difference between them is within $1 \%$.


A simple example illustrates the usefulness of analytical solutions even in this day and age.  In second order cosmological perturbation theory, the governing equations include terms of the form $h_{ij}h^{ij}$.
For example \rfcite{2016JCAP...02..021C} shows that different forms of the curvature perturbation at second order differ by terms proportional to $h_{ij}h^{ij}$. To relate this to the expressions derived in the sections above, we can use that for a gravitational wave travelling in the $z$--direction, for example,
\bea
h_{ij}h^{ij} = 2\left[\left(h^{\times}\right)^2+\left(h^{+}\right)^2\right]\,,
\eea
where $h^{\times}$ and $h^{+}$ are the two independent polarisations of the gravitational waves defined in \eref{eq:dof}. Then substituting in the solutions of the model we are interested in, we obtain an expression for how much the different curvature perturbations disagree in terms of, say, the wavenumber and the neutrino-radiation ratio $f_\nu$. 

%% file: tex/chapter_4.tex
\chapter{Galaxy Number Counts}
\label{chapter:gnc}

\section{Introduction}
\label{introdg}

Recent years have witnessed the beginning of the era of precision cosmology with future surveys such as BOSS \cite{BOSS}, eBOSS \cite{eBOSS}, Euclid \cite{euclid}, and WFIRST \cite{WFIRST} improving and tightening constraints on observable cosmological parameters. Additionally, great theoretical advancements have been made in tackling nonlinear regimes to test cosmological models and general relativity.

For theoretical cosmological models, probing the relation between redshift and angular diameter or luminosity distance of a source is of significant value.
This relation determines the parameters of the cosmological model, but when perturbations due to structure are included, new effects are revealed. One of these effects is lensing, which is observed along the line of sight. Another effect is the distortion in redshift space due to velocities and motion of the sources, giving rise to `Doppler lensing'. Doppler lensing is the apparent change in object size and magnitude due to peculiar velocities. Objects falling into an overdensity appear larger on its near side, and smaller on its far side, than typical objects at the same redshifts. This effect dominates over the usual gravitational lensing magnification at low redshift. Doppler lensing gives a new window into the peculiar velocity field in addition to the usual redshift space distortion measurements. The integrated Sachs-Wolfe (ISW) effect which arises from integrating along the full line of sight between the source and the observer.

Most of the known effects on the distance-redshift relation are calculated at linear order in cosmological perturbation theory in Refs.~\cite{jeong1, durrer1, antony, chen, zalda}. However, at second order other general relativistic effects must be considered. When structure is evolving, nonlinear modes come into play, and many of these are go beyond Newtonian theory.

One of the main observables directly affected by the angular and luminosity distance estimation is the galaxy number density (often dubbed number counts). Important examples of these effects are demonstrated in Refs.~\cite{cc1,cc2, durrer1, durrer2, yoo1, yoo2, yoo3, marozzi1, marozzi2, marozzi3}. The dominating terms of the full second order calculations have been reviewed in Ref.~\cite{nielsen}. More recently in Ref.~\cite{yoo5}, the authors present second order relativistic corrections to the observable redshift. A ``pedagogical'' approach to the lengthy calculations is provided in Ref.~\cite{yoo4} to try to ease the tension between the different groups.

In this work, we present a new path to compute the second-order galaxy number count in a Friedmann-Lema\^{i}tre-Robertson-Walker (FLRW) universe. This follows the volume determination as defined in Ref.~\cite{ellis} instead of computing the luminosity distance as in Ref.~\cite{yoo2}. We identify key effects, some of which will be observable with the next generation of cosmological surveys. To check the robustness of our results, we confirm the consistency for the first order expressions with previous works.

In section \ref{sec:whatwemeasure}, we give all the definitions needed for the linear and nonlinear calculations in the context of cosmological perturbation theory (CPT). In section \ref{sec:effects}, we compute the linear and nonlinear parts of the null geodesic equation, the observed redshift and show the geometrical effects present at this level. In section \ref{sec:distances}, we compute the angular diameter distance and the physical volume that the galaxy survey spans; we make a conformal transformation that maps null geodesics from the perturbed FLRW metric to a perturbed Minkowski spacetime $\hat{g}_{\mu \nu}\rightarrow g_{\mu \nu}$, and the way quantities transform with this map is discussed in further detail within this section.
We compute our main result of the galaxy number density in section \ref{sec:counts}.
In section \ref{sec:comparison} we compare our calculation with other results in the literature at linear order, and find an exact agreement with all those pertaining to the right interpretation of variables. In section \ref{sec:secondcomparison} we make a similar comparison as in the previous section for the leading terms in our second order galaxy number density, finding a good agreement and only some slight differences with other results in the literature. Finally, in section \ref{sec:discussion} we give a discussion of our result, some conclusions and future work.

We use the notation
\be
\label{eq:not1}
\left(X\right)^{s}_{o} = X\big|^{s}_{o} = X_{s}-X_{o} = X(\lambda_{s})-X(\lambda_{o}).
\ee
The derivative with respect to the affine parameter is  
\begin{equation}
\label{eq:der-affine}
\frac{\dd X}{\dd \lambda} = X' + n^{i}X_{,i},
\end{equation}
where $n^{i}$ represents the direction of observation. This last equation implies, for a scalar function $X$,
\begin{equation}
\label{eq:double-der-affine}
n^{i}n^{j}X_{,ij}= n^{i}n^{j}\p_{i}\p_{j}X = \frac{\dd^{2} X}{\dd \lambda^{2}}-2 \frac{\dd X'}{\dd \lambda} +X'',
\end{equation}
where $\nabla_{i}X = \partial_{i}X = X_{,i}$ is the spatial part of the covariant derivative. 

Following our discussion in Section \ref{section:flrw} and the definitions presented in subsection \ref{sub:gauge}, the perturbed FLRW spacetime in the longitudinal gauge is then described by \cite{malik}
\be
\label{eq:metric}
\dd s^{2} = a^{2}\left[ -\left(1+2\ff+\fs \right) \dd \eta^{2} + \left( 1 - 2 \pf - \ps \right) \delta_{ij} \dd x^{i}\dd x^{j}\right],
\ee
making use of Eq.~\eqref{eq:poisson-scalar}, $\eta$ is the conformal time, $a=a(\eta)$ is the scale factor and $\delta_{ij}$ is the flat spatial metric, and we have neglected the vector and tensor modes, we also allow for first and second order anisotropic stresses. Hereinafter, we consider perturbations around a FLRW metric to second-order.

\section{Basics for the definition of the galaxy number density}
\label{sec:whatwemeasure}

\subsection{Photon wavevector}

\begin{figure}
\centering
\includegraphics[scale=1]{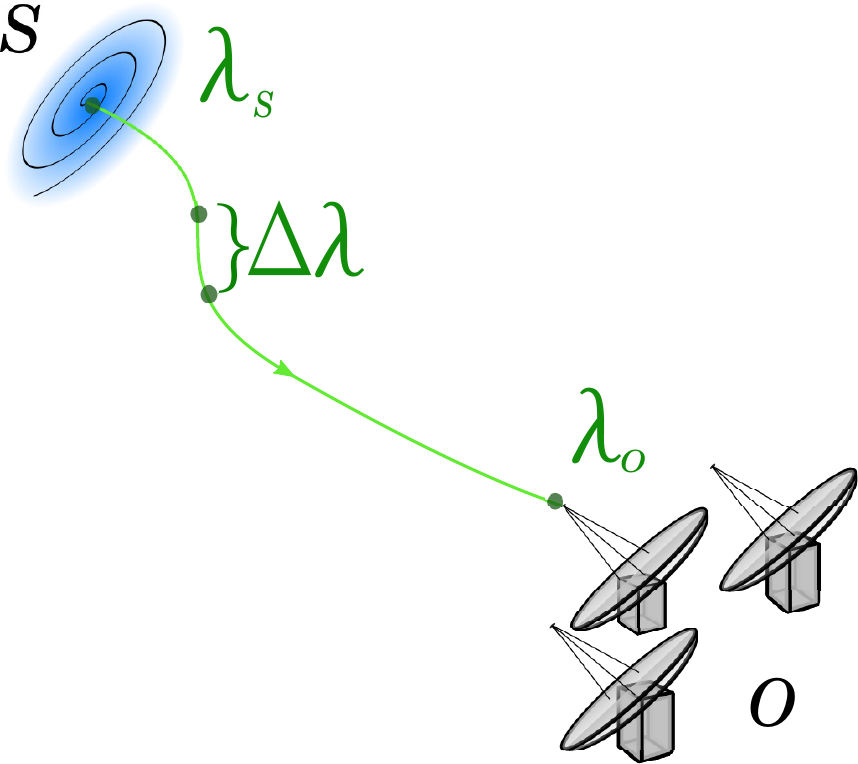}
\caption{Affine parameter convention of a light-ray in a radio observation. $S$ denotes the source, $O$ is the observer and $\lambda$ is the affine parameter.}
\label{fig:affine}
\end{figure}

In a redshift survey, galaxy positions are identified by measuring photons from the sources. As discussed in Chapter \ref{chapter:cpt}, the geodesic equation, given in Eq.~\eqref{eq:geointro}, is useful for computing dynamical evolution and for analysing the geometry of a spacetime through the trajectories of test particles. In a general spacetime, we consider a light-ray with tangent vector $k^{\mu}$ and affine parameter $\lambda$, that parametrises the curve that light-ray follows, with given values for the source, $\lambda_{s}$, and observer, $\lambda_{o}$, points, illustrated in Fig.~\ref{fig:affine}. The components of the photon wavevector can be written as
\be
\label{eq:kbg}
\bar{k}^{\mu} = \frac{\dd x^{\mu}}{\dd \lambda} = a^{-1}\Big[1,n^{i}\Big],
\ee
where the overbar denotes background quantities, $n^{i}$ is the direction of observation\footnote{Some authors define $n^{i}$ with the opposite sign. See, for example, Refs.~\cite{durrer1,cc1,cc2}.} pointing from the observer to the source, and follows the normalisation condition: $n^{i}n_{i} = 1$.

The tangent vector is \textit{null} 
\be
\label{eq:null}
k_{\mu}k^{\mu} = 0,
\ee
and \textit{geodesic} 
\be
\label{eq:geodesic}
k^{\nu}\nabla_{\nu}k^{\mu} =0.
\ee
where $\nabla_{\nu}$ is the covariant derivative defined by the metric given in \eq{eq:metric}.
In general, the perturbed wavevector can be written as
\be
\delta^{(n)} k^{\mu}=a^{-1}\Big[\delta^{(n)} \nu, \delta^{(n)} n^{i}\Big]. 
\ee
where $\delta^{(n)}$ gives the $n\text{-th}$ order perturbation as defined in Eq.~\eqref{eq:deltat}, and we are following the usual notation for the temporal component, that is $k^{0}\equiv\nu$ \cite{yoo4}.

The affine parameter that parametrises the geodesic equation, is also related to the comoving distance ($\chi$) by
\be
\label{eq:comovingd}
\chi = \lambda_{o}-\lambda_{s},
\ee
and in terms of the redshift as
\be
\label{eq:comd}
\chi(z) = \int_{0}^{z} \frac{\dd \tilde{z}}{(1+\tilde{z})\H (\tilde{z})}.
\ee

\subsection{Observed redshift}

The photon energy measured by an observer with 4-velocity $u^{\mu}$ is 
\be
\label{eq:photon-energy}
\E = - g_{\mu \nu}u^{\mu}k^{\nu}.
\ee
From \eq{eq:photon-energy} the observed redshift of a source (e.g.~a galaxy) can be defined as
\be
\label{eq:redshift}
1+z = \frac{\E_{s}}{\E_{o}} ,
\ee
where the `$s$' denotes the source and `$o$' the observer. From this definition, there will be a Doppler effect on the redshift due to the velocities $u^{\mu}$ and the observed redshift is in fact a function of the velocity and the wavevector, i.e.~$z=z(k^{\mu},u^{\mu})$.

\subsection{Angular diameter distance}
\label{daintro}

\begin{figure}
\centering
\includegraphics[trim=40 440 70 75,clip,scale=0.8]{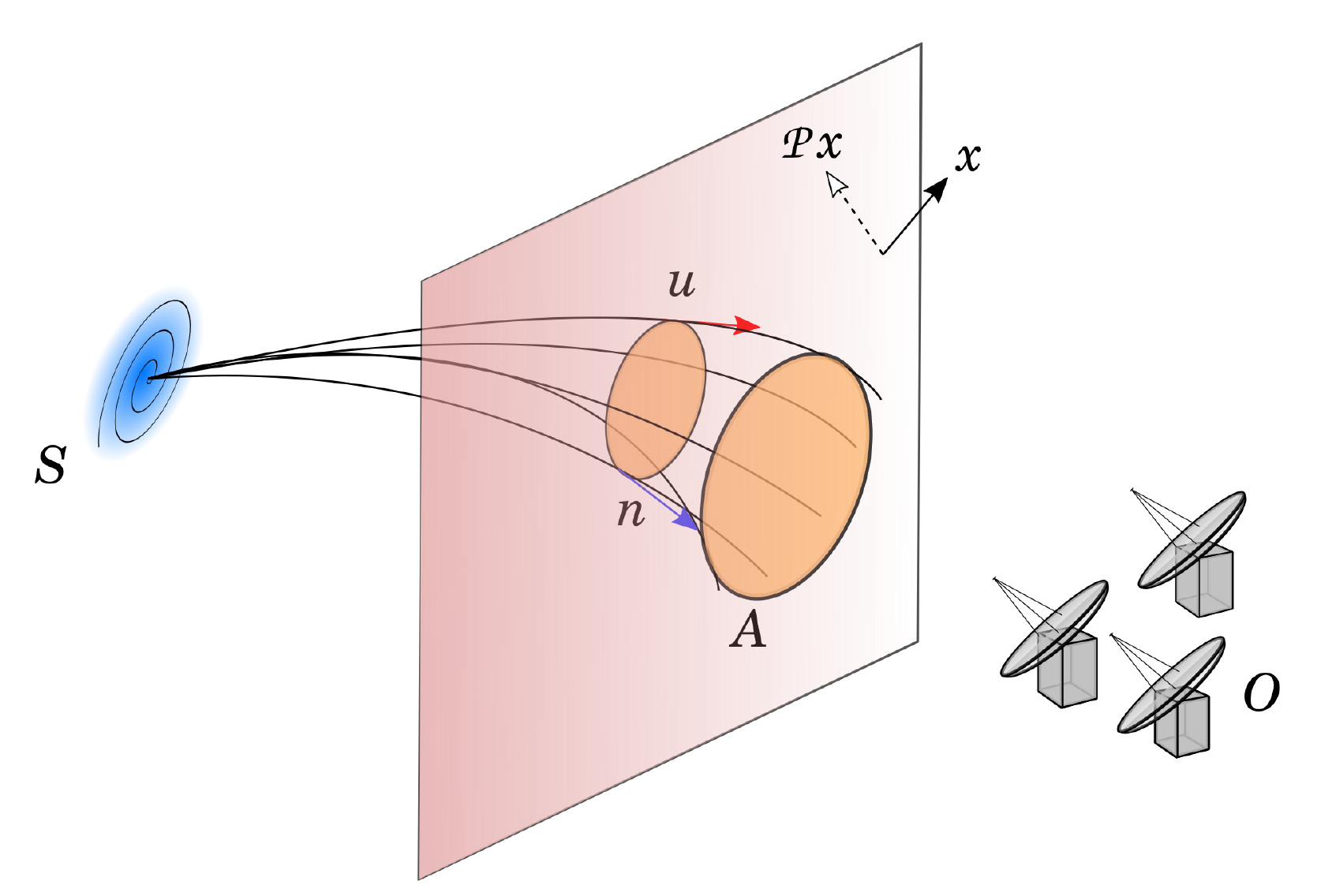}
\caption{The light-ray bundle going from the source, $S$, to the observer, $O$, the cross sectional area created by the infinitesimal separation of the light-rays and the screen space, $\mathcal{A}$, which is orthogonal to the 4-velocity, $u^{\mu}$, and the direction of observation, $n^{\mu}$. For simplicity, it is shown $n^{i}$ pointing from the source to the observer, i.e.~ the opposite from the one using in all our calculations. We present a general 4-vector $x^{\nu}$ and its projection onto screen space $\P_{\mu \nu}x^{\nu}$.}
\label{fig:screen}
\end{figure}

\begin{figure}
\centering
\includegraphics[trim=80 420 120 140,clip,scale=0.8]{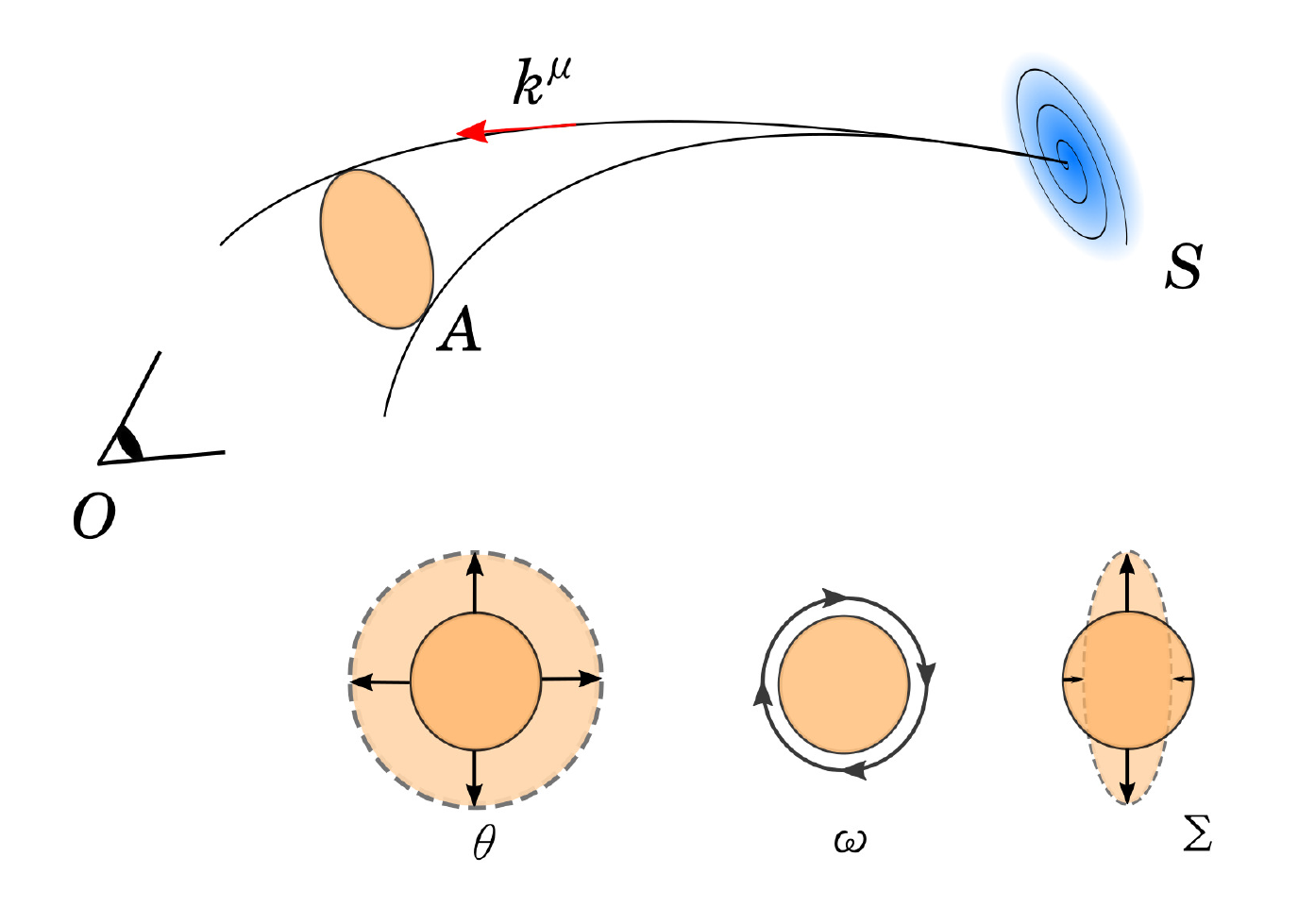}
\caption{ When a light-ray bundle traveling from a source, $S$, to an observer, $O$, passes close to matter, the cross section that the different null geodesics generate, $A$, gets distorted in different ways, and those are explained by an an expansion $\theta$, a vorticity $\omega$, and shear $\Sigma$, defined in Eqs.~\eqref{eq:deftheta} and \eqref{eq:defsigma}. The circles represent the area of the bundle's cross section and the arrows show how it is distorted.}
\label{fig:da}
\end{figure}

For a given bundle of light-rays leaving a source, the bundle will invariantly expand and create an area in between the light-rays. This area can be projected onto a screen space, perpendicular to the trajectories of the photons and the 4-velocity of the observer, as illustrated in Figs.~\ref{fig:screen} and \ref{fig:da}.

The area of a bundle in screen space, $\mathcal{A}$, defines the angular diameter distance $d_{A}$, and is directly related to the null expansion $\theta$ \cite{cc4} defined in section \ref{sec:distances},
\begin{equation}
\label{eq:daa}
\frac{1}{\sqrt{\mathcal{A}}}\frac{\dd \sqrt{\mathcal{A}}}{\dd \lambda} = \frac{\dd \ln d_{A}}{\dd \lambda} = \frac{1}{2} \theta.
\end{equation}
where $\lambda$ is the affine parameter defined in \eq{eq:kbg}. Using \eq{eq:daa} we can compute how the area of the bundle changes along the geodesic trajectory that the photons are following from the source towards the observer.

\subsection{Physical Volume}

Number counts relate to the number of sources detected in a bundle of rays, for a small affine parameter displacement $\lambda$ to $\lambda+\dd\lambda$ at an event $P$. This corresponds to a physical distance 
\be
\label{eq:dl}
\dd \ell = (k^{\mu}u_{\mu}) \dd \lambda,
\ee 
in the rest frame of a comoving galaxy at said point in space $P$, if $k^{\mu}$ is a tangent vector to the past directed null geodesics (so that $k^{\mu}u_{\mu}>0$). 

\begin{figure}
\centering
\includegraphics[scale=0.7]{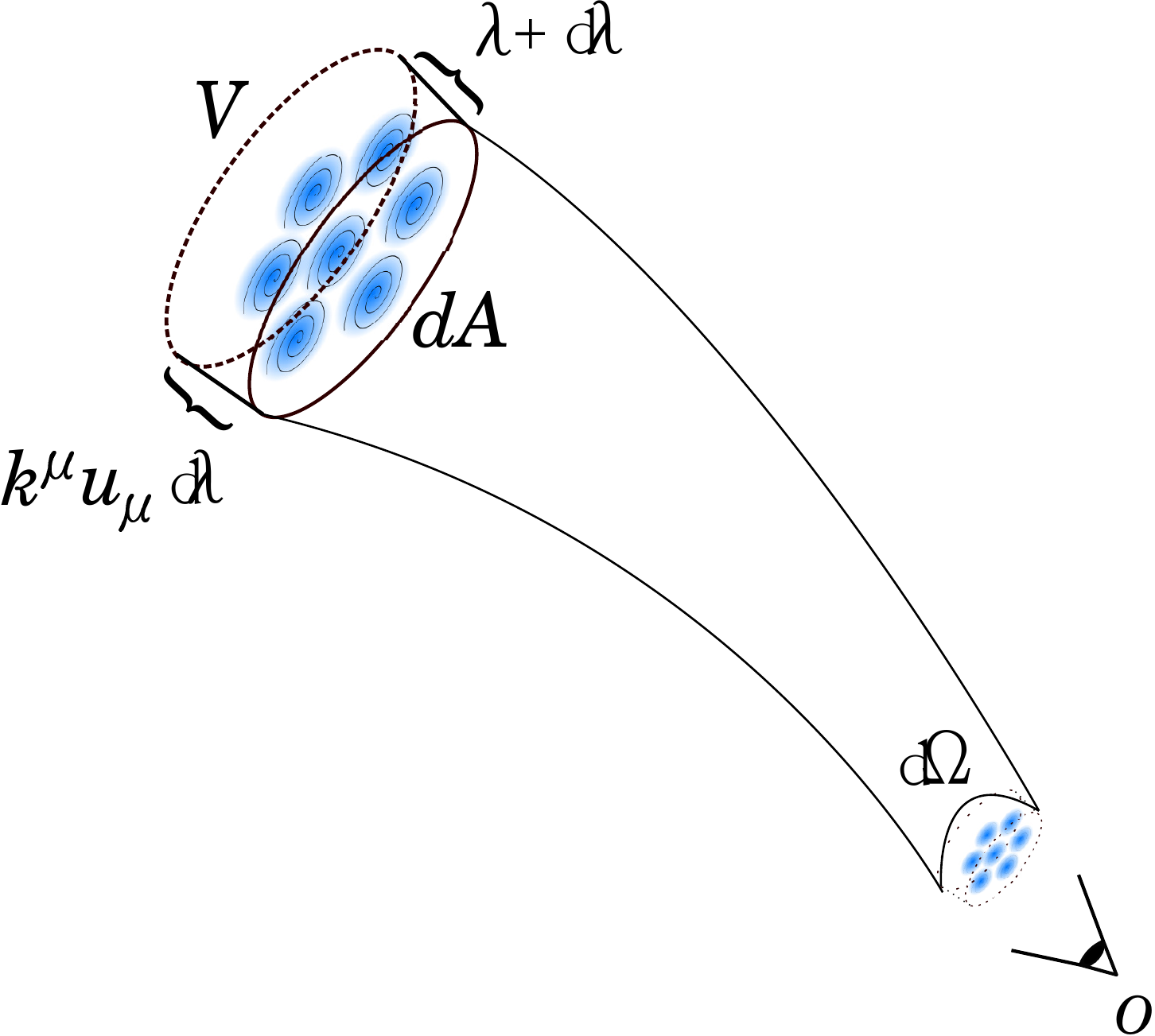}
\caption{Volume corresponding to an infinitesimal change in the affine parameter from $\lambda$ to $\lambda+\dd\lambda$.}
\label{fig:volume}
\end{figure}

The cross-sectional area of the bundle is 
\be
\label{eq:cross-a}
\dd \mathcal{A} = d_{A}^{2}(\lambda)\dd \Omega,
\ee 
if the geodesics subtend a solid angle $\dd \Omega$ to the observer, this is shown in Fig.~\ref{fig:volume}.

From \eqs{eq:dl} and \eqref{eq:cross-a} the corresponding volume element at a point $P$ in space is (see e.g. Ref.~\cite{ellis})
\begin{equation}
\label{eq:dldo}
\dd V = \dd \ell \dd \mathcal{A} = (k^{\mu}u_{\mu})d_{A}^{2}(\lambda) \dd \lambda \dd \Omega = -\E d^{2}_{A}(\lambda) \dd \lambda \dd \Omega.
\end{equation}

These covariant definitions lead to the expressions we compute in the following sections at first and second order in cosmological perturbation theory. 

\section{Perturbed null geodesics and redshift}
\label{sec:effects}
\subsection{Geodesic equation}

Let us now look at solutions to the geodesic equation. First, from \eq{eq:null} and the normalisation of $n^{i}$ we obtain the null condition of the photon wavevector
\begin{align}
\label{eq:null-conditions}
0 =& 2\Big[ n^{i}\dnf_{i} - \dnuf - \ff - \pf \Big] \notag \\
\qquad &+ \Bigg[ n^{i} \dns_{i} - \dnus - \frac{1}{2}\fs - \frac{1}{2} \ps - \left(\dnuf\right)^{2} + \dnf_{i} \dnf^{i} \notag\\
&\qquad\qquad\qquad\qquad\qquad\qquad - 4 \left(\ff+\pf\right) \dnuf - 4 \pf \left(\ff+\pf\right)\Bigg].
\end{align}
For the geodesic equation, \eq{eq:geodesic}, we obtain the propagation equations for the temporal and spatial perturbations 
\begin{align}
\label{eq:geo0}
k^{\mu}\nabla_{\mu}\delta^{(n)}\nu &= \frac{\dd \delta^{(n)}\nu}{\dd \lambda} + \Gamma^{0}_{\alpha \beta}k^{\alpha}k^{\beta}=0, \\
\label{eq:geoi}
k^{\mu}\nabla_{\mu}\delta^{(n)}n^{i} &= \frac{\dd \delta^{(n)}n^{i}}{\dd \lambda} + \Gamma^{i}_{\alpha \beta}k^{\alpha}k^{\beta}=0,
\end{align}
where $\Gamma^{\mu}_{\nu\sigma}$ are the connection coefficients at any given order $(n)$. The first order connection coefficients are given in Appendix \ref{connection}.

Substituting and rearranging, we find the geodesic equations in general at first order \cite{durrer3},
\begin{align}
\label{eq:dnu1}
\frac{\dd \dnuf}{\dd \lambda} &= - 2 \frac{\dd \ff}{\dd \lambda} + \ff' + \pf',\\
\label{eq:dn1}
\frac{\dd \dnf^{i}}{\dd \lambda} &= 2 \frac{\dd \pf }{\dd \lambda}n^{i} -\left[\pf_{,}^{i} + \ff_{,}^{i}\right],
\end{align}
where \eq{eq:dnu1} there is a term related to the Integrated Sachs-Wolfe effect defined below. The expressions for the equations at second order are given in Appendix \ref{metriccomp}.
\subsection{Observed redshift}

We now expand the photon energy $\E = - g_{\mu \nu}u^{\mu}k^{\nu}$ to second order,
\be
\E = \bar{\E} + \Ef + \frac{1}{2} \Es.
\ee
Using \eqs{eq:metricsec2}, \eq{eq:pertu0}, \eqref{eq:pertui}, \eqref{eq:dnu1}, \eqref{eq:dn1}, \eqref{eq:dnu2}  and \eqref{eq:dni2}, we find that $\bar{\E}=1$, and
\begin{align}
\label{eq:dE1}
\Ef &= \dnuf + \ff - \left(v_{1 i}n^{i}\right), \\
\label{eq:dE2}
\Es &= \dnus + \fs - \left( v_{2 i}n^{i}\right) + 2 \ff \dnuf - 2v_{1i} \dnf^{i} + \ff^{2} + \left(v_{1 k}v_{1}^{k}\right) + 4 \pf \left( v_{1i}n^{i}\right) .
\end{align}

The perturbations to the photon energy given in \eqs{eq:dE1} and \eqref{eq:dE2} are given explicitly in terms of the metric potentials in Appendix \ref{metriccomp}, in \eqs{eq:dE1-metric} and \eqref{eq:dE2-metric}. Integrating \eqs{eq:dnu1}, \eqref{eq:dn1}, \eqref{eq:dnu2} and \eqref{eq:dni2}, with respect to the affine parameter, $\lambda$, from the observer to the source, we find the perturbed photon vector, $\delta k^{\mu}$. Note that $\delta k^{\mu}|_{o}=0$, what can be seen explicitly in \eq{eq:dE1-metric} and \eq{eq:dE2-metric},  so that
\begin{align}
\label{eq:dE1o}
\Ef\big|_{o} &= \ff |_{o} - (v_{1 i}n^{i})_{o}, \\
\label{eq:dE2o}
\Es\big|_{o} &= \fs |_{o} - \left( v_{2 i}n^{i}\right)_{o} + \left(\ff|_{o}\right)^{2} + \left(v_{1 k}v_{1}^{k}\right)_{o} + 4 \pf|_{o} \left( v_{1i}n^{i}\right)_{o}. 
\end{align}
The observed redshift defined by \eq{eq:redshift}, then up to second order is
\be
\label{eq:redshift1}
1+z = 1 +\bar{z}+ \dzf+ \frac{1}{2} \dzs= \frac{\E_{s}}{\E_{o}} = \frac{(\bar{\E} + \Ef + \frac{1}{2} \Es)_{s}}{(\bar{\E} + \Ef + \frac{1}{2} \Es)_{o}}.
\ee
Note that in general, the expansion of the perturbed quotient up to second order is
\be
\label{eq:quotient}
\frac{A+\delta^{(1)} A + \frac{1}{2} \delta^{(2)}A}{B+\delta^{(1)} B + \frac{1}{2}\delta^{(2)}B} = \frac{A}{B}\left( 1 + \frac{\delta^{(1)} A}{A} - \frac{\delta^{(1)}B}{B} + \frac{\delta^{(2)}A}{2A} -\frac{\delta^{(2)}B}{2B} - \frac{\delta^{(1)}A}{A} \frac{\delta^{(1)} B}{B} + \left[\frac{\delta^{(1)}B}{B}\right]^{2}\right).
\ee
Using \eqs{eq:redshift1} and \eqref{eq:quotient}, and ignoring the background redshift, we obtain
\be
\label{eq:rfe}
1+z = 1+ \left( \Ef\big|_{s}-\Ef\big|_{o} \right) - \Ef\big|_{s}\Ef\big|_{o} + \left(\Ef\big|_{o}\right)^{2} + \frac{1}{2}\left( \Es\big|_{s}-\Es\big|_{o} \right).
\ee
where we defined the perturbed redshift $\delta z$ relative to the flat background, where the background value is zero.
From \eq{eq:rfe} the redshift of a source $s$ is, at first order,
\be
\label{eq:dz1}
\dzf= -\left( v_{1i}n^{i} + \ff\right)\big|^{s}_{o} + \int_{\lambda_{o}}^{\lambda_{s}}  \left[\ff' +\pf'\right]\dd\lambda,
\ee
where the integral runs with respect to~the affine parameter $\lambda$ along the line of sight. In \eq{eq:dz1} we identify the following elements:
\begin{itemize}
\item[a)] \textit{Doppler redshift}, which depends on the difference between the peculiar velocities of the source and the observer
\be
\dzf_{\text{Doppler}}=\left(v_{1 i}n^{i}\right)_{o} - \left(v_{1 i}n^{i}\right)_{s}.
\ee
\item[b)] \textit{Gravitational redshift}, which describes the change of energy the photon experiences when it travels from a region with potential $\ff|_{s}$ to a region with potential $\ff|_{o}$
\be
\dzf_{\text{gravitational}}=\ff |_{o} - \ff|_{s}.
\ee
\item[c)] The Integrated Sachs-Wolfe (ISW) effect, which describes the change in energy when a photon travels through a potential well between the source and the observer. This effect is only non-zero when the gravitational potential evolves during and along the photon's trajectory, so that the energy gained by going down the gravitational potential does not cancel out with the energy lost by climbing out the potential at the other end,
\be
\dzf_{\text{ISW}}=\int_{\lambda_{o}}^{\lambda_{s}}\left[\ff' +\pf'\right]\dd\lambda.
\ee
\end{itemize}

At second order the redshift \eqref{eq:dz2-metric} can also be decomposed as above
\begin{align}
\label{eq:dz2}
\dzs&= \dzs_{\text{Doppler}} + \dzs_{\text{gravitational}} + \dzs_{\text{ISW}} + \dzs_{\text{NL}},
\end{align}
where the nonlinear contribution from squared first order quantities denoted by `NL' is given by
\begin{align}
\dzs_{\text{NL}} &= \dzs_{\text{gravitational}\times\text{gravitational}} + \dzs_{\text{gravitational}\times\text{ISW}} + \dzs_{\text{IISW}}\\
&\quad + \dzs_{\text{gravitational}\times\text{Doppler}}+ \dzs_{\text{Doppler}\times\text{Doppler}}+ \dzs_{\text{Doppler}\times\text{ISW}}. \notag
\end{align}

Here, just as in the linear case, we have
\begin{itemize}
\item[(a)] Nonlinear doppler redshift
\begin{align}
\dzs_{\text{Doppler}} &= \frac{1}{2} \Big[\left(v_{2i}n^{i}\right)_{o} - \left(v_{2i}n^{i}\right)_{s} \Big] \notag,
\end{align}
\item[(b)] Nonlinear gravitational redshift
\begin{align}
\dzs_{\text{gravitational}} &=  \frac{1}{2} \Big[ \fs|_{o} - \fs|_{s} \Big] , \notag 
\end{align}
\item[(c)] Nonlinear ISW
\begin{align}
\dzs_{\text{ISW}} &= \frac{1}{2}\int_{\lambda_{o}}^{\lambda_{s}} \Big( \fs'+\ps' \Big) \dd\lambda ,\notag
\end{align}
\end{itemize}
Here the second order contribution to the effects described above are evident, plus the product of first order contributions:
\begin{itemize}
\item[(1)] Gravitational redshift squared
\be
\dzs_{\text{gravitational} \times \text{gravitational}} = - \frac{3}{2} \left( \ff|_{s} - \ff|_{o}\right)^{2}+6 \ff|_{s}\ff|_{o}, \notag
\ee
\item[(2)] Gravitational redshift $\times$ ISW 
\be
\dzs_{\text{gravitational} \times \text{ISW}} = 2\left(\ff|_{s}-\ff|_{o}\right)\int_{\lambda_{o}}^{\lambda_{s}}\dd\lambda\left( \ff' + \pf' \right), \notag
\ee
\item[(3)] Gravitational redshift $\times$ Doppler
\be
\dzs_{\text{gravitational} \times \text{Doppler}} = \ff|_{o}\left( v_{1i}n^{i} \right)_{s} -\ff|_{s}\left( v_{1i}n^{i} \right)_{o} - 2 \pf|_{o}\left( v_{1i}n^{i} \right)_{o},\notag
\ee
\item[(4)] Doppler squared
\be
\dzs_{\text{Doppler} \times \text{Doppler}} = \frac{1}{2}\Big[\left( v_{1k}v^{k}_{1}\right)_{s}-\left( v_{1k}v^{k}_{1}\right)_{o}\Big] -\left( v_{1i}n^{i} \right)_{o}\Big[\left( v_{1i}n^{i}\right)_{s}-\left( v_{1i}n^{i}\right)_{o}\Big], \notag
\ee
\item[(5)] Doppler $\times$ ISW
\be
\dzs_{\text{Doppler} \times \text{ISW}} = \left( v_{1i}n^{i} \right)_{o}\int_{\lambda_{o}}^{\lambda_{s}}\dd\lambda\left( \ff' + \pf' \right) +  2v_{1 i}\int_{\lambda_{o}}^{\lambda_{s}}\dd \lambda\left( {\ff_{,}}^{i} + {\pf_{,}}^{i} \right), \notag
\ee
\item[(6)] Integrated ISW
\begin{align}
\dzs_{\text{IISW}} &= 2\int_{\lambda_{o}}^{\lambda_{s}}\dd\lambda\left[\pf' \left( \ff + \pf \right)\right] + 2\int_{\lambda_{o}}^{\lambda_{s}}\dd\lambda \left[ \ff \left( \frac{\dd \ff}{\dd \lambda} - 2\frac{\dd \pf}{\dd \lambda}\right)\right] \notag\\
& \quad -4 \int_{\lambda_{o}}^{\lambda_{s}}\dd\lambda\left[\left( \pf \ff_{,i} - \ff \pf_{,i} \right)n^{i}\right] + 4\int_{\lambda_{o}}^{\lambda_{s}}\dd\lambda\left[ n^{i}\ff_{,i}\left( \ff - \pf \right)\right]  \notag \\
& \quad + \int_{\lambda_{o}}^{\lambda_{s}}\dd\tilde{\lambda}\Bigg\{2 \left( \ff' - \pf'\right)\int_{\lambda_{o}}^{\lambda_{s}}\dd\lambda\left( \ff' + \pf' \right) \notag \\
&\quad- n^{i} \left( \ff_{,i} + \pf_{,i} \right) \int_{\lambda_{o}}^{\lambda_{s}}\dd\lambda\left( \ff' + \pf'\right) +  \left( \frac{\dd\ff}{\dd \lambda} - \frac{\dd \pf}{\dd \lambda}\right)\int_{\lambda_{o}}^{\lambda_{s}}\dd\lambda \left( \ff' + \pf' \right) \notag \\
& \quad +2\left( \ff-\pf\right)n^{i}\int_{\lambda_{o}}^{\lambda_{s}}\dd\lambda\left( \ff' + \pf'\right)_{,i} \notag \\
&\quad +\left( \int_{\lambda_{o}}^{\lambda_{s}}\dd\lambda{\left( \ff + \pf\right)_{,}}^{i}\right)\left( \int_{\lambda_{o}}^{\lambda_{s}}\dd\lambda\left( \ff' + \pf'\right)_{,i}\right)\notag \\
&\quad - n^{i}\left( \int_{\lambda_{o}}^{\lambda_{s}}\dd\lambda\left( \ff' + \pf' \right)_{,i} \right)\left( \int_{\lambda_{o}}^{\lambda_{s}}\dd\lambda\left( \ff' + \pf' \right) \right)\Bigg\}. \notag
\end{align}
\end{itemize}

\section{Distance determinations and the observed volume}
\label{sec:distances}

\subsection{Angular Diameter Distance}

To measure the angular diameter distance ($d_{A}$) introduced in Section \ref{daintro}, we must define a projector into the screen space perpendicular to the light-ray as shown in Fig. \ref{fig:da}. The screen space is orthogonal to the light-ray {\textit{and}} to the observer 4-velocity. In fact, the tensor 
\be
\P_{\mu \nu} = g_{\mu\nu}+ u_{\mu}u_{\nu}-n_{\mu}n_{\nu},
\ee
where $g_{\mu\nu}$ is the metric, $u_{\mu}$ is the 4-velocity and $n_{\mu}$ is the 4-direction of observation\footnote{Note that in the background, $n^{\mu}= a^{-1}[0,n^i]$.}, projects 4-vectors onto screen space, as can be seen in Fig.~\ref{fig:screen}. The projector tensor satisfies the relations
\be
{\P_{\mu}}^{\mu}=2, \qquad \P_{\mu\alpha}{\P^{\alpha}}_{\nu}=\P_{\mu\nu}, \qquad \P_{\mu\nu}k^{\nu} =\P_{\mu\nu}u^{\nu} = \P_{\mu\nu}n^{\nu} = 0.
\ee

The null expansion, $\theta$, and null shear, $\Sigma_{\mu \nu}$, are optical properties given in terms of the tangent vector $k^{\mu}$ by \cite{malik,sachs3,cc4}
\begin{align}
\label{eq:deftheta}
\theta &= \P^{\mu\nu}\nabla_{\mu}k_{\nu}, \\
 \label{eq:defsigma}
 \Sigma_{\mu \nu}& = {\P_{(\mu}}^{\sigma}{\P_{\nu)}}^{\rho}\nabla_{\sigma}k_{\rho}-\frac{1}{2} \theta \P_{\mu \nu}.
\end{align}
Here, $\theta$ describes the rate of expansion of the projected area of a bundle of light-rays and $\Sigma_{\mu\nu}$ describes its rate of shear illustrated in Fig.~\ref{fig:da}. Note that the wavevector can be obtained from a scalar potential ($S$), i.e.~$k_{\mu}=\nabla_{\mu}S$, and thus there is no null vorticity, that is $\omega \equiv \nabla_{[\mu}k_{\nu]} = 0$ \cite{cc7}.

To understand how the null expansion and the null shear evolve along the trajectory parametrised with the affine parameter, one can use the Sachs propagation equations, so the ``null evolution'' is given by (see e.g.~Ref. \cite{sachs3} for full derivation) 
\begin{align}
\label{eq:dth}
\frac{\dd \theta}{\dd \lambda} &= -\frac{1}{2} \theta^{2} - \Sigma_{\mu \nu}\Sigma^{\mu \nu} - R_{\mu \nu}k^{\mu}k^{\nu},\\
\label{eq:shr}
\frac{\dd \Sigma_{\mu \nu}}{\dd \lambda} &= -\Sigma_{\mu \nu} \theta + C_{\mu \rho \nu \sigma} k^{\rho} k^{\sigma},
\end{align}
where $C_{\mu \rho \nu \sigma}$ is the Weyl tensor. 

\eqs{eq:dth} and \eqref{eq:shr} allow us to compute the angular diameter distance as a parametric function depending only on the affine parameter $\lambda$, in contrast to previous works where the dependency is on the redshift  \cite{durrer2,cc1,cc2} or the conformal time \cite{yoo2}. The advantages of maintaining this dependency are discussed in section \ref{sec:discussion}.

From \eqs{eq:daa} and (\ref{eq:dth}) we obtain a second order differential equation for the area distance,
\begin{equation}
\label{eq:da}
\frac{\dd^{2}d_{A}}{\dd \lambda^{2}} = -\frac{1}{2} \left( R_{\mu \nu} k^{\mu}k^{\nu} +\Sigma_{\mu \nu} \Sigma^{\mu \nu} \right) d_{A}.
\end{equation}

We require appropriate initial conditions to solve \eqref{eq:da}. These can be found from the series expansion of the squared distance given by Kristian and Sachs in \cite{sachs2}:
\be
d_{A}^{2}(\lambda_{s}) = (u_{\mu}k^{\mu})_{o}^{2}(\lambda_{o}-\lambda_{s})\left[ 1 - \frac{1}{6}\left( R_{\mu\nu}k^{\mu}k^{\nu}\right)_{o}(\lambda_{o}-\lambda_{s})^{2}+\cdots \right],
\ee
from where we obtain the boundary conditions at the observer
\be
\label{eq:init}
d_{A}(\lambda_{o}) = 0, \quad \text{and}, \quad \frac{\dd d_{A}}{\dd \lambda}\Big|_{o} = -\E_{o}.
\ee

In this section, we define a conformal metric $g_{\mu \nu} = a^{-2} \hat{g}_{\mu \nu}$ useful to compute the angular diameter distance. In our notation, a hat ($ \hat{ \_ } $) denotes quantities on the physical spacetime, while quantities on the conformal spacetime have no hat. The background of the metric $g_{\mu \nu}$ is Minkowski spacetime, which simplifies both the equations and the calculations.

Conformal maps preserve both angles and shapes of infinitesimally small figures, but not their overall size \cite{conformal}. The conformal transformation $\hat{g}_{\mu\nu} \rightarrow g_{\mu\nu}$ maps the null geodesic equation of the perturbed FLRW metric $\hat{g}_{\mu \nu}$ to a null geodesic on the perturbed Minkowski metric $g_{\mu\nu}$ \cite{conformalnotes} and the angular diameter distance transforms as 
$
\hat{d}_{A}= a d_{A}.
$ 
The affine parameter transforms as 
$
\dd \lambda = a^{-2}\dd \hat{\lambda}, 
$
so that the photon ray vector transforms as 
$
\hat{k}^{\mu} = a^{-2}k^{\mu} \iff \hat{k}_{\mu}=k_{\mu}
$ \cite{cc3}. For the 4-velocity we have 
$
\hat{u}_{\mu} = a u_{\mu}.
$
Finally, the energy transforms as 
$
\hat{\E}=-\hat{u}_{\nu}\hat{k}^{\nu} = -a^{-1}u_{\nu}k^{\nu}=a^{-1}\E.
$ 

Hereafter, and until the end of this section, we will be working in a perturbed Minkowski spacetime, in order to finally conformally transform our result back to a FLRW spacetime.

In the Minkowski background, \eq{eq:da} simplifies to 
\be 
\frac{\dd^{2}\bar{d}_{A} }{\dd \lambda^{2}}= 0,
\ee
since $\bar{R}_{\mu \nu}$ and the shear vanish in the background. The solution is then 
\be
\bar{d}_{A}=C_{1}+\lambda C_{2}.
\ee 
The initial conditions given in \eq{eq:init} give $C_{1}=0$ and $C_{2}=-1$, so that
\be
\label{eq:bgda-mink}
\bar{d}_{A}(\lambda_{s}) = \lambda_{o} - \lambda_{s}.
\ee
Mapping this into the FLRW background we get for the angular diameter distance
\be
\hat{d}_{A}(\hat{\lambda}_{s}) = a(\hat{\lambda}_{s})\left( \hat{\lambda}_{o}-\hat{\lambda}_{s}\right),
\ee
which can be expressed in terms of the comoving distance \eqref{eq:comovingd} as
\be
\label{eq:dacd}
\hat{d}_{A}(\hat{z}) = \frac{\chi(\hat{z})}{1+\hat{z}},
\ee
where we have used the definition of the scale factor $a(\hat{z})=1/(1+\hat{z})$ and used the fact that the comoving distance depends on the redshift as given in Eq.~\eqref{eq:comd}. Eqs.~\eqref{eq:dacd} and \eqref{eq:comd} will be useful when we are comparing our work with results in the literature.

In general, at first order \eq{eq:da} takes the form
\be
\label{eq:da2}
\frac{\dd^{2} \daf}{\dd \lambda^{2}} = -\frac{1}{2}\left[ 2 \bar{R}_{\mu \nu} \bar{k}^{\mu} \dkf^{\nu} \bar{d}_{A} + \dRf_{\mu \nu} \bar{k}^{\mu} \bar{k}^{\nu} \bar{d}_{A} + \bar{R}_{\mu \nu} \bar{k}^{\mu} \bar{k}^{\nu} \daf\right] - \bar{d}'_{A} \frac{\dd\dnuf}{\dd \lambda}-2 \bar{d}''_{A} \dnuf,
\ee
where we use that in the background the affine parameter is related to the conformal time like $\dd \lambda = \dd \eta$, so that
\be
\frac{\dd \bar{d}_{A}}{\dd \eta} = \frac{\dd \bar{d}_{A}}{\dd \lambda}.
\ee 
This relation is only fulfilled in the background, once perturbations are introduced the relation between the affine parameter and time becomes non-trivial.

In Minkowski spacetime, \eq{eq:da2} simplifies to
\begin{align}
\label{eq:dda1}
\frac{\dd^{2} \daf}{\dd \lambda^{2}} &= -\frac{1}{2}\bar{d}_{A}\left[2 \left( \frac{\dd^{2}\pf}{\dd\lambda^{2}}\right) +\nabla^{2}\left( \ff + \pf \right)-n^{i}n^{j}\left( \ff + \pf\right)_{,ij}-\frac{2}{\bar{d}_{A}}\frac{\dd \dnuf}{\dd \lambda}\right],
\end{align}
where we used the background solution for $d_{A}$ \eqref{eq:bgda-mink}, the first order perturbation of the Ricci tensor $\dRf_{\mu\nu}$ given in Appendix \ref{riccipert}, and \eqs{eq:der-affine} and \eqref{eq:double-der-affine}.

The solution to \eqref{eq:dda1} is, upon several integrations by parts, 
\begin{align}
\label{eq:da-linear}
\frac{\daf(\lambda_{s})}{\bar{d}_{A}(\lambda_{s})} &=   \ff|_{o}-\pf|_{o} - \pf|_{s} - \left(v_{1i}n^{i} \right)_{o} - \frac{1}{\lambda_{o}-\lambda_{s}}\Bigg\{2\int_{\lambda_{o}}^{\lambda_{s}}\dd\lambda \pf \\
& \quad +\frac{1}{2} \int_{\lambda_{o}}^{\lambda_{s}}\dd\lambda\left(\lambda_{s}-\lambda\right)\left( \lambda_{o}-\lambda\right)\left[ \nabla^{2}\left( \ff+\pf \right) - n^{i}n^{j}\left( \ff+\pf \right)_{,ij} \right.\notag \\
& \left. \qquad \qquad \qquad \qquad \qquad \qquad \qquad \qquad \qquad \qquad \qquad \qquad -\frac{2}{\lambda_{o}-\lambda_{s}}\frac{\dd\dnuf}{\dd \lambda} \right]\Bigg\}. \notag
\end{align}
From \eq{eq:init}, in general
\be
\label{eq:da-linear-no-slip}
\delta^{(n)} d_{A}(\lambda_{o}) = 0, \quad \text{and}, \quad \frac{\dd \delta^{(n)}d_{A}}{\dd \lambda}\Big|_{o} = - \delta^{(n)}\E_{o}.
\ee
In the absence of anisotropic stress, $\ff = \pf$ (see, e.g.~Ref.~\cite{malik}), we recover in \eq{eq:da-linear} the fully relativistic lensing convergence, usually denoted as $\kappa$ \cite{durrer1,cc5,cc7,yoo2}, at first order, which includes Sachs-Wolfe (SW), Integrated Sachs-Wolfe (ISW) and Doppler terms in addition to the standard lensing integral

\begin{align}
\label{eq:da-noan}
\frac{\daf(\lambda_{s})}{\bar{d}_{A}(\lambda_{s})} &=  - \ff|_{s} - \left(v_{1i}n^{i} \right)_{o} - \frac{1}{\lambda_{o}-\lambda_{s}}\Bigg\{2\int_{\lambda_{o}}^{\lambda_{s}}\dd\lambda \ff \\
& \quad + \int_{\lambda_{o}}^{\lambda_{s}}\dd\lambda\left(\lambda_{s}-\lambda\right)\left( \lambda_{o}-\lambda\right)\left[ \nabla^{2}\left( \ff \right) - n^{i}n^{j}\left( \ff \right)_{,ij} -\frac{1}{\lambda_{o}-\lambda_{s}}\frac{\dd\dnuf}{\dd \lambda} \right]\Bigg\}. \notag
\end{align}
In Ref.~\cite{cc2}, to simplify notation the authors define the Laplacian operator transverse to the radial direction by
\be
\label{eq:umeh}
\nabla^{2}_{\perp}X = \nabla^{2}\left( X \right) - n^{i}n^{j}\left( X \right)_{,ij} -\frac{2}{\lambda_{o}-\lambda_{s}}\frac{\dd X}{\dd \lambda}.
\ee
Using Eqs.~\eqref{eq:geo0} and \eqref{eq:umeh}, Eq.~\eqref{eq:da-noan} can be further simplified into
\be
\frac{\daf(\lambda_{s})}{\bar{d}_{A}(\lambda_{s})} =  - \ff|_{s} - \left(v_{1i}n^{i} \right)_{o} - \frac{1}{\lambda_{o}-\lambda_{s}}\Bigg\{2\int_{\lambda_{o}}^{\lambda_{s}}\dd\lambda \ff + \int_{\lambda_{o}}^{\lambda_{s}}\dd\lambda\left(\lambda_{s}-\lambda\right)\left( \lambda_{o}-\lambda\right) \nabla^{2}_{\perp} \ff \Bigg\},
\ee
which is exactly what the authors obtain in Ref.~\cite{cc2}. In this thesis, we will not use \eqref{eq:umeh} further as this expansion was for comparison with Ref.~\cite{cc2} only.

At second order, \eq{eq:da} takes the form
\begin{align}
\label{eq:da-two}
\frac{\dd^{2} \das}{\dd \lambda^{2}} &=  -\Bigg[ 2\bar{k}^{\mu}\dkf^{\nu} \dRf_{\mu\nu} +\frac{1}{2} \bar{k}^{\mu}\bar{k}^{\nu}\dRs_{\mu\nu}+\dS_{\mu\nu}\dS^{\mu\nu} \Bigg]\bar{d}_{A} \\
&\quad - \Big[ \bar{k}^{\mu}\bar{k}^{\nu}\dRf_{\mu\nu} \Big]\daf -2 \Big[ \dkf^{\mu}\nabla_{\mu}\dnuf + 3 \dkf^{\mu} \bar{k}^{\alpha}\Gamma_{\mu\alpha}^{0} \Big]\bar{d}_{A}' \notag \\
&\quad - 2\Big[ \dnus + \left( \dnuf \right)^{2} \Big]\bar{d}_{A}'' - \left( \frac{\dd \dnus}{\dd \lambda} \right)\bar{d}_{A}' -4 \left(\dnuf\right)\left( \daf ''\right) \notag \\
&\quad -2\Bigg[ \bar{k}^{\mu}\bar{k}^{\alpha}\Gamma_{\mu\alpha} + \frac{\dd \dnuf}{\dd \lambda} \Bigg]\daf' \notag,
\end{align}
where $\dS_{\mu \nu}$ is the linear perturbation to the shear. Using \eq{eq:defsigma} we obtain
\be
\label{eq:ddS}
\frac{\dd \dS_{ij}}{\dd \lambda} = \frac{1}{2}\delta_{ij}\nabla^{2}\left(\ff + \pf\right) - \left( \ff + \pf\right)_{,ij}. 
\ee
Without loss of generality, we set the perturbation of the shear at the observer $\dS^{\mu\nu}|_{o} = 0$, and integrating along the line of sight from the observer to the source ($\lambda_{o}$ to $\lambda_{s}$), we obtain
\be
\label{eq:dSij}
\dS_{ij} = \int_{\lambda_{o}}^{\lambda_{s}}\dd \lambda \Bigg[ \frac{1}{2}\delta_{ij}\nabla^{2}\left(\ff + \pf\right) - \left( \ff + \pf\right)_{,ij} \Bigg] . 
\ee
The contraction $\dS_{ij}\dS^{ij}$ is given by
\begin{align}
\label{eq:contracted-null-shear}
\dS_{ij} \dS^{ij} &= \Bigg[  \int_{\lambda_{o}}^{\lambda_{s}}\dd \lambda \Bigg[ \frac{1}{2}\delta_{ij}\nabla^{2}\left(\ff + \pf\right) - \left( \ff + \pf\right)_{,ij} \Bigg] \Bigg]\times \\ 
&\qquad\qquad\qquad \Bigg[  \int_{\lambda_{o}}^{\lambda_{s}}\dd \lambda \Bigg[ \frac{1}{2}\delta^{ij}\nabla^{2}{\left(\ff + \pf\right) - \left( \ff + \pf\right)_{,}}^{ij} \Bigg] \Bigg] , \notag\\
&= \left(\int_{\lambda_{o}}^{\lambda_{s}}\dd\lambda\left( \ff + \pf \right)_{,ij}\right)\left( \int_{\lambda_{o}}^{\lambda_{s}}\dd\lambda{\left( \ff + \pf\right)_{,}}^{ij}\right) -\frac{1}{4}\left[ \int_{\lambda_{o}}^{\lambda_{s}}\dd\lambda\nabla^{2}\left( \ff + \pf \right)\right]^{2}. \notag
\end{align}

In Eq.~\eqref{eq:da-second-metric} we find the second order part of the diameter distance, using the background solution for $\bar{d}_{A}$, and the full expression $\dRs_{\mu \nu}$, the second order perturbation of the Ricci tensor given in Appendix \ref{riccipert}.
 
Thus, the total area distance as a function of the affine parameter in a perturbed FLRW spacetime is given by
\be
\label{eq:daaff}
\hat{d}_{A}(\lambda_{s}) = a(\lambda_{s})(\lambda_{o}-\lambda_{s})\left[ 1 + \frac{\daf(\lambda_{s})}{\bar{d}_{A}(\lambda_{s})}+\frac{1}{2}\frac{\das(\lambda_{s})}{\bar{d}_{A}(\lambda_{s})} \right],
\ee
where the solutions for $\bar{d}_{A}(\lambda_{s})$, $\daf(\lambda_{s})$ and $\das(\lambda_{s})$ are given in \eqs{eq:bgda-mink}, \eqref{eq:da-linear} and \eqref{eq:da-second-metric}, respectively. From here onwards, we abandon the conformal Minkowski spacetime and return to FLRW.

\subsection{Area distance as a function of observed redshift}
\label{dafredshift}

In order to compare with previous work done in the literature, we can convert the angular diameter distance in terms of the affine parameter to a function of the observed redshift. To do so, we need to perturbatively invert $z(\lambda)$ into $\lambda(z)$ and substitute this into Eq.~\eqref{eq:daaff}. This means we need $d_{A}$ on surfaces of constant observed redshift $z$ rather than on surfaces of constant affine parameter $\lambda$, which is not observable.

We expand the affine parameter in perturbation theory as 
\be
\label{eq:pertl}
\lambda = \varsigma + \delta^{(1)} \lambda + \frac{1}{2} \delta^{(2)} \lambda,
\ee
where $\varsigma$ is the affine parameter in redshift space corresponding to the redshift $\hat{z}$, \textit{as if there were no perturbations} \cite{cc2}. We define $\varsigma$ using as an anchor the background relation
\be
\label{eq:anu}
a(\varsigma) = \frac{1}{1+\hat{z}},
\ee
with this relation we can fix $\delta^{(1)}\lambda$ and $\delta^{(2)}\lambda$, since it should always hold, and if there are any perturbations, they should cancel since Eq.~\eqref{eq:anu} is only valid in the background. To begin with, we see that at any redshift $\hat{z}$, the derivatives of $a$ with respect to $\varsigma$ are
\begin{align}
\frac{1}{a} \frac{\dd a}{\dd \varsigma} &= \H(\varsigma), \\
\frac{1}{a} \frac{\dd^{2} a}{\dd \varsigma^{2}} &= \left[ \frac{\dd \H(\varsigma)}{\dd \varsigma} + \H^{2}(\varsigma) \right]. 
\end{align}
We now expand the scale factor $a$ about $\lambda$ up to and including second order perturbations as defined in Eq.~\eqref{eq:pertl}, we have
\be
\label{eq:anutaylor}
a(\lambda) = a(\varsigma) \left[ 1 +\H \delta^{(1)} \lambda + \frac{1}{2}\H \delta^{(2)} \lambda +\frac{1}{2} \left( \frac{\dd \H}{\dd \lambda} +\H^{2} \right)\left(\delta^{(1)}\lambda\right)^{2} + \mathcal{O}\left(\delta^{(3)}\lambda\right) \right].
\ee
Using Eqs.~\eqref{eq:redshift1} and \eqref{eq:anutaylor} we find that, 
\begin{align}
\label{eq:tayloranu}
\frac{1}{1+\hat{z}} &= \frac{a(\varsigma)}{a(\varsigma_{o})}\Bigg[ 1+ \left( \H \delta^{(1)} \lambda - \dzf \right)   \\
& \qquad +\frac{1}{2}\left\{ \H \delta^{(2)}\lambda - \dzs + 2\left( \dzf\right)^{2} - 2\H \delta^{(1)}\lambda \dzf + \left( \frac{\dd \H}{\dd \varsigma} + \H^{2} \right)\left(\delta^{(1)}\lambda\right)^{2}\right\} \Bigg]. \notag
\end{align}
From the background relation given in Eq.~\eqref{eq:anu} and \eqref{eq:tayloranu} we then find that the perturbations to the affine parameter must follow the following relations:
\begin{align}
\label{eq:pertlz1}
\delta^{(1)}\lambda &= \frac{\dzf}{\H}, \\
\label{eq:pertlz2}
\delta^{(2)}\lambda &= \frac{1}{\H}\left[ \dzs - \left( \dzf \right)^{2}\left( 1 + \frac{1}{\H^{2}}\frac{\dd \H}{\dd \varsigma} \right) \right].
\end{align}
Finally, using these relations to substitute for $a(\lambda_{s})\left( \lambda_{o}-\lambda_{s}\right)$, we find that the area distance \eqref{eq:daaff} becomes
\begin{align}
\label{eq:danu1}
\hat{d}_{A} (\varsigma)&= a(\varsigma)\left( \varsigma_{o}-\varsigma\right)\Bigg\{ 1 + \left[ \frac{\daf}{\bar{d}_{A}}(\lambda)+\left( 1-\frac{1}{\varsigma_{o}-\varsigma}\dzf(\lambda)\right) \right] \\
& \quad + \frac{1}{2}\left[ \frac{\das}{\bar{d}_{A}}(\lambda) +\left( 1-\frac{1}{\varsigma_{o}-\varsigma}\dzs(\lambda)\right)  + \frac{\H'-\H^{2}}{\H^{3}(\varsigma_{o}-\varsigma)}\left( \dzf \right)^{2} \right. \notag \\
& \quad \left. +2\left( 1 - \frac{1}{\H(\varsigma_{o}-\varsigma)}\right) \frac{\daf}{\bar{d}_{A}}\dzf \right] \Bigg\}. \notag
\end{align}
Up until here we have corrected the scale factor from the affine parameter $\lambda$ to $\varsigma$. Now we need to convert the first order contributions because they give additional second order contributions. We introduce that, for a general first order quantity $\delta^{(1)}X$, converting to $\varsigma$ gives
\be
\delta^{(1)}X(\lambda) = \delta^{(1)}X(\varsigma) + \frac{\partial \delta^{(1)} X}{\partial \lambda}\Big|_{\varsigma} \frac{\dzf (\varsigma)}{\H},
\ee
where $\delta^{(1)}X(\varsigma)$ is to be understood as substituting $\varsigma$ in the expression for $\delta^{(1)}X(\lambda)$, i.e. $\delta^{(1)} X(\lambda \to \varsigma)$. For $\partial_{\lambda}\delta^{(1)}X|_{\varsigma}$ we are multiplying for a first order quantity, so it is taken to be a derivative evaluated in the background. Thus, we can write $\dd_{\varsigma}\delta^{(1)}X$.
With this, Eq.~\eqref{eq:danu1} finally becomes
\begin{align}
\label{eq:da-sigma}
\hat{d}_{A}(\hat{z}) &= \frac{\varsigma_{o}-\varsigma}{1+\hat{z}} \Bigg\{ 1 + \left[ \frac{\daf}{\bar{d}_{A}} + \left( 1-\frac{1}{\H(\varsigma_{o}-\varsigma)} \right)\dzf \right] \\
&\quad + \frac{1}{2}\left[ \frac{\das}{\bar{d}_{A}} + \left( 1-\frac{1}{\H(\varsigma_{o}-\varsigma)} \right)\dzs  \right. \notag \\
&\quad  +2\left( 1-\frac{1}{\H(\varsigma_{o}-\varsigma)} \right) \left( \frac{\daf}{\bar{d}_{A}} + \frac{1}{\H}\frac{\dd \dzf}{\dd \varsigma} \right)\dzf \notag \\
&\quad\left.  + 2\frac{\dd}{\dd \varsigma}\left(\frac{\daf}{\bar{d}_{A}}\right) \frac{\dzf}{\H} +\frac{\H'-\H^{2}}{\H^{3}(\varsigma_{o}-\varsigma)} \left( \dzf \right)^{2} \right]\Bigg\}. \notag
\end{align}
We can now write the diameter distance as a function of the redshift, although it is written in terms of integrals over the comoving distance $\chi = \varsigma_{o}-\varsigma$, which depends on the redshift itself by Eq.~\eqref{eq:comd}.

Using Eq.~\eqref{eq:da-sigma} written in terms of observable quantities such as the observable redshift and comoving distance, the angular diameter distance will, in principle, be measured with great accuracy by the upcoming surveys mentioned in Chapter \ref{chapter:intro}, and should complement to the known luminosity distance measurements quite well. Alternatively, it can be use to compute the spherical harmonic expansion of $\hat{d}_{A}(\hat{z})$, which, then gives the deviation from the background geometry expected in the standard model. This latter `backreaction' effect has been postulated to be both large and small, and is very important to calculate accurately, in order to correctly determine the background geometry and parameters \cite{cc2}.

Combining Eqs.~\eqref{eq:dz1}, \eqref{eq:da-noan} with \eqref{eq:da-sigma}, we have that at linear order the diameter distance as a function of redshift is given by,
\begin{align}
\label{eq:dazf}
\delta^{(1)}\hat{d}_{A}(\hat{z}_{s}) &= \frac{\chi_{s}}{1+\hat{z}_{s}}\Bigg\{ -\pf|_{s} - \pf|_{o} - \left( 1 - \frac{1}{\H \chi_{s}}\right)\ff|_{s} + \left( 2-\frac{2}{\H \chi_{s}} \right)\left( v_{1i}n^{i}\right)_{o} \notag \\
& \quad\quad + \left( 1-\frac{1}{\H \chi_{s}}\right)\left( v_{1i}n^{i}\right)_{s} + \left( 1 - \frac{1}{\H\chi_{s}}\right)\int_{0}^{\chi_{s}}\left( \ff'+\pf'\right)\dd\chi -\frac{2}{\chi_{s}}\int_{0}^{\chi_{s}}\pf\dd\chi \notag \\
& \quad \quad -\frac{1}{2 \chi_{s}} \int_{0}^{\chi_{s}}\dd\chi \left( \chi - \chi_{s} \right)\chi \left[ \nabla^{2}\left( \ff + \pf\right) - n^{i}n^{j}\left( \ff + \pf \right)_{,ij} - \frac{2}{\chi} \frac{\dd \dnuf}{\dd \varsigma} \right] \Bigg\}. 
\end{align}

The full expression for $\delta^{(2)}\hat{d}_{A}(\hat{z}_{s})$ in terms of the metric potentials is given in Appendix \ref{metriccomp}.

\break
\subsection{Physical Volume}

The area distance the light-ray bundle creates, changes along the line of sight as seen in Fig.~\ref{fig:screen}, and we are interested in computing the volume that these hypersurfaces enclose, since therein lie the overdensities we are accounting for.

The volume element \eqref{eq:dldo} can be rewritten in terms of the quantities we have computed in the previous sections; the angular diameter distance in \eqs{eq:da-noan} and \eqref{eq:da-second-metric}, and the energy in \eqs{eq:dE1} and \eqref{eq:dE2}. It is given up to second order by,
\begin{align}
\label{eq:volume}
\dd V &= -\E d_{A}^{2}(\lambda) \dd \lambda \dd \Omega,  \\
&= - \bar{\E}\bar{d}_{A}^{2} \Bigg[ 1+ 2 \frac{\daf}{\bar{d}_{A}}+\frac{\Ef}{\bar{\E}} \notag \\ 
&\qquad\qquad\qquad\qquad+ \left(\frac{\daf}{\bar{d}_{A}}\right)^{2} + \left( \frac{\Ef}{\bar{\E}}\right) \left( \frac{\daf}{\bar{d}_{A}}\right) +\frac{\das}{\bar{d}_{A}} +\frac{1}{2}\frac{\Es}{\bar{\E}} \Bigg]\dd \lambda \dd \Omega. \notag
\end{align}
The volume element is given in terms of the affine parameter $\lambda$, but we need to express our result in terms of the observed redshift $z$, and so we need to take the volume in bins of $\dd z$ instead of $\dd \lambda$. To do so we use the fact that we can write the affine parameter as a function of redshift, i.e.~$\lambda(z)$, and using \eqs{eq:pertl}, \eqref{eq:pertlz1} and \eqref{eq:pertlz2}, we obtain the relation
\begin{align}
\frac{\dd\lambda}{\dd z} &= -\frac{a}{\H}\left[ 1 + \left(\frac{1}{\H}+\frac{1}{\H\left( 1+\bar{z}\right)}\right) \frac{\dd \dzf}{\dd \lambda} - \frac{\H'}{\H^{2}} \dzf\right] \\
&\qquad -\frac{a}{2\H}\Bigg[ \left(\frac{1}{\H}+\frac{1}{\H\left( 1+\bar{z}\right)}\right) \frac{\dd\dzs}{\dd\lambda} -\frac{\H'}{\H^{2}}\dzs -\frac{\H'}{\H^{3}\left( 1+\bar{z} \right)}\left( \frac{\dd\dzf}{\dd\lambda}\right)\dzf\notag \\
&\qquad +\frac{1}{\H^{2}\left( 1+\bar{z}\right)}\left( 2+\frac{1}{1+\bar{z}}  \right)\left( \frac{\dd\dzf}{\dd\lambda} \right)^{2} + \left( \frac{\H'}{\H^{2}} \right)\left( 1+\frac{\H'}{\H^{2}} \right)\left( \dzf \right)^{2} \notag \\
&\qquad -\frac{1}{\H} \left( 2 \frac{\left( \H' \right)^{2}}{\H^{2}} - \frac{\H''}{\H}\right)\left( \dzf \right)^{2} - \frac{1}{\H}\left( 1+\frac{\H'}{\H^{2}}\right)\frac{\dd}{\dd \lambda}\left[\left( \dzf \right)^{2}\right] \Bigg],\notag 
\end{align}
modifying Eq.~\eqref{eq:volume} into
\begin{align}
\label{eq:volumedz}
\dd V(z) &= -\E(z) d_{A}^{2}(z) \left(\frac{\dd \lambda}{\dd z}\right) \dd z \dd \Omega, \notag \\
&= \frac{\bar{\E}\bar{d}_{A}^{2}}{\H (1+z)} \Bigg[ 1+ \frac{2}{\H}\frac{\dd \dzf}{\dd \lambda} - \frac{\H'}{\H^{2}}\dzf +2 \frac{\daf}{\bar{d}_{A}}+\frac{\Ef}{\bar{\E}} \\
&\qquad - \frac{1}{\H} \frac{\dd\dzs}{\dd\lambda} +\frac{1}{2}\frac{\H'}{\H^{2}}\dzs -\frac{3}{2\H^{2}}\left( \frac{\dd\dzf}{\dd\lambda} \right)^{2} - \frac{1}{2}\left( \frac{\H'}{\H^{2}} \right)\left( 1+\frac{\H'}{\H^{2}} \right)\left( \dzf \right)^{2} \notag \\
&\qquad +\frac{1}{2\H^{2}} \left( 2 \frac{\left( \H' \right)^{2}}{\H^{2}} - \frac{\H''}{\H}\right)\left( \dzf \right)^{2} + \frac{1}{2\H}\left( 1+2\frac{\H'}{\H^{2}}\right)\left( \frac{\dd\dzf}{\dd\lambda}\right)\dzf\notag  \\
& \qquad  + \left(\frac{\daf}{\bar{d}_{A}}\right)^{2} + \left( \frac{\Ef}{\bar{\E}}\right) \left( \frac{\daf}{\bar{d}_{A}}\right) +\frac{\das}{\bar{d}_{A}} +\frac{1}{2}\frac{\Es}{\bar{\E}} \Bigg]\dd z \dd \Omega. \notag
\end{align}

We now give an expression for the volume element order by order. In the background we have
\begin{align}
\label{eq:bgV}
\dd\bar{V} &= - \bar{\E} \bar{d}_{A}^{2} \dd \lambda \dd \Omega = a^{2}\left(\lambda_{s}\right) [\lambda_{s}-\lambda_{o}]^{2}\dd \lambda \dd \Omega, \\
&= \frac{\bar{\E}\bar{d}_{A}^{2}}{\H(1+z)} \dd z \dd \Omega= \frac{\chi^{2}}{\H(1+z)} \dd z \dd \Omega.
\end{align}

From \eq{eq:volumedz} and using \eqs{eq:dE1-metric} and \eqref{eq:dazf}, we have that the first order perturbation to the physical volume is
\begin{align}
\label{eq:dvv1}
\dd \dVf &= \frac{\bar{\E} \bar{d}^{2}_{A}}{\H(1+z)}\Bigg[ \frac{2}{\H}\frac{\dd \dzf}{\dd \varsigma} - \frac{\H'}{\H^{2}}\dzf + 2 \frac{\daf}{\bar{d}_{A}} + \frac{\delta^{(1)}\E}{\bar{\E}} \Bigg] \dd z \dd \Omega, \notag \\
&= \frac{\chi^{2}}{\H(1+z)}\Bigg[ \frac{2}{\H}\left( \pf' - \partial_{\chi}\ff + \frac{\dd \left( v_{1i}n^{i}\right)}{\dd \varsigma}\right) - 2\left( \ff + \pf \right) -3 \left( v_{1i}n^{i}\right) \notag \\
&\quad +\left( \frac{\H'}{\H^{2}} + \frac{2}{\H \chi} \right)\left( \ff - \left( v_{1i}n^{i} \right) + \int_{0}^{\chi}\dd \tilde{\chi}\left( \ff' + \pf'\right) \right) - \frac{4}{\chi}\int_{0}^{\chi}\dd\tilde{\chi}\pf \notag \\ 
&\quad  -\frac{1}{\chi}\int_{0}^{\chi}\dd \tilde{\chi}\left( \tilde{\chi}-\chi \right)\tilde{\chi} \left\{ \nabla^{2}\left( \ff + \pf \right) - n^{i}n^{j}\left( \ff + \pf\right)_{,ij} - \frac{2}{\tilde{\chi}} \frac{\dd \dnuf}{\dd \varsigma} \right\} \notag \\ 
&\quad + 3\int_{0}^{\chi}\dd \tilde{\chi}\left( \ff' + \pf' \right)\Bigg] \dd z \dd \Omega,
\end{align}
and using \eqs{eq:dE2-metric} and \eqref{eq:da-second-metric} we find that the second order perturbation to the physical volume is 
\begin{align}
\label{eq:dvv2}
\dd \dVs &= \frac{\bar{\E}\bar{d}_{A}^{2}}{\H (1+z)} \Bigg[- \frac{1}{\H} \frac{\dd\dzs}{\dd\lambda} +\frac{1}{2}\frac{\H'}{\H^{2}}\dzs -\frac{3}{2\H^{2}}\left( \frac{\dd\dzf}{\dd\lambda} \right)^{2} - \frac{1}{2}\left( \frac{\H'}{\H^{2}} \right)\left( 1+\frac{\H'}{\H^{2}} \right)\left( \dzf \right)^{2} \notag \\
&\qquad +\frac{1}{2\H^{2}} \left( 2 \frac{\left( \H' \right)^{2}}{\H^{2}} - \frac{\H''}{\H}\right)\left( \dzf \right)^{2} + \frac{1}{2\H}\left( 1+2\frac{\H'}{\H^{2}}\right)\left( \frac{\dd\dzf}{\dd\lambda}\right)\dzf\notag  \\
& \qquad  + \left(\frac{\daf}{\bar{d}_{A}}\right)^{2} + \left( \frac{\Ef}{\bar{\E}}\right) \left( \frac{\daf}{\bar{d}_{A}}\right) +\frac{\das}{\bar{d}_{A}} +\frac{1}{2}\frac{\Es}{\bar{\E}} \Bigg]\dd z \dd \Omega. 
\end{align}
and the full expression in terms of the metric potentials is given in Appendix \ref{metriccomp}. With this expansion at hand, we have all the necessary quantities to compute our main result of the galaxy number density up to second order in the next section.

\section{Galaxy number density}
\label{sec:counts}
%
%

In this section, we compute our main result; the galaxy number overdensity at second order. Here, $V(n^{i},z)$ is the physical survey volume density per redshift bin per solid angle given by (\ref{eq:dldo}), where $n^{i}$ is the direction of observation and $z=z(\lambda_{s})$. The volume is a perturbed quantity since the solid angle of observation as well as the redshift bin are distorted between the source and the observer
\be
V(n^{i},z) = \bar{V}(z)+\dVf(n^{i},z)+\frac{1}{2}\dVs(n^{i},z).
\ee 
In \eqs{eq:dvv1} and \eqref{eq:dvv2} we provide the first and second order perturbations to the volume, respectively. \textbf{Note.} We use $\delta (\dd  V) /\dd \bar{V}$ where other authors in the literature use $\delta V / \bar{V}$ (see, e.g. Ref~\cite{durrer1}).

In a galaxy redshift survey, we measure the number of galaxies in direction $n^{i}$ at redshift $z$. Let us call this $N(n^{i},z) \dd \Omega_{n}\dd z$, where $\dd \Omega_{n}$ is the solid angle the survey spans. Then one must average over the angles to obtain their redshift distribution, $\langle N \rangle(z)\dd z$, where the angle brackets correspond to this angular average \cite{campos}
\be
\langle N\rangle(z) \dd z = \dd z \int_{\Omega_{n}} N(n^{i},z)\dd \Omega,
\ee
where the integral is over the solid angle the survey spans.

We can then build the matter density perturbation, number density contrast, in redshift space, i.e. the perturbation variable \cite{durrer1}
\be
\delta_{z}(n^{i}, z) \equiv \frac{\rho(n^{i},z)-\langle \rho \rangle(z)}{\langle \rho \rangle(z)}.
\ee
and expand it up to second order as
\begin{align}
\delta_{z}(n^{i}, z) = \delta^{(1)}_{z}(n^{i},z) + \frac{1}{2}\delta^{(2)}_{z}(n^{i},z).
\end{align}

Our aim in this chapter is to compute the observed matter number density perturbation since the density of sources is proportional to the number of the sources within a given volume, i.e.
\be
\label{eq:rho}
\rho(n^{i},z) = \frac{N(n^{i},z)}{V(n^{i},z)},
\ee
and expanding \eq{eq:rho} we show that at any order
\begin{equation}
\delta_{z}(n^{i}, z) = \frac{N(n^{i},z)-\langle N \rangle(z)}{\langle N \rangle(z)} - \frac{\delta V(n^{i}, z)}{V(z)}.
\end{equation}

The observed quantity is the perturbation in the number density of galaxies, $\Delta$, and it is defined as
\begin{equation}
\label{eq:ng}
\Delta(n^{i},z) \equiv \frac{N(n^{i},z)-\langle N \rangle(z)}{\langle N \rangle(z)} =\delta_{z}(n^{i}, z) + \frac{\delta V(n^{i}, z)}{V(z)},
\end{equation}
and so we have that
\begin{align}
\label{eq:defdg1}
\Delta_{g}^{(1)}(n^{i},z) &= \delta_{z}^{(1)}(n^{i},z) + \frac{\dVf(n^{i},z)}{\bar{V}(z)}, \\
\label{eq:defdg2}
\Delta_{g}^{(2)} (n^{i},z)&= \delta_{z}^{(2)}(n^{i},z) + \frac{\dVs(n^{i},z)}{\bar{V}(z)} + \delta_{z}^{(1)}(n^{i},z)\frac{\dVf(n^{i},z)}{\bar{V}(z)}.
\end{align}

In order to compute the above, let us first relate $\delta_{z}(n^{i},z)$ to the matter density quantity $\delta(x^{i},\eta)$ and the perturbations on the redshift computed in Section \ref{sec:whatwemeasure}. The redshift density up to second order in redshift space is
\begin{align}
\label{eq:redshift-density}
\delta_{z}(n^{i},z) &= \frac{\rho(n^{i},z)-\bar{\rho}(z)}{\bar{\rho}(z)} = \frac{\bar{\rho}(z)+\drf(n^{i},z)+\frac{1}{2}\drs(n^{i},z)-\bar{\rho}(z)}{\bar{\rho}(z)} \\
&= \frac{\bar{\rho}(\bar{z}+\dzf+\frac{1}{2}\dzs)+\drf(n^{i},z)+\frac{1}{2}\drs(n^{i},z)-\bar{\rho}(z)}{\bar{\rho}(z)} \notag\\
&= \frac{\drf(n^{i},z)}{\bar{\rho}(z)} + \frac{\dd \bar{\rho}}{\dd \bar{z}} \frac{\dzf(n^{i},z)}{\bar{\rho}(\bar{z})} \notag \\
&\qquad +\frac{1}{2}\frac{\drs(n^{i},z)}{\bar{\rho}(z)} + \frac{1}{2}\frac{\dd \bar{\rho}}{\dd \bar{z}} \frac{\dzs(n^{i},z)}{\bar{\rho}(\bar{z})}+\frac{1}{2}\frac{\dd^{2} \bar{\rho}}{\dd \bar{z}^{2}} \frac{\left[\dzf(n^{i},z)\right]^{2}}{\bar{\rho}(\bar{z})} + \frac{\dd \drf}{\dd \bar{z}}\frac{\dzf(n^{i},z)}{\bar{\rho}(\bar{z})}.\notag 
\end{align}
Structure in the universe is formed mainly from dark matter and this pressureless component evolves with redshift as
\be
\label{eq:bg-rho}
\bar{\rho}(z)\approx \rho_{0}(1+z)^{3}.
\ee
Thus we have that
\be
\label{eq:der1-rho}
\frac{\dd \bar{\rho}}{\dd \bar{z}} = 3 \frac{\bar{\rho}}{1+\bar{z}},
\ee
so using \eq{eq:dz1}, the redshift density perturbation at first order is given by
\be
\label{eq:ddz1}
\delta^{(1)}_{z}(n^{i},z) = \frac{\drf(n^{i},z)}{\bar{\rho}(z)} + \frac{3}{1+\bar{z}}\left[\left( v_{1i}n^{i} + \ff\right)\big|^{s}_{o} + \int_{0}^{\chi_{s}}\dd\chi  \left\{\ff' +\pf'\right\}\right].
\ee

Combining (\ref{eq:dvv1})
and (\ref{eq:ddz1}) we find that the galaxy number density fluctuation in redshift space as defined in Eq.~(\ref{eq:defdg1}) is, at first order, 
\begin{align}
\label{eq:Dg1}
\Delta^{(1)}_{g}(n^{i},z) &= \left[\frac{\drf(n^{i},z)}{\bar{\rho}(z)} + 3\ff\right] -2\left(\ff + \pf\right) + \frac{1}{\H}\left( \pf' - \partial_{\chi}\ff + \frac{\dd \left(v_{1i}n^{i}\right)}{\dd \varsigma} \right) \\
& \quad +\left( \frac{\H'}{\H^{2}} +\frac{2}{\H \chi}\right)\left[ \ff - \left( v_{1i}n^{i}\right) + \int_{0}^{\chi}\dd\tilde{\chi}\left( \ff' + \pf' \right)\right] - \frac{4}{\chi}\int_{0}^{\chi}\dd \tilde{\chi}\pf\notag \\
& \quad - \frac{1}{\chi}\int_{0}^{\chi}\dd \tilde{\chi}\left( \tilde{\chi}-\chi\right)\tilde{\chi}\left[ \nabla^{2}\left( \ff + \pf\right) - n^{i}n^{j}\left( \ff + \pf \right)_{,ij} - \frac{2}{\tilde{\chi}}\frac{\dd \dnuf}{\dd \varsigma} \right]. \notag
\end{align}
From \eq{eq:bg-rho}, we have that the second derivative of the background density is
\be
\label{eq:der2-rho}
\frac{\dd^{2} \bar{\rho}}{\dd \bar{z}^{2}} = 6 \frac{\bar{\rho}}{(1+\bar{z})^{2}},
\ee  
so using \eqs{eq:dz1}, \eqref{eq:dz2} and \eqref{eq:der2-rho} in \eq{eq:redshift-density}  we find that the redshift density perturbation at second order is given by
\begin{align}
\delta^{(2)}_{z}(n^{i},z) &= \frac{1}{2} \frac{\drs(n^{i},z)}{\bar{\rho}(\bar{z})} + \frac{3}{2(1+\bar{z})} \dzs(n^{i},z) \\
&\qquad\qquad\qquad+ \frac{3}{(1+\bar{z})^{2}}\left[ \dzf(n^{i},z) \right]^{2}+\frac{\dd\delta^{(1)}\rho}{\dd\bar{z}}\frac{\dzf\left(n^{i},z\right)}{\bar{\rho}}, \notag
\end{align}
where the full expression in terms of the metric potentials is given in Appendix \ref{metriccomp}, Eq.~\eqref{eq:ddz2}. Finally, combining \eqs{eq:dvv2} and \eqref{eq:ddz2} we find the galaxy number density fluctuation at second order as defined in \eq{eq:defdg2} is 
\begin{align}
\label{eq:deltag2}
&\Delta^{(2)}_{g}(n^{i},z) =\frac{1}{2} \frac{\drs(n^{i},z)}{\bar{\rho(z)}} + \frac{3}{2}\Bigg[ -\frac{1}{2}\fs|^{s}_{o} - \frac{1}{2}\left( v_{2i} n^{i}\right)^{s}_{o}   \\
& + \frac{1}{2}\int_{0}^{\chi_{s}}\dd\chi \left( \fs' + \ps' \right)+\frac{1}{2}\left( v_{1k}v^{k}_{1}\right)^{s}_{o} - \frac{3}{2} \left( \ff|^{s}_{o}\right)^{2}+6 \ff|_{s}\ff|_{o} +\ff|_{o}\left( v_{1i}n^{i} \right)_{s} \notag \\
&-\ff|_{s}\left( v_{1i}n^{i} \right)_{o} -\left( v_{1i}n^{i} \right)_{o}\left( v_{1i}n^{i}\right)^{s}_{o}  - 2 \pf|_{o}\left( v_{1i}n^{i} \right)_{o}+ \ff|^{s}_{o}\int_{0}^{\chi_{s}}\dd\chi\left( \ff' + \pf' \right)  \notag \\
& + \left( v_{1i}n^{i} \right)_{o}\int_{0}^{\chi_{s}}\dd\chi\left( \ff' + \pf' \right)+ \ff \int_{0}^{\chi_{s}}\dd\chi\left( \ff' + \pf' \right) +  2v_{1 i}\int_{0}^{\chi_{s}}\dd \chi\left( {\ff_{,}}^{i} + {\pf_{,}}^{i} \right) \notag \\
& + 2\int_{0}^{\chi_{s}}\dd\chi\left[\pf' \left( \ff + \pf \right)\right] + 2\int_{0}^{\lambda_{s}}\dd\chi \left[ \ff \left( \frac{\dd \ff}{\dd \varsigma} - 2\frac{\dd \pf}{\dd \varsigma}\right)\right] \notag\\
& -4 \int_{0}^{\chi_{s}}\dd\chi\left[\left( \pf \ff_{,i} - \ff \pf_{,i} \right)n^{i}\right] + 4\int_{0}^{\chi_{s}}\dd\chi\left[ n^{i}\ff_{,i}\left( \ff - \pf \right)\right]  \notag \\
& + \int_{0}^{\chi_{s}}\dd\tilde{\chi}\Bigg\{2 \left( \ff' - \pf'\right)\int_{0}^{\chi_{s}}\dd\chi\left( \ff' + \pf' \right) - n^{i} \left( \ff_{,i} + \pf_{,i} \right) \int_{0}^{\chi_{s}}\dd\chi\left( \ff' + \pf'\right) \notag \\
& +  \left( \frac{\dd\ff}{\dd \varsigma} - \frac{\dd \pf}{\dd \varsigma}\right)\int_{0}^{\chi_{s}}\dd\chi \left( \ff' + \pf' \right)+2\left( \ff-\pf\right)n^{i}\int_{0}^{\chi_{s}}\dd\chi\left( \ff' + \pf'\right)_{,i} \notag \\
& +\left( \int_{0}^{\chi_{s}}\dd\chi{\left( \ff + \pf\right)_{,}}^{i}\right)\left( \int_{0}^{\chi_{s}}\dd\chi\left( \ff' + \pf'\right)_{,i}\right)\notag \\
& - n^{i}\left( \int_{0}^{\chi_{s}}\dd\chi\left( \ff' + \pf' \right)_{,i} \right)\left( \int_{0}^{\chi_{s}}\dd\chi\left( \ff' + \pf' \right) \right)\Bigg\}\Bigg]\notag \\
&+3\Bigg[ -\left( v_{1i}n^{i} + \ff\right)\big|^{s}_{o} + \int_{0}^{\chi_{s}}  \left[\ff' +\pf'\right]\dd\chi \Bigg]^{2}\notag \\
&+\frac{\dd\delta^{(1)}\rho}{\dd\bar{z}}\frac{1}{\bar{\rho}}\Bigg[ -\left( v_{1i}n^{i} + \ff\right)\big|^{s}_{o} + \int_{0}^{\chi_{s}}  \left[\ff' +\pf'\right]\dd\chi  \Bigg] \notag \\
&+\frac{1}{\H}\Bigg[\frac{1}{2}\left( \ps' -\partial_{\chi}\fs - \frac{\dd\left( v_{2i} n^{i}\right)}{\dd\varsigma}+\frac{\dd\left( v_{1k}v^{k}_{1}\right)}{\dd\varsigma}\right) \notag 
\end{align}
\begin{align}
& - 3\ff \left( \frac{\dd\ff}{\dd\varsigma}\right)+\frac{\dd\ff}{\dd\varsigma}\left( v_{1i}n^{i} \right)+\ff\frac{\dd\left( v_{1i}n^{i} \right)}{\dd\varsigma}-\left( v_{1i}n^{i} \right)\frac{\dd\left( v_{1i}n^{i}\right)}{\dd\varsigma} - 2 \frac{\dd\pf}{\dd\varsigma}\left( v_{1i}n^{i} \right) \notag\\ 
&-2\pf\frac{\dd\left( v_{1i}n^{i} \right)}{\dd\varsigma} + \frac{\dd\ff}{\dd\varsigma}\int_{0}^{\chi_{s}}\dd\chi\left( \ff' + \pf' \right) + \ff\left( \ff'+\pf'\right)\notag \\
& + \frac{\dd\left( v_{1i}n^{i} \right)}{\dd\varsigma}\int_{0}^{\chi_{s}}\dd\chi\left( \ff' + \pf' \right)+\left(v_{1i}n^{i}\right)\left( \ff'+\pf'\right) \notag \\
& + \frac{\dd\ff}{\dd\varsigma} \int_{0}^{\chi_{s}}\dd\chi\left( \ff' + \pf' \right)+\ff\left( \ff'+\pf' \right)+  2\frac{\dd v_{1 i}}{\dd\varsigma}\int_{0}^{\chi{s}}\dd \chi\left( {\ff_{,}}^{i} + {\pf_{,}}^{i} \right) + 2v_{1i}\left( {\ff_{,}}^{i} + {\pf_{,}}^{i} \right)   \notag \\
& + 2\left[\pf' \left( \ff + \pf \right)\right] + 2 \left[ \ff \left( \frac{\dd \ff}{\dd \varsigma} - 2\frac{\dd \pf}{\dd \varsigma}\right)\right]-4 \left[\left( \pf \ff_{,i} - \ff \pf_{,i} \right)n^{i}\right] \notag\\
&+ 4\left[ n^{i}\ff_{,i}\left( \ff - \pf \right)\right]+2 \left( \ff' - \pf'\right)\int_{0}^{\chi_{s}}\dd\chi\left( \ff' + \pf' \right) - n^{i} \left( \ff_{,i} + \pf_{,i} \right) \int_{0}^{\chi_{s}}\dd\chi\left( \ff' + \pf'\right) \notag \\
&  +  \left( \frac{\dd\ff}{\dd \varsigma} - \frac{\dd \pf}{\dd \varsigma}\right)\int_{0}^{\chi_{s}}\dd\chi \left( \ff' + \pf' \right)+2\left( \ff-\pf\right)n^{i}\int_{0}^{\chi_{s}}\dd\chi\left( \ff' + \pf'\right)_{,i} \notag \\
& +\left( \int_{0}^{\chi_{s}}\dd\chi{\left( \ff + \pf\right)_{,}}^{i}\right)\left( \int_{0}^{\chi_{s}}\dd\chi\left( \ff' + \pf'\right)_{,i}\right)\notag \\
& - n^{i}\left( \int_{0}^{\chi_{s}}\dd\chi\left( \ff' + \pf' \right)_{,i} \right)\left( \int_{0}^{\chi_{s}}\dd\chi\left( \ff' + \pf' \right) \right) \Bigg]\notag\\
&-\frac{1}{2}\frac{\H'}{\H^{2}}\Bigg[ -\frac{1}{2}\fs|^{s}_{o} - \frac{1}{2}\left( v_{2i} n^{i}\right)^{s}_{o} + \frac{1}{2}\int_{0}^{\chi_{s}}\dd\chi \left( \fs' + \ps' \right) +\frac{1}{2}\left( v_{1k}v^{k}_{1}\right)^{s}_{o} \notag  \\
& - \frac{3}{2} \left( \ff|^{s}_{o}\right)^{2}+6 \ff|_{s}\ff|_{o} +\ff|_{o}\left( v_{1i}n^{i} \right)_{s} -\ff|_{s}\left( v_{1i}n^{i} \right)_{o} -\left( v_{1i}n^{i} \right)_{o}\left( v_{1i}n^{i}\right)^{s}_{o}  \notag \\
& - 2 \pf|_{o}\left( v_{1i}n^{i} \right)_{o}+ \ff|^{s}_{o}\int_{0}^{\chi_{s}}\dd\chi\left( \ff' + \pf' \right) + \left( v_{1i}n^{i} \right)_{o}\int_{0}^{\chi_{s}}\dd\chi\left( \ff' + \pf' \right) \notag \\
& + \ff \int_{0}^{\chi_{s}}\dd\chi\left( \ff' + \pf' \right) +  2v_{1 i}\int_{0}^{\chi_{s}}\dd \chi\left( {\ff_{,}}^{i} + {\pf_{,}}^{i} \right) \notag \\
& + 2\int_{0}^{\chi_{s}}\dd\chi\left[\pf' \left( \ff + \pf \right)\right] + 2\int_{0}^{\lambda_{s}}\dd\chi \left[ \ff \left( \frac{\dd \ff}{\dd \varsigma} - 2\frac{\dd \pf}{\dd \varsigma}\right)\right] \notag\\
& -4 \int_{0}^{\chi_{s}}\dd\chi\left[\left( \pf \ff_{,i} - \ff \pf_{,i} \right)n^{i}\right] + 4\int_{0}^{\chi_{s}}\dd\chi\left[ n^{i}\ff_{,i}\left( \ff - \pf \right)\right]  \notag \\
& + \int_{0}^{\chi_{s}}\dd\tilde{\chi}\Bigg\{2 \left( \ff' - \pf'\right)\int_{0}^{\chi_{s}}\dd\chi\left( \ff' + \pf' \right) - n^{i} \left( \ff_{,i} + \pf_{,i} \right) \int_{0}^{\chi_{s}}\dd\chi\left( \ff' + \pf'\right) \notag \\
& +  \left( \frac{\dd\ff}{\dd \varsigma} - \frac{\dd \pf}{\dd \varsigma}\right)\int_{0}^{\chi_{s}}\dd\chi \left( \ff' + \pf' \right)+2\left( \ff-\pf\right)n^{i}\int_{0}^{\chi_{s}}\dd\chi\left( \ff' + \pf'\right)_{,i} \notag \\
& +\left( \int_{0}^{\chi_{s}}\dd\chi{\left( \ff + \pf\right)_{,}}^{i}\right)\left( \int_{0}^{\chi_{s}}\dd\chi\left( \ff' + \pf'\right)_{,i}\right)\notag \\
& - n^{i}\left( \int_{0}^{\chi_{s}}\dd\chi\left( \ff' + \pf' \right)_{,i} \right)\left( \int_{0}^{\chi_{s}}\dd\chi\left( \ff' + \pf' \right) \right)\Bigg\} \Bigg]+\frac{3}{2\H^{2}}\Bigg[ \pf' -\partial_{\chi}\ff -\frac{\dd\left( v_{1i}n^{i}\right)}{\dd\varsigma} \Bigg]^{2}\notag
\end{align}
\begin{align}
&+\frac{1}{2}\left( \frac{\H'}{\H^{2}} \right)\left( 1+\frac{\H'}{\H^{2}}\right)\Bigg[ \left(v_{1i}n^{i}+\ff+\int_{0}^{\chi_{s}}\left( \ff'+\pf'\right)\dd\chi\right) \Bigg]^{2}\notag\\
&+\frac{1}{2\H^{2}}\left(2\left[\frac{\H'}{\H}\right]^{2}+\frac{\H''}{\H} \right)\Bigg[ \left(v_{1i}n^{i}+\ff+\int_{0}^{\chi_{s}}\left( \ff'+\pf'\right)\dd\chi\right)  \Bigg]^{2} \notag \\
&+\frac{1}{2\H}\left( 1+2\frac{\H'}{\H^{2}} \right)\left( \pf' - \partial_{\chi}\ff - \frac{\dd\left( v_{1i}n^{i} \right)}{\dd\varsigma} \right)\Bigg[ v_{1i}n^{i}+\ff+\int_{0}^{\chi_{s}}\left( \ff'+\pf'\right)\dd\chi \Bigg] \notag \\
& +\Bigg[-\pf|_{s} - \pf|_{o} - \left( 1 - \frac{1}{\H \chi_{s}}\right)\ff|_{s} + \left( 2-\frac{2}{\H \chi_{s}} \right)\left( v_{1i}n^{i}\right)_{o} + \left( 1-\frac{1}{\H \chi_{s}}\right)\left( v_{1i}n^{i}\right)_{s} \notag\\
&+ \left( 1 - \frac{1}{\H\chi_{s}}\right)\int_{0}^{\chi_{s}}\left( \ff'+\pf'\right)\dd\chi -\frac{2}{\chi_{s}}\int_{0}^{\chi_{s}}\pf\dd\chi \notag \\
& -\frac{1}{2 \chi_{s}} \int_{0}^{\chi_{s}}\dd\chi \left( \chi - \chi_{s} \right)\chi \left[ \nabla^{2}\left( \ff + \pf\right) - n^{i}n^{j}\left( \ff + \pf \right)_{,ij} - \frac{2}{\chi} \frac{\dd \dnuf}{\dd \varsigma} \right]\Bigg]^{2} \notag\\
& +\Bigg[ -\ff|^{s}_{o}+\int_{0}^{\chi_{s}}\left( \ff'+\pf' \right)\dd\chi+\ff-\left( v_{1i}n^{i} \right) \Bigg]\Bigg[ -\pf|_{s} - \pf|_{o} - \left( 1 - \frac{1}{\H \chi_{s}}\right)\ff|_{s} \notag \\
& + \left( 2-\frac{2}{\H \chi_{s}} \right)\left( v_{1i}n^{i}\right)_{o}+ \left( 1-\frac{1}{\H \chi_{s}}\right)\left( v_{1i}n^{i}\right)_{s} + \left( 1 - \frac{1}{\H\chi_{s}}\right)\int_{0}^{\chi_{s}}\left( \ff'+\pf'\right)\dd\chi  \notag \\
& -\frac{2}{\chi_{s}}\int_{0}^{\chi_{s}}\pf\dd\chi -\frac{1}{2 \chi_{s}} \int_{0}^{\chi_{s}}\dd\chi \left( \chi - \chi_{s} \right)\chi \left[ \nabla^{2}\left( \ff + \pf\right) - n^{i}n^{j}\left( \ff + \pf \right)_{,ij} - \frac{2}{\chi} \frac{\dd \dnuf}{\dd \varsigma} \right] \Bigg] \notag\\
&+\frac{1}{2}\Bigg[ \dnus + \fs - \left( v_{2 i}n^{i}\right) + 2 \ff \dnuf - 2v_{1i} \dnf^{i} + \ff^{2} + \left(v_{1 k}v_{1}^{k}\right) + 4 \pf \left( v_{1i}n^{i}\right) \Bigg]\notag\\
&+\frac{\das}{\bar{d}_{A}} +\frac{\drf(n^{i},z)}{\bar{\rho}(z)} + 3\left[\left( v_{1i}n^{i} + \ff\right)\big|^{s}_{o} + \int_{0}^{\chi_{s}}\dd\chi  \left\{\ff' +\pf'\right\}\right]\times \notag \\
&\quad\Bigg[ \frac{2}{\H}\left( \pf' - \partial_{\chi}\ff + \frac{\dd \left( v_{1i}n^{i}\right)}{\dd \varsigma}\right) - 2\left( \ff + \pf \right) -3 \left( v_{1i}n^{i}\right) \notag \\
&\quad +\left( \frac{\H'}{\H^{2}} + \frac{2}{\H \chi} \right)\left( \ff - \left( v_{1i}n^{i} \right) + \int_{0}^{\chi}\dd \tilde{\chi}\left( \ff' + \pf'\right) \right) - \frac{4}{\chi}\int_{0}^{\chi}\dd\tilde{\chi}\pf \notag \\ 
&\quad  -\frac{1}{\chi}\int_{0}^{\chi}\dd \tilde{\chi}\left( \tilde{\chi}-\chi \right)\tilde{\chi} \left\{ \nabla^{2}\left( \ff + \pf \right) - n^{i}n^{j}\left( \ff + \pf\right)_{,ij} - \frac{2}{\tilde{\chi}} \frac{\dd \dnuf}{\dd \varsigma} \right\} \notag \\ 
&\quad + 3\int_{0}^{\chi}\dd \tilde{\chi}\left( \ff' + \pf' \right)\Bigg]. \notag
\end{align}
Which is the main result of this chapter. In the following section we make a comparison of our result with others in the literature \cite{cc1,cc2,durrer2}. The second-order effects that we derive, especially those involving integrals along the line of sight, may make a non-negligible contribution to the observed number counts. This will be important for removing potential biases on parameter estimation in precision cosmology with galaxy surveys. It will also be important for an accurate analysis of the `contamination' of primordial non-Gaussianity by relativistic projection effects \cite{cc1,cc2}. 

\section{Comparison with previous work}
\label{sec:comparison}

In this section we compare our linear result given in \eq{eq:Dg1} with those in the literature given in Refs.~\cite{cc1, cc2, yoo2} which also compute second order corrections and in particular with Di Dio, et al.~\cite{durrer2}. 
We compare to ensure that our result is correct, since, since the number counts are well established to linear order with Refs.~\cite{durrer1, antony}. In Section \ref{sec:secondcomparison} we perform a comparison of the leading terms of the second order expansion of the galaxy number counts.

Our result, as given in \eq{eq:ng} is
\begin{align}
\Delta^{(1)}_{g}(n^{i},z) &= \left[\frac{\drf(n^{i},z)}{\bar{\rho}(z)} + 3\ff\right] -2\left(\ff + \pf\right) + \frac{1}{\H}\left( \pf' - \partial_{\chi}\ff + \frac{\dd \left(v_{1i}n^{i}\right)}{\dd \varsigma} \right) \notag \\
& \quad +\left( \frac{\H'}{\H^{2}} +\frac{2}{\H \chi}\right)\left[ \ff - \left( v_{1i}n^{i}\right) + \int_{0}^{\chi}\dd\tilde{\chi}\left( \ff' + \pf' \right)\right] - \frac{4}{\chi}\int_{0}^{\chi}\dd \tilde{\chi}\pf\notag \\
& \quad - \frac{1}{\chi}\int_{0}^{\chi}\dd \tilde{\chi}\left( \tilde{\chi}-\chi\right)\tilde{\chi}\left[ \nabla^{2}\left( \ff + \pf\right) - n^{i}n^{j}\left( \ff + \pf \right)_{,ij} - \frac{2}{\tilde{\chi}}\frac{\dd \dnuf}{\dd \varsigma} \right]. 
\end{align}

\subsection{Di Dio, et al.}

Rewriting the result from Ref.~\cite{durrer2}, in Poisson gauge, allowing for anisotropic stress. At first order, Ref.~\cite{durrer2} have
\begin{align}
\label{eq:didio}
\Delta^{(1)}_{\text{Di Dio}}&= \delta^{(1)}_{\rho} + \left( \frac{2}{\H r}+\frac{\H'}{\H^{2}}\right)\left[ \left( v_{1i}n^{i}\right) + \psi^{I}-\psi^{A}+2\int_{\eta_{s}}^{\eta_{o}} \dd\eta' \partial_{\eta'}\psi^{I}\right] -\psi^{I} \\
&\quad +\frac{4}{r}\int_{\eta_{s}}^{\eta_{o}}\dd\eta' \psi^{I} - \frac{2}{r}\int_{\eta_{s}}^{\eta_{o}}\dd\eta'\frac{\eta'-\eta_{s}}{\eta_{o}-\eta'}\Delta_{2}\psi^{I} + \frac{1}{\H}\left[ \partial_{\eta}\psi^{I} + \partial_{r}\left( v_{1i}n^{i} \right) \right] \notag\\
&\quad -3\psi^{A} + \frac{1}{\H}\partial_{\eta}\psi^{A}, \notag
\end{align}
where ${\cal{H}} = a'(\eta)/a(\eta)$ is the Hubble parameter, the `s' denotes {\textit{source}} and the `o' denotes {\textit{observer}}, and
\be
\psi^{I} = \frac{\psi+\phi}{2}, \qquad \text{and}, \qquad \psi^{A}=\frac{\psi - \phi}{2},
\ee
which rewriting in our notation is
\begin{align}
\Delta^{(1)}_{\text{Di Dio}}&= \delta^{(1)}_{\rho} + \left( \frac{2}{\H r}+\frac{\H'}{\H^{2}}\right)\left[ \left( v_{1i}n^{i}\right) + \ff+\int_{\eta_{s}}^{\eta_{o}} \dd\eta' \left( \ff'+\pf' \right)\right] \\
&\quad+\left( 2\pf-\ff\right)+\frac{2}{r}\int_{\eta_{s}}^{\eta_{o}}\dd\eta' \left( \ff+\pf \right) - \frac{1}{r}\int_{\eta_{s}}^{\eta_{o}}\dd\eta'\frac{\eta'-\eta_{s}}{\eta_{o}-\eta'}\Delta_{2}\left( \ff+\pf\right) \notag \\
&\quad+ \frac{1}{\H}\left[ \pf' + \partial_{r}\left( v_{1i}n^{i} \right) \right] . \notag
\end{align}

Computing the difference between \eq{eq:didio} and \eq{eq:Dg1}, we have
\begin{align}
\Delta_{g}^{(1)}-\Delta_{\text{Di Dio}}^{(1)} &\approx \Bigg[\frac{\drf(n^{i},z)}{\bar{\rho}(z)} + 3\ff\Bigg] -\Bigg[\delta^{(1)}_{\rho}\Bigg]  \\
& \quad - \frac{1}{\chi}\int_{0}^{\chi}\dd \tilde{\chi}\left( \tilde{\chi}-\chi\right)\tilde{\chi}\left[ \nabla^{2}\left( \ff + \pf\right) - n^{i}n^{j}\left( \ff + \pf \right)_{,ij} \right] \notag \\
&\quad +\Bigg[ \frac{1}{r}\int_{\eta_{s}}^{\eta_{o}}\dd\eta'\frac{\eta'-\eta_{s}}{\eta_{o}-\eta'}\Delta_{2}\left( \ff+\pf\right) \Bigg] \notag
\end{align}
where the first line cancels out from the definition of the comoving density perturbation ($\delta_{\rho}^{(1)}$), and the last integral cancels out from the definition of the angular operator $\Delta_{2}$, both given in Ref.~\cite{durrer2}, and we find,
\begin{align}
\Delta_{g}^{(1)}-\Delta_{\text{Di Dio}}^{(1)} &=  0.
\end{align}

\subsection{Bertacca, et al.}

In Refs.~\cite{cc1,cc2}, their result is written in terms of ``cosmic rulers'' and it is given by
\be
\Delta_{\text{Bertacca}}^{(1)} = \delta_{g}^{(1)} +\frac{1}{2}\hat{g}_{\mu}^{\mu (1)} + b_{e} \Delta \ln a^{(1)} + \partial_{\parallel}\Delta x_{\parallel}^{(1)} + \frac{2}{\bar{\chi}}\Delta x_{\parallel}^{(1)} - 2\kappa^{(1)} + E_{\hat{0}}^{0(1)} + E_{\hat{0}}^{\parallel (1)},
\ee
which in Poisson gauge, translates into
\begin{align}
\label{eq:bertacca}
\Delta^{(1)}_{\text{Bertacca}}&= \frac{\drf(n^{i},z)}{\bar{\rho}(n^{i},z)}  - \left(  \frac{{\cal{H}}'}{{\cal{H}}^{2}}+\frac{2}{\bar{\chi} {\cal{H}}} \right)\left[ (v_{1i}n^{i} - \ff)_{o}^{s} - 2 \int_{0}^{\bar{\chi}}\ff'\dd\tilde{\chi} \right] \\
&\qquad -\ff + \frac{\ff'}{{\cal{H}}} + \frac{4}{\bar{\chi}}\int_{0}^{\bar{\chi}}\ff \dd\tilde{\chi}-\frac{1}{{\cal{H}}}\frac{\dd}{\dd \chi} (v_{1i}n^{i})- \frac{1}{{\cal{H}}}\frac{\dd\ff}{\dd \chi} \notag\\
&\qquad-2\int_{0}^{\bar{\chi}}\dd\tilde{\chi}(\bar{\chi}-\tilde{\chi})\frac{\tilde{\chi}}{\bar{\chi}}\Big[\nabla^{2}\ff + \frac{\dd^{2} \ff}{\dd \tilde{\chi}^{2}} + \ff'' - 2 \frac{\dd \ff'}{\dd \tilde{\chi}} - \frac{2}{\bar{\chi}}\left(\frac{\dd \ff}{\dd \tilde{\chi}} - \ff' \right)\Big], \notag
\end{align}
where we omitted the terms with the evolution bias $b_{e}$. We must rewrite our own result taking $\pf = \ff$ in \eq{eq:Dg1} to make the comparison, so we have that
\begin{align}
\Delta^{(1)}_{g}(n^{i},z) &= \left[\frac{\drf(n^{i},z)}{\bar{\rho}(z)} + 3\ff\right] -4\ff + \frac{1}{\H}\left( \ff' - \partial_{\chi}\ff + \frac{\dd \left(v_{1i}n^{i}\right)}{\dd \varsigma} \right) \\
& \quad +\left( \frac{\H'}{\H^{2}} +\frac{2}{\H \chi}\right)\left[ \ff - \left( v_{1i}n^{i}\right) + 2\int_{0}^{\chi}\dd\tilde{\chi} \ff' \right] - \frac{4}{\chi}\int_{0}^{\chi}\dd \tilde{\chi}\ff\notag \\
& \quad - \frac{2}{\chi}\int_{0}^{\chi}\dd \tilde{\chi}\left( \tilde{\chi}-\chi\right)\tilde{\chi}\left[ \nabla^{2}\ff - n^{i}n^{j} \ff_{,ij} - \frac{1}{\tilde{\chi}}\frac{\dd \dnuf}{\dd \varsigma} \right]. \notag
\end{align}

Computing the difference between \eq{eq:bertacca} and \eq{eq:Dg1}, we have
\begin{align}
\Delta_{g}^{(1)} - \Delta_{\text{Bertacca}}^{(1)} &\approx \frac{1}{\H}\left( \frac{\dd \left(v_{1i}n^{i}\right)}{\dd \varsigma} \right)+\frac{1}{{\cal{H}}}\frac{\dd}{\dd \chi} (v_{1i}n^{i}),
\end{align}
where both are derivatives of first order terms with respect to background quantities, and in the background $\dd \chi = \dd\varsigma$. Note that the direction in the sky, $n^{i}$, is $\left(-n^{i}\right)$ from Refs.~\cite{cc1,cc2}, so
\be
\Delta_{g}^{(1)}-\Delta_{\text{Bertacca}}^{(1)} =0.
\ee

\subsection{Yoo \& Zaldarriaga}
The galaxy overdensity in \cite{yoo2}, is given by
\be
\delta_{g}^{\text{obs}(1)} = \delta_{g}^{\text{int}(1)} +3\delta z + \delta g + 2 \frac{\delta r}{\bar{r}_{z}} - 2 \kappa + H_{z}\frac{\partial \delta r}{\partial z} + \delta u^{0} + V_{\parallel} - e_{1} \delta z_{t_{p}} - t_{1} \delta \mathcal{D}_{L},
\ee

which in Poisson gauge, allowing for anisotropic stress takes the form
\begin{align}
\label{eq:yoo}
\Delta^{(1)}_{\text{Yoo}}&= \delta_{g}^{\text{int}(1)} + 3{\cal{H}}_{o}\delta\tau_{o} -3\ff|^{z}_{o} - 3\int_{0}^{\bar{r}_{z}}\dd \bar{r} \left( \ff' + \pf'\right) + 3\ff + 3v_{1i}n^{i}+\ff + 3\pf \notag\\
&\qquad +\frac{2}{r_{z}}\Bigg[ \delta\tau_{o} - \frac{1}{{\cal{H}}_{z}}\Big( {\cal{H}}_{o}\delta\tau_{o} -\ff|^{z}_{o}-\int_{0}^{\bar{r}_{z}}\dd \bar{r}\left( \ff' + \pf'\right) + (v_{1i}n^{i})_{o}^{z}\Big)\notag \\
&\qquad +\int_{0}^{\bar{r}_{z}}\left( \ff - \pf\right)\dd \bar{r} - 2 \kappa + H_{z}\frac{\p}{\p z}\Big( \delta\tau_{o} - \frac{1}{{\cal{H}}_{z}} \Big\{ {\cal{H}}_{o}\delta\tau_{o} -\ff|^{z}_{o}-\int_{0}^{\bar{r}_{z}}\dd \bar{r}\left( \ff'+\pf'\right) \notag \\
&\qquad  +(v_{1i}n^{i})^{z}_{o}\Big\}+\int_{0}^{\bar{r}_{z}}\dd \bar{r}\left( \ff-\pf\right)  \Big)-\ff+v_{1i}n^{i}\Bigg], 
\end{align}
where we did not use the evolution bias or the running and slope of the luminosity.

The difference between \eq{eq:yoo} and \eq{eq:Dg1} is then,
\begin{align}
\Delta_{g}^{(1)}-\Delta_{\text{Yoo}}^{(1)} &\approx \left[\frac{\drf(n^{i},z)}{\bar{\rho}(z)} + 3\ff\right]-\delta_{g}^{\text{int}(1)} -3{\cal{H}}_{o}\delta\tau_{o} -\frac{2}{r_{z}} \delta\tau_{o}  \\
& \quad - \frac{1}{\chi}\int_{0}^{\chi}\dd \tilde{\chi}\left( \tilde{\chi}-\chi\right)\tilde{\chi}\left[ \nabla^{2}\left( \ff + \pf\right) - n^{i}n^{j}\left( \ff + \pf \right)_{,ij}  \right] +\frac{4}{r_{z}}  \kappa, \notag
\end{align}
where the first line is zero from the definition of $\delta_{g}^{\text{int}(1)}$, and without loss of generality we take $\delta \tau_{o} = 0$, and the integrals in the second line cancel from the definition of $\kappa$ in Ref.~\cite{yoo2}, so
\be
\Delta_{g}^{(1)}-\Delta_{\text{Yoo}}^{(1)} =  0.
\ee

\section{Second order comparison with literature}
\label{sec:secondcomparison}

\subsection{Introduction}

In 2014, almost simultaneously, three different derivations of the second order number counts were published \cite{durrer2, cc1,cc2, yoo2}.

Even though these works are interesting on their own, it is very hard to compare them as the final expressions for the galaxy number density occupy several pages. In particular, the result in Ref.~\cite{yoo2} has to be put together from a multitude of different contributions. On top of this, the notation and the break-up into separate terms is very different throughout all the derivations. An additional difficulty is that terms can be converted into each other by integration by parts, in a non-trivial way, making the comparison of even partial results tricky at best.

Because of these difficulties, there is still a general question of whether or not all these derivations coincide. In Ref.~\cite{nielsen} the authors concentrate on the terms which dominate on sub-horizon scales, considering only the terms of the order of $\left( k/\H \right)^{4}\Psi^{2}$ and neglect smaller contributions to the second order number count. Here, $k$ is the comoving wave number, $\H$ the conformal Hubble parameter and $\Psi$ a scalar metric perturbation. In Ref.~\cite{nielsen} the authors use a straight-forward derivation to reproduce the result of Ref.~\cite{durrer2}, finding disagreements concerning lensing terms and a double counting of volume distortion effects in Refs.~\cite{cc1} and \cite{yoo2}, respectively.

In this chapter, we use our independent derivation of the second order number counts given in \eq{eq:deltag2} and follow a similar approach as the one made in Ref.~\cite{nielsen} comparing leading terms in the aforementioned literature.

\subsection{Dominant terms at second order}

To determine the leading terms of the second order number counts, we first need to analyse our result given in Eq.~\eqref{eq:deltag2}. From \eqs{eq:defdg1} and \eqref{eq:defdg2} we consider that the general expression for the dominating terms is of the form
\be
\label{eq:dleading}
\Delta_{\text{Leading}} = \left( 1 + \delta \right)\left( 1 + \delta V \right) \simeq \left(1+\delta\right)\left(1+\text{RSD}\right)\left(1+\kappa\right),
\ee
where $\delta$ is the matter overdensity, $\delta V$ is the perturbed volume, and we further split the volume perturbation into its dominant components, being the lensing contribution ($\kappa$), and redshift space distortions (RSD).

Perturbing Eq.~\eqref{eq:dleading} up to second order, we find that the nonlinear contribution to the leading terms is given by
\begin{align}
\label{eq:leadingdg}
\Delta^{(2)}_{\text{Leading}} &= \delta_{g}^{(2)} + \delta_{g}^{(1)} \dVf + \dVs, \\
&= \delta_{g}^{(2)} + \left[\text{RSD}\right]^{(2)}+ \kappa^{(2)} + \delta_{g}^{(1)}\left[ \text{RSD}\right]^{(1)} + \delta_{g}^{(1)}\kappa^{(1)} + \left[ \text{RSD} \right]^{(1)}\kappa^{(1)}. \notag
\end{align}

We now present the leading terms of our calculation to second order given in \eq{eq:deltag2}, which is
\begin{align}
&\Delta^{(2)}_{\text{Leading}} \simeq \delta_{g}^{(2)} -\frac{1}{2\H}\frac{\dd\left( v_{2i}n^{i}\right)}{\dd\varsigma} + \frac{1}{2\H}\frac{\dd\left(v_{1k}v_{1}^{k}\right)}{\dd\varsigma}-\left(v_{1i}n^{i}\right)\frac{\dd\left(v_{1i}n^{i}\right)}{\dd\varsigma}\\
& +\frac{\dd\left(v_{1i}n^{i}\right)}{\dd\varsigma}\int_{0}^{\chi_{s}}\dd\chi\left( \ff'+\pf' \right)-\frac{\H''}{\H^{3}}\frac{\dd\left(v_{1i}n^{i}\right)}{\dd\varsigma}\int_{0}^{\chi_{s}}\dd\chi\left( \ff'+\pf' \right) \notag\\
& +\frac{2}{\H}\frac{\dd\left(v_{1i}n^{i}\right)}{\dd\varsigma} \int_{0}^{\chi_{s}}\dd\chi\left( \ff'+\pf' \right) +\frac{\delta_{g}^{(1)}}{\H}\frac{\dd \left(v_{1i}n^{i}\right)}{\dd\varsigma}\notag\\
&+ \frac{1}{\H}\frac{\dd\left(v_{1i}n^{i}\right)}{\dd\varsigma}\frac{1}{\chi}\int_{0}^{\chi_{s}}\dd\chi\left[ \nabla^{2}\left( \ff+\ff \right) + n^{i}n^{j}\left( \ff+\pf \right)_{,ij}+\frac{2}{\chi}\frac{\dd\dnuf}{\dd\varsigma}\right] \notag\\
& +\frac{\delta_{g}^{(1)}}{\chi}\int_{0}^{\chi_{s}}\dd\chi\left[ \nabla^{2}\left( \ff+\pf \right) + n^{i}n^{j}\left( \ff+\pf \right)_{,ij}+\frac{2}{\chi}\frac{\dd\dnuf}{\dd\varsigma}\right] \notag\\
& +\frac{2}{\chi} \int_{0}^{\chi_{s}}\dd\chi\left[ \nabla^{2}\left( \ps+\fs \right) + n^{i}n^{j}\left( \fs+\ps \right)_{,ij}+\frac{2}{\chi}\frac{\dd\dnus}{\dd\varsigma}\right]. \notag 
\end{align}

All these terms come from density fluctuations, 
\begin{align}
&\Delta^{(2)}_{\text{Leading--}\delta} \simeq \delta_{g}^{(2)},
\end{align}
redshift space distortions 
\begin{align}
&\Delta^{(2)}_{\text{Leading--RSD}} \simeq -\frac{1}{2\H}\frac{\dd\left( v_{2i}n^{i}\right)}{\dd\varsigma} + \frac{1}{2\H}\frac{\dd\left(v_{1k}v_{1}^{k}\right)}{\dd\varsigma}-\left(v_{1i}n^{i}\right)\frac{\dd\left(v_{1i}n^{i}\right)}{\dd\varsigma}\\
& \qquad\qquad\qquad\qquad +\left[1-\frac{\H''}{\H^{3}}+\frac{2}{\H}\right]\frac{\dd\left(v_{1i}n^{i}\right)}{\dd\varsigma} \int_{0}^{\chi_{s}}\dd\chi\left( \ff'+\pf' \right), \notag
\end{align}
lensing terms
\begin{align}
&\Delta^{(2)}_{\text{Leading--}\kappa} \simeq \frac{2}{\chi} \int_{0}^{\chi_{s}}\dd\chi\left[ \nabla^{2}\left( \ps+\fs \right) + n^{i}n^{j}\left( \fs+\ps \right)_{,ij}+\frac{2}{\chi}\frac{\dd\dnus}{\dd\varsigma}\right],  
\end{align}
and crossed terms
\begin{align}
&\Delta^{(2)}_{\text{Leading--}\delta\times \text{RSD}} \simeq \frac{\delta_{g}^{(1)}}{\H}\frac{\dd \left(v_{1i}n^{i}\right)}{\dd\varsigma}, \\
&\Delta^{(2)}_{\text{Leading--}\delta\times\kappa} \simeq \frac{\delta_{g}^{(1)}}{\chi}\int_{0}^{\chi_{s}}\dd\chi\left[ \nabla^{2}\left( \ff+\pf \right) + n^{i}n^{j}\left( \ff+\pf \right)_{,ij}+\frac{2}{\chi}\frac{\dd\dnuf}{\dd\varsigma}\right], \\
&\Delta^{(2)}_{\text{Leading--RSD}\times \kappa} \simeq \frac{1}{\H}\frac{\dd\left(v_{1i}n^{i}\right)}{\dd\varsigma}\frac{1}{\chi}\int_{0}^{\chi_{s}}\dd\chi\left[ \nabla^{2}\left( \ff+\ff \right) + n^{i}n^{j}\left( \ff+\pf \right)_{,ij}+\frac{2}{\chi}\frac{\dd\dnuf}{\dd\varsigma}\right].
\end{align}

\subsection{Comparison with Di Dio, et al.}

In the following, we identify and match terms from Ref.~\cite{durrer2} to \eq{eq:deltag2}. We start by relating the notation of one to the other, the main notational differences between our work and Ref.~\cite{durrer2} are:
\begin{itemize}
\item Work in the geodesic light-cone gauge (GLC) to obtain their solution which is then expanded in a more conventional gauge, Poisson gauge, to second order.
\item Latin indices in the GLC take only the values 1, 2.
\item Maintain their integrals in terms of the conformal time $\eta$.
\item Separate the final result into an isotropic and anisotropic part.
\item Projected notation is used $v_{\parallel} = n^{i}v_{i}$.
\item There is a factor of $(-1)$ in the definition of the observation vector $n^{i}$.
\end{itemize}

To second order, the main result is given by Eq.~(4.41) in their paper,
\be
\Delta^{(2)} = \Sigma - \langle \Sigma \rangle,
\ee
where 
\be
\Sigma = \Sigma_{\text{IS}} + \Sigma_{\text{AS}},
\ee
and the isotropic, $\Sigma_{\text{IS}}$, and anisotropic, $\Sigma_{\text{AS}}$, parts are given in Eqs.~(4.42) and (4.43) of Ref.~\cite{durrer2}, respectively.

The translation of the majority of the terms is straightforward. $r$ is the comoving distance we call $\chi$ and $\partial_{\eta} = \dd/\dd\eta = \dd/\dd\varsigma - n^{i}\partial_{i} = -\dd/\dd\chi$. Thankfully, the authors of Ref.~\cite{durrer2} include a \textit{leading} order second order contribution in their paper, which is
\begin{align}
&{}^{(2)}\Delta^{\text{Di Dio}}_{\text{Leading}} \simeq \delta_{\rho}^{(2)} + \frac{1}{\H}\frac{\dd \left( v_{2i}n^{i}\right)}{\dd\varsigma} -\frac{1}{2 r} \int_{\eta_{s}}^{\eta_{o}}\dd\eta'\frac{\eta'-\eta_{s}}{\eta_{o}-\eta'} \Delta_{2}\left( \ps + \fs \right) \\
&+\frac{1}{\H}\left[ \left( v_{1i}n^{i} \right)\frac{\dd^{2}\left( v_{1i}n^{i} \right)}{\dd \varsigma^{2}} + \left( \frac{\dd\left( v_{1i}n^{i} \right)}{\dd\varsigma} \right)^{2} \right] - \frac{2}{\H}\frac{\dd\left( v_{1i}n^{i}\right)}{\dd\varsigma} \frac{1}{r}\int_{\eta_{s}}^{\eta_{o}}\dd\eta'\frac{\eta'-\eta_{s}}{\eta_{o}-\eta'}\Delta_{2}\psi^{I} \notag \\
& -\frac{2}{\H}\partial_{a}\left( \frac{\dd\left( v_{1i}n^{i} \right)}{\dd\varsigma} \right)\int_{\eta_{s}}^{\eta_{o}}\dd\eta' \gamma_{0}^{ab}\partial_{b}\int_{\eta_{s}}^{\eta_{s}}\dd\eta''\psi^{I} +2\left( \frac{1}{r}\int_{\eta_{s}}^{\eta_{o}}\dd\eta'\frac{\eta'-\eta_{s}}{\eta_{o}-\eta'}\Delta_{2}\psi^{I} \right)^{2} \notag\\
& +\frac{4}{r}\int_{\eta_{s}}^{\eta_{o}}\dd\eta'\frac{\eta'-\eta_{s}}{\eta_{o}-\eta'}\Bigg\{\partial_{b}\left[\Delta_{2}\psi^{I}\right]\int_{\eta_{s}}^{\eta_{o}}\dd\eta''\gamma_{0}^{ab}\partial_{a}\int_{\eta''}^{\eta_{o}}\dd\eta'''\psi^{I}  \notag \\
& +\Delta_{2}\left[ -\frac{1}{2}\gamma_{0}^{ab}\partial_{a}\left( \int_{\eta'}^{\eta_{o}}\dd\eta''\psi^{I} \right)\partial_{b}\left( \int_{\eta'}^{\eta_{o}}\dd\varsigma''\psi^{I} \right) \right]\Bigg\} \notag\\
& -4\int_{\eta_{s}}^{\eta_{o}}\dd\eta'\Big\{ \gamma_{0}^{ab}\partial_{b}\left( \int_{\eta'}^{\eta_{o}}\dd\eta''\psi^{I} \right)\frac{1}{\eta_{o}-\eta'}\int_{\eta'}^{\eta_{o}}\dd\eta''\frac{\eta''-\eta'}{\eta_{o}-\eta''}\partial_{a}\Delta_{2}\psi^{I} \Big\}\notag\\
& +\left[ \frac{1}{\H}\left( \frac{\dd\left( v_{1i}n^{i} \right)}{\dd\varsigma}\right) - \frac{2}{r}\int_{\eta_{s}}^{\eta_{o}}\dd\eta'\frac{\eta'-\eta_{s}}{\eta_{o}-\eta'}\Delta_{2}\psi^{I} \right]\delta_{\rho}^{(1)} +\frac{1}{\H}\left( v_{1i}n^{i} \right)\frac{\dd \delta_{\rho}^{(1)}}{\dd\varsigma} \notag \\
& -2\partial_{a}\delta_{\rho}^{(1)}\int_{\eta_{s}}^{\eta_{o}}\dd\eta'\gamma_{0}^{ab}\partial_{b}\int_{\eta'}^{\eta_{o}}\dd\eta\psi^{I}. \notag
\end{align}
After some integrations by parts to rewrite some terms, and rewriting what the angular Laplacian is in our notation, the only difference between our leading terms and the ones from Ref.~\cite{durrer2} only comes from the different definition of the observation vector $n^{i}$, and a numerical factor between the four velocities, one half.
\begin{align}
\Delta^{(2)}_{\text{Leading}} - {}^{(2)}\Delta^{\text{Di Dio}}_{\text{Leading}} &\simeq \Big[\delta_{g}^{(2)} - \delta_{\rho}^{(2)}\Big] - \frac{1}{\H}\frac{\dd\left( v_{2i}n^{i}\right)}{\dd\varsigma}\left[ \frac{1}{2} + 1 \right], \\
&\simeq - \frac{1}{\H}\frac{\dd\left( v_{2i}n^{i}\right)}{\dd\varsigma}\left[ \frac{1}{2} + 1 \right], \notag
\end{align}
where it can be seen explicitly that the difference comes from the different sign in the definition of the observation vector $n^{i}$, and a $(1/2)$-factor in the the definition of our second order velocity perturbation.

\subsection{Comparison with Bertacca, et al.}

We now proceed to identify the terms of Refs.~\cite{cc1,cc2}, the main remarks about the notation here are:
\begin{itemize}
\item They work with cosmic rulers, then change to Poisson gauge.
\item There is a factor of $(-1)$ in the definition of the observation vector $n^{i}$.
\item Projected derivatives, meaning $\partial_{\parallel}=n^{i}\partial_{i}$ and $\partial_{\perp}^{i} = r^{-1}\partial^{i}$.
\end{itemize}

The leading terms in the expression for the number counts, would be
\begin{align}
\label{eq:bert2}
&{}^{(2)}\Delta^{\text{Bertacca}}_{\text{Leading}} \simeq \delta_{g}^{(2)} - \frac{1}{\H} \frac{\dd^{2} \left( v_{2i}n^{i}\right)}{\dd\varsigma^{2}} - 2\kappa^{(2)} + 4\left[ \kappa^{(1)} \right]^{2} - 4\delta_{g}^{(1)}\kappa^{(1)} \\
&- 2 \frac{\delta_{g}^{(1)}}{\H} \frac{\dd^{2}\left( v_{1i}n^{i} \right)}{\dd\varsigma^{2}} + 4\frac{\kappa^{(1)}}{\H}\frac{\dd^{2}\left( v_{1i}n^{i} \right)}{\dd\varsigma^{2}} + \frac{2}{\H^{2}}\left[ \frac{\dd^{2}\left( v_{1i}n^{i} \right)}{\dd\varsigma^{2}} \right]^{2} + \frac{2}{\H^{2}}\left[\frac{\dd\left( v_{1i}n^{i} \right)}{\dd\varsigma}\right]\left[ \frac{\dd^{3}\left( v_{1i}n^{i} \right)}{\dd\varsigma^{3}}\right] \notag\\
& -\frac{2}{\H}\frac{\dd \delta_{g}^{(1)}}{\dd\chi}\Delta\ln a^{(1)} + \frac{2}{\chi}\left[ \left( \delta_{g}^{(1)}\right)_{,i} -\frac{1}{\H}\frac{\dd^{2}\left(v_{1k}n^{k}\right)_{,i}}{\dd\varsigma^{2}} \right]\int_{0}^{\chi_{s}}\dd\chi\left( {\ff_{,}}^{i} + {\pf_{,}}^{i} \right) \notag \\
& -2\left[ \left( \delta_{g}^{(1)}\right)_{,i} -\frac{1}{\H}\frac{\dd^{2}\left(v_{1k}n^{k}\right)_{,i}}{\dd\varsigma^{2}} \right] \int_{0}^{\chi_{s}}\dd\chi\frac{1}{\chi}\left( {\ff_{,}}^{i} + {\pf_{,}}^{i} \right)\notag\\
&-4\left( \int_{0}^{\chi_{s}}\dd\tilde{\chi}\frac{\tilde{\chi}}{\chi}\left( \bar{\chi}-\tilde{\chi} \right)\mathcal{P}^{n}_{i}\mathcal{P}^{jm}\partial_{m}\partial_{n}\ff \right)\left( \int_{0}^{\chi_{s}}\dd\tilde{\chi}\frac{\tilde{\chi}}{\chi}\left( \bar{\chi}-\tilde{\chi} \right)\mathcal{P}^{p}_{i}\mathcal{P}^{jq}\partial_{p}\partial_{q}\ff \right). \notag
\end{align}

The translation of the majority of the terms is straightforward, $a^{(1)}$ is the first order perturbation of the scale factor taken as $1/(1+z)$. To leading order, we can substitute $\Delta \ln a^{(1)} = -\partial_{\chi}\left( v_{1i}n^{i}\right)$. 

After several integrations by parts and translations between the leading expression in our notation and the authors from Ref.~\cite{cc1}, the main difference comes in the definition of the convergence, where there is a numerical value factor of difference. Note that we are not taking into account the so called `post-Born' contributions.
\begin{align}
&\Delta^{(2)}_{\text{Leading}} - {}^{(2)}\Delta^{\text{Bertacca}}_{\text{Leading}} \simeq  \\
&\qquad\qquad \frac{\delta_{g}^{(1)}}{\chi}\int_{0}^{\chi_{s}}\dd\chi\left[ \nabla^{2}\left( \ff+\pf \right) + n^{i}n^{j}\left( \ff+\pf \right)_{,ij}+\frac{2}{\chi}\frac{\dd\dnuf}{\dd\varsigma}\right]  - 4\delta_{g}^{(1)}\kappa^{(1)}\notag\\
&\qquad\qquad +\frac{2}{\chi} \int_{0}^{\chi_{s}}\dd\chi\left[ \nabla^{2}\left( \ps+\fs \right) + n^{i}n^{j}\left( \fs+\ps \right)_{,ij}+\frac{2}{\chi}\frac{\dd\dnus}{\dd\varsigma}\right]- 2\kappa^{(2)}, \notag \\
&\approx 2 \delta_{g}^{(1)}\kappa^{(1)} - 4\delta_{g}^{(1)}\kappa^{(1)} + \kappa^{(2)} -2\kappa^{(2)}, \notag \\
& = - 2 \delta_{g}^{(1)}\kappa^{(1)}-\kappa^{(2)},
\end{align}
where in the second equality we used the definition given in Ref.~\cite{cc1} [see Eqs. (209) and (217) therein], to rewrite our notation into their convergence to explicitly see the difference that arises from the convergence, where it appears to be adding twice as much to the galaxy overdensity.

\subsection{Comparison with Yoo and Zaldarriaga}
\label{subsec:yoozal2}

We now proceed to identify the terms of Ref.~\cite{yoo2}. Since the full final result is not written down in closed form, we have to perform a \textit{stitching-together} by going backwards through the paper. The way the results are presented is one by one, so the main result -- the galaxy overdensity -- is not written explicitly; it is only shown in terms of previously found expressions, that the reader needs to find and then ``stitch-together'' in order to obtain the full expression. The main differences between our notation and Ref.~\cite{yoo2} are:
\begin{itemize}
\item Latin indices go from 0 to 3, Greek indices go from 1 to 3 (the opposite of our notation).
\item Metric perturbations are called $\mathcal{A}$ and $\mathcal{C}_{\alpha\beta}$, where $\mathcal{C}_{\alpha\beta} = \Psi \delta_{\alpha\beta}$ in our notation, for purely scalar perturbations. The second order perturbations are defined with no factor 2 at second order.
\item All perturbation orders are left implicit.
\item Work in a different basis that depends on angles, and leave some factors of $\left(\sin\theta\right)$ between expressions, which are not present in our derivation.
\end{itemize}
From \eq{eq:leadingdg}, we have that for Ref.~\cite{yoo2} the leading terms are
\be
{}^{(2)}\Delta_{\text{Leading}}^{\text{Yoo}} = \delta_{g}^{(2)} + \dVs + \delta_{g}^{(1)}\dVf.
\ee
The first order volume perturbation in Ref.~\cite{yoo2} is given by
\be
\dVf = -2 \kappa^{(1)} + H_{z}\partial_{z}\delta r^{(1)} = -2\kappa^{(1)} + \frac{1}{\H}\frac{\dd\left( v_{1i}n^{i} \right)}{\dd\varsigma},
\ee
where $\partial_{z} = H_{z}^{-1}\partial_{\chi}$. In Ref.~\cite{yoo2} the second order perturbation of the volume is
\be
\dVs = \delta \mathcal{D}_{L}^{(2)} + H_{z}\partial_{z}\delta r^{(2)} -2H_{z}\kappa^{(1)}\partial_{z} \delta r^{(1)} + \Delta x^{(1)b}\partial_{b}\dVf,
\ee
where $\delta\mathcal{D}_{L}^{(2)}$ is the luminosity distance, which is related to $\das$ through the \textit{Etherington's reciprocity theorem}, which is expressed as $d_{L} = \left( 1+z \right)^{2}d_{A}$. Lastly we need to expand 
\be
\Delta x^{(1)b}\partial_{b}\dVf = -2\nabla^{a}\ff\nabla_{a}\kappa^{(1)} + \frac{1}{\H}\nabla^{a}\ff\nabla_{a}\frac{\dd\left( v_{1i}n^{i} \right)}{\dd\varsigma} + \frac{1}{\H^{2}}\left(\frac{\dd\left( v_{1i}n^{i} \right)}{\dd\varsigma}\right)\left(\frac{\dd^{2}\left( v_{1i}n^{i} \right)}{\dd\varsigma^{2}}\right).
\ee
All these terms are already accounted for in $\delta\mathcal{D}_{L}^{(2)}$ and in $\delta r^{(2)}$. To avoid the cumbersome rewriting of the full term, we refrain from rewriting the terms here. For reference they are given in Eqs.~(78) and (50) in Ref.~\cite{yoo2}, respectively. Here, in agreement with Ref.~\cite{nielsen}, it can be seen that with these two terms, there is a double-counting effect, so this is the only difference between our leading terms, if we ignore this, then we are in agreement.
\begin{align}
\Delta^{(2)}_{\text{Leading}} - {}^{(2)}\Delta^{\text{Yoo}}_{\text{Leading}} &\simeq -2\nabla^{a}\ff\nabla_{a}\kappa^{(1)} + \frac{1}{\H}\nabla^{a}\ff\nabla_{a}\frac{\dd\left( v_{1i}n^{i} \right)}{\dd\varsigma} \\
&\qquad\qquad\qquad+ \frac{1}{\H^{2}}\left(\frac{\dd\left( v_{1i}n^{i} \right)}{\dd\varsigma}\right)\left(\frac{\dd^{2}\left( v_{1i}n^{i} \right)}{\dd\varsigma^{2}}\right), \notag
\end{align}
where we show explicitly that the leading terms are in complete agreement, except for those terms which are accounted for twice inside the luminosity distance and the radial perturbation.

\section{Discussion}
\label{sec:discussion}

In this chapter we presented a new and independent approach to calculate the galaxy number overdensity. We have shown that through an independent approach we could reproduce the leading terms that have only been computed by three groups before, in Refs.~\cite{durrer2,cc1,yoo2}. We present the galaxy number counts in a general form depending only on the affine parameter, which allows for simple plotting along the line of sight if the potentials are known. The potentials can be calculated either using the field equations discussed in Section \ref{section:flrw} or N-body simulations. We also present the second order galaxy number counts in terms of observable quantities such as the redshift, so future surveys can potentially measure this quantity and provide sufficient information on large and small scales. Our results will aid in the analysis of the data and enable comparisons of the theoretical number counts with observed quantities.

This result can be used to study the impact of relativistic effects on parameters estimation. Since we expect that the relativistic effects do not significantly shift the position of the BAO peak we can focus on the bias parameters and astrophysical systematics in the cross-correlations.
The main result in this Chapter is presented in \eq{eq:deltag2} -- the galaxy number counts up to and including second order in cosmological perturbation theory. We used scalar perturbations in longitudinal gauge allowing for non-zero anisotropic stress. We assumed a flat FLRW background universe filled with a pressureless fluid throughout the whole chapter, except where otherwise stated.

As discussed in Section \ref{introdg} we are not the first group to perform this calculation. We performed a comparison of our main result with others in the literature. We find that at linear order we are in complete agreement. Nevertheless, the approaches taken by different groups lead to slight differences at higher order. We provide a comparison of the leading terms at second order in Section \ref{sec:secondcomparison}, finding that our approach and the previous works differ by numerical factors with one group, and that other groups double count some of the effects, leading to a more significant difference with our expression. Although we do not provide a full comparison of the complete expression, we will return to this issue in the future. Our result is in agreement with the previous comparison made in Ref.~\cite{nielsen}. 



The contribution to the bispectrum from the dominant terms discussed in Section \ref{sec:secondcomparison} has been computed numerically in \cite{didio2015}. There, it is shown that for equal redshifts, the result is dominated by standard Newtonian terms. However, for different redshifts, the Newtonian terms rapidly decay and the result is soon dominated by new lensing contributions. 

The dominant projection effects are the redshift space distortions, which contribute up to 30\% of the total signal at equal redshifts, both in the power spectrum and in the bispectrum, and the lensing terms which dominate the signal (99\% of the total) for widely separated redshifts \cite{didio2015}.

%% file: tex/chapter_5.tex
\chapter{Conclusions and Outlook}
\label{chapter:conclusions}

In this thesis, we have investigated instances in which nonlinear cosmological perturbation theory is needed to understand the Universe. We focused particularly on the late Universe and derived new analytic solutions to the gravitational wave or tensor evolution equation at linear order. We derived in an independent approach, the galaxy number overdensity up to second order and computed the differences with other groups in the literature, demonstrating the equivalence between the different approaches at linear order and for the leading terms at second order in perturbation theory. In spite of the emphasis of this thesis being only on the late Universe, many of the techniques used can be applied in far more general settings in cosmology.

We started with a review of cosmological perturbation theory in Chapter \ref{chapter:cpt}, by discussing the main aspects of general relativity. We presented the governing equations for the evolution of spacetime, Einstein's field equations, both for the geometry as well as that of matter. We then reviewed relativistic perturbation theory in detail, accounting for the gauge issue and establishing general expressions for the later parts of the thesis. Finally, we applied the results to the case of perturbations around a FLRW spacetime and we wrote down the perturbed field equations up to second order.\\

Chapter \ref{chapter:wkb} included the research published in Ref.~\cite{fuentes1}, where we derived new analytical solutions to the gravitational wave evolution equation at linear order in cosmological perturbation theory using the \textit{double power series method for approximating cosmological perturbations} presented in \rfcite{DPS} and including neutrino anisotropic stress. 
A key objective of this study was to find an approximate solution that would provide as much information as the numerical solution would, without having to perform the full numerical simulation, thus introducing a faster way to obtain solutions.
We succeed in obtaining an approximation that is accurate to within $1\%$ and is demonstrated in Figure~\ref{fig:Fighappr3k10error}. We then apply our analytical solutions to the governing equations for primordial gravitational waves in a radiation--matter model, presenting our main result in \eq{eq:match} which is a good approximation on both sub--horizon ($\tau\lesssim \infrac{1}{k}$) and super--horizon ($\tau\gtrsim \infrac{1}{k}$) scales for wavenumbers $k$ over a wide range of values relevant for structure formation in the Universe. The best approximation we have found in the literature for sub--horizon primordial gravitational waves is, in fact, the WKB solution. This is accurate to within $0.8\%$, which compares with the accuracies for the two approximations of Figure~\ref{fig:Fighappr3k10error} of $\sim0.25\%$ for the real part $(\cos)$ and $1\%$ for the imaginary part $(\sin)$, but is still considerably less accurate than our matching approximation for the radiation--matter mixture when started outside the horizon.

Future work on this topic could take different directions. LIGO is detecting gravitational wave signals more often than ever and so it is important to take into account the existence of a gravitational wave background. Our approximation will be useful 
since it does not require the computational time demanded by numerical simulation and does not compromise on accuracy.
For example, in second order cosmological perturbation theory, the governing equations include terms of the form $h_{ij}h^{ij}$ and different forms of the curvature perturbation, $\zeta$, at second order differ by terms proportional to this $h_{ij}h^{ij}$ (see e.g.~\rfcite{cm2016}). Another effect our approximation allows us to study, is the dependence of the curvature perturbation at second order on the wavenumber and the neutrino--radiation ratio $f_{\nu}$, since the existence of neutrinos should be taken into account if we want accurate results. \\

In Chapter \ref{chapter:gnc}, we presented an independent approach to computing the galaxy number counts, presenting the work done in Refs.~\cite{fuentes3, fuentes4}. We  have reviewed the research undertaken in the literature, from the first papers on cosmological observables by Kristian and Sachs in \rfcite{sachs2} in 1966 to one of the latest theoretical studies of the relativistic redshift in \rfcite{yoo5}. 
We then introduce definitions to provide an understanding of the galaxy number density, defining the photon wavevector, observed redshift, the angular diameter distance and the physical volume in which galaxies reside.
We then proceed to apply the results from cosmological perturbation theory discussed in Chapter \ref{chapter:cpt} and obtain the perturbed expressions for the null geodesics, the photon energy, the observed redshift, the angular diameter distance and the physical volume, all of which all of which define the galaxy number counts. We present all of our results in terms of the affine parameter $\lambda$ which parametrises the geodesic equation and makes our result easy to program, making it easy to compare with numerical simulations. We also provide an expression for the angular diameter distance in terms of the observed redshift, which can in principle be measured by the next generation of galaxy surveys and can be complementary to the luminosity distance through \textit{Etherington's reciprocity theorem} mentioned in Section \ref{subsec:yoozal2}.

One of the main results of Chapter \ref{chapter:gnc}, presented in \eq{eq:deltag2}, is the galaxy number density fluctuation at second order. We proceed to compare this result to expressions derived in the literature, starting with the linear perturbation and finding that our result is in complete agreement with those in previous works. Finally, we present a comparison of the second order leading terms of the galaxy number overdensity that had been computed previously in Refs.~\cite{durrer2, cc1,cc2, yoo2}, finding that there is a slight difference arising from our definition of the direction of observation, the definition of convergence and double counting 
of some terms in \rfcite{cc1,cc2,yoo2} which would, in principle, over-estimate the galaxy number counts.

We are still working to perform the full comparison term by term, including those that are only important at ultra large scales, where normally different observables tend to take a different shape depending on the gauge choice, similar to the power spectrum. We will also include vector and tensor perturbations up to and including second order, for completeness. We shall also include the effects of biasing and magnification bias which have been neglected in this first discussion. The calculation of its bispectrum and the comparison with others, will include the nonlinear effects of gravity, with a possible primordial bispectrum, which is necessary for any conclusions on primordial non-Gaussianities.

Future work also includes the calculation of the Cartesian Fourier-space galaxy bispectrum and the galaxy 3-point correlation function including relativistic corrections using our model to probe non-Gaussianity. We could also perform a similar comparison as the one made in Section \ref{sec:secondcomparison} instead using numerical simulations to get the differences between the different results studied in \ref{sec:secondcomparison} and then plot a percentage difference, not only taking the leading terms but the full second order expressions. The main issue with this comparison would be that since it is numerical we would not be able to discern from which term (or terms) the difference comes from; integrated effects, pseudo-RSDs, or new unknown effects. Another possible issue is that the precision of upcoming galaxy surveys may not be able to measure these small differences. However, there will be more precise surveys in the future and this approach should not be neglected.

The formalism followed in this chapter can also be used to study the impact of relativistic effects in the 3-dimensional cross-correlation between Lyman-$\alpha$ forest and quasars, or other combination of observables, taking into account the density parameters that go with each one of them.\\

Throughout the thesis, we worked in standard unmodified theory of gravity, that is general relativity. However we have also studied a modified version of gravity which has recently become popular whilst writing this thesis, and our results are included and presented in Appendix \ref{chapter:mgb}; we present the calculation of the equations of motion in a modified Gauss-Bonnet theory of gravity at linear order in cosmological perturbation theory. Future work has been planned for this project in \rfcite{FGM2}, since there has been a surge in interest in these modified theories of gravity in the past year (see, e.g. Refs.~\cite{PGS1,PGS2}). Another interesting avenue of research to take would be the study of observables in such modified theories of gravity, and the possibility of these observables to act as a test for other gravity formulations. \\

Finally, it is important to mention that during the research for this thesis, there were many challenging moments in which we encountered many setbacks. However, we overcame these challenges and proceeded to devise new solutions.
One of the main problems that occurred whilst working on the galaxy number density was the problem of the notation;  different groups used different notations and surprisingly, the notation seen in papers coming from the same group was also different (see Section \ref{sec:secondcomparison}).
To overcome this problem it was needed for me to write all the other results in our own notation and that took some time, but in the end, that made the comparison easier, at least up to the leading terms in second order cosmological perturbation theory.

The most challenging and frustrating problem we stumbled upon was the fact that the codes \texttt{xAct} and \texttt{xPand} can only work with derivatives of perturbed and unperturbed quantities, but do not work with integrated equations. For example, \texttt{xPand} can be used to derive \eqs{eq:dnu1} and \eqref{eq:dn1}, however, it would be impossible to compute \eq{eq:dE1} solely with functions defined within the code itself, since $\dnuf$ is obtained by integrating \eq{eq:dnu1} along the line of sight. This problem became much more cumbersome whilst working at second order in cosmological perturbation theory, and was a major obstacle for the calculation of the galaxy number counts since almost every variable needs to be integrated along the line of sight. We were able to find a workaround to this problem by partially integrating some terms using the code and solving some other integrals by hand, and after writing everything back up in the code as like a normal equation, it was a long process, but it worked.

%% file: tex/appendix.tex
\begin{appendices}

\chapter{Matching conditions and anisotropic stress calculations}
\label{app1}

\section{Calculation of the matching conditions for approximating primordial gravitational waves}
\label{sec:calculation-of-the-matching-conditions-for-approximating-primordial-gravitational-waves}

In this appendix, we calculate the matching conditions needed to obtain  the solution in \eref{eq:match} of Section~\ref{sec:modelling-gravitational-waves-through-the-whole-radiation--matter-epoch}.  We start by matching the values of the functions $\hr$ of \eref{eq:flat-rad-only-soln3} and $h_{(2)}$ of \ref{eq:second-order-trig2} at the conformal time $\tau=\infrac{1}{k}.$ This gives
\bas \label{eq:fval}
\sin(1)&=\frac{B_\text{s}\sin \left(\lambda\left(k,\frac{1}{k}\right)\right)+B_\text{c}\cos \left(\lambda\left(k,\frac{1}{k}\right)\right)}{\frac{1}{k}  (\frac{1}{k} +4)}\\
&=
\frac{k^2}{4k+1}
\left\lbrace B_\text{s}\sin\left(\lambda\left(k,\frac{1}{k}\right)\right) +B_\text{c}\cos\left(\lambda\left(k,\frac{1}{k}\right)\right)\right\rbrace,
\eas
where
\be \label{eq:cts}
\lambda\left(k,\frac{1}{k}\right)=1+\frac{1}{4 k}\ln \left(1+4k\right).
\ee

Note that
\be \label{eq:d1}
\frac{\partial \left(\lambda(k,\tau)\right)}{\partial \tau}\bigg|_{\tau=\frac{1}{k}}
=k-\frac{1}{k \tau ^2+4 k \tau }\bigg|_{\tau=\frac{1}{k}}
=k-\frac{k}{4k+1}
=\frac{4k^2}{4k+1},
\ee
and 
\be \label{eq:d2}
\frac{\partial \Big(\frac{1}{\tau\left(\tau+4\right)}\Big)}{\partial \tau}\Bigg|_{\tau=\frac{1}{k}}
=-\frac{2 (\tau +2)}{\tau ^2 (\tau +4)^2}\Bigg|_{\tau=\frac{1}{k}}
=-\frac{2 k^3 (2 k+1)}{(4 k+1)^2}.
\ee 
Using  Eqs.~\ref{eq:d1} and \ref{eq:d2} to calculate $h'_{(2)}\left(\infrac{1}{k}\right),$ we match the first derivatives of with respect to $\tau$ of $\hr$ and $h_{(2)}$ at $\tau=\infrac{1}{k}$ to get
\bml \label{eq:fderiv}
k\cos(1)-k\sin(1)
=
\frac{4k^4}{\left(4k+1\right)^2}
\Bigg\lbrace B_\text{s}\cos\left(\lambda\left(k,\frac{1}{k}\right)\right)-B_\text{c}\sin\left(\lambda\left(k,\frac{1}{k}\right)\right)\Bigg\rbrace \\
-\frac{2 k^3 (2 k+1)}{(4 k+1)^2}\Bigg\lbrace B_\text{s}\sin\left(\lambda\left(k,\frac{1}{k}\right)\right)+B_\text{c}\cos\left(\lambda\left(k,\frac{1}{k}\right)\right)\Bigg\rbrace.
\eml\\

We can solve Eqs.~\ref{eq:fval} and \ref{eq:fderiv} simultaneously to find
\begin{align}
\label{eq:b-s}
B_\text{s}
&=
\frac{4 k+1}{4 k^3} \Bigg\{ 4 k\left[
\cos(1)\cos \left(\lambda\left(k,\frac{1}{k}\right)\right) +\sin (1)\sin\left(\lambda\left(k,\frac{1}{k}\right)\right)\right] \notag \\
&\qquad\qquad\qquad\qquad\qquad\qquad\qquad\qquad\qquad+\left[\sin(1)+\cos (1)\right]\cos\left(\lambda\left(k,\frac{1}{k}\right)\right)\Bigg\} \notag\\
&=
\frac{4 k+1}{4 k^3} \left\lbrace 4 k\cos \left(1-\lambda\left(k,\frac{1}{k}\right)\right) +\left[\sin(1)+\cos (1)\right]\cos\left(\lambda\left(k,\frac{1}{k}\right)\right)\right\rbrace \notag\\
&=
\frac{4 k+1}{4 k^3} \left\lbrace 4 k\cos \left(\frac{1}{4k}\ln\left(1+4k\right)\right) +\left[\sin(1)+\cos (1)\right]\cos\left(1+\frac{1}{4k}\ln\left(1+4k\right)\right)\right\rbrace
\end{align}
and
\begin{align}
\label{eq:b-c}
B_\text{c}
&=
\frac{4 k+1}{4 k^3} \Bigg\{ 4 k\left[
\sin(1)\cos \left(\lambda\left(k,\frac{1}{k}\right)\right) -\cos (1)\sin\left(\lambda\left(k,\frac{1}{k}\right)\right)\right] \notag \\
&\qquad\qquad\qquad\qquad\qquad\qquad\qquad\qquad\qquad -\left[\sin(1)+\cos (1)\right]\sin\left(\lambda\left(k,\frac{1}{k}\right)\right)\Bigg\} \notag\\
&=
\frac{4 k+1}{4 k^3} \left\lbrace 4 k\sin \left(1-\lambda\left(k,\frac{1}{k}\right)\right) -\left[\sin(1)+\cos (1)\right]\sin\left(\lambda\left(k,\frac{1}{k}\right)\right)\right\rbrace \notag\\
&=
\frac{4 k+1}{4 k^3} \left\lbrace - 4 k\sin \left(\frac{1}{4k}\ln\left(1+4k\right)\right) -\left[\sin(1)+\cos (1)\right]\sin\left(1+\frac{1}{4k}\ln\left(1+4k\right)\right)\right\rbrace,
\end{align}
where we recalled the definition of $\lambda(k,\tau)$ from \eref{eq:lamdba}.
In going from the first to second lines of Eqs.~\ref{eq:b-s} and \ref{eq:b-c}, we have used standard trigonometric addition formulae.\\

From \eref{eq:second-order-trig2}, this gives us
\begin{subequations}
\begin{align} 
\begin{split}\label{eq:trigs-a}
h_{(2)}(\tau)
&=
\frac{4 k+1}{4 k^3 \tau  (4+\tau)}
\left[
4k\left\lbrace
\cos\left(\frac{1}{4k}\ln\left(1+4k\right)\right)\sin \left(\lambda(k,t)\right)\right.\right.\\
&\qquad\qquad\qquad\qquad\qquad\qquad\quad\left.\left.
-\sin\left(\frac{1}{4k}\ln\left(1+4k\right)\right)\cos \left(\lambda(k,t)\right)
\right\rbrace\right.
\\
&\left.\qquad+\big[\sin (1)+\cos (1)\big]
\bigg\lbrace\cos\left(1+\frac{1}{4k}\ln\left(1+4k\right)\right)\sin \left(\lambda(k,t)\right)\right.
\\
&\qquad\qquad\qquad\qquad\qquad\qquad\left.
-\sin\left(1+\frac{1}{4k}\ln\left(1+4k\right)\right)\cos \left(\lambda(k,t)\right)\bigg\rbrace
\right]
\end{split}
\\[9pt]
\begin{split}\label{eq:trigs-b}
&=
\frac{4 k+1}{4 k^3 \tau  (4+\tau)}
\bigg[
4k \sin \Big(k\tau + L(k,t)\Big)+\left[\sin (1)+\cos (1)\right]
\sin \Big(k\tau + L(k,t)-1\Big)
\bigg],
\end{split}
\end{align}
\end{subequations}
where again we recalled the definition of $\lambda(k,\tau)$ from \eref{eq:lamdba}, and we have defined
\be
L(k,\tau)=\frac{1}{4 k}\ln \left(\frac{1+4\,\tau^{-1} }{1+4\,k\hfill}\right).
\ee
We used standard trigonometric addition formulae to go from \eref{eq:trigs-a} to \eref{eq:trigs-b}.\\

From \eref{eq:flat-rad-only-soln3} and \eref{eq:trigs-b}, the matching approximation is therefore
\begin{equation}\label{eq:match-a}
\hmatch(\tau)=
\begin{dcases}
\frac{\sin(k\tau)}{k\tau}
&\mbox{if } \tau\le\frac{1}{k}
\\[9pt]
\frac{4 k+1}{4 k^3 \tau  (4+\tau)}
\bigg[
\,4 k\, \sin \Big(k\tau + L(k,t)\Big)
+ \mu
\sin \Big(k\tau + L(k,\tau)-1\Big)
\bigg]
 &\mbox{if } \tau\ge\frac{1}{k},
\end{dcases}
\end{equation}
where $
\mu
=
\sin\left(1\right)+\cos\left(1\right)
=
1.38177...\, .
$

\section{Calculation of the anisotropic stress solution}
\label{sec:calculation-of-the-anisotropic-stress-solution}

\nt The method described in \rfcite{DPS} works best for equations like \ent 
\ba
\mathbf{A} \mathbf{f}''(\tau) + \mathbf{C} \mathbf{f}'(\tau) + \mathbf{B} \mathbf{f}(\tau) = \mathbf{0}. 
\ea
\nt so we take \eref{eq:gov-tau-stress} and write it as a system of ordinary second order differential equations to get\footnote{ \textbf{Note.} We are neglecting the term $\int_{0}^{\tau}K'(\tau-t)h'(t)dt$ that arises from differentiating~\eref{eq:gov-tau-stress} because numerically the value is \textit{negligible}.} \ent 
\ba
\mathbf{A} \mathbf{X}''(\tau) + \mathbf{C} \mathbf{X}'(\tau) + \mathbf{B} \mathbf{X}(\tau) = \mathbf{0}. 
\ea
\nt where \ent 
\[
   \mathbf{X}(\tau)=
  \left[ {\begin{array}{cc}
   x_{1}(\tau) \\
   x_{2}(\tau) \\
  \end{array} } \right] =
    \left[ {\begin{array}{cc}
   h(\tau) \\
   h'(\tau) \\
  \end{array} } \right],
\]
\nt and \ent 
\[
   \mathbf{A}=
  \left[ {\begin{array}{cc}
   1 & 0\\
   0 & 1\\
  \end{array} } \right], \quad
     \mathbf{C}(\tau)=
  \left[ {\begin{array}{cc}
   0 & -1\\
   0 & f_{2}(\tau)/f_{1}(\tau)\\
  \end{array} } \right], \quad
     \mathbf{B}(\tau)=
  \left[ {\begin{array}{cc}
   0 & 0\\
   f_{4}(\tau)/f_{1}(\tau) & f_{3}(\tau)/f_{1}(\tau)\\
  \end{array} } \right] ,
\]
\nt where we have defined \ent
\begin{align}
f_{1}(\tau) &= -\frac{1}{96 f_{\nu}(0)}\frac{\tau^{2}(1+\tau)(4+\tau)^{2}}{(2+\tau)^{2}}, \\
f_{2}(\tau) &= -\frac{1}{96 f_{\nu}(0)}\frac{\tau(4+\tau)\{32+\tau(64+\tau[36+7\tau])\}}{(2+\tau)^{3}}, \\
f_{3}(\tau) &= -\frac{1}{96 f_{\nu}(0)}\frac{32+\tau\left\{80+\tau[44+8\tau+k^{2}(1+\tau)(4+\tau)^{2}]\right\}}{(2+\tau)^{2}} - \frac{1}{15}, \\
f_{4}(\tau) &= -\frac{1}{96 f_{\nu}(0)}\frac{k^{2} \tau \left\{4+\tau [16+\tau (16+3\tau) ]\right\}}{121569(2+\tau)^{3}}.
\end{align}
\nt Following \rfcite{DPS} we need to further decompose $\mathbf{B}(\tau) = \mathbf{B_{0}}(\tau) + \mathbf{B_{-2}}(\tau) k^{2}$ and those matrices are \ent
\[
     \mathbf{B_{-2}}(\tau)=
  \left[ {\begin{array}{cc}
   0 & 0\\
   0 & b_{0}(\tau)\
  \end{array} } \right], \quad \text{\nt and \ent} \quad
     \mathbf{B_{0}}(\tau)=
  \left[ {\begin{array}{cc}
   0 & 0\\
   b_{-2}(\tau)& 1\\
  \end{array} } \right] .
\]
\nt where \ent
\begin{align}
b_{0}(\tau) &= \frac{4\left\{ 8f_{\nu}(0)(2+\tau)^{2} + 5 (8+\tau[20+\tau(11+2\tau)]) \right\}}{5 \tau^{2}(1+\tau)(4+\tau)^{2}}, \\
b_{-2}(\tau) &= \frac{2}{\tau} + \frac{1}{1+\tau} - \frac{2}{2+\tau} + \frac{2}{4+\tau}. 
\end{align}

\nt Finally, we use de \textit{Double Power Series Method} in \rfcite{GWGithub} to get Eqs.~(\ref{eq:flat-rad-only-soln-stress}) and (\ref{eq:third-order-stress}). \ent


\chapter{Perturbed quantities to second order}
\label{app2}

\section{Connection coefficients}
\label{connection}

The connection coefficients in a FLRW spacetime, in longitudinal gauge, up to second order are
\begin{align}
\label{eq:coeff1}
\Gamma^{0}_{00} &= {\cal H} + \ff'+\frac{1}{2}\fs'-2\ff\ff', \\
\label{eq:coeff2}
\Gamma^{0}_{0i} &= \ff_{,i} + \frac{1}{2} \fs_{,i} -2\ff\ff_{,i}, \\
\label{eq:coeff3}
\Gamma^{i}_{00} &= {\ff_{,}}^{i} + \frac{1}{2} {\fs_{,}}^{i} + 2 \pf\ff_{,}^{i},  \\
\label{eq:coeff4}
\Gamma^{i}_{j0} &= \Big[{\cal H} - \pf'-\frac{1}{2}\ps' -2 \pf\pf'\Big]\delta^{i}_{j}, \\
\label{eq:coeff5}
\Gamma^{0}_{ij} &= \Bigg[ {\cal H} - 2{\cal H} \left(\ff +\pf+ \frac{1}{2} \fs+\frac{1}{2}\ps -2 \ff\pf -2\ff^{2}  \right)    \\
& \quad  -\pf' - \frac{1}{2}\ps' +2\ff\pf'  \Bigg]\delta_{ij} ,\notag \\
\label{eq:coeff6}
\Gamma^{i}_{jk} &= -{\delta^{i}_{k}}\pf_{,j} - {\delta^{i}_{j}}\pf_{,k} + {\delta_{j k}}\pf_{,}^{i}- \frac{1}{2} \left( \delta_{k}^{i}\ps_{,j} + \delta_{j}^{i}\ps_{,k} - {\delta_{jk}}\ps_{,}^{i} \right)  \\
& \quad -2 \pf \left( \delta^{i}_{k}\pf_{, j} + \delta^{i}_{j}\pf_{, k} - \delta_{j k}\pf_{,}^{i} \right),\notag 
\end{align}
including only scalar perturbations. To translate the FLRW coefficients into Minkowski spacetime we just set ${\cal H}=0$.

\section{Perturbed Ricci Tensor $R_{\mu \nu}$}
\label{riccipert}

The perturbed Ricci tensor components in a FLRW spacetime, in longitudinal gauge, up to second order are
\begin{align}
\label{eq:Ricci00}
R_{00} &= -3{\cal H}' +3\pf'' + \nabla^{2}\ff +3{\cal H} \left( \ff'+\pf' \right) +{\cal H}\Big[ \frac{3}{2}\left( \fs'+\ps' \right) + 6\left( \pf\pf' - \ff\ff' \right)\Big] \notag\\
&\quad + 2\pf\left( 3\pf''+\nabla^{2}\ff \right) +\frac{1}{2}\Big[ 6\left( \pf' \right)^{2}-6\ff' \pf'+3\ps''+\nabla^{2}\fs \Big] - \left( \ff+\pf\right)_{,i}\ff_{,}^{i}, \\
\label{eq:Ricci0i}
R_{0j} &= 2\left( \pf' +{\cal H} \ff\right)_{,j} +{\cal H}\left( \fs_{,j} -4 \ff \ff_{,j} \right) +2 \pf' \left( 2\pf-\ff \right)_{,j} +4 \pf \pf_{,j}' + \ps'_{,j}, \notag\\
&\quad \\
\label{eq:Ricci-ij}
R_{ij} &=  \left(2 {\cal H}^{2}+{\cal H}'\right) \delta_{ij} +\Big[ \nabla^{2}\pf -\pf'' - 2\left( 2{\cal H}^{2}+{\cal H}'\right)\left(\ff+\pf\right)-{\cal H}\left( \ff'+5\pf' \right) \Big] \delta_{ij}  \notag\\
&\quad + \left(\pf-\ff\right)_{,ij} +\Big\{ \ff'\pf' + \left( 2{\cal H}^{2}-{\cal H}' \right)\left[ 4\left(\ff\right)^{2}+4\ff\pf-\fs-\ps \right]  \notag\\
&\quad +{\cal H}\left[ \ff' \left( 4\ff + 2\pf\right) +10 \ff \pf' -\frac{1}{2}\left( \fs'+5\ps' \right)\right] -\frac{1}{2}\left( \ps'' - \nabla^{2}\ps\right) \notag \\
&\quad +\left( \pf '\right)^{2} + 2\ff \pf'' + 2 \pf \nabla^{2}\pf+{\left( \ff+\pf \right)_{,}}^{i}\pf_{,i}\Big\}\delta_{ij} + \left( 3\pf -\ff \right)_{,i} \pf_{,j}\notag \\
&\quad +\left( \ff-\pf\right)_{,i}\ff_{,j} + 2 \ff \ff_{,ij} + 2 \pf \pf_{,ij} +\frac{1}{2}\left( \ps - \fs \right)_{,ij} ,
\end{align}
including only scalar perturbations. To translate the FLRW components into Minkowski spacetime we just set ${\cal H}=0$.

%

\section{Second order quantities in terms of the metric potentials}
\label{metriccomp}

\subsection{Geodesic Equation}

Solving \eq{eq:geodesic} at second order gives
\begin{align}
\label{eq:dnu2}
\frac{\dd \dnus}{\dd \lambda} &= -2 \frac{\dd \fs}{\dd \lambda} + \left( \fs' +\ps' \right) + 4\left( \pf\pf' - \ff\ff'\right)+ 4 \ff \left( \frac{\dd \ff}{\dd \lambda}\right)  \\
& \quad - 4 \dnuf\left( \frac{\dd \ff}{\dd \lambda} - \frac{\dd \pf}{\dd \lambda} \right)- 4 \dnf^{i}\ff_{,i} - 4 \dnuf \left( n^{i}\pf_{,i} \right)  \notag\\
&\quad -2 \dnuf\left( \frac{\dd \dnuf}{\dd \lambda}\right) - 2 \dnf^{i}\left(\dnuf\right)_{,i} + 2\dnuf\left( n^{i}\dnuf_{,i}\right), \notag \\
%
\label{eq:dni2}
\frac{\dd \dns^{i}}{\dd \lambda} &=  2\frac{\dd\ps}{\dd \lambda}n^{i} - \left( {\fs_{,}}^{i} + {\ps_{,}}^{i}\right) +8\pf \left( \frac{\dd \pf}{\dd \lambda}\right)n^{i} + 4 \left( \frac{\dd\pf}{\dd \lambda}\right)\dnf^{i} \\
&\quad -4\pf\left( {\ff_{,}}^{i} + {\pf_{,}}^{i}\right) - 4\left( \ff + \pf\right){\pf_{,}}^{i} - 4\dnuf \left( {\ff_{,}}^{i} + {\pf_{,}}^{i} \right) \notag \\
& \quad +4\left( \dnuf \pf' + \pf_{,k}\dnf^{k} \right)n^{i} - 2 \dnuf \left( \frac{\dd \dnf^{i}}{\dd \lambda} \right) \notag \\
&\quad -2 \left(\dnf^{k}\right)\left( \dnf^{i} \right)_{,k} + 2 \left[ n^{k}\left( \dnf^{i} \right)_{,k}\right]. \notag
\end{align}
Using \eq{eq:null-conditions}, and the integrated version of \eqs{eq:dnu1} and \eqref{eq:dn1}, we rewrite \eqs{eq:dnu2} and \eqref{eq:dni2} purely in terms of the metric potentials, 
\begin{align}
\label{eq:dnu2-metric}
\frac{\dd \dnus}{\dd \lambda} &= -2 \frac{\dd \fs}{\dd \lambda} + \left( \fs' + \ps' \right) + 4\pf' \left( \ff + \pf \right) + 4\ff \left( \frac{\dd \ff}{\dd \lambda} - 2\frac{\dd \pf}{\dd \lambda}\right) \\
& \quad -8 \left( \pf \ff_{,i} - \ff \pf_{,i} \right)n^{i} + 8 n^{i}\ff_{,i}\left( \ff - \pf \right)  \notag \\
& \quad + 4 \left( \ff' - \pf'\right)\int_{\lambda_{o}}^{\lambda_{s}}\dd\lambda\left( \ff' + \pf' \right) - 2n^{i} \left( \ff_{,i} + \pf_{,i} \right) \int_{\lambda_{o}}^{\lambda_{s}}\dd\lambda\left( \ff' + \pf'\right) \notag \\
& \quad + 2 \left( \frac{\dd\ff}{\dd \lambda} - \frac{\dd \pf}{\dd \lambda}\right)\int_{\lambda_{o}}^{\lambda_{s}}\dd\lambda \left( \ff' + \pf' \right)+4\left( \ff-\pf\right)n^{i}\int_{\lambda_{o}}^{\lambda_{s}}\dd\lambda\left( \ff' + \pf'\right)_{,i} \notag \\
&\quad +2\left( \int_{\lambda_{o}}^{\lambda_{s}}\dd\lambda{\left( \ff + \pf\right)_{,}}^{i}\right)\left( \int_{\lambda_{o}}^{\lambda_{s}}\dd\lambda\left( \ff' + \pf'\right)_{,i}\right)\notag \\
&\quad -2 n^{i}\left( \int_{\lambda_{o}}^{\lambda_{s}}\dd\lambda\left( \ff' + \pf' \right)_{,i} \right)\left( \int_{\lambda_{o}}^{\lambda_{s}}\dd\lambda\left( \ff' + \pf' \right) \right),\notag 
\end{align}
\begin{align}
\label{eq:dni2-metric}
\frac{\dd \dns^{i}}{\dd \lambda} &=  2 \frac{\dd \ps}{\dd \lambda} n^{i} - \left( {\fs_{,}}^{i}+{\ps_{,}}^{i}\right) +4 \left( \frac{\dd \pf}{\dd \lambda} \right)\left( 2\ff + 3\pf\right)n^{i} \\
&\quad +8\left( \ff {\ff_{,}}^{i} - \pf {\pf_{,}}^{i} \right) + 4\left( \ff {\pf_{,}}^{i} - \pf {\ff_{,}}^{i}\right) -8\ff\left( \pf' - n^{k}\pf_{,k}\right)n^{i}\notag \\
&\quad +4 \left( \pf - \ff\right)n^{k}\int_{\lambda_{o}}^{\lambda_{s}}\dd\lambda\left( {\ff_{,}}^{i} + {\pf_{,}}^{i} \right)_{,k} +4 \left( \pf'-n^{k}\pf_{,k}\right)n^{i}\int_{\lambda_{o}}^{\lambda_{s}}\dd\lambda\left( \ff' + \pf'\right) \notag\\
&\quad -4\left( \frac{\dd\pf}{\dd\lambda}\right)\int_{\lambda_{o}}^{\lambda_{s}}\dd\lambda\left( {\ff_{,}}^{i}+{\pf_{,}}^{i}\right) - 4n^{i}\left( \frac{\dd \pf}{\dd \lambda}\right)\int_{\lambda_{o}}^{\lambda_{s}}\dd\lambda\left( \ff' + \pf'\right) \notag \\
&\quad - 4\left( {\ff_{,}}^{i}+{\pf_{,}}^{i}\right)\int_{\lambda_{o}}^{\lambda_{s}}\dd\lambda\left( \ff' + \pf'\right) - 4 \ff \int_{\lambda_{o}}^{\lambda_{s}}\dd\lambda\left( {\ff_{,}}^{i} + {\pf_{,}}^{i} \right) \notag \\
& \quad +2\left( \int_{\lambda_{o}}^{\lambda_{s}}\dd\lambda\left( \ff' + \pf' \right) \right)\left( \int_{\lambda_{o}}^{\lambda_{s}}\dd\lambda\left( {\ff_{,}}^{i} + {\pf_{,}}^{i}\right)  \right)\notag \\
& \quad - 2\left( \int_{\lambda_{o}}^{\lambda_{s}}\dd\lambda\left( {\ff_{,}}^{k} + {\pf_{,}}^{k} \right) \right) \left( \int_{\lambda_{o}}^{\lambda_{s}}\dd\lambda\left( {\ff_{,}}^{i} + {\pf_{,}}^{i} \right)_{,k} \right) \notag \\
& \quad -2 n^{k}\left( \int_{\lambda_{o}}^{\lambda_{s}}\dd\lambda\left( {\ff_{,}}^{i}+{\pf_{,}}^{i} \right)_{,k}\right) \left( \int_{\lambda_{o}}^{\lambda_{s}}\dd\lambda\left( \ff' + \pf'\right) \right), \notag
\end{align}
where we integrate along the line of sight from $\lambda_{o}$ to $\lambda_{s}$.

\subsection{Energy}

\begin{align}
\label{eq:dE1-metric}
\Ef &= \left(-2 \ff \Big|_{o}^{s}+\int_{\lambda_{o}}^{\lambda_{s}}\left( \ff' +\pf' \right)\dd\lambda\right) + \ff - \left(v_{1i}n^{i}\right),  \\
\label{eq:dE2-metric}
\Es &=  \left( -2 \fs\Big|_{o}^{s} + \int_{\lambda_{o}}^{\lambda_{s}}\dd\lambda\left( \fs' + \ps' \right)\right)+\fs - \left( v_{2i} n^{i}\right) - 3 \pf^{2} + \left( v_{1k}v_{1}^{k} \right)  \\
&\quad + 2\ff \int_{\lambda_{o}}^{\lambda_{s}}\dd\lambda\left( \ff' + \pf' \right) + 2 v_{1 i}\int_{\lambda_{o}}^{\lambda_{s}}\dd \lambda\left( {\ff_{,}}^{i} + {\pf_{,}}^{i} \right) \notag \\
&\quad + 4\int_{\lambda_{o}}^{\lambda_{s}}\dd\lambda\left[\pf' \left( \ff + \pf \right)\right] + 4\int_{\lambda_{o}}^{\lambda_{s}}\dd\lambda \left[ \ff \left( \frac{\dd \ff}{\dd \lambda} - 2\frac{\dd \pf}{\dd \lambda}\right)\right] \notag\\
& \quad -8 \int_{\lambda_{o}}^{\lambda_{s}}\dd\lambda\left[\left( \pf \ff_{,i} - \ff \pf_{,i} \right)n^{i}\right] + 8\int_{\lambda_{o}}^{\lambda_{s}}\dd\lambda\left[ n^{i}\ff_{,i}\left( \ff - \pf \right)\right]  \notag \\
& \quad + \int_{\lambda_{o}}^{\lambda_{s}}\dd\tilde{\lambda}\Bigg\{4 \left( \ff' - \pf'\right)\int_{\lambda_{o}}^{\lambda_{s}}\dd\lambda\left( \ff' + \pf' \right) - 2n^{i} \left( \ff_{,i} + \pf_{,i} \right) \int_{\lambda_{o}}^{\lambda_{s}}\dd\lambda\left( \ff' + \pf'\right) \notag \\
& \quad + 2 \left( \frac{\dd\ff}{\dd \lambda} - \frac{\dd \pf}{\dd \lambda}\right)\int_{\lambda_{o}}^{\lambda_{s}}\dd\lambda \left( \ff' + \pf' \right)+4\left( \ff-\pf\right)n^{i}\int_{\lambda_{o}}^{\lambda_{s}}\dd\lambda\left( \ff' + \pf'\right)_{,i} \notag \\
&\quad +2\left( \int_{\lambda_{o}}^{\lambda_{s}}\dd\lambda{\left( \ff + \pf\right)_{,}}^{i}\right)\left( \int_{\lambda_{o}}^{\lambda_{s}}\dd\lambda\left( \ff' + \pf'\right)_{,i}\right)\notag \\
&\quad -2 n^{i}\left( \int_{\lambda_{o}}^{\lambda_{s}}\dd\lambda\left( \ff' + \pf' \right)_{,i} \right)\left( \int_{\lambda_{o}}^{\lambda_{s}}\dd\lambda\left( \ff' + \pf' \right) \right)\Bigg\}. \notag
\end{align}

\subsection{Observed redshift}

\begin{align}
\label{eq:dz2}
\dzs&= \dnuf\Big[ \left( v_{1 i}n^{i}\right)_{o} - \ff|_{o} \Big] - \left(\ff|_{s}\right)\left(\ff|_{o}\right) + \ff|_{o}\left( v_{1i}n^{i}\right) + \ff|_{s}\left( v_{1i}n^{i} \right)_{o} \\
&\quad - \left( v_{1i}n^{i} \right)_{o}\left( v_{1i}n^{i} \right)_{s} + \left( \ff|_{o}\right)^{2} - 2\ff|_{o}\left( v_{1i}n^{i}\right)_{o} + \left( v_{1i}n^{i}\right)_{o}^{2} \notag \\
&\quad +\frac{1}{2}\Bigg\{ \dnus +\fs|^{s}_{o} - \left( v_{2i}n^{i}\right)^{s}_{o} + \left( \ff|^{s}_{o}\right)^{2} + \left( v_{1k}v_{1}^{k}\right)^{s}_{o}+\left[ 4\pf \left( v_{1i}n^{i}\right) \right]^{s}_{o} \notag \\
&\quad\quad\quad\quad\quad\quad\quad\quad\quad\quad\quad\quad\quad\quad\quad\quad\quad\quad\quad\quad\quad +2 \ff|_{s} \dnuf - 2 v_{1i} \dnf^{i} \Bigg\}, \notag
\end{align}
and using \eqs{eq:dnu1}, \eqref{eq:dn1}, \eqref{eq:dnu2} and \eqref{eq:dE1-metric}, in terms of the metric potentials is
\begin{align}
\label{eq:dz2-metric}
\dzs&= -\frac{1}{2}\fs|^{s}_{o} - \frac{1}{2}\left( v_{2i} n^{i}\right)^{s}_{o} + \frac{1}{2}\int_{\lambda_{o}}^{\lambda_{s}}\dd\lambda \left( \fs' + \ps' \right) +\frac{1}{2}\left( v_{1k}v^{k}_{1}\right)^{s}_{o}  \\
&\quad - \frac{3}{2} \left( \ff|^{s}_{o}\right)^{2}+6 \ff|_{s}\ff|_{o} +\ff|_{o}\left( v_{1i}n^{i} \right)_{s} -\ff|_{s}\left( v_{1i}n^{i} \right)_{o} -\left( v_{1i}n^{i} \right)_{o}\left( v_{1i}n^{i}\right)^{s}_{o}  \notag \\
&\quad - 2 \pf|_{o}\left( v_{1i}n^{i} \right)_{o}+ \ff|^{s}_{o}\int_{\lambda_{o}}^{\lambda_{s}}\dd\lambda\left( \ff' + \pf' \right) + \left( v_{1i}n^{i} \right)_{o}\int_{\lambda_{o}}^{\lambda_{s}}\dd\lambda\left( \ff' + \pf' \right) \notag \\
&\quad + \ff \int_{\lambda_{o}}^{\lambda_{s}}\dd\lambda\left( \ff' + \pf' \right) +  2v_{1 i}\int_{\lambda_{o}}^{\lambda_{s}}\dd \lambda\left( {\ff_{,}}^{i} + {\pf_{,}}^{i} \right) \notag \\
&\quad + 2\int_{\lambda_{o}}^{\lambda_{s}}\dd\lambda\left[\pf' \left( \ff + \pf \right)\right] + 2\int_{\lambda_{o}}^{\lambda_{s}}\dd\lambda \left[ \ff \left( \frac{\dd \ff}{\dd \lambda} - 2\frac{\dd \pf}{\dd \lambda}\right)\right] \notag\\
& \quad -4 \int_{\lambda_{o}}^{\lambda_{s}}\dd\lambda\left[\left( \pf \ff_{,i} - \ff \pf_{,i} \right)n^{i}\right] + 4\int_{\lambda_{o}}^{\lambda_{s}}\dd\lambda\left[ n^{i}\ff_{,i}\left( \ff - \pf \right)\right]  \notag \\
& \quad + \int_{\lambda_{o}}^{\lambda_{s}}\dd\tilde{\lambda}\Bigg\{2 \left( \ff' - \pf'\right)\int_{\lambda_{o}}^{\lambda_{s}}\dd\lambda\left( \ff' + \pf' \right) - n^{i} \left( \ff_{,i} + \pf_{,i} \right) \int_{\lambda_{o}}^{\lambda_{s}}\dd\lambda\left( \ff' + \pf'\right) \notag \\
& \quad +  \left( \frac{\dd\ff}{\dd \lambda} - \frac{\dd \pf}{\dd \lambda}\right)\int_{\lambda_{o}}^{\lambda_{s}}\dd\lambda \left( \ff' + \pf' \right)+2\left( \ff-\pf\right)n^{i}\int_{\lambda_{o}}^{\lambda_{s}}\dd\lambda\left( \ff' + \pf'\right)_{,i} \notag \\
&\quad +\left( \int_{\lambda_{o}}^{\lambda_{s}}\dd\lambda{\left( \ff + \pf\right)_{,}}^{i}\right)\left( \int_{\lambda_{o}}^{\lambda_{s}}\dd\lambda\left( \ff' + \pf'\right)_{,i}\right)\notag \\
&\quad - n^{i}\left( \int_{\lambda_{o}}^{\lambda_{s}}\dd\lambda\left( \ff' + \pf' \right)_{,i} \right)\left( \int_{\lambda_{o}}^{\lambda_{s}}\dd\lambda\left( \ff' + \pf' \right) \right)\Bigg\}. \notag
\end{align}

We will need the derivative of the redshift perturbation w.r.t. the affine parameter $\nu$ for the \textit{volume density}, and that is given by
\begin{align}
\label{eq:dz2dnu}
\frac{\dd\dzs}{\dd\nu}&= \frac{1}{2}\left[ \ps' -\partial_{\chi}\fs - \frac{\dd\left( v_{2i} n^{i}\right)}{\dd\nu}+\frac{\dd\left( v_{1k}v^{k}_{1}\right)}{\dd\nu}\right] - 3\ff \left( \frac{\dd\ff}{\dd\nu}\right)\\
&\quad +\frac{\dd\ff}{\dd\nu}\left( v_{1i}n^{i} \right)+\ff\frac{\dd\left( v_{1i}n^{i} \right)}{\dd\nu}-\left( v_{1i}n^{i} \right)\frac{\dd\left( v_{1i}n^{i}\right)}{\dd\nu} - 2 \frac{\dd\pf}{\dd\nu}\left( v_{1i}n^{i} \right)-2\pf\frac{\dd\left( v_{1i}n^{i} \right)}{\dd\nu}\notag\\
&\quad + \frac{\dd\ff}{\dd\nu}\int_{0}^{\chi_{s}}\dd\chi\left( \ff' + \pf' \right) + \ff\left( \ff'+\pf'\right)\notag \\
&\quad + \frac{\dd\left( v_{1i}n^{i} \right)}{\dd\nu}\int_{0}^{\chi_{s}}\dd\chi\left( \ff' + \pf' \right)+\left(v_{1i}n^{i}\right)\left( \ff'+\pf'\right) \notag \\
&\quad + \frac{\dd\ff}{\dd\nu} \int_{0}^{\chi_{s}}\dd\chi\left( \ff' + \pf' \right)+\ff\left( \ff'+\pf' \right) \notag \\
&\quad +  2\frac{\dd v_{1 i}}{\dd\nu}\int_{0}^{\chi{s}}\dd \chi\left( {\ff_{,}}^{i} + {\pf_{,}}^{i} \right) + 2v_{1i}\left( {\ff_{,}}^{i} + {\pf_{,}}^{i} \right)   \notag \\
&\quad + 2\left[\pf' \left( \ff + \pf \right)\right] + 2 \left[ \ff \left( \frac{\dd \ff}{\dd \nu} - 2\frac{\dd \pf}{\dd \nu}\right)\right] \notag\\
& \quad -4 \left[\left( \pf \ff_{,i} - \ff \pf_{,i} \right)n^{i}\right] + 4\left[ n^{i}\ff_{,i}\left( \ff - \pf \right)\right]  \notag \\
& \quad +2 \left( \ff' - \pf'\right)\int_{0}^{\chi_{s}}\dd\chi\left( \ff' + \pf' \right) - n^{i} \left( \ff_{,i} + \pf_{,i} \right) \int_{0}^{\chi_{s}}\dd\chi\left( \ff' + \pf'\right) \notag \\
& \quad +  \left( \frac{\dd\ff}{\dd \nu} - \frac{\dd \pf}{\dd \nu}\right)\int_{0}^{\chi_{s}}\dd\chi \left( \ff' + \pf' \right)+2\left( \ff-\pf\right)n^{i}\int_{0}^{\chi_{s}}\dd\chi\left( \ff' + \pf'\right)_{,i} \notag \\
&\quad +\left( \int_{0}^{\chi_{s}}\dd\chi{\left( \ff + \pf\right)_{,}}^{i}\right)\left( \int_{0}^{\chi_{s}}\dd\chi\left( \ff' + \pf'\right)_{,i}\right)\notag \\
&\quad - n^{i}\left( \int_{0}^{\chi_{s}}\dd\chi\left( \ff' + \pf' \right)_{,i} \right)\left( \int_{0}^{\chi_{s}}\dd\chi\left( \ff' + \pf' \right) \right). \notag
\end{align}

\subsection{Angular diameter distance}

Using \eqs{eq:dnu1}, \eqref{eq:dn1}, \eqref{eq:dnu2-metric}, \eqref{eq:dni2-metric}, \eqref{eq:dE2-metric}, \eqref{eq:da-linear}, \eqref{eq:Ricci00}, \eqref{eq:Ricci0i}, \eqref{eq:Ricci-ij} and \eqref{eq:contracted-null-shear} we find that the second order perturbation to the angular diameter distance becomes
\begin{align}
\label{eq:da-second-metric}
\frac{\das(\lambda_{s})}{\bar{d}_{A}(\lambda_{s})} &= \fs|_{o}-\ps|_{o}-\ps |_{s} - \left( v_{2 i}n^{i}\right)_{o} + \left(\ff|_{o}\right)^{2} + \left(v_{1 k}v_{1}^{k}\right)_{o} \\
&\quad + 4 \pf|_{o} \left( v_{1i}n^{i}\right)_{o}  -\frac{1}{\lambda_{o}-\lambda_{s}} \int_{\lambda_{o}}^{\lambda_{s}}\dd\lambda\left( \fs + \ps \right) \notag\\
&\quad - \frac{1}{2\left( \lambda_{o}-\lambda_{s}\right)}\int_{\lambda_{o}}^{\lambda_{s}}\dd\lambda\left(\lambda_{s}-\lambda\right)\left( \lambda_{o}-\lambda\right)\Bigg[\frac{1}{2}\nabla^{2}\left( \fs + \ps\right) \notag \\
&\quad +\frac{1}{2}\left( \ff + \pf \right)_{,ij}n^{i}n^{j} - \frac{2}{\lambda_{o}-\lambda}\left( \frac{\dd \dnus}{\dd\lambda} \right) \Bigg] \notag \\
&\quad - \int_{\lambda_{o}}^{\lambda_{s}}\dd\tilde{\lambda}\Bigg\{ \left(\int_{\lambda_{o}}^{\lambda_{s}}\dd\lambda\left( \ff + \pf \right)_{,ij}\right)\left( \int_{\lambda_{o}}^{\lambda_{s}}\dd\lambda{\left( \ff + \pf\right)_{,}}^{ij}\right) \notag \\
&\quad -\frac{1}{4}\left[ \int_{\lambda_{o}}^{\lambda_{s}}\dd\lambda\nabla^{2}\left( \ff + \pf \right)\right]^{2}\Bigg\}+\frac{1}{\lambda_{o}-\lambda_{s}}\int_{\lambda_{o}}^{\lambda_{s}}\dd\tilde{\lambda}\Bigg[\frac{1}{2}\ff'\left( 5\pf'-3\pf \right) \notag  \\
&\quad +\frac{1}{2}\left( \ff'-\pf'-\frac{\dd \ff}{\dd \lambda}+\frac{\dd \pf}{\dd \lambda}\right)^{2}+\pf\nabla^{2}\left( \ff + \pf \right) + \frac{1}{2}{\pf_{,}}^{k} \left( \ff_{,k}+\pf_{,k}\right)\notag\\
&\quad +\pf\left( 3\pf'' + \pf_{,ij}n^{i}n^{j} \right) + \ff\left( \pf'' + \ff_{,ij}n^{i}n^{j} \right) +4 \pf \left( \pf' - \frac{\dd \pf}{\dd \lambda}\right) \notag \\
&\quad -2\pf'' \left( \frac{\dd \ff}{\dd\lambda} \right) + 2\left( \frac{\dd\pf}{\dd \lambda} \right)^{2} + 2\dnuf \nabla^{2}\left( \ff + \pf \right) + 4\dnuf \left( \frac{\dd \pf'}{\dd \lambda} \right) \notag\\
&\quad +2\left( \ff+\pf \right)\left( \nabla^{2}\pf -\pf'' \right) + 4\dnf^{i}\pf_{,i} + 2 n^{i}\dnf^{j}\left( \pf-\ff \right)_{,ij} \Bigg] \notag \\
&\quad -\int_{\lambda_{o}}^{\lambda_{s}}\dd\tilde{\lambda}\Bigg\{ \left[\frac{\dd^{2}\pf}{\dd\lambda^{2}} + \nabla^{2}\left( \ff + \pf \right) - \left( \ff+\pf\right)_{,ij}n^{i}n^{j}\right]\daf \Bigg\} \notag\\
&\quad -\int_{\lambda_{o}}^{\lambda_{s}}\dd\tilde{\lambda}\Bigg\{ \dnuf\left( \dnuf\right)' + \dnf^{i}\left( \dnuf \right)_{,i} +3\Big[ \dnuf\left( \frac{\dd \ff}{\dd \lambda} -\pf \right)  \notag \\
&\quad - \left( \ff+\pf \right)\pf'  \Big] +4\dnuf\left( \frac{\dd^{2}\das}{\dd\lambda^{2}} \right)+ 4\left( \frac{\dd\dnuf}{\dd\lambda} \right)\left( \frac{\dd\daf}{\dd\lambda} \right)  \Bigg\}. \notag 
\end{align}
And the angular distance in terms of the observed redshift is given by
\begin{align}
\label{eq:da2dz}
\delta^{(2)}\hat{d}_{A}\left(\hat{z}_{s}\right) &= \frac{\chi_{s}}{2\left(1+\hat{z}_{s}\right)}\Bigg\{ \frac{\das}{\bar{d}_{A}} + \left( 1-\frac{1}{\H\chi_{s}}\right)\dzs \\
&\quad +2\left( 1-\frac{1}{\H\chi_{s}}\right) \left( \frac{\daf}{\bar{d}_{A}}+\frac{1}{\H}\frac{\dd\dzf}{\dd\nu}\right)\dzf \notag\\
&\quad +2\frac{\dd}{\dd\nu}\left( \frac{\daf}{\bar{d}_{A}} \right)\frac{\dzf}{\H} + \frac{\H' - \H^{2}}{\H^{3}\chi_{s}}\left(\dzf\right)^{2}\Bigg\} \notag \\
&= \frac{\chi_{s}}{2\left(1+\hat{z}_{s}\right)}\Bigg[ -\ps|_{s} - \ps|_{o} - \left(1-\frac{1}{\H \chi_{s}}\right)\fs|_{s} +2\left(1-\frac{1}{\H\chi_{s}}\right)\left( v_{2i}n^{i} \right)_{o} \notag \\  
&\quad + \left( 1-\frac{1}{\H \chi_{s}}\right)\left( v_{2i}n^{i} \right)_{s} + \left( 1-\frac{1}{\H\chi_{s}}\right)\int_{0}^{\chi_{s}}\dd\chi\left( \fs'+\ps' \right) - \frac{1}{\chi}\int_{0}^{\chi_{s}}\left( \fs+\ps \right)\dd\chi\notag \\  
&\quad -\frac{1}{2\chi_{s}}\int_{0}^{\chi_{s}}\dd\chi\left(\chi-\chi_{s}\right)\chi\left[ \nabla^{2}\left( \fs+\ps \right) - n^{i}n^{j}\left( \fs + \ps \right)_{,ij} -\frac{2}{\chi}\frac{\dd \dnus}{\dd\nu}\right]\notag \\  
&\quad +\frac{1}{2}\left( 1-\frac{1}{\H\chi_{s}}\right)\left(v_{1k}v_{1}^{k}\right)_{s}+\frac{1}{2}\left( 1+\frac{1}{\H\chi_{s}}\right)\left( v_{1k}v_{1}^{k}\right)_{o} -\frac{1}{2}\left( 1-\frac{3}{\H\chi_{s}}\right)\left(\ff|_{s}\right)^{2} \notag \\  
&\quad -\frac{1}{2}\left( 1+\frac{1}{\H\chi_{s}}\right)\left( \ff|_{o}\right)^{2} +2\left( 1+\frac{2}{\H\chi_{s}}\right)\pf|_{o}\left( v_{1i}n^{i} \right)_{o} \notag \\  
&\quad +\left( 1-\frac{1}{\H\chi_{s}}\right)\Big[ \ff|_{o}\left( v_{1i}n^{i}\right)_{s} - \ff|_{s}\left( v_{1i}n^{i}\right)_{o} + 6\ff|_{s}\ff|_{o} -\left( v_{1i}n^{i} \right)_{o}\left( v_{1i}n^{i} \right)_{o}^{s} \Big]\notag \\  
&\quad - \int_{\lambda_{o}}^{\lambda_{s}}\dd\tilde{\lambda}\Bigg\{ \left(\int_{\lambda_{o}}^{\lambda_{s}}\dd\lambda\left( \ff + \pf \right)_{,ij}\right)\left( \int_{\lambda_{o}}^{\lambda_{s}}\dd\lambda{\left( \ff + \pf\right)_{,}}^{ij}\right) \notag \\
&\quad -\frac{1}{4}\left[ \int_{\lambda_{o}}^{\lambda_{s}}\dd\lambda\nabla^{2}\left( \ff + \pf \right)\right]^{2}\Bigg\}+\frac{1}{\lambda_{o}-\lambda_{s}}\int_{\lambda_{o}}^{\lambda_{s}}\dd\tilde{\lambda}\Bigg[\frac{1}{2}\ff'\left( 5\pf'-3\pf \right) \notag  \\
&\quad +\frac{1}{2}\left( \ff'-\pf'-\frac{\dd \ff}{\dd \lambda}+\frac{\dd \pf}{\dd \lambda}\right)^{2}+\pf\nabla^{2}\left( \ff + \pf \right) + \frac{1}{2}{\pf_{,}}^{k} \left( \ff_{,k}+\pf_{,k}\right)\notag\\
&\quad +\pf\left( 3\pf'' + \pf_{,ij}n^{i}n^{j} \right) + \ff\left( \pf'' + \ff_{,ij}n^{i}n^{j} \right) +4 \pf \left( \pf' - \frac{\dd \pf}{\dd \lambda}\right) \notag \\
&\quad -2\pf'' \left( \frac{\dd \ff}{\dd\lambda} \right) + 2\left( \frac{\dd\pf}{\dd \lambda} \right)^{2} + 2\dnuf \nabla^{2}\left( \ff + \pf \right) + 4\dnuf \left( \frac{\dd \pf'}{\dd \lambda} \right) \notag\\
&\quad +2\left( \ff+\pf \right)\left( \nabla^{2}\pf -\pf'' \right) + 4\dnf^{i}\pf_{,i} + 2 n^{i}\dnf^{j}\left( \pf-\ff \right)_{,ij} \Bigg] \notag \\
&\quad -\int_{\lambda_{o}}^{\lambda_{s}}\dd\tilde{\lambda}\Bigg\{ \left[\frac{\dd^{2}\pf}{\dd\lambda^{2}} + \nabla^{2}\left( \ff + \pf \right) - \left( \ff+\pf\right)_{,ij}n^{i}n^{j}\right]\daf \Bigg\} \notag\\
&\quad -\int_{\lambda_{o}}^{\lambda_{s}}\dd\tilde{\lambda}\Bigg\{ \dnuf\left( \dnuf\right)' + \dnf^{i}\left( \dnuf \right)_{,i} +3\Big[ \dnuf\left( \frac{\dd \ff}{\dd \lambda} -\pf \right)  \notag \\
&\quad - \left( \ff+\pf \right)\pf'  \Big] +4\dnuf\left( \frac{\dd^{2}\daf}{\dd\lambda^{2}} \right)+ 4\left( \frac{\dd\dnuf}{\dd\lambda} \right)\left( \frac{\dd\daf}{\dd\lambda} \right)  \Bigg\}\notag
\end{align}
\begin{align}
&\quad + \left( 1-\frac{1}{\H \chi_{s}}\right)\ff|^{s}_{o}\int_{\lambda_{o}}^{\lambda_{s}}\dd\lambda\left( \ff' + \pf' \right) + \left( 1-\frac{1}{\H \chi_{s}}\right)\left( v_{1i}n^{i} \right)_{o}\int_{\lambda_{o}}^{\lambda_{s}}\dd\lambda\left( \ff' + \pf' \right) \notag \\
&\quad + \left( 1-\frac{1}{\H \chi_{s}}\right)\ff \int_{\lambda_{o}}^{\lambda_{s}}\dd\lambda\left( \ff' + \pf' \right) +  2\left( 1-\frac{1}{\H \chi_{s}}\right)v_{1 i}\int_{\lambda_{o}}^{\lambda_{s}}\dd \lambda\left( {\ff_{,}}^{i} + {\pf_{,}}^{i} \right) \notag \\
&\quad + 2\left( 1-\frac{1}{\H \chi_{s}}\right)\int_{\lambda_{o}}^{\lambda_{s}}\dd\lambda\left[\pf' \left( \ff + \pf \right)\right] + 2\left( 1-\frac{1}{\H \chi_{s}}\right)\int_{\lambda_{o}}^{\lambda_{s}}\dd\lambda \left[ \ff \left( \frac{\dd \ff}{\dd \lambda} - 2\frac{\dd \pf}{\dd \lambda}\right)\right] \notag\\
& \quad -4 \left( 1-\frac{1}{\H \chi_{s}}\right)\int_{\lambda_{o}}^{\lambda_{s}}\dd\lambda\left[\left( \pf \ff_{,i} - \ff \pf_{,i} \right)n^{i}\right] + 4\left( 1-\frac{1}{\H \chi_{s}}\right)\int_{\lambda_{o}}^{\lambda_{s}}\dd\lambda\left[ n^{i}\ff_{,i}\left( \ff - \pf \right)\right]  \notag \\
& \quad + \int_{\lambda_{o}}^{\lambda_{s}}\dd\tilde{\lambda}\Bigg\{2 \left( \ff' - \pf'\right)\int_{\lambda_{o}}^{\lambda_{s}}\dd\lambda\left( \ff' + \pf' \right) - n^{i} \left( \ff_{,i} + \pf_{,i} \right) \int_{\lambda_{o}}^{\lambda_{s}}\dd\lambda\left( \ff' + \pf'\right) \notag \\
& \quad +  \left( \frac{\dd\ff}{\dd \lambda} - \frac{\dd \pf}{\dd \lambda}\right)\int_{\lambda_{o}}^{\lambda_{s}}\dd\lambda \left( \ff' + \pf' \right)+2\left( \ff-\pf\right)n^{i}\int_{\lambda_{o}}^{\lambda_{s}}\dd\lambda\left( \ff' + \pf'\right)_{,i} \notag \\
&\quad +\left( \int_{\lambda_{o}}^{\lambda_{s}}\dd\lambda{\left( \ff + \pf\right)_{,}}^{i}\right)\left( \int_{\lambda_{o}}^{\lambda_{s}}\dd\lambda\left( \ff' + \pf'\right)_{,i}\right)\notag \\
&\quad - n^{i}\left( \int_{\lambda_{o}}^{\lambda_{s}}\dd\lambda\left( \ff' + \pf' \right)_{,i} \right)\left( \int_{\lambda_{o}}^{\lambda_{s}}\dd\lambda\left( \ff' + \pf' \right) \right)\Bigg\}\left( 1-\frac{1}{\H \chi_{s}}\right)\notag \\
&\quad -2\left( 1-\frac{1}{\H \chi_{s}}\right)\left[ \left(v_{1i}n^{i}\right)_{o}^{s}+\ff|^{s}_{o} \int_{0}^{\chi_{s}}\left( \ff'+\pf'\right)\dd\chi \right]\times\notag \\
&\quad \Big[ -\pf|_{s} - \pf|_{o} - \left( 1 - \frac{1}{\H \chi_{s}}\right)\ff|_{s} + \left( 2-\frac{2}{\H \chi_{s}} \right)\left( v_{1i}n^{i}\right)_{o}+ \left( 1-\frac{1}{\H \chi_{s}}\right)\left( v_{1i}n^{i}\right)_{s} \notag\\
&\quad +\frac{1}{\H}\left( \pf'-\partial_{\chi}\ff + \frac{\dd\left(v_{1i}n^{i}\right)}{\dd\nu}\right)+ \left( 1 - \frac{1}{\H\chi_{s}}\right)\int_{0}^{\chi_{s}}\left( \ff'+\pf'\right)\dd\chi -\frac{2}{\chi_{s}}\int_{0}^{\chi_{s}}\pf\dd\chi \notag \\
& \quad \quad -\frac{1}{2 \chi_{s}} \int_{0}^{\chi_{s}}\dd\chi \left( \chi - \chi_{s} \right)\chi \left[ \nabla^{2}\left( \ff + \pf\right) - n^{i}n^{j}\left( \ff + \pf \right)_{,ij} - \frac{2}{\chi} \frac{\dd \dnuf}{\dd \nu} \right] \Big] \notag\\
&\quad -\frac{2}{\H} \Big[ -\pf|_{s} - \pf|_{o} - \left( 1 - \frac{1}{\H \chi_{s}}\right)\ff|_{s} + \left( 2-\frac{2}{\H \chi_{s}} \right)\left( v_{1i}n^{i}\right)_{o}\notag\\
&\quad + \left( 1-\frac{1}{\H \chi_{s}}\right)\left( v_{1i}n^{i}\right)_{s} + \left( 1 - \frac{1}{\H\chi_{s}}\right)\int_{0}^{\chi_{s}}\left( \ff'+\pf'\right)\dd\chi -\frac{2}{\chi_{s}}\int_{0}^{\chi_{s}}\pf\dd\chi \notag \\
& \quad \quad -\frac{1}{2 \chi_{s}} \int_{0}^{\chi_{s}}\dd\chi \left( \chi - \chi_{s} \right)\chi \left[ \nabla^{2}\left( \ff + \pf\right) - n^{i}n^{j}\left( \ff + \pf \right)_{,ij} - \frac{2}{\chi} \frac{\dd \dnuf}{\dd \nu} \right] \Big]\times \notag \\
&\quad \left[ \left(v_{1i}n^{i}\right)^{s}_{o}+\ff|^{s}_{o} + \int_{0}^{\chi_{s}}\left( \ff'+\pf'\right)\dd\chi \right]\notag \\
&\quad +\frac{\H'-\H^{2}}{\H^{3}\chi_{s}}\left[ \left(v_{1i}n^{i}\right)^{s}_{o}+\ff|^{s}_{o} + \int_{0}^{\chi_{s}}\left( \ff'+\pf'\right)\dd\chi \right]^{2} \Bigg]. \notag 
\end{align}

\subsection{Physical volume}

The second order perturbation to the physical volume is 
\begin{align}
\label{eq:dvv2}
&\dd \dVs = \frac{\chi^{2}}{\H\left( 1+z \right)}\Bigg\{\frac{1}{\H}\Bigg[\frac{1}{2}\left( \ps' -\partial_{\chi}\fs - \frac{\dd\left( v_{2i} n^{i}\right)}{\dd\nu}+\frac{\dd\left( v_{1k}v^{k}_{1}\right)}{\dd\nu}\right) \\
& - 3\ff \left( \frac{\dd\ff}{\dd\nu}\right)+\frac{\dd\ff}{\dd\nu}\left( v_{1i}n^{i} \right)+\ff\frac{\dd\left( v_{1i}n^{i} \right)}{\dd\nu}-\left( v_{1i}n^{i} \right)\frac{\dd\left( v_{1i}n^{i}\right)}{\dd\nu} - 2 \frac{\dd\pf}{\dd\nu}\left( v_{1i}n^{i} \right) \notag\\ 
&-2\pf\frac{\dd\left( v_{1i}n^{i} \right)}{\dd\nu} + \frac{\dd\ff}{\dd\nu}\int_{0}^{\chi_{s}}\dd\chi\left( \ff' + \pf' \right) + \ff\left( \ff'+\pf'\right)\notag \\
& + \frac{\dd\left( v_{1i}n^{i} \right)}{\dd\nu}\int_{0}^{\chi_{s}}\dd\chi\left( \ff' + \pf' \right)+\left(v_{1i}n^{i}\right)\left( \ff'+\pf'\right) \notag \\
& + \frac{\dd\ff}{\dd\nu} \int_{0}^{\chi_{s}}\dd\chi\left( \ff' + \pf' \right)+\ff\left( \ff'+\pf' \right)+  2\frac{\dd v_{1 i}}{\dd\nu}\int_{0}^{\chi{s}}\dd \chi\left( {\ff_{,}}^{i} + {\pf_{,}}^{i} \right) + 2v_{1i}\left( {\ff_{,}}^{i} + {\pf_{,}}^{i} \right)   \notag \\
& + 2\left[\pf' \left( \ff + \pf \right)\right] + 2 \left[ \ff \left( \frac{\dd \ff}{\dd \nu} - 2\frac{\dd \pf}{\dd \nu}\right)\right]-4 \left[\left( \pf \ff_{,i} - \ff \pf_{,i} \right)n^{i}\right] \notag\\
&+ 4\left[ n^{i}\ff_{,i}\left( \ff - \pf \right)\right]+2 \left( \ff' - \pf'\right)\int_{0}^{\chi_{s}}\dd\chi\left( \ff' + \pf' \right) - n^{i} \left( \ff_{,i} + \pf_{,i} \right) \int_{0}^{\chi_{s}}\dd\chi\left( \ff' + \pf'\right) \notag \\
&  +  \left( \frac{\dd\ff}{\dd \nu} - \frac{\dd \pf}{\dd \nu}\right)\int_{0}^{\chi_{s}}\dd\chi \left( \ff' + \pf' \right)+2\left( \ff-\pf\right)n^{i}\int_{0}^{\chi_{s}}\dd\chi\left( \ff' + \pf'\right)_{,i} \notag \\
& +\left( \int_{0}^{\chi_{s}}\dd\chi{\left( \ff + \pf\right)_{,}}^{i}\right)\left( \int_{0}^{\chi_{s}}\dd\chi\left( \ff' + \pf'\right)_{,i}\right)\notag \\
& - n^{i}\left( \int_{0}^{\chi_{s}}\dd\chi\left( \ff' + \pf' \right)_{,i} \right)\left( \int_{0}^{\chi_{s}}\dd\chi\left( \ff' + \pf' \right) \right) \Bigg]\notag\\
&-\frac{1}{2}\frac{\H'}{\H^{2}}\Bigg[ -\frac{1}{2}\fs|^{s}_{o} - \frac{1}{2}\left( v_{2i} n^{i}\right)^{s}_{o} + \frac{1}{2}\int_{0}^{\chi_{s}}\dd\chi \left( \fs' + \ps' \right) +\frac{1}{2}\left( v_{1k}v^{k}_{1}\right)^{s}_{o} \notag  \\
& - \frac{3}{2} \left( \ff|^{s}_{o}\right)^{2}+6 \ff|_{s}\ff|_{o} +\ff|_{o}\left( v_{1i}n^{i} \right)_{s} -\ff|_{s}\left( v_{1i}n^{i} \right)_{o} -\left( v_{1i}n^{i} \right)_{o}\left( v_{1i}n^{i}\right)^{s}_{o}  \notag \\
& - 2 \pf|_{o}\left( v_{1i}n^{i} \right)_{o}+ \ff|^{s}_{o}\int_{0}^{\chi_{s}}\dd\chi\left( \ff' + \pf' \right) + \left( v_{1i}n^{i} \right)_{o}\int_{0}^{\chi_{s}}\dd\chi\left( \ff' + \pf' \right) \notag \\
& + \ff \int_{0}^{\chi_{s}}\dd\chi\left( \ff' + \pf' \right) +  2v_{1 i}\int_{0}^{\chi_{s}}\dd \chi\left( {\ff_{,}}^{i} + {\pf_{,}}^{i} \right) \notag \\
& + 2\int_{0}^{\chi_{s}}\dd\chi\left[\pf' \left( \ff + \pf \right)\right] + 2\int_{0}^{\lambda_{s}}\dd\chi \left[ \ff \left( \frac{\dd \ff}{\dd \nu} - 2\frac{\dd \pf}{\dd \nu}\right)\right] \notag\\
& -4 \int_{0}^{\chi_{s}}\dd\chi\left[\left( \pf \ff_{,i} - \ff \pf_{,i} \right)n^{i}\right] + 4\int_{0}^{\chi_{s}}\dd\chi\left[ n^{i}\ff_{,i}\left( \ff - \pf \right)\right]  \notag \\
& + \int_{0}^{\chi_{s}}\dd\tilde{\chi}\Bigg\{2 \left( \ff' - \pf'\right)\int_{0}^{\chi_{s}}\dd\chi\left( \ff' + \pf' \right) - n^{i} \left( \ff_{,i} + \pf_{,i} \right) \int_{0}^{\chi_{s}}\dd\chi\left( \ff' + \pf'\right) \notag \\
& +  \left( \frac{\dd\ff}{\dd \nu} - \frac{\dd \pf}{\dd \nu}\right)\int_{0}^{\chi_{s}}\dd\chi \left( \ff' + \pf' \right)+2\left( \ff-\pf\right)n^{i}\int_{0}^{\chi_{s}}\dd\chi\left( \ff' + \pf'\right)_{,i} \notag \\
& +\left( \int_{0}^{\chi_{s}}\dd\chi{\left( \ff + \pf\right)_{,}}^{i}\right)\left( \int_{0}^{\chi_{s}}\dd\chi\left( \ff' + \pf'\right)_{,i}\right)\notag
\end{align}
\begin{align}
& - n^{i}\left( \int_{0}^{\chi_{s}}\dd\chi\left( \ff' + \pf' \right)_{,i} \right)\left( \int_{0}^{\chi_{s}}\dd\chi\left( \ff' + \pf' \right) \right)\Bigg\} \Bigg]+\frac{3}{2\H^{2}}\Bigg[ \pf' -\partial_{\chi}\ff -\frac{\dd\left( v_{1i}n^{i}\right)}{\dd\nu} \Bigg]^{2}\notag\\
&+\frac{1}{2}\left( \frac{\H'}{\H^{2}} \right)\left( 1+\frac{\H'}{\H^{2}}\right)\Bigg[ \left(v_{1i}n^{i}+\ff+\int_{0}^{\chi_{s}}\left( \ff'+\pf'\right)\dd\chi\right) \Bigg]^{2}\notag\\
&+\frac{1}{2\H^{2}}\left(2\left[\frac{\H'}{\H}\right]^{2}+\frac{\H''}{\H} \right)\Bigg[ \left(v_{1i}n^{i}+\ff+\int_{0}^{\chi_{s}}\left( \ff'+\pf'\right)\dd\chi\right)  \Bigg]^{2} \notag \\
&+\frac{1}{2\H}\left( 1+2\frac{\H'}{\H^{2}} \right)\left( \pf' - \partial_{\chi}\ff - \frac{\dd\left( v_{1i}n^{i} \right)}{\dd\nu} \right)\Bigg[ v_{1i}n^{i}+\ff+\int_{0}^{\chi_{s}}\left( \ff'+\pf'\right)\dd\chi \Bigg] \notag \\
& +\Bigg[-\pf|_{s} - \pf|_{o} - \left( 1 - \frac{1}{\H \chi_{s}}\right)\ff|_{s} + \left( 2-\frac{2}{\H \chi_{s}} \right)\left( v_{1i}n^{i}\right)_{o} + \left( 1-\frac{1}{\H \chi_{s}}\right)\left( v_{1i}n^{i}\right)_{s} \notag\\
&+ \left( 1 - \frac{1}{\H\chi_{s}}\right)\int_{0}^{\chi_{s}}\left( \ff'+\pf'\right)\dd\chi -\frac{2}{\chi_{s}}\int_{0}^{\chi_{s}}\pf\dd\chi \notag \\
& -\frac{1}{2 \chi_{s}} \int_{0}^{\chi_{s}}\dd\chi \left( \chi - \chi_{s} \right)\chi \left[ \nabla^{2}\left( \ff + \pf\right) - n^{i}n^{j}\left( \ff + \pf \right)_{,ij} - \frac{2}{\chi} \frac{\dd \dnuf}{\dd \nu} \right]\Bigg]^{2} \notag\\
& +\Bigg[ -\ff|^{s}_{o}+\int_{0}^{\chi_{s}}\left( \ff'+\pf' \right)\dd\chi+\ff-\left( v_{1i}n^{i} \right) \Bigg]\Bigg[ -\pf|_{s} - \pf|_{o} - \left( 1 - \frac{1}{\H \chi_{s}}\right)\ff|_{s} \notag \\
& + \left( 2-\frac{2}{\H \chi_{s}} \right)\left( v_{1i}n^{i}\right)_{o}+ \left( 1-\frac{1}{\H \chi_{s}}\right)\left( v_{1i}n^{i}\right)_{s} + \left( 1 - \frac{1}{\H\chi_{s}}\right)\int_{0}^{\chi_{s}}\left( \ff'+\pf'\right)\dd\chi  \notag \\
& -\frac{2}{\chi_{s}}\int_{0}^{\chi_{s}}\pf\dd\chi -\frac{1}{2 \chi_{s}} \int_{0}^{\chi_{s}}\dd\chi \left( \chi - \chi_{s} \right)\chi \left[ \nabla^{2}\left( \ff + \pf\right) - n^{i}n^{j}\left( \ff + \pf \right)_{,ij} - \frac{2}{\chi} \frac{\dd \dnuf}{\dd \nu} \right] \Bigg] \notag\\
&+\frac{1}{2}\Bigg[ \dnus + \fs - \left( v_{2 i}n^{i}\right) + 2 \ff \dnuf - 2v_{1i} \dnf^{i} + \ff^{2} + \left(v_{1 k}v_{1}^{k}\right) + 4 \pf \left( v_{1i}n^{i}\right) \Bigg]\notag\\
&+\frac{\das}{\bar{d}_{A}}\Bigg\}. \notag
\end{align}

\subsection{Redshift density}

Using \eqs{eq:dz1}, \eqref{eq:dz2} and \eqref{eq:der2-rho} in \eq{eq:redshift-density}  we find that the redshift density perturbation at second order is given by
\begin{align}
\label{eq:ddz2}
\delta^{(2)}_{z}(n^{i},z) &= \frac{1}{2} \frac{\drs(n^{i},z)}{\bar{\rho(z)}} + \frac{3}{2(1+\bar{z})} \dzs(n^{i},z) + \frac{3}{(1+\bar{z})^{2}}\left[ \dzf(n^{i},z) \right]^{2} + \frac{\dd\delta^{(1)}\rho}{\dd\bar{z}}\frac{\dzf(n^{i},z)}{\bar{\rho}}  \notag \\
&=\frac{1}{2} \frac{\drs(n^{i},z)}{\bar{\rho(z)}} + \frac{3}{2}\Bigg[ -\frac{1}{2}\fs|^{s}_{o} - \frac{1}{2}\left( v_{2i} n^{i}\right)^{s}_{o} + \frac{1}{2}\int_{0}^{\chi_{s}}\dd\chi \left( \fs' + \ps' \right) +\frac{1}{2}\left( v_{1k}v^{k}_{1}\right)^{s}_{o} \notag  \\
& - \frac{3}{2} \left( \ff|^{s}_{o}\right)^{2}+6 \ff|_{s}\ff|_{o} +\ff|_{o}\left( v_{1i}n^{i} \right)_{s} -\ff|_{s}\left( v_{1i}n^{i} \right)_{o} -\left( v_{1i}n^{i} \right)_{o}\left( v_{1i}n^{i}\right)^{s}_{o}  \notag \\
& - 2 \pf|_{o}\left( v_{1i}n^{i} \right)_{o}+ \ff|^{s}_{o}\int_{0}^{\chi_{s}}\dd\chi\left( \ff' + \pf' \right) + \left( v_{1i}n^{i} \right)_{o}\int_{0}^{\chi_{s}}\dd\chi\left( \ff' + \pf' \right) \notag \\
& + \ff \int_{0}^{\chi_{s}}\dd\chi\left( \ff' + \pf' \right) +  2v_{1 i}\int_{0}^{\chi_{s}}\dd \chi\left( {\ff_{,}}^{i} + {\pf_{,}}^{i} \right) \notag \\
& + 2\int_{0}^{\chi_{s}}\dd\chi\left[\pf' \left( \ff + \pf \right)\right] + 2\int_{0}^{\lambda_{s}}\dd\chi \left[ \ff \left( \frac{\dd \ff}{\dd \nu} - 2\frac{\dd \pf}{\dd \nu}\right)\right] \notag\\
& -4 \int_{0}^{\chi_{s}}\dd\chi\left[\left( \pf \ff_{,i} - \ff \pf_{,i} \right)n^{i}\right] + 4\int_{0}^{\chi_{s}}\dd\chi\left[ n^{i}\ff_{,i}\left( \ff - \pf \right)\right]  \notag \\
& + \int_{0}^{\chi_{s}}\dd\tilde{\chi}\Bigg\{2 \left( \ff' - \pf'\right)\int_{0}^{\chi_{s}}\dd\chi\left( \ff' + \pf' \right) - n^{i} \left( \ff_{,i} + \pf_{,i} \right) \int_{0}^{\chi_{s}}\dd\chi\left( \ff' + \pf'\right) \notag \\
& +  \left( \frac{\dd\ff}{\dd \nu} - \frac{\dd \pf}{\dd \nu}\right)\int_{0}^{\chi_{s}}\dd\chi \left( \ff' + \pf' \right)+2\left( \ff-\pf\right)n^{i}\int_{0}^{\chi_{s}}\dd\chi\left( \ff' + \pf'\right)_{,i} \notag \\
& +\left( \int_{0}^{\chi_{s}}\dd\chi{\left( \ff + \pf\right)_{,}}^{i}\right)\left( \int_{0}^{\chi_{s}}\dd\chi\left( \ff' + \pf'\right)_{,i}\right)\notag \\
& - n^{i}\left( \int_{0}^{\chi_{s}}\dd\chi\left( \ff' + \pf' \right)_{,i} \right)\left( \int_{0}^{\chi_{s}}\dd\chi\left( \ff' + \pf' \right) \right)\Bigg\}\Bigg]\notag\\
&+3\Bigg[ -\left( v_{1i}n^{i} + \ff\right)\big|^{s}_{o} + \int_{0}^{\chi_{s}}  \left[\ff' +\pf'\right]\dd\chi \Bigg]^{2}\notag \\
&+\frac{\dd\delta^{(1)}\rho}{\dd\bar{z}}\frac{1}{\bar{\rho}}\Bigg[ -\left( v_{1i}n^{i} + \ff\right)\big|^{s}_{o} + \int_{0}^{\chi_{s}}  \left[\ff' +\pf'\right]\dd\chi  \Bigg] 
\end{align}



\chapter{Modified Gauss-Bonnet}
\label{chapter:mgb}

\section{Introduction}
\label{intro}

Recent observations show that the universe is expanding at an
accelerated rate, at the moment the cause of this is still unknown. In
the cosmological standard model, one can assume different mechanisms
to describe dark energy (DE) such as the cosmological constant
$\Lambda$, or scalar fields: quintessence, $k$-essence, and many other
alternatives \cite{b,c,Nojiri:2017ncd,LDS2007}. Current surveys are being planned
like DES \cite{DES}, DESI \cite{Guy:2016zel}, Euclid
\cite{Tereno:2015hja}, and LSST \cite{LSST} to probe large scales in
order to find an answer to this problem.

Dark energy in general relativity (GR) is usually considered as a
change in the energy momentum tensor, $T_{\mu \nu}$, however one can
change the left hand side of the Einstein field equations and take the
accelerated expansion as an effect coming from the geometry of
spacetime, this is usually called modified gravity. Research has
focused until recently in models with $f(R)$, a function of the Ricci
scalar \cite{f,8f,PJ2007,CLF2010,SS2003}, but one can also focus on
more complex models like $f(R,T)$, where $T$ is the trace of the
energy momentum tensor \cite{TFS2011}. Among this modified theories of
gravity, Gauss-Bonnet (GB) gravity has been widely studied in its
$f(G)$ approach \cite{SSD2009,ERV2010}, where gravity is required to
couple with some scalar field $G$ \cite{c,BI2007,SSS2002}. Recent
work \cite{HD2013} has focused on the study of ``almost
scale-invariant theories'' where $f(R,G)$ is a function of the Ricci
scalar and the GB term
\begin{equation}
\label{eq:gb}
G=R^2-4R_{\mu\nu}R^{\mu\nu}+R_{\mu\nu\rho\sigma}R^{\mu\nu\rho\sigma}\,,
\end{equation}
where $R_{\mu \nu}$ is the Ricci tensor and $R_{\mu \nu \rho \sigma}$
is the Riemann tensor. This model was initially proposed as a
gravitational alternative for DE and inflation in Ref.~\cite{g} and
its applications to late-time cosmology have been studied in
\cite{g,g1,g2}.

Modified theories of gravity do have their problems, such as
Ostrogradsky instabilities \cite{Becker:2017tcx,Motohashi:2014opa} and
ghosts \cite{Himmetoglu:2009qi}, but with a particular choice of
a Lagrangian one can avoid this typical problems and find cosmological
solutions as power-law inflation and local attractors
\cite{HD2013}. In Ref.~\cite{HD2013} the authors prove that it is possible to get interesting cosmological solutions for a particular choice of Lagrangian, moreover in Ref.~\cite{nojiri:2018} it is shown that ghosts can be removed from $f(R,\G)$ theories of gravity. Therefore, motivated by the work presented in Refs.~\cite{HD2013,nojiri:2018}, our analysis is focused in the special
choice of $f(R,G)$ that gives
\begin{equation}
\label{eq:lagrangiangb}
\mathcal{L}=\frac{1}{2}m_p^{2}\sqrt{-g}
\left( \alpha R^2+\beta G \log G \right)\,,
\end{equation}
for the Lagrangian with constants $\alpha, \beta$, where $m_p$ is
Planck's mass, $g$ is determinant of the metric tensor $g_{\mu\nu}$,
$R$ is the Ricci scalar, and $G$ is the Gauss-Bonnet invariant defined above in Eq.~\eqref{eq:gb}.

The action is, as usual, defined as
\begin{equation}
\label{eq:action}
\mathcal{S}=\int{\mathcal{L} dx^4}\,.
\end{equation}

The paper is organised as follows: In Section \ref{background} we give
a comparison on how the Einstein tensor is written in GR, $f(R)$ and
$f(R,G)$ in general, then we present the background field equations
for the Lagrangian \eqref{eq:lagrangiangb}. In Section \ref{perturbed}
we present the linear order perturbed Einstein-like tensor that
describes the geometry of the universe, in longitudinal gauge for
scalar perturbations.
In Section \ref{conclusion} we conclude and give an outlook on future
work. The expression for the Einstein-like tensor, derived from a
general Lagrangian as a function $f(R,G)$ is given in Appendix
\ref{einstein}.

For simplicity we work with a flat background spatial metric which is
compatible with current observations.

\section{Governing equations}
\label{background}

In standard GR, the Einstein-Hilbert action is given by integrating
the Lagrangian,
\begin{equation}
\label{eq:einstein-hilbert}
\mathcal{L}_{GR} = \frac{1}{2}m_{p}^{2} \sqrt{-g} R\,,
\end{equation}
and the field equations are obtained by varying the action with
respect to the metric tensor $g_{\mu \nu}$, in GR one has only the
Einstein tensor, given in \eq{eq:ETensor}, 
and one can then relate the geometry with the matter content of the
universe by making it equal to the energy momentum tensor and
obtaining \eq{eq:efield}, the usual equations of motion (EOM).
%
Similarly, in the case of the most popular versions of modified gravity, one can
replace the Ricci tensor, $R$, in the Einstein-Hilbert action
\eqref{eq:einstein-hilbert} with a function $f(R)$ and get a modified
Einstein tensor,
\begin{align}
\label{eq:fr}
\hat{G}_{\mu \nu} &= R_{\mu \nu} \p_{R}f +g_{\mu \nu}\Bigg[
  \left(\nabla^{2}R\right) \p^{2}_{R}f +
  \left(\nabla_{\alpha}R\right)\left(\nabla^{\alpha}R\right)\p^{3}_{R}f
  -\frac{1}{2}f \Bigg] \\
  &\qquad\qquad\qquad\qquad\qquad\qquad
  - \left(\nabla_{\mu}\nabla_{\nu}R\right)
\p^{2}_{R}f - \left(\nabla_{\mu}R\right) \left(\nabla_{\nu}R\right)
\p^{3}_{R}f\, \notag,
\end{align}
where we dropped the dependency on $R$ on $f(R)$ to keep the equation
more compact. We can see that \eqref{eq:fr} reduces to the usual
Einstein tensor if we set $f(R)=R$.\\

In principle one can use a function $f(R)$ as complex as one likes,
however, in this paper we concentrate on the modified gravity function
that includes the Gauss-Bonnet term $G$ defined in the previous
section in Eq.~\eqref{eq:gb}, and therefore we work with $f(R,G)$. The
full expression for the general Einstein-like tensor is rather lengthy
and is given in Appendix \ref{einstein}. Once more, if one drops the
dependency on $G$ and makes $f(R,G)=R$ only, one recovers the usual
Einstein tensor from GR. In this paper we use a particular function
$f(R,G)$ studied in Ref.~\cite{HD2013} given by
\begin{equation}
\label{eq:frg}
f(R,G)=\alpha R^2+\beta G \log G\,,
\end{equation}
where $\alpha$ and $\beta$ are arbitrary constants. This function
leads to a particular Einstein-like tensor
\begin{align}
\label{eq:eom2}
\G_{\mu \nu} &= 2 \alpha \left(R_{\mu \nu} R + g_{\mu \nu} \nabla^{2} R -\nabla_{\mu}\nabla_{\nu}R\right)-\frac{1}{2} \left( \alpha R^2+\beta G \log G \right)g_{\mu \nu}\\
&\qquad\qquad+\frac{\beta}{2}\left(1+\log G\right)\Big[\mathcal{C}\indices{^1_{\mu\nu}}\Big]+\frac{\beta}{2G}\Big[\mathcal{C}\indices{^2_{\mu\nu}}\Big]-\frac{\beta}{2G^{2}}\Big[\mathcal{C}\indices{^3_{\mu\nu}}\Big]\,, \notag
\end{align}
where the $\mathcal{C}\indices{^i_{\mu\nu}}$'s are the coefficients
corresponding to the derivatives $\p^{i}_{G}f$ from
Eq.~\eqref{eq:eom} \footnote{Note that $\p_{R}\p_{G}f(R,G)=\p_{G}\p_{R}f(R,G)=0$ with the choice for $f(R,G)$ made in Eq.~\eqref{eq:frg}.}, we will
be working with this description of the universe geometry.

By modifying the Lagrangian and adding the Gauss-Bonnet term, the governing
equations are also modified,
\begin{equation}
\label{mod_ein_equ}  
\G_{\mu \nu} = \kappa^{2}T_{\mu \nu}\,,
\end{equation}
%
analogous to \eq{eq:efield} where the right hand side stays the
same, since we are only modifying the way we describe the geometry of
the universe.\\

\subsection{Background}
In standard Einstein gravity, \eq{eq:efield},
the governing equations in component form can also be rewritten
as the Friedmann equations,
\begin{align}
\label{fried1}  
\H^{2} &= \frac{\kappa^{2}}{3}a^{2}\rho_{0}\,,\\
\label{fried2}
\H' &= -\frac{\kappa^{2}}{6}a^{2}\left( \rho_{0}+3 P_{0}\right)\,,
\end{align}
where $\H = a'/a$ is the Hubble parameter. 

Using the definition of the modified governing equations,
Eq.~\eqref{mod_ein_equ} we can also find Friedmann-like equations for
$f(R,G)$ gravity. In the background these equations of motion are
given, for our choice of theory and hence $\G_{\mu \nu}$, by
\bea
\label{eq:g00bg}
&&\frac{18}{a^{2}}\Bigg[ \alpha\left(\H\H''-{\H'}^{2}-\H^{4}\right)+\frac{2}{3}\beta \left(66\H\H''-20{\H'}^{2}-44\H^{4}-31\H^{2}\H'\right.\\
&&\qquad\left.+\frac{1}{\H'}\left[35\H^{3}\H''-52\H^{6}-6{\H'}^{2}\right]-\frac{1}{\H^{2}}\left[4{\H'}^{3}+18{\H''}^{2}\right]+\frac{32}{\H}\H'\H''\right. \nonumber\\
&&\qquad\qquad\qquad\qquad\qquad\qquad\qquad\qquad\left.+\frac{16}{\H^{3}}{\H'}^{2}\H''-\frac{13}{\H^{4}}\H' {\H''}^{2}\right) \Bigg]=\kappa^2 T_{00}\,,\nonumber \\
\label{eq:gijbg}
&&\frac{6}{a^{2}} \delta_{ij}\Bigg[\alpha\left( \H\H''-\H'''-{\H'}^{2}+6\H^{2}\H'-\H^{4} \right) +\frac{2}{3}\beta\left(9\H^{2}\H'-2\H\H''+\frac{10}{3}{\H'}^{2}\right.\nonumber\\
&&\qquad\qquad\left.-\frac{2}{\H}\H'\H''+\frac{1}{\H'}\left[\H^{3}\H''-\H^{2}\H'''-\frac{4}{3}\H^{6}\right]+\frac{1}{{\H'}^{2}}\H^{2}{\H''}^{2}\right) \Bigg]=\kappa^2 T_{ij}\,. \nonumber\\
\eea
The off-diagonal components of the field equations vanish in the
background. Also note that the logarithmic dependence $\log G$ cancels
out. 
Already at the background level in this theory the field equations,
(\ref{eq:g00bg}) and (\ref{eq:gijbg}), are more complicated than the
ones in the standard GR case, Eqs.~(\ref{fried1}) and
(\ref{fried2}). In particular the equations now contain time
derivatives up to fourth-order, instead of up to second-order in the
standard case.
The above equations have been previously derived in Ref.~\cite{nojiri:2018}, for a Ghost-free $f(\G)$ theory and in general for any $F(R,\G)$ in their Section V.~Ghost-free $F(R,\G)$ gravity, and our results agree with the ones found in the latter after taking $\Phi(R,\G) \approx \alpha R$ and $\Theta(R,\G) \approx \beta \log\G$ and neglecting the potential $V(\Phi,\Theta)$ used in their analysis.

\section{Perturbed governing equations}
\label{perturbed}

Due to the complexity of the governing equations in fourth-order
gravity, we derived the components of the Einstein-like tensor in
longitudinal gauge, instead of leaving the gauge unspecified. We can
easily reconstruct the tensor components for an arbitrary gauge, by
substituting in the definitions of the variables in longitudinal gauge
defined in Chapter \ref{chapter:cpt}.

%
%


\subsection{Scalar perturbations}
The scalar perturbations give the biggest contribution to the
components of the Einstein-like tensor, and this is given by
\begin{align}
\label{eq:g00n}
&\delta\G_{00} = -\frac{2}{a^{2}}\Bigg[ \alpha\Big\{18\left(\H^{4}+{\H'}^{2}-\H\H''\right)\phi-9\H\H'\phi'-9\H^{2}\phi''+9\left(4\H^{3}-\H\H'-\H''\right)\psi'\notag\\
& +9\left(2\H'-\H^{2}\right)\psi''-9\H\psi'''+12\left(\H^{2}+\H'\right)\nabla^{2}\phi+15\H\nabla^{2}\psi'+3\nabla^{2}\psi''+\nabla^{2}\nabla^{2}\left(\phi-2\psi\right) \Big\}\notag\\
&+\beta\Big\{ \left( 24\H^{2}\left[26\H^{8}+19\H^{6}\H'+20\H^{4}{\H'}^{2}+7\H^{2}{\H'}^{3}+2{\H'}^{4}\right]-6\H\left[73\H^{6}+126\H^{4}\H'\right.\right.\notag\\
&\left.\left.+64\H^{2}{\H'}^{2}+32{\H'}^{3}\right]\H''+12\left[6\H^{4}+18\H^{2}\H'+13{\H'}^{2}\right]{\H''}^{2} \right)\frac{\phi}{\H^{4}\H'}\notag\\
&+\left(6\H\left[ 13{\H'}^{2}-6\H^{4} \right]{\H''}^{2} -3\H\left[104\H^{10}+4\H^{8}\H'+139\H^{6}{\H'}^{2}+322\H^{4}{\H'}^{3}+168\H^{2}{\H'}^{4}\right.\right.\notag\\
&\left.\left.+96{\H'}^{5}\right]+6\left[ 35\H^{8}+\H^{6}\H'+4\H^{4}{\H'}^{2}+76\H^{2}{\H'}^{3}+78{\H'}^{4} \right]\H'' \right)\frac{\phi'}{\H^{4}{\H'}^{2}}\notag\\
&-\left( 71\H^{7}+130\H^{5}\H'+64\H^{3}{\H'}^{2}+32\H{\H'}^{3}-24\H^{4}\H''-72\H^{2}\H'\H'' -52{\H}^{2}\H'' \right)\frac{3\phi''}{\H^{3}\H'}\notag\\
&-\left( 104\H^{12}-620\H^{10}\H'-189\H^{8}{\H'}^{2}+174\H^{6}{\H'}^{3}+176\H^{4}{\H'}^{4}+112\H^{2}{\H'}^{5} -70\H^{9}\H''\right.\notag\\
& +213\H^{7}\H'\H''+118\H^{5}{\H'}^{2}\H'' -216\H^{3}{\H'}^{3}\H''-252\H{\H'}^{4}\H''+12\H^{6}{\H''}^{2}+46\H^{2}{\H'}^{2}{\H''}^{2}\notag\\
&\left.+104{\H'}^{3}{\H''}^{2}\right)\frac{3\psi'}{\H^{5}{\H'}^{2}}\notag\\
&-\left( 104\H^{10}+75\H^{8}\H'+56\H^{6}{\H'}^{2}-4\H^{4}{\H'}^{3}+8\H^{2}{\H'}^{4}-70\H^{7}\H''-26\H^{5}\H'\H''-8\H^{3}{\H'}^{2}\H''\right.\notag\\
&\left. +12\H{\H'}^{3}\H''+12\H^{4}{\H''}^{2}-26{\H'}^{2}{\H''}^{2} \right)\frac{3\psi''}{\H{\H'}^{2}}\notag\\
&-\left( 71\H^{7}+130\H^{5}\H'+64\H^{3}{\H'}^{2}+32\H{\H'}^{3}-24\H^{4}\H''-72\H^{2}\H'\H'' -52{\H}^{2}\H'' \right)\frac{3\psi'''}{\H^{4}\H'}\notag\\
&+\left( 4\H^{2}\left[ 18{\H'}^{4}+44\H^{2}{\H'}^{3}+77\H^{4}{\H'}^{2}+51\H^{6}\H'-26^{8} \right] +2\left[ 13{\H'}^{2}-6\H^{4} \right]{\H''}^{2} \right.\notag\\
&\left. +2\H\left[ 35\H^{6}-34\H^{4}\H' -109\H^{2}{\H'}^{2}-68{\H'}^{3}\right]\H'' \right)\frac{\nabla^{2}\phi}{\H^{4}{\H'}^{2}}\notag\\
&-\left( 70\H^{7}+130\H^{5}\H' +66\H^{3}{\H'}^{2}+32\H{\H'}^{3}-24\H^{4}\H''-73\H^{2}\H'\H''-52{\H'}^{2}\H''\right)\frac{\nabla^{2}\phi'}{\H^{4}\H'}\notag\\
&-\left( 104\H^{10}+42\H^{8}\H' +2\H^{6}{\H'}^{2}-20\H^{4}{\H'}^{3}-8\H^{2}{\H'}^{4}-36\H^{7}\H''+\H^{5}\H'\H''+32\H^{3}{\H'}^{2}\H''\right.\notag\\
&\left.+32\H{\H'}^{3}\H''-18\H^{2}\H'{\H''}^{2}-26{\H'}^{2}{\H''}^{2} \right)\frac{4\nabla^{2}\psi}{\H^{6}\H'}\notag\\
&+\left( 209\H^{7}+236\H^{5}\H'+100\H^{3}{\H'}^{2}+48\H{\H'}^{3}-70\H^{4}\H''-152\H^{2}\H'\H''-72{\H'}^{2}\H'' \right)\nabla^{2}\frac{\psi'}{\H^{4}\H'}\notag\\
&+\left( \frac{\H^{2}}{\H'} \right)\nabla^{2}\psi''+\left( \frac{\H^{2}}{3\H'} \right)\nabla^{2}\nabla^{2}\phi-\left( \frac{2}{3} \right)\nabla^{2}\nabla^{2}\psi  \Big\} \Bigg]\,,
\end{align}
\begin{align}
\label{eq:g0in}
&\delta \G_{0i} = \delta\G_{i0} = \frac{2}{a^{2}}\Bigg[\alpha\Big\{3\left(3\H''+2\H\H'-4\H^{3}\right)\phi-3\left( \H^{2}-3\H' \right)\phi'+3\H\phi''+3\left(7\H'-5\H^{2}\right)\psi'\notag\\
&+3\psi'''-3\H\nabla^{2}\left(\phi-2\psi\right)+\nabla^{2}\left(\phi'-2\psi'\right)\Big\}+\beta\Big\{ \left(60{\H'}^{3}\H''+171\H^{4}\H'\H'' +172\H^{4}\H'\H'' \right.\notag\\
&\left. +44\H^{6}\H''-72\H{\H'}^{4} -134\H^{3}{\H'}^{3}-220\H^{5}{\H'}^{2}-282\H^{7}\H'-192\H^{9}\right)\frac{\nabla^{2}\phi}{6\H^{4}{\H'}^{2}} \notag\\
&+\left( \right. 96\H^{9}+156\H^{7}\H' +84\H^{5}{\H'}^{2}+83\H^{3}{\H'}^{3}+24\H{\H'}^{4}-24\H^{6}\H''-82\H^{4}\H'\H'' \notag\\
&\left. -90\H^{2}{\H'}^{2}\H''-24{\H'}^{3}\H'' \right)\frac{2\nabla^{2}\psi}{3\H^{4}{\H'}^{2}}\notag\\
&-\left( 1880\H^{9}\H'+2156\H^{7}{\H'}^{2} +1710\H^{5}{\H'}^{3}+992\H^{3}{\H'}^{4}+432\H{\H'}^{5}+576\H^{8}\H''+178\H\H'\H''\right. \notag\\
&-1231\H^{4}{\H'}^{2}\H''-984\H^{2}{\H'}^{3}\H'' -360{\H'}^{4}\H''-144\H^{5}{\H''}^{2}-444\H^{3}\H'{\H''}^{2} \notag\\
&\left.-288\H{\H'}^{2}{\H''}^{2}\right)\frac{\phi}{6\H^{4}{\H'}^{2}}\notag\\
&-\left( 288\H^{9}+419\H^{7}\H'+323\H^{5}{\H'}^{2} +176\H^{3}{\H'}^{3}+108\H{\H'}^{4}-66\H^{6}\H''-258\H^{4}\H'\H''\right.\notag\\
&\left. -246\H^{2}{\H'}^{2}\H''-90{\H'}^{3}\H'' \right)\frac{\phi'}{3\H^{3}{\H'}^{2}}\notag\\
&-\left( 576\H^{10}+1566\H\H'+1126\H^{6}{\H'}^{2}+734\H^{4}{\H'}^{3}+504\H^{2}{\H'}^{4}+444\H^{7}\H''-56\H^{5}\H'\H'' \right. \notag\\
&\left. -691\H^{3}{\H'}^{2}\H'' -180\H{\H'}^{3}\H'' -144\H^{4}{\H''}^{2}-444\H^{2}\H'{\H''}^{2}-288{\H'}^{2}{\H''}^{2} \right)\frac{\psi'}{6\H^{4}{\H'}^{2}}\notag\\
&-\left( 144\H^{9}+208\H^{7}\H'+166\H^{5}{\H'}^{2}+88\H^{3}{\H'}^{3}+54\H{\H'}^{4}-33\H^{6}\H''-129\H^{4}\H'\H'' \right. \notag\\
&\left. -123\H^{2}{\H'}^{2}\H''-45{\H'}^{3}\H'' \right)\frac{2\psi''}{3\H^{4}{\H'}^{2}}+\left( \frac{\H^{2}}{3\H'} \right)\nabla^{2}\phi'-\left(\frac{2}{3}\right)\nabla^{2}\psi'\notag\\
&+\left( \frac{\H^{3}}{\H'} \right)\phi''+\left( \frac{\H^{2}}{\H'} \right)\psi'''  \Big\} \Bigg]_{,i}\,, 
\end{align}
\begin{align}
\label{eq:gijn}
&\delta \G_{ij} = \frac{2}{a^{2}}\Bigg[ \alpha\Big\{\delta_{ij}\bigg[ 12\left(\H^{4}-6\H^{2}\H'+{\H'}^{2}-\H\H''+\H'''\right)\phi-3\left( 6\H^{3}+\H\H'-6\H'' \right)\phi' \notag\\
&-3\left(\H^{2}-4\H'\right)\phi''+3\H\phi'''+6\left(\H^{4}-6\H^{2}\H'+{\H'}^{2}-\H\H''+\H'''\right)\psi \notag \\ 
&-3\left( 2\H^{3}+13\H\H'-5\H'' \right)\psi'+3\psi''''-3\left(7\H^{2}-6\H'\right)\psi'' -2\left( \H^{2}+2\H' \right)\nabla^{2}\phi\notag\\
&-3\H\nabla^{2}\left(2\phi'+\psi'\right)+\nabla^{2}\left(\phi''-5\psi''\right)-2\left(\H^{2}-\H'\right)\nabla^{2}\psi -\nabla^{2}\nabla^{2}\left(\phi-2\psi\right)\bigg]\notag\\
&+\bigg[ 3\H\left(\phi'+3\psi'\right) + 6\left( \H^{2}+\H' \right)\psi \bigg]_{,ij}+\bigg[ 3\psi''+\nabla^{2}\nabla^{2}\left( \phi-2\psi \right) \bigg]_{,ij} \Big\} \notag\\
&+ \beta\Big\{ \delta_{ij}\bigg[ \left( 8\H^{3}\H'\H'' -16 \H^{4}{\H'}^{2}-4{\H'}^{4}-4\H{\H'}^{2}\H''+\H^{2}\left[13{\H'}^{3}-{\H''}^{2}\right] \right)\frac{8\phi}{{\H'}^{2}}\notag\\
&-\left( 8\H^{8}\H' +48\H^{6} {\H'}^{2}+64\H^{2}{\H'}^{4} -18{\H'}^{5}-6\H^{5}\H'\H''+36\H^{3}{\H'}^{2}\H''\right.\notag\\
&\left. -3\H^{4}\left[ 47{\H'}^{3}+4{\H''}^{2}-2\H'\H''' \right] \right)\frac{\phi'}{3\H{\H'}^{3}}+\left( 7\H^{4}\H'+2\H^{2}{\H'}^{2}+2{\H'}^{3}-4\H^{3}\H'' \right)\frac{\phi''}{{\H'}^{2}}\notag\\
&+\left( 4\H^{7}\H'-48\H^{5}{\H'}^{2}-22\H{\H'}^{4}+21\H^{4}\H'\H''-6\H^{2}{\H'}^{2}\H''+6{\H'}^{3}\H''\right.\notag\\
&\left. +3\H^{3}\left[ 4{\H'}^{3}-2{\H''}^{2}+\H'\H'' \right] \right)\frac{2\psi}{3\H{\H'}^{2}}\notag\\
&-\left( 8\H^{9}\H' +32\H^{7}{\H'}^{2}-14\H{\H'}^{5}-6\H^{6}\H'\H'' -81\H^{4}{\H'}^{2}\H'' + 42\H^{2}{\H'}^{3}\H'' +18{\H'}^{4}\H''\right.\notag\\
& \left. -\H^{3}\left[ 92{\H'}^{4}-12\H'{\H''}^{2} \right]+3\H^{5}\left[ 49{\H'}^{3}-4{\H''}^{2}+2\H'\H'' \right] \right)\frac{\psi'}{3\H^{2}{\H'}^{3}}\notag\\
&-\frac{\psi''}{3{\H'}^{3}}\left( 8\H^{6}\H'+27\H^{4}{\H'}^{2}+40{\H'}^{4}+6\H^{3}\H'\H''+18\H{\H'}^{2}\H''\right.\notag\\
&\left.-6\H^{2}\left[ 14{\H'}^{3}+2{\H''}^{2}-\H'\H''' \right] \right)+\left( 4\H^{4}\H'-\H^{2}{\H'}^{2}+{\H'}^{3}-2\H^{3}\H'' \right)\frac{2\psi'''}{\H{\H'}^{2}}\notag\\
&-\left( 16\H^{7}\H' +30\H^{5}{\H'}^{2}+131\H{\H'}^{4}-12\H^{4}\H'\H''+6\H^{2}{\H'}^{2}\H''-6{\H'}^{3}\H''\right. \notag\\
&\left.-3\H^{3}\left[7{\H'}^{3}+{\H''}^{2}-4\H'\H'''\right] \right)\frac{\nabla^{2}\phi}{18\H{\H'}^{3}}+\left( 7\H^{4}\H' -7\H^{2}{\H'}^{2} +{\H'}^{3} -4\H^{3}\H'' \right)\frac{\nabla^{2}\phi'}{3\H{\H'}^{2}}\notag\\
&+\left( 2\H^{7}\H'+453\H^{5}{\H'}^{2}-68\H{\H'}^{4} -234\H^{4}\H'\H''+78\H^{2}{\H'}^{2}\H'' +48{\H'}^{3}\H''\right. \notag \\
&\left. -3\H^{3}\left[ 69{\H'}^{3}-12{\H''}^{2}+2\H'\H''' \right] \right)\frac{\nabla^{2}\psi}{18\H^{{\H'}^{2}}}+\left( 4\H^{4}-21\H^{2}\H'+6{\H'}^{2}+4\H\H'' \right)\frac{\nabla^{2}\psi'}{3\H\H'}\notag\\ 
&+\left( \frac{\H^{2}}{\H'} \right)\psi''''+\left( \frac{\H^{3}}{\H'} \right)\phi'''+\left( \frac{\H^{2}}{3\H'} \right)\nabla^{2}\phi''-\left( \frac{5}{3} \right)\nabla^{2}\psi''-\left( \frac{1}{3} \right)\nabla^{2}\nabla^{2}\phi+\left( \frac{2\H'}{3\H^{2}} \right)\nabla^{2}\nabla^{2}\psi \bigg] \notag\\
&+\bigg[\left( 11\H^{3}\H'-10\H^{5}+23\H{\H'}^{2}-2\H^{2}\H''-6\H'\H'' \right)\frac{\phi}{2\H\H'}\notag\\
&+\frac{\psi}{6\H^{2}{\H'}^{2}}\left( 2\H^{6}\H'-79\H^{4}{\H'}^{2}-36{\H'}^{4}+42\H^{3}\H'\H''-6\H{\H'}^{2}\H''\right.\notag\\
&\left.+\H^{2}\left[ 53{\H'}^{3}-12{\H''}^{2}+6\H'\H'' \right] \right)-\left( 4\H^{5}+\H^{3}\H' -26\H{\H'}^{2}+6\H'\H''\right)\frac{\psi'}{3\H^{2}\H'}\notag\\
&+\left( \H \right)\phi' +\psi'' +\left( \frac{1}{3} \right)\nabla^{2}\phi-\left( \frac{2\H'}{\H^{2}} \right)\nabla^{2}\psi \bigg]_{,ij} \Big\}\Bigg]\,.
\end{align}

We now have the Einstein-like tensor, and following the prescription given in Ref.~\cite{MFB}, the usual way of dealing with long equations for perturbations is to decompose them into trace-free and trace. The idea is to reduce everything into only scalar quantities so we have to compute the trace of the Einstein-like tensor and substract that from Eq.~\eqref{eq:tracegn}.

The trace of the spatial part of the Einstein-like tensor is given by
\begin{align}
\label{eq:tracegn}
&\delta {\G^{k}}_{k} = \frac{2}{a^{2}}\Bigg[ \alpha\Big\{ 36\left(\H^{4}-6\H^{2}\H'+{\H'}^{2}-\H\H''+\H'''\right)\phi-9\left( 6\H^{3}+\H\H' -6\H'' \right)\phi'\notag\\
&-9\left( \H^{2}-4\H' \right)\phi''+9 \H \phi''' +18\left( \H^{4}-6\H^{2}\H'+{\H'}^{2}-\H\H''+\H''' \right)\psi\notag\\
&-9\left( 2\H^{3}+13\H\H'-5\H'' \right)\psi'-9\left( 7\H^{2}-6\H' \right)\psi'' +9\psi''''-6\left( \H^{2}+2\H' \right)\nabla^{2}\phi \notag\\
&-15\H\nabla^{2}\phi'+3\nabla^{2}\phi''+12 \H' \nabla^{2}\psi-12\nabla^{2}\psi''-2\nabla^{2}\nabla^{2}\phi+4\nabla^{2}\nabla^{2}\psi \Big\}\notag\\
&+\beta\Big\{\left( 8\H^{3}\H'\H''-16\H^{4}{\H'}^{2}-4{\H'}^{4}-4\H{\H'}^{2}\H''+\H^{2}\left[ 13{\H'}^{3}-{\H''}^{2} \right] \right)\frac{24\phi}{{\H'}^{2}}\notag\\ 
&-\left( 8\H^{8}\H'+48\H^{6}{\H'}^{2}+64\H^{2}{\H'}^{4}-18{\H'}^{5}-6\H^{5}\H'\H''+36\H^{3}{\H'}^{2}\H'' \right.\notag\\
&\left.-3\H^{4}\left[ 47{\H'}^{3}+4{\H''}^{2} -2\H'\H''' \right]\right)\frac{\phi'}{\H{\H'}^{3}}\notag\\
&+\left( 7\H\H'+2\H^{2}{\H'}^{2}+2{\H'}^{3}-4\H^{3}\H'' \right)\frac{3\phi''}{{\H'}^{2}}\notag\\
&+\left( 4\H^{7}\H'+48\H^{5}{\H'}^{2}-22\H{\H'}^{4}+21\H^{4}\H'\H'' -6\H^{2}{\H'}^{2}+6{\H'}^{3}\H''\right.\notag\\
&\left. +3\H^{3}\left[ 4{\H'}^{3}-2{\H''}^{2}+\H'\H''' \right] \right)\frac{2\psi}{\H{\H'}^{2}}\notag\\
&-\left( 8\H^{6}\H'+27\H^{4}{\H'}^{2}+40{\H'}^{4}+6\H^{3}\H'\H''+18\H{\H'}^{2}\H''\right.\notag\\
&\left.-6\H^{2}\left[ 14{\H'}^{3}+2{\H''}^{2}-\H'\H''' \right] \right)\frac{\psi''}{{\H'}^{3}}+\left( 4\H^{4}\H' -\H^{2}{\H'}^{2}+{\H'}^{3}-2\H^{3}\H'' \right)\frac{6\psi'''}{\H{\H'}^{2}}\notag\\
&-\left( 8\H^{7}\H' +30\H^{5}{\H'}^{2}-27\H^{3}{\H'}^{3}+31\H{H'}^{4}+6\H'\left[ {\H'}^{2}+\H^{2}\H'-\H^{4} \right]\right.\notag\\
&\left. -12\H^{3}{\H''}^{2}+6\H^{3}\H'\H'''  \right)\frac{\nabla^{2}\phi}{3\H{\H'}^{3}}+\left( \H'\left[ 7\H^{4}-6\H^{2}\H'+2{H'}^{2} \right]-4\H^{3}\H'' \right)\nabla^{2}\phi'\notag\\
&+\left( 2\H^{7}\H' +187\H^{5}{\H'}^{2}-52\H{\H'}^{4}-96\H^{4}\H'\H''+36\H^{2}{\H'}^{2}\H'' +24{\H'}^{3}\H''\right.\notag\\
&\left. +\H^{3}\left[ 12{\H''}^{2}-77{\H'}^{3} \right]  \right)\frac{\nabla^{2}\psi}{3\H^{3}{\H'}^{2}}\notag\\
&+\left( 2\H\left[ 2\H^{4}-16\H^{2}\H' +11{\H'}^{2} \right]+\left[ 6\H^{2}-3\H' \right]\H'' \right)\frac{2\nabla^{2}\psi'}{3\H^{2}\H'}\notag\\
&+\left(\frac{\H^{3}}{\H'}\right)\phi'''+\left(\frac{3\H^{2}}{\H'}\right)\psi''''+\left( \frac{\H^{2}}{\H'} \right)\nabla^{2}\phi''-4\nabla^{2}\psi''-\left(\frac{2}{3}\right)\nabla^{2}\nabla^{2}\phi+\left( \frac{4\H'}{3\H^{2}} \right)\nabla^{2}\nabla^{2}\psi\Big\} \Bigg]\,.
\end{align}

To construct the traceless Einstein-like tensor we need to subtract
the trace from the spatial part given by Eq.~\eqref{eq:tracegn}. In
order to do this, we apply two spatial derivatives to $\delta \G_{ij}$
(see e.g.~Ref.~\cite{malik}). We then use the linearity of the equation
to simplify it even more, and we are left with single Laplacian
operators instead of double Laplacians, simplifying the calculations,
%
\begin{align}
\label{eq:iintg}
&\iint \left(\p^{i}\p^{j}\left[\delta \G_{ij}\right]\right) = \frac{2}{a^{2}}\Bigg[  \alpha\Big\{ \left(\H^{4}-6\H^{2}\H'{\H}^{2}-\H\H''+\H'''\right)\phi-3\left( 6\H^{3}+\H\H'-6\H'' \right)\phi'  \notag\\ 
&-3\left( \H^{2}-4\H' \right)\phi''+3\H\phi'''+6\left( \H^{4}-6\H^{2}\H'+{\H}^{2}-\H\H'' \right) \psi-3\left( 2\H^{3}+13\H\H'-5\H'' \right)\psi'\notag\\ 
&-3\left( 7\H^{2}-6\H' \right)\psi'' +3\psi''''-2\left(\H^{2}+2\H'\right)\nabla^{2}\left(\phi-2\psi\right)-3\H\nabla^{2}\left(\phi' -2\psi' \right)+\nabla^{2}\left(\phi''-2\psi''\right) \Big\} \notag\\ 
&+ \beta\Big\{\left(8\H^{3}\H'\H''-16\H^{4}{\H'}^{2}-4{\H'}^{4}-4\H{\H'}^{2}\H''+\H^{2}\left[ 13{\H'}^{3}-{\H''}^{2} \right]\right)\frac{8\phi}{{\H'}^{2}}\notag\\ 
&-\left( 8\H^{8}\H'+48\H^{6}{\H'}^{2}+64\H^{2}{\H'}^{4}-18{\H'}^{5}-6\H^{5}\H'\H''+36\H^{3}{\H'}^{2}\H''\right.\notag\\
&\left. -3\H^{4}\left[ 47{\H'}^{3}+4{\H''}^{2}-2\H'\H''' \right] \right)\frac{\phi'}{3\H {\H'}^{3}}+\left( 7\H^{4}\H' +2\H^{2}{\H'}^{2}+2{\H'}^{3} - 4\H^{3}\H'' \right)\frac{\phi''}{{\H'}^{2}}\notag\\ 
&+\left( 4\H^{7}\H' - 48\H^{5}{\H'}^{2} - 22\H{\H'}^{4} + 21\H^{4}\H'\H''-6{\H'}^{3}\H''\right.\notag \\
&\left. +3\H^{3}\left[ 4{\H'}^{3}-2{\H''}^{2}+\H'\H''' \right]  \right)\frac{2\psi}{3\H{\H'}^{2}}\notag\\ 
&-\left( 8\H^{9}\H' + 32\H^{7}{\H'}^{2} -14\H{\H'}^{5} -14\H{\H'}^{5} -6\H^{6}\H'\H''-81\H^{4}{\H'}^{2}\H'' +42\H^{2}{\H'}^{2}\H''\right.\notag\\
&\left. +\H^{3}\left[ 12\H'{\H''}^{2} -92{\H'}^{4} \right] +3\H^{5}\left[ 49{\H'}^{3}-4{\H''}^{2} +2\H'\H''' \right] \right)\frac{\psi'}{3\H^{2}{\H'}^{3}}\notag\\ 
&-\left( 8\H^{6}\H' +27\H^{4}{\H'}^{2}+40{\H'}^{4}+6\H^{3}\H'\H'' +18\H {\H'}^{2} \H'' \right.\notag\\
&\left. -6\H^{2}\left[ 14{\H'}^{3} +2{\H''}^{2} -\H'\H''' \right] \right)\frac{\psi''}{3{\H'}^{3}}+\left( 4\H^{4}\H' -\H^{2}{\H'}^{2}+{\H'}^{3}-2\H^{3}\H'' \right)\frac{2\psi'''}{\H{\H'}^{2}}\notag\\ 
&-\left( 4\H^{7}\H'+30\H^{5}{\H'}^{2}-19\H{\H'}^{4}-3\H^{4}\H'\H''+6\H^{2}{\H'}^{2}\H''+12{\H'}^{3}\H'' \right.\notag\\
&\left. +\H^{3}\left[ 3\H'\H'''-6{\H''}^{2}-30{\H'}^{3} \right] \right)\frac{2\nabla^{2}\phi}{9\H{\H'}^{3}}+\left( 7\H^{4}\H'-4\H^{2}{\H'}^{2}+2{\H'}^{3}-4\H^{3}\H'' \right)\frac{ \nabla^{2}\phi'}{3\H{\H'}^{2}}\notag\\ 
&+\left( 2^{7}+54^{5}\H' -44\H{\H'}^{3} -27\H^{4}\H'' +15\H^{2}\H'\H'' + 12{\H'}^{2}\H''\right.\notag\\
&\left. +3\H^{3}\left[ \H'''-4{\H'}^{2} \right] \right)\frac{\nabla^{2}\psi}{81\H^{6}{\H'}^{4}} -\left(11\H^{3}\H' -16\H{\H'}^{2} - 2\H^{2}\H'' + 3\H' \H'' \right) \frac{\nabla^{2}\psi'}{27\H^{5}{\H'}^{4}} \notag\\ 
&+\left( \frac{\H^{3}}{\H'} \right)\phi'''+\left( \frac{\H^{2}}{{\H'}^{2}} \right)\psi''''+\left( \frac{\H^{2}}{3\H'}\right)\nabla^{2}\phi''-\frac{\nabla^{2}\psi''}{27\H^{3}{\H'}^{3}}\Big\} \Bigg]\,.
\end{align}

Subtracting \eqref{eq:iintg} and \eqref{eq:tracegn} from Eq.~\eqref{eq:gijn} one gets the trace free Einstein-like tensor. Using Eqs.~\eqref{mod_ein_equ},~\eqref{eq:g00n},~\eqref{eq:g0in} and~\eqref{eq:gijn} we can get the evolution equations, these are in \ref{evoeqns}.

\section{Einstein-like tensor}
\label{einstein}

The choice of a function of the Ricci scalar is arbitrary, so we keep the function only as $f(R,G)$ without any assumption, we do take into account the fact that the Gauss-Bonnet term $G$, given in Eq.~\eqref{eq:gb}, also has a dependency on the metric, so it has to be varied with respect of the metric itself, from where the equation becomes rather lengthy, this general equation, reduces to the known cases of GR and $f(R)$ when we choose $f(R)=R$ and we drop the dependance of $G$ respectively.

In general, the Einstein-like tensor, without choosing any particular $f(R,G)$, is given by
\begin{align}
\label{eq:eom}
&\tilde{G}_{\mu \nu} = g_{\mu \nu} \Big\{ \p_{R}f\Big[0\Big]+\p^{2}_{R}f\Big[ \nabla^{2}R \Big]+\p^{3}_{R}f\Big[ \left( \nabla_{\alpha}R\right)\left( \nabla^{\alpha}R \right)\Big] +\p_{G}f\Big[2 \nabla^{2}R-4\nabla_{\alpha}\nabla_{\beta}R^{\alpha \beta }\Big] -\frac{1}{2}f \notag \\
&+\p^{2}_{G}f\Big[ 4R^{2}\left( \nabla^{2}R \right)-32R^{\beta\sigma}\left( \nabla_{\alpha}R_{\beta \sigma} \right)\left( \nabla^{\alpha}R \right)+12R\left(\nabla_{\alpha}R\right)\left(\nabla^{\alpha}R\right) + 8R^{\beta\sigma\lambda\rho}\left(\nabla_{\alpha}R_{\beta\sigma\lambda\rho}\right)\left(\nabla^{\alpha}R\right) \notag \\
&-16R \left(\nabla^{\alpha}R\right)\left(\nabla_{\beta}R\indices{_\alpha^\beta}\right) + 32 R^{\alpha \beta}\left(\nabla_{\alpha}R^{\sigma \lambda}\right)\left(\nabla_{\beta}R_{\sigma\lambda}\right)-16R^{\sigma\lambda\rho\tau}\left(\nabla_{\alpha}R^{\alpha\beta}\right)\left(\nabla_{\beta}R_{\sigma\lambda\rho\tau}\right) \notag \\
& -8R^{\alpha \beta} \left( \nabla_{\alpha}R^{\sigma\lambda\rho\tau} \right)\left(\nabla_{\beta}R_{\sigma\lambda\rho\tau}\right) -8 R^{\alpha\beta}R \left(\nabla_{\alpha}\nabla_{\beta}R\right)-8R^{\alpha \beta}\left(R^{\sigma\lambda\rho\tau}\right)\left(\nabla_{\alpha}\nabla_{\beta}R_{\sigma\lambda\rho\tau}\right)  \notag \\
&-8R_{\alpha\beta}\left(\nabla^{\alpha}R\right)\left(\nabla^{\beta}R\right)-16R^{\alpha\beta} R \left( \nabla^{2}R_{\alpha \beta} \right)-16R \left(\nabla_{\sigma}R_{\alpha\beta}\right)\left(\nabla^{\sigma}R^{\alpha\beta}  \right)\notag \\
& +64R^{\alpha\beta}\left(\nabla^{\sigma}R_{\alpha\beta}\right)\left(\nabla_{\lambda}R\indices{_\sigma^\lambda}\right)+32R^{\alpha\beta}R^{\sigma\lambda}\left(\nabla_{\lambda}\nabla_{\sigma}R_{\alpha\beta}\right)+4R^{\alpha\beta\sigma\lambda}R\left(\nabla^{2}R_{\alpha\beta\sigma\lambda}\right) \notag\\
&+4 R \left(\nabla_{\rho}R_{\alpha\beta\sigma\lambda}\right)\left(\nabla^{\rho}R^{\alpha\beta\sigma\lambda}\right)\Big] \notag \\
&+\p^{3}_{G}f\Big[ 8R^{3}\left(\nabla_{\alpha}R\right)\left(\nabla^{\alpha}R\right)-64R^{\sigma\beta}R^{2}\left(\nabla_{\alpha}R_{\beta\sigma}\right)\left(\nabla^{\alpha}R\right)+16R^{2}R^{\beta\sigma\lambda\rho}\left(\nabla_{\alpha}R_{\beta\sigma\lambda\rho}\right)\left(\nabla^{\alpha}R\right) \notag \\
&+128 R\indices{_\alpha^\beta}R^{\sigma\lambda}R\left(\nabla^{\alpha}R\right)\left(\nabla_{\beta}R_{\sigma\lambda}\right)-32R\indices{_\alpha^\beta}R^{\sigma\lambda\rho\tau}R\left(\nabla^{\alpha}R\right)\left(\nabla_{\beta}R_{\sigma\lambda\rho\tau}\right) \notag \\
&-16R^{\alpha\beta}R^{\sigma\lambda\rho\tau}R^{\xi\chi\omega\kappa}\left(\nabla_{\alpha}R_{\sigma\lambda\rho\tau}\right)\left(\nabla_{\beta}R_{\xi\chi\omega\kappa}\right)-16R_{\alpha\beta}R^{2}\left(\nabla^{\alpha}R\right)\left(\nabla^{\beta}R\right) \notag\\
&-64R^{\alpha\beta}R^{\lambda\rho\xi\chi}R\left(\nabla_{\sigma}R_{\lambda\rho\xi\chi}\right)\left(\nabla^{\sigma}R_{\alpha\beta}\right)-256R^{\alpha\beta}R^{\sigma\lambda}R^{\rho\tau}\left(\nabla_{\sigma}R_{\alpha\beta}\right)\left(\nabla_{\lambda}R_{\rho\tau}\right)\notag \\
&+128R^{\alpha \beta}R^{\sigma\lambda}R^{\rho\tau\xi\chi}\left(\nabla_{\sigma}R_{\alpha \beta}\right)\left(\nabla_{\lambda}R_{\rho\tau\xi\chi}\right) +128 R^{\alpha \beta}R^{\sigma\lambda}R\left(\nabla_{\rho}R_{\sigma\lambda}\right)\left(\nabla^{\rho}R_{\alpha\beta}\right)\notag\\
&+8R^{\alpha\beta\sigma\lambda}R^{\rho\tau\xi\chi}R\left(\nabla_{\kappa}R_{\rho\tau\xi\chi}\right)\left(\nabla^{\kappa}R_{\alpha\beta\sigma\lambda}\right) \Big] \notag \\
&+\p_{R}\p_{G}f\Big[4 R\left( \nabla^{2}R \right)+6 \left( \nabla_{\alpha}R\right)\left( \nabla^{\alpha}R \right)-8\left( \nabla^{\alpha}R\right)\left(\nabla_{\beta}R\indices{_\alpha^\beta} \right)-4R^{\alpha \beta}\left( \nabla_{\alpha}\nabla_{\beta}R\right)-8R^{\alpha \beta}\left( \nabla^{2}R_{\alpha \beta}\right) \notag \\
&-8\left( \nabla_{\sigma}R_{\alpha \beta} \right)\left( \nabla^{\sigma} R^{\alpha \beta}\right)+2R^{\alpha \beta \lambda \rho} \left( \nabla^{2}R_{\alpha \beta \lambda \rho}\right)+2\left( \nabla_{\sigma}R_{\alpha \beta \lambda \rho}\right)\left( \nabla^{\sigma}R^{\alpha \beta \lambda \rho} \right)\Big] \notag \\
&+\p^{2}_{R}\p_{G}f\Big[ 6R\left( \nabla_{\alpha}R\right)\left( \nabla^{\alpha}R \right) -16R^{\beta \sigma}\left( \nabla_{\alpha}R_{\beta \sigma}\right)\left( \nabla^{\alpha} R\right) + 4R^{\beta \sigma \lambda \rho}\left( \nabla_{\alpha}R_{\beta \sigma\lambda \rho} \right)\left( \nabla^{\alpha}R\right) \notag\\
&- 4R_{\alpha \beta}\left( \nabla^{\alpha}R\right)\left( \nabla^{\beta}R\right)\Big] \notag 
\end{align}
\begin{align}
&+\p_{R}\p^{2}_{G}f\Big[ 12 R^{2}\left( \nabla_{\alpha}R\right)\left( \nabla^{\alpha}R\right) -64R^{\beta \sigma}R \left( \nabla_{\alpha}R_{\beta \sigma} \right)\left( \nabla^{\alpha}R \right)+16 R^{\beta \sigma \lambda \rho}R\left( \nabla_{\alpha}R_{\beta \sigma \lambda \rho}\right)\left( \nabla^{\alpha}R \right)\notag\\ 
&+64R\indices{_\alpha^\beta}R^{\sigma \lambda} \left( \nabla^{\alpha}R \right)\left( \nabla_{\beta}R_{\sigma \lambda} \right) -16 R\indices{_\alpha^\beta}R^{\sigma\lambda\rho\tau}\left( \nabla^{\alpha}R\right)\left( \nabla_{\beta}R_{\sigma\lambda\rho\tau}\right)-16R_{\alpha \beta}R \left( \nabla^{\alpha}R\right)\left( \nabla^{\beta}R\right) \notag \\
&-32R^{\alpha \beta} R^{\lambda\rho\tau\xi}\left( \nabla_{\sigma}R_{\lambda\rho\tau\xi}\right)\left( \nabla^{\sigma}R_{\alpha \beta} \right)+64R^{\alpha \beta} R^{\sigma \lambda}\left( \nabla_{\rho}R_{\sigma\lambda}\right)\left( \nabla^{\rho}R_{\alpha \beta}\right)\notag\\
&+4R^{\alpha \beta \sigma \lambda}R^{\rho \tau\xi\kappa}\left( \nabla_{\chi}R_{\rho\tau\xi\kappa} \right)\left( \nabla^{\chi}R_{\alpha\beta\sigma\lambda} \right)\Big]\Big\} \notag \\
&+ \p_{R}f\Big[ R_{\mu \nu} \Big]-\p^{2}_{R}f\Big[ \left(\nabla_{\mu}\nabla_{\nu}R\right) \Big]-\p^{3}_{R}f\Big[ \left(\nabla_{\mu}R\right)\left(\nabla_{\nu}R\right) \Big]\notag\\ 
&+\p_{G}f\Big[ 2R_{\mu\nu}R-8R\indices{_\mu^\alpha}R_{\nu\alpha}+2R\indices{_\mu^\alpha^\beta^\sigma}R_{\nu\alpha\beta\sigma} -4\left(\nabla^{2}R_{\mu\nu}\right)+2\left(\nabla_{\alpha}\nabla_{\beta}R\indices{_\mu^\alpha_\nu^\beta}\right)+4\left(\nabla_{\alpha}\nabla_{\mu}R\indices{_\nu^\alpha}\right) \notag \\
&+4\left(\nabla_{\alpha}\nabla_{\nu}R\indices{_\mu^\alpha}\right)+2\left(\nabla_{\beta}\nabla_{\alpha}R\indices{_\mu^\alpha_\nu^\beta}\right)-2\left(\nabla_{\nu}\nabla_{\mu}R\right)\Big]\notag \\
&+\p_{R}\p_{G}f\Big[ 4\left(\nabla^{\alpha}R\right)\left(\nabla_{\beta}R\indices{_\mu_\alpha_\nu^\beta}\right)-4R_{\mu\nu}\left(\nabla^{2}R\right) -8\left(\nabla_{\alpha}R_{\mu\nu}\right)\left(\nabla^{\alpha}R\right)+4\left(\nabla^{\alpha}R\right)\left(\nabla_{\beta}R\indices{_\mu^\beta_\nu_\alpha}\right)\notag \\
&+4R_{\mu\alpha\nu\beta}\left(\nabla^{\beta}\nabla^{\alpha}R\right) +4\left(\nabla^{\alpha}R\right)\left(\nabla_{\mu}R_{\nu\alpha}\right)+4\left(\nabla_{\alpha}R\indices{_\nu^\alpha}\right)\left(\nabla_{\mu}R\right)\notag\\
&- R^{\alpha\beta\sigma\lambda}\left(\nabla_{\mu}\nabla_{\nu}R_{\alpha\beta\sigma\lambda}\right)+8\left(\nabla_{\mu}R^{\alpha\beta}\right)\left(\nabla_{\nu}R_{\alpha\beta}\right)+4\left(\nabla^{\alpha}R\right)\left(\nabla_{\nu}R_{\mu\alpha}\right)+4\left(\nabla_{\alpha}R\indices{_\mu^\alpha}\right)\left(\nabla_{\nu}R\right)\notag\\
&-6\left(\nabla_{\mu}R\right)\left(\nabla_{\nu}R\right)-2\left(\nabla_{\mu}R^{\alpha\beta\sigma\lambda}\right)\left(\nabla_{\nu}R_{\alpha\beta\sigma\lambda}\right)+4R\indices{_\mu^\alpha}\left(\nabla_{\nu}\nabla_{\alpha}R\right)+4R^{\alpha\beta}\left(\nabla_{\mu}\nabla_{\nu}R_{\alpha\beta}\right)\notag\\
&-4R\left(\nabla_{\mu}\nabla_{\mu}R\right) -R^{\alpha\beta\sigma\lambda}\left(\nabla_{\mu}\nabla_{\nu}R_{\alpha\beta\sigma\lambda}\right)+4R\indices{_\nu^\alpha}\left(\nabla_{\mu}\nabla_{\alpha}R\right)+4R^{\alpha\beta}\left(\nabla_{\mu}\nabla_{\nu}R_{\alpha\beta}\right)\Big]\notag\\ 
&+\p^{2}_{R}\p_{G}f\Big[4 R_{\mu\alpha\nu\beta}\left(\nabla^{\alpha}R\right)\left(\nabla^{\beta}R\right)-4R_{\mu\nu}\left(\nabla_{\alpha}R\right)\left(\nabla^{\alpha}R\right)+4R_{\nu\alpha}\left(\nabla^{\alpha}R\right)\left(\nabla_{\mu}R\right)\notag\\
&+8R^{\alpha\beta}\left(\nabla_{\mu}R\right)\left(\nabla_{\nu}R_{\alpha\beta}\right) +4R_{\mu\alpha}\left(\nabla^{\alpha}R\right)\left(\nabla_{\nu}R\right) +8R^{\alpha \beta}\left(\nabla_{\mu}R_{\alpha\beta}\right)\left(\nabla_{\nu}R\right)-6R\left(\nabla_{\mu}R\right)\left(\nabla_{\nu}R\right)\notag\\
&-2R^{\alpha\beta\sigma\lambda}\left(\nabla_{\mu}R_{\alpha\beta\sigma\lambda}\right)\left(\nabla_{\nu}R\right) -2R^{\alpha\beta\sigma\lambda}\left(\nabla_{\mu}R\right)\left(\nabla_{\nu}R_{\alpha\beta\sigma\lambda}\right)\Big]\notag\\
&+\p^{2}_{G}f\Big[ 4 R^{\sigma\lambda\rho\tau}R\indices{_\mu^\alpha_\nu^\beta}\left(\nabla_{\alpha}\nabla_{\beta}R_{\sigma\lambda\rho\tau}\right)-8R_{\mu\nu}R\left(\nabla^{2}R\right) -32R^{\beta\sigma}R\indices{_\nu^\alpha}\left(\nabla_{\alpha}\nabla_{\mu}R_{\beta\sigma}\right)\notag\\
&+8R\indices{_\nu^\alpha}R^{\beta\sigma\lambda\rho}\left(\nabla_{\alpha}\nabla_{\mu}R_{\beta\sigma\lambda\rho}\right)-32R^{\beta\sigma}R\indices{_\mu^\alpha}\left(\nabla_{\alpha}\nabla_{\nu}R_{\beta\sigma}\right)+8R\indices{_\mu^\alpha}R^{\beta\sigma\lambda\rho}\left(\nabla_{\alpha}\nabla_{\nu}R_{\beta\sigma\lambda\rho}\right)\notag\\
&-16R^{\beta\sigma\lambda\rho}\left(\nabla_{\alpha}R_{\beta\sigma\lambda\rho}\right)\left(\nabla^{\alpha}R_{\mu\nu}\right)-16 R \left(\nabla_{\alpha}R_{\mu\nu}\right)\left(\nabla^{\alpha}R\right)-8R_{\mu\nu}\left(\nabla_{\alpha}R\right)\left(\nabla^{\alpha}R\right)\notag \\
& +8R\indices{_\mu^\alpha_\nu^\beta}\left(\nabla_{\alpha}R^{\sigma\lambda\rho\tau}\right)\left(\nabla_{\beta}R_{\sigma\lambda\rho\tau}\right)+8R \left(\nabla^{\alpha}R\right)\left(\nabla_{\beta}R\indices{_{\mu\alpha\nu}^\beta}\right)+8R \left(\nabla^{\alpha}R\right)\left(\nabla_{\beta}R\indices{_\mu^\beta_{\nu\alpha}}\right)\notag \\
& +4R^{\sigma\lambda\rho\tau}R\indices{_\mu^\alpha_\nu^\beta}\left(\nabla_{\beta}\nabla_{\alpha}R_{\sigma\lambda\rho\tau}\right)+8R_{\mu\alpha\nu\beta}\left(\nabla^{\alpha}R\right)\left(\nabla^{\beta}R\right)+8R_{\mu\alpha\nu\beta}\left(\nabla^{\beta}\nabla^{\alpha}R\right)\notag\\
&+32R^{\alpha\beta}R_{\mu\nu}\left(\nabla^{2}R_{\alpha\beta}\right)+32R_{\mu\nu}\left(\nabla_{\sigma}R_{\alpha\beta}\right)\left(\nabla^{\sigma}R^{\alpha\beta}\right)+64R^{\alpha\beta}\left(\nabla_{\sigma}R_{\alpha\beta}\right)\left(\nabla^{\sigma}R_{\mu\nu}\right)\notag\\
&-32R^{\alpha\beta}\left(\nabla^{\sigma}R_{\alpha\beta}\right)\left(\nabla_{\lambda}R\indices{_{\mu\sigma\nu}^\lambda}\right)-32R^{\alpha\beta}\left(\nabla^{\sigma}R_{\alpha\beta}\right)\left(\nabla_{\lambda}R\indices{_\mu^\lambda_\nu_\sigma}\right)\notag\\
&-32R_{\mu\sigma\nu\lambda}\left(\nabla^{\sigma}R^{\alpha\beta}\right)\left(\nabla^{\lambda}R_{\alpha\beta}\right)-16R^{\alpha\beta}R_{\mu\sigma\nu\lambda}\left(\nabla^{\lambda}\nabla^{\sigma}R_{\alpha\beta}\right)-16R^{\alpha\beta}R_{\mu\lambda\nu\sigma}\left(\nabla^{\lambda}\nabla^{\sigma}R_{\alpha\beta}\right) \notag \\
& -32R^{\alpha\beta}\left(\nabla_{\sigma}R\indices{_\nu^\sigma}\right)\left(\nabla_{\mu}R_{\alpha\beta}\right)-32R\indices{_\nu^\alpha}\left(\nabla_{\alpha}R_{\beta\sigma}\right)\left(\nabla_{\mu}R^{\beta\sigma}\right)+8R\left(\nabla^{\alpha}R\right)\left(\nabla_{\mu}R_{\nu\alpha}\right)\notag \\ 
& +8R^{\beta\sigma\lambda\rho}\left(\nabla_{\alpha}R_{\beta\sigma\lambda\rho}\right)\left(\nabla_{\mu}R\indices{_\nu^\alpha}\right)-32R^{\alpha\beta}\left(\nabla_{\sigma}R_{\alpha\beta}\right)\left(\nabla_{\mu}R\indices{_\nu^\sigma}\right)+8R\left(\nabla_{\alpha}R\indices{_\nu^\alpha}\right)\left(\nabla_{\mu}R\right)\notag\\
&+8R_{\nu\alpha}\left(\nabla^{\alpha}R\right)\left(\nabla_{\mu}R\right)+8R^{\beta\sigma\lambda\rho}\left(\nabla_{\alpha}R\indices{_\nu^\alpha}\right)\left(\nabla_{\mu}R_{\beta\sigma\lambda\rho}\right)+8R\indices{_\nu^\alpha}\left(\nabla_{\alpha}R_{\beta\sigma\lambda\rho}\right)\left(\nabla_{\mu}R^{\beta\sigma\lambda\rho}\right)\notag\\
&+8R\indices{_\nu^\alpha}R\left(\nabla_{\mu}\nabla_{\alpha}R\right)+8R^{\alpha\beta}R\left(\nabla_{\mu}\nabla_{\nu}R_{\alpha\beta}\right)-2R^{\alpha\beta\sigma\lambda}R\left(\nabla_{\mu}\nabla_{\nu}R_{\alpha\beta\sigma\lambda}\right) \notag \\
&-32R^{\alpha\beta}\left(\nabla_{\sigma}R\indices{_\mu^\sigma}\right)\left(\nabla_{\nu}R_{\alpha\beta}\right)+16R\left(\nabla_{\mu}R^{\alpha\beta}\right)\left(\nabla_{\nu}R_{\alpha\beta}\right)+16R^{\alpha\beta}\left(\nabla_{\mu}R\right)\left(\nabla_{\nu}R_{\alpha\beta}\right)\notag \\
&-32R\indices{_\mu^\alpha}\left(\nabla_{\alpha}R_{\beta\sigma}\right)\left(\nabla_{\nu}R^{\beta\sigma}\right)+8R\left(\nabla^{\alpha}R\right)\left(\nabla_{\nu}R_{\mu\alpha}\right)+8R^{\beta\sigma\lambda\rho}\left(\nabla_{\alpha}R_{\beta\sigma\lambda\rho}\right)\left(\nabla_{\nu}R\indices{_\mu^\alpha}\right)\notag \\
&-32R^{\alpha\beta}\left(\nabla_{\sigma}R_{\alpha\beta}\right)\left(\nabla_{\nu}R\indices{_\mu^\sigma}\right) +8R \left(\nabla_{\alpha}R\indices{_\mu^\alpha}\right)\left(\nabla_{\nu}R\right)+8R_{\mu\alpha}\left(\nabla^{\alpha}R\right)\left(\nabla_{\nu}R\right)\notag\\
&+16R^{\alpha\beta}\left(\nabla_{\mu}R_{\alpha\beta}\right)\left(\nabla_{\nu}R\right)-12R\left(\nabla_{\mu}R\right)\left(\nabla_{\nu}R\right)-4R^{\alpha\beta\sigma\lambda}\left(\nabla_{\mu}R_{\alpha\beta\sigma\lambda}\right)\left(\nabla_{\nu}R\right)\notag
\end{align}
\begin{align}
&-4R^{\alpha\beta\sigma\lambda}\left(\nabla_{\mu}R\right)\left(\nabla_{\nu}R_{\alpha\beta\sigma\lambda}\right)-4R\left(\nabla_{\mu}R^{\alpha\beta\sigma\lambda}\right)\left(\nabla_{\nu}R_{\alpha\beta\sigma\lambda}\right)-8R_{\mu\nu}\left(\nabla_{\rho}R_{\alpha\beta\sigma\lambda}\right)\left(\nabla^{\rho}R^{\alpha\beta\sigma\lambda}\right)\notag\\
&+8R^{\beta\sigma\lambda\rho}\left(\nabla_{\alpha}R\indices{_\mu^\alpha}\right)\left(\nabla_{\nu}R_{\beta\sigma\lambda\rho}\right)+8R\indices{_\mu^\alpha}\left(\nabla_{\alpha}R_{\beta\sigma\lambda\rho}\right)\left(\nabla_{\nu}R^{\beta\sigma\lambda\rho}\right)+8R\indices{_\mu^\alpha}R\left(\nabla_{\nu}\nabla_{\alpha}R\right)\notag \\
&+8R^{\alpha\beta}R\left(\nabla_{\nu}\nabla_{\mu}R_{\alpha\beta}\right)-4R^{2}\left(\nabla_{\nu}\nabla_{\mu}R\right)-2R^{\alpha\beta\sigma\lambda}R\left(\nabla_{\mu}\nabla_{\nu}R_{\alpha\beta\sigma\lambda}\right)-8R_{\mu\nu}R^{\alpha\beta\sigma\lambda}\left(\nabla^{2}R_{\alpha\beta\sigma\lambda}\right)\notag \\
& +8R^{\alpha\beta\sigma\lambda}\left(\nabla_{\rho}R\indices{_\mu^\rho_\nu^\tau}\right)\left(\nabla_{\tau}R_{\alpha\beta\sigma\lambda}\right)+8R^{\alpha\beta\sigma\lambda}\left(\nabla_{\rho}R_{\alpha\beta\sigma\lambda}\right)\left(\nabla^{\rho}R\indices{_\mu^\rho_\nu^\tau}\right)  \Big]\notag\\ 
&+\p^{3}_{G}f\Big[ 128R^{\beta\sigma}R_{\mu\nu}R\left(\nabla_{\alpha}R_{\beta\sigma}\right)\left(\nabla^{\alpha}R\right) -16R_{\mu\nu}R^{2}\left(\nabla_{\alpha}R\right)\left(\nabla^{\alpha}R\right)\notag\\
&-32R_{\mu\nu}R^{\beta\sigma\lambda\rho}R\left(\nabla_{\alpha}R_{\beta\sigma\lambda\rho}\right)\left(\nabla^{\alpha}R\right)+16R^{\sigma\lambda\rho\tau}R\indices{_{\mu\alpha\nu}^\beta}R\left(\nabla^{\alpha}R\right)\left(\nabla_{\beta}R_{\sigma\lambda\rho\tau}\right)\notag \\
& +16R^{\sigma\lambda\rho\tau}R\indices{_\mu^\alpha_\nu^\beta}R^{\xi\chi\kappa\omega}\left(\nabla_{\alpha}R_{\sigma\lambda\rho\lambda}\right)\left(\nabla_{\beta}R_{\xi\chi\kappa\omega}\right)+16R^{2}R_{\mu\alpha\nu\beta}\left(\nabla^{\alpha}R\right)\left(\nabla^{\beta}R\right) \notag\\
&+128R^{\alpha\beta}R_{\mu\nu}R^{\lambda\rho\tau\xi}\left(\nabla_{\sigma}R_{\lambda\rho\tau\xi}\right)\left(\nabla^{\sigma}R_{\alpha\beta}\right)+16R^{\sigma\lambda\rho\tau}R\indices{_\mu^\beta_{\nu\alpha}}R\left(\nabla^{\alpha}R\right)\left(\nabla_{\beta}R_{\sigma\lambda\rho\tau}\right)\notag \\
& -64R^{\alpha\beta}R\indices{_{\mu\sigma\nu}^\lambda}R^{\rho\tau\xi\chi}\left(\nabla^{\sigma}R_{\alpha\beta}\right)\left(\nabla_{\lambda}R_{\rho\tau\xi\chi}\right)-64R^{\alpha\beta}R\indices{_\mu^\lambda_{\nu\sigma}}R^{\rho\tau\xi\chi}\left(\nabla^{\sigma}R_{\alpha\beta}\right)\left(\nabla_{\lambda}R_{\rho\tau\xi\chi}\right)\notag \\
& -64R^{\beta\sigma} R_{\mu\alpha\nu\lambda}R\left(\nabla^{\alpha}R\right)\left(\nabla^{\lambda}R_{\beta\sigma}\right)-64R^{\beta\sigma}R_{\mu\lambda\nu\alpha}R\left(\nabla^{\alpha}R\right)\left(\nabla^{\lambda}R_{\beta\sigma}\right)\notag\\
&+256R^{\beta\sigma}R^{\lambda\rho}R\indices{_\nu^\alpha}\left(\nabla_{\alpha}R_{\lambda\rho}\right)\left(\nabla_{\mu}R_{\beta\sigma}\right)-64R^{\beta\sigma}R\indices{_\nu^\alpha}R^{\lambda\rho\tau\xi}\left(\nabla_{\alpha}R_{\lambda\rho\tau\xi}\right)\left(\nabla_{\mu}R_{\beta\sigma}\right)\notag\\
&-64R^{\beta\sigma}R_{\nu\alpha}R\left(\nabla^{\alpha}R\right)\left(\nabla_{\mu}R_{\beta\sigma}\right)-64R^{\beta\sigma}R\indices{_\nu^\alpha}R\left(\nabla_{\alpha}R_{\beta\sigma}\right)\left(\nabla_{\mu}R\right)\notag \\
&+16R\indices{_\nu^\alpha}R^{\beta\sigma\lambda\rho}R\left(\nabla_{\alpha}R_{\beta\sigma\lambda\rho}\right)\left(\nabla_{\mu}R\right)+16R_{\nu\alpha}R^{2}\left(\nabla^{\alpha}R\right)\left(\nabla_{\mu}R\right)\notag\\
&+16R\indices{_\nu^\alpha}R^{\beta\sigma\lambda\rho}R^{\tau\xi\chi\kappa}\left(\nabla_{\alpha}R_{\tau\xi\chi\kappa}\right)\left(\nabla_{\mu}R_{\beta\sigma\lambda\rho}\right)+16R_{\nu\alpha}R^{\beta\sigma\lambda\rho}R\left(\nabla^{\alpha}R\right)\left(\nabla_{\mu}R_{\beta\sigma\lambda\rho}\right)\notag\\
& -64R^{\beta\sigma}R\indices{_\nu^\alpha}R^{\lambda\rho\tau\xi}\left(\nabla_{\alpha}R_{\beta\sigma}\right)\left(\nabla_{\mu}R_{\lambda\rho\tau\xi}\right)+32R^{\alpha\beta}R^{2}\left(\nabla_{\mu}R\right)\left(\nabla_{\nu}R_{\alpha\beta}\right)\notag\\
&+32R^{\alpha\beta}R^{\sigma\lambda\rho\tau}R\left(\nabla_{\mu}R_{\sigma\lambda\rho\tau}\right)\left(\nabla_{\nu}R_{\alpha\beta}\right)+256R^{\beta\sigma}R^{\lambda\rho}R\indices{_\mu^\alpha}\left(\nabla_{\alpha}R_{\lambda\rho}\right)\left(\nabla_{\nu}R_{\beta\sigma}\right)\notag\\
&-64R^{\beta\sigma}R\indices{_\mu^\alpha}R^{\lambda\rho\tau\xi}\left(\nabla_{\alpha}R_{\lambda\rho\tau\xi}\right)\left(\nabla_{\nu}R_{\beta\sigma}\right)-64R^{\beta\sigma}R_{\mu\alpha}R\left(\nabla^{\alpha}R\right)\left(\nabla_{\nu}R_{\beta\sigma}\right)\notag\\
&-128R^{\alpha\beta}R^{\sigma\lambda}R\left(\nabla_{\mu}R_{\alpha\beta}\right)\left(\nabla_{\nu}R_{\sigma\lambda}\right)-64 R^{\beta\sigma}R\indices{_\mu^\alpha}R\left(\nabla_{\alpha}R_{\beta\sigma}\right)\left(\nabla_{\nu}R\right)\notag\\
&+16R\indices{_\mu^\alpha}R^{\beta\sigma\lambda\rho}R\left(\nabla_{\alpha}R_{\beta\sigma\lambda\rho}\right)\left(\nabla_{\nu}R\right)+16R_{\mu\alpha}R^{2}\left(\nabla^{\alpha}R\right)\left(\nabla_{\nu}R\right)+32R^{\alpha\beta}R^{2}\left(\nabla_{\mu}R_{\alpha\beta}\right)\left(\nabla_{\nu}R\right)\notag\\
&-8R^{3}\left(\nabla_{\mu}R\right)\left(\nabla_{\nu}R\right)-8R^{2}R^{\alpha\beta\sigma\lambda}\left(\nabla_{\mu}R_{\alpha\beta\sigma\lambda}\right)\left(\nabla_{\nu}R\right)-8R^{2}R^{\alpha\beta\sigma\lambda}\left(\nabla_{\mu}R\right)\left(\nabla_{\nu}R_{\alpha\beta\sigma\lambda}\right)\notag\\
&+16R\indices{_\mu^\alpha}R^{\beta\sigma\lambda\rho}R^{\tau\xi\chi\kappa}\left(\nabla_{\alpha}R_{\tau\xi\chi\kappa}\right)\left(\nabla_{\nu}R_{\beta\sigma\lambda\rho}\right) +16R_{\mu\alpha}R^{\beta\sigma\lambda\rho}R\left(\nabla^{\alpha}R\right)\left(\nabla_{\nu}R_{\beta\sigma\lambda\rho}\right)\notag\\
&+32R^{\alpha\beta}R^{\sigma\lambda\rho\tau}R\left(\nabla_{\mu}R_{\alpha\beta}\right)\left(\nabla_{\nu}R_{\sigma\lambda\rho\tau}\right)-64R^{\beta\sigma}R\indices{_\mu^\alpha}R^{\lambda\rho\tau\xi}\left(\nabla_{\alpha}R_{\beta\sigma}\right)\left(\nabla_{\nu}R_{\lambda\rho\tau\xi}\right)\notag\\
&-8R^{\alpha\beta\sigma\lambda}R^{\rho\xi\chi\kappa}R\left(\nabla_{\mu}R_{\alpha\beta\sigma\lambda}\right)\left(\nabla_{\nu}R_{\rho\xi\chi\kappa}\right)-256R^{\alpha\beta}R^{\sigma\lambda}R_{\mu\nu}\left(\nabla_{\rho}R_{\sigma\lambda}\right)\left(\nabla^{\rho}R_{\alpha\beta}\right)\notag\\
&+256R^{\alpha\beta}R^{\sigma\lambda}R_{\mu\rho\nu\tau}\left(\nabla^{\rho}R_{\alpha\beta}\right)\left(\nabla^{\tau}R_{\sigma\lambda}\right)-16R_{\mu\nu}R^{\alpha\beta\sigma\lambda}R^{\rho\tau\xi\chi}\left(\nabla_{\kappa}R_{\rho\tau\xi\chi}\right)\left(\nabla^{\kappa}R_{\alpha\beta\sigma\lambda}\right) \Big]\notag \\
&+\p_{R}\p^{2}_{G}f\Big[ 64R^{\beta\sigma}R_{\mu\nu}\left(\nabla_{\alpha}R_{\beta\sigma}\right)\left(\nabla^{\alpha}R\right)-16R_{\mu\nu}\left(\nabla_{\alpha}R\right)\left(\nabla^{\alpha}R\right)-16R_{\mu\nu}R^{\beta\sigma\lambda\rho}\left(\nabla_{\alpha}R_{\beta\sigma\lambda\rho}\right)\left(\nabla^{\alpha}R\right)\notag\\
&+8R^{\sigma\lambda\rho\tau}R\indices{_\mu_\alpha_\nu^\beta}\left(\nabla^{\alpha}R\right)\left(\nabla_{\beta}R_{\sigma\lambda\rho\tau}\right)+8R^{\sigma\lambda\rho\tau}R\indices{_\mu^\beta_{\nu\alpha}}\left(\nabla^{\alpha}R\right)\left(\nabla_{\beta}R_{\sigma\lambda\rho\tau}\right)+16R_{\mu\alpha\nu\beta}R\left(\nabla^{\alpha}R\right)\left(\nabla^{\beta}R\right)\notag \\
&-32R^{\beta\sigma}R_{\mu\alpha\nu\lambda}\left(\nabla^{\alpha}R\right)\left(\nabla^{\lambda}R_{\beta\sigma}\right)-32R^{\beta\sigma}R_{\mu\lambda\nu\alpha}\left(\nabla^{\alpha}R\right)\left(\nabla^{\lambda}R_{\beta\sigma}\right)-32R^{\beta\sigma}R_{\nu\alpha}\left(\nabla^{\alpha}R\right)\left(\nabla_{\mu}R_{\beta\sigma}\right)\notag\\
&-32R^{\beta\sigma}R\indices{_\nu^\alpha}\left(\nabla_{\alpha}R_{\beta\sigma}\right)\left(\nabla_{\mu}R\right)+8R\indices{_\nu^\alpha}R^{\beta\sigma\lambda\rho}\left(\nabla_{\alpha}R_{\beta\sigma\lambda\rho}\right)\left(\nabla_{\mu}R\right)+16R_{\nu\alpha}R\left(\nabla^{\alpha}R\right)\left(\nabla_{\mu}R\right)\notag \\
&+8R_{\nu\alpha}R^{\beta\sigma\lambda\rho}\left(\nabla^{\alpha}R\right)\left(\nabla_{\mu}R_{\beta\sigma\lambda\rho}\right)+32R^{\alpha\beta}R\left(\nabla_{\mu}R\right)\left(\nabla_{\nu}R_{\alpha\beta}\right)+16R^{\alpha\beta}R^{\sigma\lambda\rho\tau}\left(\nabla_{\mu}R_{\sigma\lambda\rho\tau}\right)\left(\nabla_{\nu}R_{\alpha\beta}\right)\notag \\
&-32R^{\beta\sigma}R_{\mu\alpha}\left(\nabla^{\alpha}R\right)\left(\nabla_{\nu}R_{\beta\sigma}\right)-64R^{\alpha\beta}R^{\sigma\lambda}\left(\nabla_{\mu}R_{\alpha\beta}\right)\left(\nabla_{\nu}R_{\sigma\lambda}\right)-32R^{\beta\sigma}R\indices{_\mu^\alpha}\left(\nabla_{\alpha}R_{\beta\sigma}\right)\left(\nabla_{\nu}R\right)\notag 
\end{align}
\begin{align}
&+8R\indices{_\mu^\alpha}R^{\beta\sigma\lambda\rho}\left(\nabla_{\alpha}R_{\beta\sigma\lambda\rho}\right)\left(\nabla_{\nu}R\right) +16R_{\mu\alpha}R\left(\nabla^{\alpha}R\right)\left(\nabla_{\nu}R\right)+32R^{\alpha\beta}R\left(\nabla_{\mu}R_{\alpha\beta}\right)\left(\nabla_{\nu}R\right)\notag \\
&-12R^{2}\left(\nabla_{\mu}R\right)\left(\nabla_{\nu}R\right)-8R^{\alpha\beta\sigma\lambda}R\left(\nabla_{\mu}R_{\alpha\beta\sigma\lambda}\right)\left(\nabla_{\nu}R\right)-8R^{\alpha\beta\sigma\lambda}R\left(\nabla_{\mu}R_{\alpha\beta}\right)\left(\nabla_{\nu}R\right)\notag\\
&-8R^{\alpha\beta\sigma\lambda}R\left(\nabla_{\mu}R\right)\left(\nabla_{\nu}R_{\alpha\beta\sigma\lambda}\right)+8R_{\mu\alpha}R^{\beta\sigma\lambda\rho}\left(\nabla^{\alpha}R\right)\left(\nabla_{\nu}R_{\beta\sigma\lambda\rho}\right)\notag\\
&+16R^{\alpha\beta}R^{\sigma\lambda\rho\tau}\left(\nabla_{\mu}R_{\alpha\beta}\right)\left(\nabla_{\nu}R_{\sigma\lambda\rho\tau}\right)-4R^{\alpha\beta\sigma\lambda}R^{\rho\tau\xi\chi}\left(\nabla_{\mu}R_{\alpha\beta\sigma\lambda}\right)\left(\nabla_{\nu}R_{\rho\tau\xi\chi}\right)\Big]
\end{align}
where $\p_{R}$ and $\p_{G}$ are the derivatives with respect to the Ricci scalar and the Gauss-Bonnet term, respectively, i.e. $\p_{R} = \p/\p R$ and $\p_{G} = \p/\p G$. From here also one can identify the $\mathcal{C}\indices{^i_{\mu\nu}}$ present in Eq.~\eqref{eq:eom2}.

\section{Evolution equations}
\label{evoeqns}

Using Eqs.~\eqref{mod_ein_equ},~\eqref{eq:g00n},~\eqref{eq:g0in} and~\eqref{eq:gijn} we can get the evolution equations:

\begin{align}
\label{eq:t00n}
{} & -\frac{2}{a^{2}}\Bigg[ \alpha\Big\{18\left(\H^{4}+{\H'}^{2}-\H\H''\right)\phi-9\H\H'\phi'-9\H^{2}\phi''+9\left(4\H^{3}-\H\H'-\H''\right)\psi'\notag\\
& +9\left(2\H'-\H^{2}\right)\psi''-9\H\psi'''+12\left(\H^{2}+\H'\right)\nabla^{2}\phi+15\H\nabla^{2}\psi'+3\nabla^{2}\psi''+\nabla^{2}\nabla^{2}\left(\phi-2\psi\right) \Big\}\notag\\
&+\beta\Big\{ \left( 24\H^{2}\left[26\H^{8}+19\H^{6}\H'+20\H^{4}{\H'}^{2}+7\H^{2}{\H'}^{3}+2{\H'}^{4}\right]-6\H\left[73\H^{6}+126\H^{4}\H'\right.\right.\notag\\
&\left.\left.+64\H^{2}{\H'}^{2}+32{\H'}^{3}\right]\H''+12\left[6\H^{4}+18\H^{2}\H'+13{\H'}^{2}\right]{\H''}^{2} \right)\frac{\phi}{\H^{4}\H'}\notag\\
&+\left(6\H\left[ 13{\H'}^{2}-6\H^{4} \right]{\H''}^{2} -3\H\left[104\H^{10}+4\H^{8}\H'+139\H^{6}{\H'}^{2}+322\H^{4}{\H'}^{3}+168\H^{2}{\H'}^{4}\right.\right.\notag\\
&\left.\left.+96{\H'}^{5}\right]+6\left[ 35\H^{8}+\H^{6}\H'+4\H^{4}{\H'}^{2}+76\H^{2}{\H'}^{3}+78{\H'}^{4} \right]\H'' \right)\frac{\phi'}{\H^{4}{\H'}^{2}}\notag\\
&-\left( 71\H^{7}+130\H^{5}\H'+64\H^{3}{\H'}^{2}+32\H{\H'}^{3}-24\H^{4}\H''-72\H^{2}\H'\H'' -52{\H}^{2}\H'' \right)\frac{3\phi''}{\H^{3}\H'}\notag\\
&-\left( 104\H^{12}-620\H^{10}\H'-189\H^{8}{\H'}^{2}+174\H^{6}{\H'}^{3}+176\H^{4}{\H'}^{4}+112\H^{2}{\H'}^{5} -70\H^{9}\H''\right.\notag\\
&+213\H^{7}\H'\H''+118\H^{5}{\H'}^{2}\H'' -216\H^{3}{\H'}^{3}\H''-252\H{\H'}^{4}\H''+12\H^{6}{\H''}^{2}+46\H^{2}{\H'}^{2}{\H''}^{2}\notag\\
&\left.+104{\H'}^{3}{\H''}^{2}\right)\frac{3\psi'}{\H^{5}{\H'}^{2}}\notag\\
&-\left( 104\H^{10}+75\H^{8}\H'+56\H^{6}{\H'}^{2}-4\H^{4}{\H'}^{3}+8\H^{2}{\H'}^{4}-70\H^{7}\H''-26\H^{5}\H'\H''-8\H^{3}{\H'}^{2}\H''\right.\notag\\
&\left. +12\H{\H'}^{3}\H''+12\H^{4}{\H''}^{2}-26{\H'}^{2}{\H''}^{2} \right)\frac{3\psi''}{\H{\H'}^{2}}\notag\\
&-\left( 71\H^{7}+130\H^{5}\H'+64\H^{3}{\H'}^{2}+32\H{\H'}^{3}-24\H^{4}\H''-72\H^{2}\H'\H'' -52{\H}^{2}\H'' \right)\frac{3\psi'''}{\H^{4}\H'}\notag\\
&+\left( 4\H^{2}\left[ 18{\H'}^{4}+44\H^{2}{\H'}^{3}+77\H^{4}{\H'}^{2}+51\H^{6}\H'-26^{8} \right] +2\left[ 13{\H'}^{2}-6\H^{4} \right]{\H''}^{2} \right.\notag\\
&\left. +2\H\left[ 35\H^{6}-34\H^{4}\H' -109\H^{2}{\H'}^{2}-68{\H'}^{3}\right]\H'' \right)\frac{\nabla^{2}\phi}{\H^{4}{\H'}^{2}}\notag\\
&-\left( 70\H^{7}+130\H^{5}\H' +66\H^{3}{\H'}^{2}+32\H{\H'}^{3}-24\H^{4}\H''-73\H^{2}\H'\H''-52{\H'}^{2}\H''\right)\frac{\nabla^{2}\phi'}{\H^{4}\H'}\notag\\
&-\left( 104\H^{10}+42\H^{8}\H' +2\H^{6}{\H'}^{2}-20\H^{4}{\H'}^{3}-8\H^{2}{\H'}^{4}-36\H^{7}\H''+\H^{5}\H'\H''+32\H^{3}{\H'}^{2}\H''\right.\notag\\
&\left.+32\H{\H'}^{3}\H''-18\H^{2}\H'{\H''}^{2}-26{\H'}^{2}{\H''}^{2} \right)\frac{4\nabla^{2}\psi}{\H^{6}\H'}\notag\\
&+\left( 209\H^{7}+236\H^{5}\H'+100\H^{3}{\H'}^{2}+48\H{\H'}^{3}-70\H^{4}\H''-152\H^{2}\H'\H''-72{\H'}^{2}\H'' \right)\nabla^{2}\frac{\psi'}{\H^{4}\H'}\notag\\
&+\left( \frac{\H^{2}}{\H'} \right)\nabla^{2}\psi''+\left( \frac{\H^{2}}{3\H'} \right)\nabla^{2}\nabla^{2}\phi-\left( \frac{2}{3} \right)\nabla^{2}\nabla^{2}\psi  \Big\} \Bigg]=a^{2}\left( \delta \rho + 2 \rho \phi \right)\,,
\end{align}
\begin{align}
\label{eq:t0in}
{} & \frac{2}{a^{2}}\Bigg[\alpha\Big\{3\left(3\H''+2\H\H'-4\H^{3}\right)\phi-3\left( \H^{2}-3\H' \right)\phi'+3\H\phi''+3\left(7\H'-5\H^{2}\right)\psi'+3\psi'''\notag\\
&-3\H\nabla^{2}\left(\phi-2\psi\right)+\nabla^{2}\left(\phi'-2\psi'\right)\Big\}+\beta\Big\{ \left(60{\H'}^{3}\H''+171\H^{4}\H'\H'' +172\H^{4}\H'\H'' \right.\notag\\
&\left. +44\H^{6}\H''-72\H{\H'}^{4} -134\H^{3}{\H'}^{3}-220\H^{5}{\H'}^{2}-282\H^{7}\H'-192\H^{9}\right)\frac{\nabla^{2}\phi}{6\H^{4}{\H'}^{2}} \notag\\
&+\left( \right. 96\H^{9}+156\H^{7}\H' +84\H^{5}{\H'}^{2}+83\H^{3}{\H'}^{3}+24\H{\H'}^{4}-24\H^{6}\H''-82\H^{4}\H'\H'' \notag\\
&\left. -90\H^{2}{\H'}^{2}\H''-24{\H'}^{3}\H'' \right)\frac{2\nabla^{2}\psi}{3\H^{4}{\H'}^{2}}\notag\\
&-\left( 1880\H^{9}\H'+2156\H^{7}{\H'}^{2} +1710\H^{5}{\H'}^{3}+992\H^{3}{\H'}^{4}+432\H{\H'}^{5}+576\H^{8}\H''+178\H\H'\H''\right. \notag\\
& -1231\H^{4}{\H'}^{2}\H''-984\H^{2}{\H'}^{3}\H'' -360{\H'}^{4}\H''-144\H^{5}{\H''}^{2}-444\H^{3}\H'{\H''}^{2} \notag\\
&\left.-288\H{\H'}^{2}{\H''}^{2}\right)\frac{\phi}{6\H^{4}{\H'}^{2}}\notag\\
&-\left( 288\H^{9}+419\H^{7}\H'+323\H^{5}{\H'}^{2} +176\H^{3}{\H'}^{3}+108\H{\H'}^{4}-66\H^{6}\H''-258\H^{4}\H'\H''\right.\notag\\
&\left. -246\H^{2}{\H'}^{2}\H''-90{\H'}^{3}\H'' \right)\frac{\phi'}{3\H^{3}{\H'}^{2}}\notag\\
&-\left( 576\H^{10}+1566\H\H'+1126\H^{6}{\H'}^{2}+734\H^{4}{\H'}^{3}+504\H^{2}{\H'}^{4}+444\H^{7}\H''-56\H^{5}\H'\H'' \right. \notag\\
&\left. -691\H^{3}{\H'}^{2}\H'' -180\H{\H'}^{3}\H'' -144\H^{4}{\H''}^{2}-444\H^{2}\H'{\H''}^{2}-288{\H'}^{2}{\H''}^{2} \right)\frac{\psi'}{6\H^{4}{\H'}^{2}}\notag\\
&-\left( 144\H^{9}+208\H^{7}\H'+166\H^{5}{\H'}^{2}+88\H^{3}{\H'}^{3}+54\H{\H'}^{4}-33\H^{6}\H''-129\H^{4}\H'\H'' \right. \notag\\
& \left. -123\H^{2}{\H'}^{2}\H''-45{\H'}^{3}\H'' \right)\frac{2\psi''}{3\H^{4}{\H'}^{2}}\notag\\
&+\left( \frac{\H^{2}}{3\H'} \right)\nabla^{2}\phi'-\left(\frac{2}{3}\right)\nabla^{2}\psi'+\left( \frac{\H^{3}}{\H'} \right)\phi''+\left( \frac{\H^{2}}{\H'} \right)\psi'''  \Big\} \Bigg]_{,i}=-a^{2}\left[ P_{0} v_{i} + \rho\left(v_{i}+B_{i} \right)\right]\,, 
\end{align}
\begin{align}
\label{eq:tijn}
{} & \frac{2}{a^{2}}\Bigg[ \alpha\Big\{\delta_{ij}\bigg[ 12\left(\H^{4}-6\H^{2}\H'+{\H'}^{2}-\H\H''+\H'''\right)\phi-3\left( 6\H^{3}+\H\H'-6\H'' \right)\phi' -3\left(\H^{2}-4\H'\right)\phi''\notag\\
& +3\H\phi'''+6\left(\H^{4}-6\H^{2}\H'+{\H'}^{2}-\H\H''+\H'''\right)\psi -3\left( 2\H^{3}+13\H\H'-5\H'' \right)\psi'+3\psi''''\notag \\ 
&-3\left(7\H^{2}-6\H'\right)\psi'' -2\left( \H^{2}+2\H' \right)\nabla^{2}\phi-3\H\nabla^{2}\left(2\phi'+\psi'\right)+\nabla^{2}\left(\phi''-5\psi''\right)\notag\\
&-2\left(\H^{2}-\H'\right)\nabla^{2}\psi -\nabla^{2}\nabla^{2}\left(\phi-2\psi\right)\bigg]+\bigg[ 3\H\left(\phi'+3\psi'\right) + 6\left( \H^{2}+\H' \right)\psi \bigg]_{,ij}\notag\\
&+\bigg[ 3\psi''+\nabla^{2}\nabla^{2}\left( \phi-2\psi \right) \bigg]_{,ij} \Big\} \notag\\
&+ \beta\Big\{ \delta_{ij}\bigg[ \left( 8\H^{3}\H'\H'' -16 \H^{4}{\H'}^{2}-4{\H'}^{4}-4\H{\H'}^{2}\H''+\H^{2}\left[13{\H'}^{3}-{\H''}^{2}\right] \right)\frac{8\phi}{{\H'}^{2}}\notag\\
&-\left( 8\H^{8}\H' +48\H^{6} {\H'}^{2}+64\H^{2}{\H'}^{4} -18{\H'}^{5}-6\H^{5}\H'\H''+36\H^{3}{\H'}^{2}\H''\right.\notag\\
& \left. -3\H^{4}\left[ 47{\H'}^{3}+4{\H''}^{2}-2\H'\H''' \right] \right)\frac{\phi'}{3\H{\H'}^{3}}+\left( 7\H^{4}\H'+2\H^{2}{\H'}^{2}+2{\H'}^{3}-4\H^{3}\H'' \right)\frac{\phi''}{{\H'}^{2}}\notag\\
&+\left( 4\H^{7}\H'-48\H^{5}{\H'}^{2}-22\H{\H'}^{4}+21\H^{4}\H'\H''-6\H^{2}{\H'}^{2}\H''+6{\H'}^{3}\H''\right.\notag\\
&\left. +3\H^{3}\left[ 4{\H'}^{3}-2{\H''}^{2}+\H'\H'' \right] \right)\frac{2\psi}{3\H{\H'}^{2}}\notag\\
&-\left( 8\H^{9}\H' +32\H^{7}{\H'}^{2}-14\H{\H'}^{5}-6\H^{6}\H'\H'' -81\H^{4}{\H'}^{2}\H'' + 42\H^{2}{\H'}^{3}\H'' +18{\H'}^{4}\H''\right.\notag\\
&\left. -\H^{3}\left[ 92{\H'}^{4}-12\H'{\H''}^{2} \right]+3\H^{5}\left[ 49{\H'}^{3}-4{\H''}^{2}+2\H'\H'' \right] \right)\frac{\psi'}{3\H^{2}{\H'}^{3}}\notag\\
&-\left( 8\H^{6}\H'+27\H^{4}{\H'}^{2}+40{\H'}^{4}+6\H^{3}\H'\H''+18\H{\H'}^{2}\H''-6\H^{2}\left[ 14{\H'}^{3}+2{\H''}^{2}-\H'\H''' \right] \right)\frac{\psi''}{3{\H'}^{3}}\notag\\
&+\left( 4\H^{4}\H'-\H^{2}{\H'}^{2}+{\H'}^{3}-2\H^{3}\H'' \right)\frac{2\psi'''}{\H{\H'}^{2}}\notag\\
&-\left( 16\H^{7}\H' +30\H^{5}{\H'}^{2}+131\H{\H'}^{4}-12\H^{4}\H'\H''+6\H^{2}{\H'}^{2}\H''-6{\H'}^{3}\H''\right. \notag\\
&\left.-3\H^{3}\left[7{\H'}^{3}+{\H''}^{2}-4\H'\H'''\right] \right)\frac{\nabla^{2}\phi}{18\H{\H'}^{3}}+\left( 7\H^{4}\H' -7\H^{2}{\H'}^{2} +{\H'}^{3} -4\H^{3}\H'' \right)\frac{\nabla^{2}\phi'}{3\H{\H'}^{2}}\notag\\
&+\left( 2\H^{7}\H'+453\H^{5}{\H'}^{2}-68\H{\H'}^{4} -234\H^{4}\H'\H''+78\H^{2}{\H'}^{2}\H'' +48{\H'}^{3}\H''\right. \notag \\
&\left. -3\H^{3}\left[ 69{\H'}^{3}-12{\H''}^{2}+2\H'\H''' \right] \right)\frac{\nabla^{2}\psi}{18\H^{{\H'}^{2}}}+\left( 4\H^{4}-21\H^{2}\H'+6{\H'}^{2}+4\H\H'' \right)\frac{\nabla^{2}\psi'}{3\H\H'}\notag\\ 
&+\left( \frac{\H^{2}}{\H'} \right)\psi''''+\left( \frac{\H^{3}}{\H'} \right)\phi'''+\left( \frac{\H^{2}}{3\H'} \right)\nabla^{2}\phi''-\left( \frac{5}{3} \right)\nabla^{2}\psi''-\left( \frac{1}{3} \right)\nabla^{2}\nabla^{2}\phi+\left( \frac{2\H'}{3\H^{2}} \right)\nabla^{2}\nabla^{2}\psi \bigg] \notag\\
&+\bigg[\left( 11\H^{3}\H'-10\H^{5}+23\H{\H'}^{2}-2\H^{2}\H''-6\H'\H'' \right)\frac{\phi}{2\H\H'}\notag\\
&+\left( 2\H^{6}\H'-79\H^{4}{\H'}^{2}-36{\H'}^{4}+42\H^{3}\H'\H''-6\H{\H'}^{2}\H''+\H^{2}\left[ 53{\H'}^{3}-12{\H''}^{2}+6\H'\H'' \right] \right)\frac{\psi}{6\H^{2}{\H'}^{2}}\notag\\
&+\left( \H \right)\phi' -\left( 4\H^{5}+\H^{3}\H' -26\H{\H'}^{2}+6\H'\H''\right)\frac{\psi'}{3\H^{2}\H'}+\psi'' +\left( \frac{1}{3} \right)\nabla^{2}\phi-\left( \frac{2\H'}{\H^{2}} \right)\nabla^{2}\psi \bigg]_{,ij} \Big\}\Bigg] \notag\\
&\qquad\qquad\qquad\qquad\qquad\qquad\qquad\qquad\qquad\qquad\qquad\qquad\qquad  = a^{2}\left[\delta_{ij}\left( \delta P - 2 P_{0}\psi \right) + 2 P_{0} h_{ij} \right]\,.
\end{align}

\subsection{Discussion and Future Work}
\label{conclusion}

In this appendix, we have provided a derivation of the governing
equations for a fourth order $f(R,G)$ theory. Throughout we assumed a
flat FRLW universe. We rederived the governing equations in the
background, and, using cosmological perturbation theory to linear
order, derived the governing equations for the scalar perturbations.
Since the equations are rather complex, we only present them in
longitudinal gauge. However, we also provide the definitions of the
gauge-invariant variables in this gauge, both for the metric and the
matter perturbations in Chapter \ref{chapter:cpt}. One can therefore 
easily rewrite the governing equations for any other gauge.


Here we only studied scalar perturbations, leaving the discussion of
vector and tensor perturbations for future work \cite{FGM2}. The 
detection of gravitational waves and present and future
gravitational wave observatories like LIGO \cite{Abbott:2007kv},
Virgo \cite{Abbott:2009kk}, KAGRA \cite{KAGRA}, and LISA
\cite{AmaroSeoane:2012je} makes calculating the tensor perturbation
evolution equations an exciting future project.

We also need to find solutions to the governing equations, since only
then can we calculate observational signatures that can be compared to
the observational data. The equations presented in this appendix are
rather complex and hence difficult to solve. In order to make solving
the equations less time consuming, we have also provided the
Mathematica sheets in Github \cite{githubGB}.

\end{appendices}

%% file: thesis.bbl
\providecommand{\href}[2]{#2}\begingroup\raggedright\begin{thebibliography}{100}

\bibitem{2016PhRvL.116f1102A}
B.~P. {Abbott}, R.~{Abbott}, T.~D. {Abbott}, M.~R. {Abernathy}, F.~{Acernese},
  K.~{Ackley} et~al., \emph{{Observation of Gravitational Waves from a Binary
  Black Hole Merger}},
  \href{https://doi.org/10.1103/PhysRevLett.116.061102}{\emph{Physical Review
  Letters} {\bfseries 116} (2016) 061102}
  [\href{https://arxiv.org/abs/1602.03837}{{\ttfamily 1602.03837}}].

\bibitem{weinberg}
S.~Weinberg, \emph{{Gravitation and Cosmology: Principles and Applications of
  the General Theory of Relativity}}. Wiley, New York, NY, 1972.

\bibitem{TCMG}
T.~Clifton, P.~G. Ferreira, A.~Padilla and C.~Skordis, \emph{{Modified Gravity
  and Cosmology}},
  \href{https://doi.org/10.1016/j.physrep.2012.01.001}{\emph{Phys. Rept.}
  {\bfseries 513} (2012) 1} [\href{https://arxiv.org/abs/1106.2476}{{\ttfamily
  1106.2476}}].

\bibitem{CP1}
C.~Pittordis and W.~Sutherland, \emph{{Testing Modified Gravity Theories via
  Wide Binaries and GAIA}},
  \href{https://doi.org/10.1093/mnras/sty1578}{\emph{Mon. Not. Roy. Astron.
  Soc.} {\bfseries 480} (2018) 1778}
  [\href{https://arxiv.org/abs/1711.10867}{{\ttfamily 1711.10867}}].

\bibitem{CP2}
C.~Pittordis and W.~Sutherland, \emph{{Testing Modified Gravity with Wide
  Binaries in GAIA DR2}},
  \href{https://doi.org/10.1093/mnras/stz1898}{\emph{Mon. Not. Roy. Astron.
  Soc.} {\bfseries 488} (2019) 4740}
  [\href{https://arxiv.org/abs/1905.09619}{{\ttfamily 1905.09619}}].

\bibitem{Nojiri:2017ncd}
S.~Nojiri, S.~D. Odintsov and V.~K. Oikonomou, \emph{{Modified Gravity Theories
  on a Nutshell: Inflation, Bounce and Late-time Evolution}},
  \href{https://doi.org/10.1016/j.physrep.2017.06.001}{\emph{Phys. Rept.}
  {\bfseries 692} (2017) 1} [\href{https://arxiv.org/abs/1705.11098}{{\ttfamily
  1705.11098}}].

\bibitem{horn}
G.~W. Horndeski, \emph{Second-order scalar-tensor field equations in a
  four-dimensional space}, \href{https://doi.org/10.1007/BF01807638}{\emph{Int.
  J. Theor. Phys., v. 10, no. 6, pp. 363-384} }.

\bibitem{esposito}
C.~Deffayet, G.~Esposito-Far\`ese and A.~Vikman, \emph{Covariant galileon},
  \href{https://doi.org/10.1103/PhysRevD.79.084003}{\emph{Phys. Rev. D}
  {\bfseries 79} (2009) 084003}.

\bibitem{PGS1}
P.~G. Fernandes, \emph{{Charged Black Holes in AdS Spaces in $4D$ Einstein
  Gauss-Bonnet Gravity}},  \href{https://arxiv.org/abs/2003.05491}{{\ttfamily
  2003.05491}}.

\bibitem{PGS2}
P.~G. Fernandes, P.~Carrilho, T.~Clifton and D.~J. Mulryne, \emph{{Derivation
  of Regularized Field Equations for the Einstein-Gauss-Bonnet Theory in Four
  Dimensions}},  \href{https://arxiv.org/abs/2004.08362}{{\ttfamily
  2004.08362}}.

\bibitem{RevModPhys.75.559}
P.~J.~E. Peebles and B.~Ratra, \emph{The cosmological constant and dark
  energy}, \href{https://doi.org/10.1103/RevModPhys.75.559}{\emph{Rev. Mod.
  Phys.} {\bfseries 75} (2003) 559}.

\bibitem{Carroll_2001}
S.~M. Carroll, \emph{The cosmological constant},
  \href{https://doi.org/10.12942/lrr-2001-1}{\emph{Living Reviews in
  Relativity} {\bfseries 4} (2001) }.

\bibitem{euclid}
{\scshape Euclid} collaboration, R.~Scaramella et~al., \emph{{Euclid space
  mission: a cosmological challenge for the next 15 years}},
  \href{https://doi.org/10.1017/S1743921314011089}{\emph{IAU Symp.} {\bfseries
  306} (2015) 375} [\href{https://arxiv.org/abs/1501.04908}{{\ttfamily
  1501.04908}}].

\bibitem{LSST}
R.~L. Jones, C.~T. Slater, J.~Moeyens, L.~Allen, T.~Axelrod, K.~Cook et~al.,
  \emph{The large synoptic survey telescope as a near-earth object discovery
  machine}, \href{https://doi.org/10.1016/j.icarus.2017.11.033}{\emph{Icarus}
  {\bfseries 303} (2018) 181}.

\bibitem{SKA}
A.~Weltman et~al., \emph{{Fundamental Physics with the Square Kilometre
  Array}}, \href{https://doi.org/10.1017/pasa.2019.42}{\emph{Publ. Astron. Soc.
  Austral.} {\bfseries 37} (2020) e002}
  [\href{https://arxiv.org/abs/1810.02680}{{\ttfamily 1810.02680}}].

\bibitem{malik}
K.~A. Malik and D.~Wands, \emph{Cosmological perturbations},
  \href{https://doi.org/http://dx.doi.org/10.1016/j.physrep.2009.03.001}{\emph{Physics
  Reports} {\bfseries 475} (2009) 1 }.

\bibitem{kaiser}
N.~Kaiser, \emph{Elements of Astrophysics}. CreateSpace Independent Publishing
  Platform, 2014.

\bibitem{chen}
S.~Chen and D.~J. Schwarz, \emph{{Fluctuations of differential number counts of
  radio continuum sources}},
  \href{https://doi.org/10.1103/PhysRevD.91.043507}{\emph{Phys. Rev.}
  {\bfseries D91} (2015) 043507}
  [\href{https://arxiv.org/abs/1407.4682}{{\ttfamily 1407.4682}}].

\bibitem{hamilton}
A.~J.~S. Hamilton, \emph{Linear Redshift Distortions: A Review}, pp.~185--275.
\newblock Springer Netherlands, Dordrecht, 1998.
\newblock 10.1007/978-94-011-4960-0\_17.

\bibitem{kaiser2}
N.~Kaiser, \emph{Clustering in real space and in redshift space},
  \href{https://doi.org/10.1093/mnras/227.1.1}{\emph{Monthly Notices of the
  Royal Astronomical Society} {\bfseries 227} (1987) 1}.

\bibitem{saito}
S.~Saito, \emph{{Lecture notes in \textit{Galaxy Clustering in Redshift Space}
  given in Max-Planck-Institut f\"{u}r Physik}},  June, 2016.

\bibitem{yoo6}
N.~Grimm and J.~Yoo, \emph{{Jacobi Mapping Approach for a Precise Cosmological
  Weak Lensing Formalism}},  \href{https://arxiv.org/abs/1806.00017}{{\ttfamily
  1806.00017}}.

\bibitem{roulettes}
C.~Clarkson, \emph{{Roulettes: A weak lensing formalism for strong lensing -
  II. Derivation and analysis}},
  \href{https://doi.org/10.1088/0264-9381/33/24/245003}{\emph{Class. Quant.
  Grav.} {\bfseries 33} (2016) 245003}
  [\href{https://arxiv.org/abs/1603.04652}{{\ttfamily 1603.04652}}].

\bibitem{1979Nature}
D.~{Walsh}, R.~F. {Carswell} and R.~J. {Weymann}, \emph{{0957+561 A, B: twin
  quasistellar objects or gravitational lens?}},
  \href{https://doi.org/10.1038/279381a0}{\emph{\nat} {\bfseries 279} (1979)
  381}.

\bibitem{sachs}
R.~K. {Sachs} and A.~M. {Wolfe}, \emph{{Perturbations of a Cosmological Model
  and Angular Variations of the Microwave Background}},
  \href{https://doi.org/10.1086/148982}{\emph{The Astrophysical Journal}
  {\bfseries 147} (1967) 73}.

\bibitem{sachs2}
J.~{Kristian} and R.~K. {Sachs}, \emph{{Observations in Cosmology}},
  \href{https://doi.org/10.1086/148522}{\emph{The Astrophysical Journal}
  {\bfseries 143} (1966) 379}.

\bibitem{sachs3}
R.~K. Sachs, \emph{{Gravitational waves in general relativity. 6. The outgoing
  radiation condition}},
  \href{https://doi.org/10.1098/rspa.1961.0202}{\emph{Proc. Roy. Soc. Lond.}
  {\bfseries A264} (1961) 309}.

\bibitem{antony}
A.~Challinor and A.~Lewis, \emph{Linear power spectrum of observed source
  number counts}, \href{https://doi.org/10.1103/PhysRevD.84.043516}{\emph{Phys.
  Rev. D} {\bfseries 84} (2011) 043516}.

\bibitem{jeong1}
D.~Jeong, F.~Schmidt and C.~M. Hirata, \emph{Large-scale clustering of galaxies
  in general relativity},
  \href{https://doi.org/10.1103/PhysRevD.85.023504}{\emph{Phys. Rev. D}
  {\bfseries 85} (2012) 023504}.

\bibitem{durrer1}
C.~Bonvin and R.~Durrer, \emph{What galaxy surveys really measure},
  \href{https://doi.org/10.1103/PhysRevD.84.063505}{\emph{Phys. Rev. D}
  {\bfseries 84} (2011) 063505}.

\bibitem{cc1}
D.~Bertacca, R.~Maartens and C.~Clarkson, \emph{{Observed galaxy number counts
  on the lightcone up to second order: I. Main result}}, {\emph{Journal of
  Cosmology and Astroparticle Physics} {\bfseries 2014} (2014) 037}.

\bibitem{cc2}
D.~Bertacca, R.~Maartens and C.~Clarkson, \emph{{Observed galaxy number counts
  on the lightcone up to second order: II. Derivation}}, {\emph{Journal of
  Cosmology and Astroparticle Physics} {\bfseries 2014} (2014) 013}.

\bibitem{durrer2}
E.~Di~Dio, R.~Durrer, G.~Marozzi and F.~Montanari, \emph{{Galaxy number counts
  to second order and their bispectrum}},
  \href{https://doi.org/10.1088/1475-7516/2014/12/017,
  10.1088/1475-7516/2015/06/E01}{\emph{JCAP} {\bfseries 1412} (2014) 017}
  [\href{https://arxiv.org/abs/1407.0376}{{\ttfamily 1407.0376}}].

\bibitem{yoo1}
J.~Yoo, \emph{General relativistic description of the observed galaxy power
  spectrum: Do we understand what we measure?},
  \href{https://doi.org/10.1103/PhysRevD.82.083508}{\emph{Phys. Rev. D}
  {\bfseries 82} (2010) 083508}.

\bibitem{yoo2}
J.~Yoo and M.~Zaldarriaga, \emph{{Beyond the Linear-Order Relativistic Effect
  in Galaxy Clustering: Second-Order Gauge-Invariant Formalism}},
  \href{https://doi.org/10.1103/PhysRevD.90.023513}{\emph{Phys. Rev.}
  {\bfseries D90} (2014) 023513}
  [\href{https://arxiv.org/abs/1406.4140}{{\ttfamily 1406.4140}}].

\bibitem{jeong2}
F.~Schmidt and D.~Jeong, \emph{Cosmic rulers},
  \href{https://doi.org/10.1103/PhysRevD.86.083527}{\emph{Phys. Rev. D}
  {\bfseries 86} (2012) 083527}.

\bibitem{yoo5}
G.~Fanizza, J.~Yoo and S.~G. Biern, \emph{{Non-linear general relativistic
  effects in the observed redshift}},
  \href{https://arxiv.org/abs/1805.05959}{{\ttfamily 1805.05959}}.

\bibitem{didio3}
E.~Di~Dio and F.~Beutler, \emph{{The relativistic galaxy number counts in the
  weak field approximation}},
  \href{https://arxiv.org/abs/2004.07916}{{\ttfamily 2004.07916}}.

\bibitem{nielsen}
J.~T. Nielsen and R.~Durrer, \emph{{Higher order relativistic galaxy number
  counts: dominating terms}},
  \href{https://doi.org/10.1088/1475-7516/2017/03/010}{\emph{JCAP} {\bfseries
  1703} (2017) 010} [\href{https://arxiv.org/abs/1606.02113}{{\ttfamily
  1606.02113}}].

\bibitem{misner}
C.~W. {Misner}, K.~S. {Thorne} and J.~A. {Wheeler}, \emph{{Gravitation}}. 1973.

\bibitem{ellis}
G.~Ellis, R.~Maartens and M.~MacCallum, \emph{Relativistic Cosmology},
  Relativistic Cosmology. Cambridge University Press, 2012.

\bibitem{xact}
J.~M. Mart\'{i}n-Garc\'{i}a, \emph{{xAct: Efficient tensor computer algebra for
  the Wolfram Language}},  2016.

\bibitem{brizuela}
D.~Brizuela, J.~M. Martin-Garcia and G.~A. Mena~Marugan, \emph{{xPert: Computer
  algebra for metric perturbation theory}},
  \href{https://doi.org/10.1007/s10714-009-0773-2}{\emph{Gen. Rel. Grav.}
  {\bfseries 41} (2009) 2415}
  [\href{https://arxiv.org/abs/0807.0824}{{\ttfamily 0807.0824}}].

\bibitem{toolkit}
E.~Poisson, \emph{A Relativist's Toolkit: The Mathematics of Black-Hole
  Mechanics}. Cambridge University Press, 2004.

\bibitem{Lifshitz:2017aa}
E.~Lifshitz, \emph{Republication of: On the gravitational stability of the
  expanding universe},
  \href{https://doi.org/10.1007/s10714-016-2165-8}{\emph{General Relativity and
  Gravitation} {\bfseries 49} (2017) 18}.

\bibitem{lif}
E.~Lifshitz and I.~Khalatnikov, \emph{Investigations in relativistic
  cosmology}, \href{https://doi.org/10.1080/00018736300101283}{\emph{Advances
  in Physics} {\bfseries 12} (1963) 185}
  [\href{https://arxiv.org/abs/https://doi.org/10.1080/00018736300101283}{{\ttfamily
  https://doi.org/10.1080/00018736300101283}}].

\bibitem{tomita}
K.~Tomita, \emph{{Non-Linear Theory of Gravitational Instability in the
  Expanding Universe. II}},
  \href{https://doi.org/10.1143/PTP.45.1747}{\emph{Progress of Theoretical
  Physics} {\bfseries 45} (1971) 1747}
  [\href{https://arxiv.org/abs/https://academic.oup.com/ptp/article-pdf/45/6/1747/5306647/45-6-1747.pdf}{{\ttfamily
  https://academic.oup.com/ptp/article-pdf/45/6/1747/5306647/45-6-1747.pdf}}].

\bibitem{Bardeen80}
J.~M. Bardeen, \emph{Gauge-invariant cosmological perturbations},
  \href{https://doi.org/10.1103/PhysRevD.22.1882}{\emph{Phys. Rev. D}
  {\bfseries 22} (1980) 1882}.

\bibitem{KS}
H.~Kodama and M.~Sasaki, \emph{{Cosmological Perturbation Theory}},
  \href{https://doi.org/10.1143/PTPS.78.1}{\emph{Progress of Theoretical
  Physics Supplement} {\bfseries 78} (1984) 1}.

\bibitem{MFB}
V.~Mukhanov, H.~Feldman and R.~Brandenberger, \emph{Theory of cosmological
  perturbations},
  \href{https://doi.org/https://doi.org/10.1016/0370-1573(92)90044-Z}{\emph{Physics
  Reports} {\bfseries 215} (1992) 203 }.

\bibitem{MM2008}
K.~A. Malik and D.~R. Matravers, \emph{{A Concise Introduction to Perturbation
  Theory in Cosmology}},
  \href{https://doi.org/10.1088/0264-9381/25/19/193001}{\emph{Class. Quant.
  Grav.} {\bfseries 25} (2008) 193001}
  [\href{https://arxiv.org/abs/0804.3276}{{\ttfamily 0804.3276}}].

\bibitem{xpand}
C.~Pitrou, X.~Roy and O.~Umeh, \emph{{xPand: An algorithm for perturbing
  homogeneous cosmologies}},
  \href{https://doi.org/10.1088/0264-9381/30/16/165002}{\emph{Class. Quant.
  Grav.} {\bfseries 30} (2013) 165002}
  [\href{https://arxiv.org/abs/1302.6174}{{\ttfamily 1302.6174}}].

\bibitem{wald}
R.~M. Wald, \emph{{General relativity}}. Chicago Univ. Press, Chicago, IL,
  1984.

\bibitem{nakahara}
M.~Nakahara, \emph{{Geometry, topology and physics}}, Graduate student series
  in physics. Hilger, Bristol, 1990.

\bibitem{durrer3}
R.~Durrer, \emph{The Cosmic Microwave Background}. Cambridge University Press,
  2008.

\bibitem{carroll}
S.~M. Carroll, \emph{{Spacetime and Geometry}}. Cambridge University Press,
  2019.

\bibitem{hartle}
J.~B. Hartle, \emph{{Gravity: An Introduction to Einstein's General
  Relativity}}. Benjamin Cummings, illustrate~ed., Jan., 2003.

\bibitem{tong}
D.~Tong, \emph{Lectures on cosmology},  2019.

\bibitem{penrose}
J.~M. Stewart, M.~Walker and R.~Penrose, \emph{Perturbations of space-times in
  general relativity},
  \href{https://doi.org/10.1098/rspa.1974.0172}{\emph{Proceedings of the Royal
  Society of London. A. Mathematical and Physical Sciences} {\bfseries 341}
  (1974) 49}
  [\href{https://arxiv.org/abs/https://royalsocietypublishing.org/doi/pdf/10.1098/rspa.1974.0172}{{\ttfamily
  https://royalsocietypublishing.org/doi/pdf/10.1098/rspa.1974.0172}}].

\bibitem{malik2008}
K.~A. Malik and D.~R. Matravers, \emph{A concise introduction to perturbation
  theory in cosmology},
  \href{https://doi.org/10.1088/0264-9381/25/19/193001}{\emph{Classical and
  Quantum Gravity} {\bfseries 25} (2008) 193001}.

\bibitem{malik2}
K.~A. Malik and D.~R. Matravers, \emph{Comments on gauge-invariance in
  cosmology}, \href{https://doi.org/10.1007/s10714-013-1573-2}{\emph{General
  Relativity and Gravitation} {\bfseries 45} (2013) 1989}.

\bibitem{cc3}
O.~Umeh, C.~Clarkson and R.~Maartens, \emph{{Nonlinear relativistic corrections
  to cosmological distances, redshift and gravitational lensing magnification:
  I. Key results}}, {\emph{Classical and Quantum Gravity} {\bfseries 31} (2014)
  202001}.

\bibitem{pedro}
P.~Carrilho, \emph{Non-linear effects in early Universe cosmology}, Ph.D.
  thesis, Queen Mary University of London, 2018.

\bibitem{cc4}
O.~Umeh, C.~Clarkson and R.~Maartens, \emph{{Nonlinear relativistic corrections
  to cosmological distances, redshift and gravitational lensing magnification:
  II. Derivation}}, {\emph{Classical and Quantum Gravity} {\bfseries 31} (2014)
  205001}.

\bibitem{cc5}
C.~Clarkson, G.~F.~R. Ellis, A.~Faltenbacher, R.~Maartens, O.~Umeh and J.-P.
  Uzan, \emph{{(Mis-)Interpreting supernovae observations in a lumpy
  universe}},
  \href{https://doi.org/10.1111/j.1365-2966.2012.21750.x}{\emph{Mon. Not. Roy.
  Astron. Soc.} {\bfseries 426} (2012) 1121}
  [\href{https://arxiv.org/abs/1109.2484}{{\ttfamily 1109.2484}}].

\bibitem{PhysRevLett.114.101301}
{\scshape (BICEP2/Keck and Planck Collaborations)} collaboration, P.~A.~R. Ade,
  N.~Aghanim, Z.~Ahmed, R.~W. Aikin, K.~D. Alexander, M.~Arnaud et~al.,
  \emph{Joint analysis of bicep2/ \textit{Keck Array} and \textit{Planck}
  data}, \href{https://doi.org/10.1103/PhysRevLett.114.101301}{\emph{Phys. Rev.
  Lett.} {\bfseries 114} (2015) 101301}
  [\href{https://arxiv.org/abs/1502.00612}{{\ttfamily 1502.00612}}].

\bibitem{2015arXiv150201589P}
{Planck Collaboration}, P.~A.~R. {Ade}, N.~{Aghanim}, M.~{Arnaud},
  M.~{Ashdown}, J.~{Aumont} et~al., \emph{{Planck 2015 results. XIII.
  Cosmological parameters}}, {\emph{ArXiv e-prints} (2015) }
  [\href{https://arxiv.org/abs/1502.01589}{{\ttfamily 1502.01589}}].

\bibitem{1993PhRvD..48.4613T}
M.~S. {Turner}, M.~{White} and J.~E. {Lidsey}, \emph{{Tensor perturbations in
  inflationary models as a probe of cosmology}},
  \href{https://doi.org/10.1103/PhysRevD.48.4613}{\emph{\prd} {\bfseries 48}
  (1993) 4613} [\href{https://arxiv.org/abs/astro-ph/9306029}{{\ttfamily
  astro-ph/9306029}}].

\bibitem{1994PhRvD..50.3713A}
B.~{Allen} and S.~{Koranda}, \emph{{CBR anisotropy from primordial
  gravitational waves in inflationary cosmologies}},
  \href{https://doi.org/10.1103/PhysRevD.50.3713}{\emph{\prd} {\bfseries 50}
  (1994) 3713} [\href{https://arxiv.org/abs/astro-ph/9404068}{{\ttfamily
  astro-ph/9404068}}].

\bibitem{1995PhRvD..52.2112N}
K.-W. {Ng} and A.~D. {Speliotopoulos}, \emph{{Cosmological evolution of
  scale-invariant gravity waves}},
  \href{https://doi.org/10.1103/PhysRevD.52.2112}{\emph{\prd} {\bfseries 52}
  (1995) 2112} [\href{https://arxiv.org/abs/astro-ph/9405043}{{\ttfamily
  astro-ph/9405043}}].

\bibitem{1996PhRvD..53..639W}
Y.~{Wang}, \emph{{Simple analytical methods for computing the gravity-wave
  contribution to the cosmic background radiation anisotropy}},
  \href{https://doi.org/10.1103/PhysRevD.53.639}{\emph{\prd} {\bfseries 53}
  (1996) 639} [\href{https://arxiv.org/abs/astro-ph/9501116}{{\ttfamily
  astro-ph/9501116}}].

\bibitem{2004PhRvD..69b3503W}
S.~{Weinberg}, \emph{{Damping of tensor modes in cosmology}},
  \href{https://doi.org/10.1103/PhysRevD.69.023503}{\emph{\prd} {\bfseries 69}
  (2004) 023503} [\href{https://arxiv.org/abs/astro-ph/0306304}{{\ttfamily
  astro-ph/0306304}}].

\bibitem{2005AnPhy.318....2P}
J.~R. {Pritchard} and M.~{Kamionkowski}, \emph{{Cosmic microwave background
  fluctuations from gravitational waves: An analytic approach}},
  \href{https://doi.org/10.1016/j.aop.2005.03.005}{\emph{Annals of Physics}
  {\bfseries 318} (2005) 2}
  [\href{https://arxiv.org/abs/astro-ph/0412581}{{\ttfamily
  astro-ph/0412581}}].

\bibitem{2005astro.ph..5502B}
S.~{Bashinsky}, \emph{{Coupled Evolution of Primordial Gravity Waves and Relic
  Neutrinos}}, {\emph{ArXiv Astrophysics e-prints} (2005) }
  [\href{https://arxiv.org/abs/astro-ph/0505502}{{\ttfamily
  astro-ph/0505502}}].

\bibitem{2008cosm.book.....W}
S.~{Weinberg}, \emph{{Cosmology}}. Oxford University Press, 2008.

\bibitem{2013PhRvD..88h3536S}
B.~A. {Stefanek} and W.~W. {Repko}, \emph{{Analytic description of the damping
  of gravitational waves by free streaming neutrinos}},
  \href{https://doi.org/10.1103/PhysRevD.88.083536}{\emph{\prd} {\bfseries 88}
  (2013) 083536} [\href{https://arxiv.org/abs/1207.7285}{{\ttfamily
  1207.7285}}].

\bibitem{PhysRevD.50.2541}
A.~K. Rebhan and D.~J. Schwarz, \emph{Kinetic versus thermal-field-theory
  approach to cosmological perturbations},
  \href{https://doi.org/10.1103/PhysRevD.50.2541}{\emph{Phys. Rev. D}
  {\bfseries 50} (1994) 2541}.

\bibitem{FSN}
S.~F. Feshchenko, N.~I. Shkil and L.~D. Nikolenko, \emph{Asymptotic methods in
  the theory of linear differential equations, S.F. Feshchenko, N.I. Shkil, and
  L.D. Nikolenko. Translated by Scripta Technica}. American Elsevier Pub. Co
  New York, 1966.

\bibitem{DPS}
A.~J. Wren and K.~A. Malik, \emph{Double power series method for approximating
  cosmological perturbations},
  \href{https://doi.org/10.1103/PhysRevD.95.083526}{\emph{Phys. Rev. D}
  {\bfseries 95} (2017) 083526}.

\bibitem{GWGithub}
A.~J. {Wren}, K.~A. {Malik} and J.~L. {Fuentes}, \emph{{Mathematica notebooks
  for \emph{Gravitational waves in a flat radiation-matter universe including
  anisotropic stress}}},  Dec., 2017.

\bibitem{2006AIPC..843..111P}
T.~{Padmanabhan}, \emph{{Advanced Topics in Cosmology: A Pedagogical
  Introduction}},  in \emph{Graduate School in Astronomy: X} (S.~{Daflon},
  J.~{Alcaniz}, E.~{Telles} and R.~{de la Reza}, eds.), vol.~843 of
  \emph{American Institute of Physics Conference Series}, pp.~111--166, June,
  2006, \href{https://arxiv.org/abs/astro-ph/0602117}{{\ttfamily
  astro-ph/0602117}}, \href{https://doi.org/10.1063/1.2219327}{DOI}.

\bibitem{2006PASP..118.1711W}
E.~L. {Wright}, \emph{{A Cosmology Calculator for the World Wide Web}},
  \href{https://doi.org/10.1086/510102}{\emph{\pasp} {\bfseries 118} (2006)
  1711} [\href{https://arxiv.org/abs/astro-ph/0609593}{{\ttfamily
  astro-ph/0609593}}].

\bibitem{2003itc..book.....R}
B.~{Ryden}, \emph{{Intoduction to cosmology}}. Addison Wesley, 2003.

\bibitem{2009pdp..book.....L}
D.~H. {Lyth} and A.~R. {Liddle}, \emph{{The Primordial Density Perturbation}}.
  June, 2009.

\bibitem{2016JCAP...02..021C}
P.~{Carrilho} and K.~A. {Malik}, \emph{{Vector and tensor contributions to the
  curvature perturbation at second order}},
  \href{https://doi.org/10.1088/1475-7516/2016/02/021}{\emph{\jcap} {\bfseries
  2} (2016) 021} [\href{https://arxiv.org/abs/1507.06922}{{\ttfamily
  1507.06922}}].

\bibitem{BOSS}
{The BOSS Collaboration}, \emph{{The Baryon Oscillation Spectroscopic Survey of
  SDSS-III}}, \href{https://doi.org/10.1088/0004-6256/145/1/10}{\emph{The
  Astronomical Journal} {\bfseries 145} (2013) 10}
  [\href{https://arxiv.org/abs/1208.0022}{{\ttfamily 1208.0022}}].

\bibitem{eBOSS}
G.-B. Zhao et~al., \emph{{The extended Baryon Oscillation Spectroscopic Survey:
  a cosmological forecast}},
  \href{https://doi.org/10.1093/mnras/stw135}{\emph{Mon. Not. Roy. Astron.
  Soc.} {\bfseries 457} (2016) 2377}
  [\href{https://arxiv.org/abs/1510.08216}{{\ttfamily 1510.08216}}].

\bibitem{WFIRST}
J.~{Green} et~al., \emph{{Wide-Field InfraRed Survey Telescope (WFIRST) Final
  Report}}, {\emph{ArXiv e-prints} (2012) }
  [\href{https://arxiv.org/abs/1208.4012}{{\ttfamily 1208.4012}}].

\bibitem{zalda}
J.~{Yoo}, A.~L. {Fitzpatrick} and M.~{Zaldarriaga}, \emph{{New perspective on
  galaxy clustering as a cosmological probe: General relativistic effects}},
  \href{https://doi.org/10.1103/PhysRevD.80.083514}{\emph{Phys. Rev. D}
  {\bfseries 80} (2009) 083514}
  [\href{https://arxiv.org/abs/0907.0707}{{\ttfamily 0907.0707}}].

\bibitem{yoo3}
J.~Yoo and F.~Scaccabarozzi, \emph{{Unified Treatment of the Luminosity
  Distance in Cosmology}},
  \href{https://doi.org/10.1088/1475-7516/2016/09/046}{\emph{JCAP} {\bfseries
  1609} (2016) 046} [\href{https://arxiv.org/abs/1606.08453}{{\ttfamily
  1606.08453}}].

\bibitem{marozzi1}
I.~Ben-Dayan, G.~Marozzi, F.~Nugier and G.~Veneziano, \emph{{The second-order
  luminosity-redshift relation in a generic inhomogeneous cosmology}},
  \href{https://doi.org/10.1088/1475-7516/2012/11/045}{\emph{JCAP} {\bfseries
  1211} (2012) 045} [\href{https://arxiv.org/abs/1209.4326}{{\ttfamily
  1209.4326}}].

\bibitem{marozzi2}
G.~Fanizza, M.~Gasperini, G.~Marozzi and G.~Veneziano, \emph{{An exact Jacobi
  map in the geodesic light-cone gauge}},
  \href{https://doi.org/10.1088/1475-7516/2013/11/019}{\emph{JCAP} {\bfseries
  1311} (2013) 019} [\href{https://arxiv.org/abs/1308.4935}{{\ttfamily
  1308.4935}}].

\bibitem{marozzi3}
G.~Marozzi, \emph{{The luminosity distance--redshift relation up to second
  order in the Poisson gauge with anisotropic stress}},
  \href{https://doi.org/10.1088/0264-9381/32/17/179501,
  10.1088/0264-9381/32/4/045004}{\emph{Class. Quant. Grav.} {\bfseries 32}
  (2015) 045004} [\href{https://arxiv.org/abs/1406.1135}{{\ttfamily
  1406.1135}}].

\bibitem{yoo4}
J.~Yoo, \emph{{Relativistic Effect in Galaxy Clustering}},
  \href{https://doi.org/10.1088/0264-9381/31/23/234001}{\emph{Class. Quant.
  Grav.} {\bfseries 31} (2014) 234001}
  [\href{https://arxiv.org/abs/1409.3223}{{\ttfamily 1409.3223}}].

\bibitem{cc7}
C.~Clarkson, G.~F.~R. Ellis, A.~Faltenbacher, R.~Maartens, O.~Umeh and J.-P.
  Uzan, \emph{{(Mis-)Interpreting supernovae observations in a lumpy
  universe}},
  \href{https://doi.org/10.1111/j.1365-2966.2012.21750.x}{\emph{Mon. Not. Roy.
  Astron. Soc.} {\bfseries 426} (2012) 1121}
  [\href{https://arxiv.org/abs/1109.2484}{{\ttfamily 1109.2484}}].

\bibitem{conformal}
M.~Ibison, \emph{{On the conformal forms of the Robertson-Walker metric}},
  \href{https://doi.org/10.1063/1.2815811}{\emph{J. Math. Phys.} {\bfseries 48}
  (2007) 122501} [\href{https://arxiv.org/abs/0704.2788}{{\ttfamily
  0704.2788}}].

\bibitem{conformalnotes}
S.~Aretakis, \emph{{Lecture notes on general relativity. Columbia
  University.}},  2013.

\bibitem{campos}
A.~Campos, \emph{{Galaxy number counts}},
  \href{https://doi.org/10.3254/978-1-61499-217-2-123}{\emph{Proc. Int. Sch.
  Phys. Fermi} {\bfseries 132} (1996) 123}
  [\href{https://arxiv.org/abs/astro-ph/9510051}{{\ttfamily
  astro-ph/9510051}}].

\bibitem{didio2015}
E.~Di~Dio, R.~Durrer, G.~Marozzi and F.~Montanari, \emph{{The bispectrum of
  relativistic galaxy number counts}},
  \href{https://doi.org/10.1088/1475-7516/2016/01/016}{\emph{JCAP} {\bfseries
  01} (2016) 016} [\href{https://arxiv.org/abs/1510.04202}{{\ttfamily
  1510.04202}}].

\bibitem{fuentes1}
A.~J. Wren, J.~L. Fuentes and K.~A. Malik, \emph{{Gravitational waves in a flat
  radiation-matter universe including anisotropic stress}},
  \href{https://arxiv.org/abs/1712.10012}{{\ttfamily 1712.10012}}.

\bibitem{cm2016}
P.~Carrilho and K.~A. Malik, \emph{Vector and tensor contributions to the
  curvature perturbation at second order},
  \href{https://doi.org/10.1088/1475-7516/2016/02/021}{\emph{Journal of
  Cosmology and Astroparticle Physics} {\bfseries 2016} (2016) 021}.

\bibitem{fuentes3}
J.~L. Fuentes, J.~C. Hidalgo and K.~A. Malik, \emph{{Galaxy number counts at
  second order: an independent approach}},
  \href{https://arxiv.org/abs/1908.08400}{{\ttfamily 1908.08400}}.

\bibitem{fuentes4}
J.~L. Fuentes, J.~C. Hidalgo and K.~A. Malik, \emph{{Galaxy number counts at
  second order II: a leading order comparison (in preparation)}}, .

\bibitem{FGM2}
J.~L. Fuentes, U.~A. Gillani and K.~A. Malik, \emph{Vector and tensor
  perturbations in fourth order gravity (in preparation)}, .

\bibitem{b}
S.~Nojiri and S.~D. Odintsov, \emph{{Introduction to modified gravity and
  gravitational alternative for dark energy}},
  \href{https://doi.org/10.1142/S0219887807001928}{\emph{eConf} {\bfseries
  C0602061} (2006) 06} [\href{https://arxiv.org/abs/hep-th/0601213}{{\ttfamily
  hep-th/0601213}}].

\bibitem{c}
S.~Nojiri and S.~D. Odintsov, \emph{{Unified cosmic history in modified
  gravity: from F(R) theory to Lorentz non-invariant models}},
  \href{https://doi.org/10.1016/j.physrep.2011.04.001}{\emph{Phys. Rept.}
  {\bfseries 505} (2011) 59} [\href{https://arxiv.org/abs/1011.0544}{{\ttfamily
  1011.0544}}].

\bibitem{LDS2007}
L.~Amendola, D.~Polarski and S.~Tsujikawa, \emph{{Are f(R) dark energy models
  cosmologically viable ?}},
  \href{https://doi.org/10.1103/PhysRevLett.98.131302}{\emph{Phys. Rev. Lett.}
  {\bfseries 98} (2007) 131302}
  [\href{https://arxiv.org/abs/astro-ph/0603703}{{\ttfamily
  astro-ph/0603703}}].

\bibitem{DES}
{\scshape DES} collaboration, B.~Flaugher, \emph{{The Dark Energy Survey}},
  \href{https://doi.org/10.1142/S0217751X05025917}{\emph{Int. J. Mod. Phys.}
  {\bfseries A20} (2005) 3121}.

\bibitem{Guy:2016zel}
{\scshape DESI} collaboration, A.~Aghamousa et~al., \emph{{The DESI Experiment
  Part I: Science,Targeting, and Survey Design}},
  \href{https://arxiv.org/abs/1611.00036}{{\ttfamily 1611.00036}}.

\bibitem{Tereno:2015hja}
{\scshape Euclid} collaboration, I.~Tereno et~al., \emph{{Euclid Space Mission:
  building the sky survey}},
  \href{https://doi.org/10.1017/S174392131401093X}{\emph{IAU Symp.} {\bfseries
  306} (2014) 379} [\href{https://arxiv.org/abs/1502.00903}{{\ttfamily
  1502.00903}}].

\bibitem{f}
S.~Nojiri and S.~D. Odintsov, \emph{Modified gravity and its reconstruction
  from the universe expansion history},
  \href{https://doi.org/10.1088/1742-6596/66/1/012005}{\emph{Journal of
  Physics: Conference Series} {\bfseries 66} (2007) 012005}.

\bibitem{8f}
M.~Chaichian, S.~Nojiri, S.~D. Odintsov, M.~Oksanen and A.~Tureanu,
  \emph{Modifiedf(r) ho{\v r}ava--lifshitz gravity: a way to accelerating frw
  cosmology},
  \href{https://doi.org/10.1088/0264-9381/27/18/185021}{\emph{Classical and
  Quantum Gravity} {\bfseries 27} (2010) 185021}.

\bibitem{PJ2007}
P.~Zhang, \emph{Behavior of $f(r)$ gravity in the solar system, galaxies, and
  clusters}, \href{https://doi.org/10.1103/PhysRevD.76.024007}{\emph{Phys. Rev.
  D} {\bfseries 76} (2007) 024007}.

\bibitem{CLF2010}
C.~G. B\"ohmer, L.~Hollenstein and F.~S.~N. Lobo, \emph{Stability of the
  einstein static universe in $f(r)$ gravity},
  \href{https://doi.org/10.1103/PhysRevD.76.084005}{\emph{Phys. Rev. D}
  {\bfseries 76} (2007) 084005}.

\bibitem{SS2003}
S.~Nojiri and S.~D. Odintsov, \emph{Modified gravity with negative and positive
  powers of curvature: Unification of inflation and cosmic acceleration},
  \href{https://doi.org/10.1103/physrevd.68.123512}{\emph{Physical Review D}
  {\bfseries 68} (2003) }.

\bibitem{TFS2011}
T.~Harko, F.~S.~N. Lobo, S.~Nojiri and S.~D. Odintsov, \emph{$f(r,t)$ gravity},
  \href{https://doi.org/10.1103/PhysRevD.84.024020}{\emph{Phys. Rev. D}
  {\bfseries 84} (2011) 024020}.

\bibitem{SSD2009}
S.~Nojiri, S.~D. Odintsov and D.~Saez-Gomez, \emph{{Cosmological reconstruction
  of realistic modified F(R) gravities}},
  \href{https://doi.org/10.1016/j.physletb.2009.09.045}{\emph{Phys. Lett.}
  {\bfseries B681} (2009) 74}
  [\href{https://arxiv.org/abs/0908.1269}{{\ttfamily 0908.1269}}].

\bibitem{ERV2010}
E.~Elizalde, R.~Myrzakulov, V.~V. Obukhov and D.~S{\'a}ez-G{\'o}mez,
  \emph{{$\Lambda$CDM epoch reconstruction from F(R,G) and modified
  Gauss--Bonnet gravities}},
  \href{https://doi.org/10.1088/0264-9381/27/9/095007}{\emph{Classical and
  Quantum Gravity} {\bfseries 27} (2010) 095007}.

\bibitem{BI2007}
B.~M. Leith and I.~P. Neupane, \emph{Gauss--bonnet cosmologies: crossing the
  phantom divide and the transition from matter dominance to dark energy},
  \href{https://doi.org/10.1088/1475-7516/2007/05/019}{\emph{Journal of
  Cosmology and Astroparticle Physics} {\bfseries 2007} (2007) 019}.

\bibitem{SSS2002}
S.~Nojiri, S.~D. Odintsov and S.~Ogushi, \emph{{Friedmann-Robertson-Walker
  brane cosmological equations from the five-dimensional bulk (A)dS black
  hole}}, \href{https://doi.org/10.1142/S0217751X02012156}{\emph{Int. J. Mod.
  Phys.} {\bfseries A17} (2002) 4809}
  [\href{https://arxiv.org/abs/hep-th/0205187}{{\ttfamily hep-th/0205187}}].

\bibitem{HD2013}
H.-J. Schmidt and D.~Singleton, \emph{{Isotropic universe with almost
  scale-invariant fourth-order gravity}},
  \href{https://doi.org/10.1063/1.4808255}{\emph{J. Math. Phys.} {\bfseries 54}
  (2013) 062502} [\href{https://arxiv.org/abs/1212.1769}{{\ttfamily
  1212.1769}}].

\bibitem{g}
S.~Nojiri and S.~D. Odintsov, \emph{Modified gauss--bonnet theory as
  gravitational alternative for dark energy},
  \href{https://doi.org/10.1016/j.physletb.2005.10.010}{\emph{Physics Letters
  B} {\bfseries 631} (2005) 1}.

\bibitem{g1}
G.~Cognola, E.~Elizalde, S.~Nojiri, S.~D. Odintsov and S.~Zerbini, \emph{Dark
  energy in modified gauss-bonnet gravity: Late-time acceleration and the
  hierarchy problem},
  \href{https://doi.org/10.1103/physrevd.73.084007}{\emph{Physical Review D}
  {\bfseries 73} (2006) }.

\bibitem{g2}
S.~Nojiri, S.~D. Odintsov and M.~Sasaki, \emph{Gauss-bonnet dark energy},
  \href{https://doi.org/10.1103/physrevd.71.123509}{\emph{Physical Review D}
  {\bfseries 71} (2005) }.

\bibitem{Becker:2017tcx}
D.~Becker, C.~Ripken and F.~Saueressig, \emph{On avoiding ostrogradski
  instabilities within asymptotic safety},
  \href{https://doi.org/10.1007/jhep12(2017)121}{\emph{Journal of High Energy
  Physics} {\bfseries 2017} (2017) }.

\bibitem{Motohashi:2014opa}
H.~Motohashi and T.~Suyama, \emph{{Third order equations of motion and the
  Ostrogradsky instability}},
  \href{https://doi.org/10.1103/PhysRevD.91.085009}{\emph{Phys. Rev.}
  {\bfseries D91} (2015) 085009}
  [\href{https://arxiv.org/abs/1411.3721}{{\ttfamily 1411.3721}}].

\bibitem{Himmetoglu:2009qi}
B.~Himmetoglu, C.~R. Contaldi and M.~Peloso, \emph{{Ghost instabilities of
  cosmological models with vector fields nonminimally coupled to the
  curvature}}, \href{https://doi.org/10.1103/PhysRevD.80.123530}{\emph{Phys.
  Rev.} {\bfseries D80} (2009) 123530}
  [\href{https://arxiv.org/abs/0909.3524}{{\ttfamily 0909.3524}}].

\bibitem{nojiri:2018}
S.~Nojiri, S.~D. Odintsov and V.~K. Oikonomou, \emph{{Ghost-free Gauss-Bonnet
  Theories of Gravity}},
  \href{https://doi.org/10.1103/PhysRevD.99.044050}{\emph{Phys. Rev.}
  {\bfseries D99} (2019) 044050}
  [\href{https://arxiv.org/abs/1811.07790}{{\ttfamily 1811.07790}}].

\bibitem{Abbott:2007kv}
B.~P. Abbott, R.~Abbott, R.~Adhikari, P.~Ajith, B.~Allen, G.~Allen et~al.,
  \emph{{LIGO}: the laser interferometer gravitational-wave observatory},
  \href{https://doi.org/10.1088/0034-4885/72/7/076901}{\emph{Reports on
  Progress in Physics} {\bfseries 72} (2009) 076901}.

\bibitem{Abbott:2009kk}
B.~P. Abbott, R.~Abbott, F.~Acernese, R.~Adhikari, P.~Ajith, B.~Allen et~al.,
  \emph{{Search} {for} {gravitational}-{wave} {burst} {associated} {with}
  {gamma}-{ray} {bursts} {using} {data} {from} {LIGO} {science} {run} 5 {and}
  {VIRGO} {science} {run} 1},
  \href{https://doi.org/10.1088/0004-637x/715/2/1438}{\emph{The Astrophysical
  Journal} {\bfseries 715} (2010) 1438}.

\bibitem{KAGRA}
T.~A. and, \emph{Large-scale cryogenic gravitational-wave telescope in japan:
  {KAGRA}}, \href{https://doi.org/10.1088/1742-6596/610/1/012016}{\emph{Journal
  of Physics: Conference Series} {\bfseries 610} (2015) 012016}.

\bibitem{AmaroSeoane:2012je}
P.~Amaro-Seoane, S.~Aoudia, S.~Babak, P.~Bin{\'{e}}truy, E.~Berti,
  A.~Boh{\'{e}} et~al., \emph{Low-frequency gravitational-wave science with
  {eLISA}/{NGO}},
  \href{https://doi.org/10.1088/0264-9381/29/12/124016}{\emph{Classical and
  Quantum Gravity} {\bfseries 29} (2012) 124016}.

\bibitem{githubGB}
J.~L. Fuentes, \emph{{Linear Gauss-Bonnet,
  [https://github.com/jorchfv/Linear-Gauss-Bonnet]}},  2018.

\end{thebibliography}\endgroup
